\documentclass[11pt,twoside]{book}
\usepackage{fancyhdr}
\fancypagestyle{plain}{%
\fancyhf{} 
\fancyfoot[R]{\textsf{ILC Reference Design Report\hspace{0.5cm}IV-\thepage}}}
\usepackage{epsfig}
\usepackage{rotate}
\usepackage[figuresright]{rotating}
\usepackage{lscape}
\usepackage{hyperref}
\usepackage{url}
\usepackage{graphicx}
\usepackage{color}
\usepackage{epstopdf}
\usepackage{epsf}
\usepackage{hhline}
\usepackage{mytrc_style}
\usepackage{ulem}
\usepackage{verbatim}

\begin{document}
\frontmatter
%

{\sffamily\bfseries
\begin{titlepage}
\begin{center}
~
 ~ \vskip 4cm

    {\Huge I}{\huge NTERNATIONAL} 
    {\Huge L}{\huge INEAR} 
    {\Huge C}{\huge OLLIDER}
    
  \vskip 1.2cm

    {\Huge R}{\huge EFERENCE}
    {\Huge D}{\huge ESIGN}
    {\Huge R}{\huge EPORT}

  \vskip 1.2cm

\vskip 3cm

{\huge ILC Global Design Effort and} \\
    
  \vskip 0.5cm

{\huge World Wide Study }

  \vskip 3cm

    {\huge AUGUST, 2007}

\end{center}
\end{titlepage}

\newpage\thispagestyle{empty}
~
 ~ \vskip 2cm

{\LARGE Volume 1:~~~EXECUTIVE SUMMARY}
 \vskip 0.5cm
{\Large Editors:} 
 \vskip 0.25cm
{\Large James~Brau, Yasuhiro~Okada, Nicholas~J.~Walker}

  \vskip 1.5cm

{\LARGE Volume 2:~~~PHYSICS AT THE ILC}
 \vskip 0.5cm
{\Large Editors:} 
 \vskip 0.25cm
{\Large Abdelhak~Djouadi, Joseph~Lykken, Klaus~M{\"o}nig} 
 \vskip 0.25cm
{\Large Yasuhiro~Okada, Mark~Oreglia, Satoru~Yamashita}

  \vskip 1.5cm

{\LARGE Volume 3:~~~ACCELERATOR}
 \vskip 0.5cm
{\Large Editors:} 
 \vskip 0.25cm
{\Large Nan~Phinney, Nobukazu~Toge, Nicholas~Walker}

  \vskip 1.5cm

{\LARGE Volume 4:~~~DETECTORS}
 \vskip 0.5cm
{\Large Editors:} 
 \vskip 0.25cm
{\Large Ties~Behnke, Chris~Damerell, John~Jaros, Akiya~Miyamoto}

\newpage\thispagestyle{empty}
~
 ~ \vskip 5cm
~~~{\Huge Volume 4:~~~DETECTORS}
 \vskip 1cm
~~~{\LARGE Editors:} 
 \vskip 0.5cm
~~~{\LARGE Ties~Behnke, Chris~Damerell, John~Jaros} 
 \vskip 0.4cm
~~~{\LARGE Akiya~Miyamoto}

\newpage\thispagestyle{empty}

}

\cleardoublepage\setcounter{page}{1}
\chapter*{List of Contributors} 

\begin{center}

\begin{center}

Gerald~Aarons$^{203}$,
Toshinori~Abe$^{290}$,
Jason~Abernathy$^{293}$,
Medina~Ablikim$^{87}$,
Halina~Abramowicz$^{216}$,
David~Adey$^{236}$,
Catherine~Adloff$^{128}$,
Chris~Adolphsen$^{203}$,
Konstantin~Afanaciev$^{11,47}$,
Ilya~Agapov$^{192,35}$
% ...

\end{center}

\clearpage

\chapter*{List of Institutions}

\begin{center}

{\sl $^{1}$ Abdus Salam International Centre for Theoretical Physics, Strada Costriera 11, 34014 Trieste, Italy}

{\sl $^{2}$ Academy, RPR, National Institute of Physics and Nuclear Engineering `Horia Hulubei' (IFIN-HH), Str. Atomistilor no. 407, P.O. Box MG-6, R-76900 Bucharest - Magurele, Romania}

{\sl $^{3}$ AGH University of Science and Technology Akademia Gorniczo-Hutnicza im. Stanislawa Staszica w Krakowie al. Mickiewicza 30 PL-30-059 Cracow, Poland}

{\sl $^{4}$ Albert-Ludwigs Universit{\"a}t Freiburg, Physikalisches Institut, Hermann-Herder Str. 3, D-79104 Freiburg, Germany}

{\sl $^{5}$ Aligarh Muslim University, Aligarh, Uttar Pradesh 202002, India}

{\sl $^{6}$ Amberg Engineering AG, Trockenloostr. 21, P.O.Box 27, 8105 Regensdorf-Watt, Switzerland}

{\sl $^{7}$ Angstromquelle Karlsruhe (ANKA), Forschungszentrum Karlsruhe, Hermann-von-Helmholtz-Platz 1, D-76344 Eggenstein-Leopoldshafen, Germany}

{\sl $^{8}$ Argonne National Laboratory (ANL), 9700 S. Cass Avenue, Argonne, IL 60439, USA}

{\sl $^{9}$ Baylor University, Department of Physics, 101 Bagby Avenue, Waco, TX 76706, USA}

{\sl $^{10}$ Beijing University, Department of Physics, Beijing, China 100871}

{\sl $^{11}$ Belarusian State University, National Scientific \& Educational Center, Particle \& HEP Physics, M. Bogdanovich St., 153, 240040 Minsk, Belarus}

{\sl $^{12}$ Benares Hindu University, Benares, Varanasi 221005, India}

{\sl $^{13}$ Bhabha Atomic Research Centre, Trombay, Mumbai 400085, India}

{\sl $^{14}$ Birla Institute of Technology and Science, EEE Dept., Pilani, Rajasthan, India}

{\sl $^{15}$ Bogazici University, Physics Department, 34342 Bebek / Istanbul, 80820 Istanbul, Turkey}

{\sl $^{16}$ Boston University, Department of Physics, 590 Commonwealth Avenue, Boston, MA 02215, USA}

{\sl $^{17}$ Brandenburg University of Technology, Postfach 101344, D-03013 Cottbus, Germany}

{\sl $^{18}$ Brno University of Technology, Anton\'insk\'a; 548/1, CZ 601 90 Brno, Czech Republic}

{\sl $^{19}$ Brookhaven National Laboratory (BNL), P.O.Box 5000, Upton, NY 11973-5000, USA}

{\sl $^{20}$ Brown University, Department of Physics, Box 1843, Providence, RI 02912, USA}

{\sl $^{21}$ Budkar Institute for Nuclear Physics (BINP), 630090 Novosibirsk, Russia}

{\sl $^{22}$ Calcutta University, Department of Physics, 92 A.P.C. Road, Kolkata 700009, India}

{\sl $^{23}$ California Institute of Technology, Physics, Mathematics and Astronomy (PMA), 1200 East California Blvd, Pasadena, CA 91125, USA}

{\sl $^{24}$ Carleton University, Department of Physics, 1125 Colonel By Drive, Ottawa, Ontario, Canada K1S 5B6}

{\sl $^{25}$ Carnegie Mellon University, Department of Physics, Wean Hall 7235, Pittsburgh, PA 15213, USA}

{\sl $^{26}$ CCLRC Daresbury Laboratory, Daresbury, Warrington, Cheshire WA4 4AD, UK }

{\sl $^{27}$ CCLRC Rutherford Appleton Laboratory, Chilton, Didcot, Oxton OX11 0QX, UK }

{\sl $^{28}$ CEA Saclay, DAPNIA, F-91191 Gif-sur-Yvette, France}

{\sl $^{29}$ CEA Saclay, Service de Physique Th{\'e}orique, CEA/DSM/SPhT, F-91191 Gif-sur-Yvette Cedex, France}

{\sl $^{30}$ Center for High Energy Physics (CHEP) / Kyungpook National University, 1370 Sankyuk-dong, Buk-gu, Daegu 702-701, Korea}

{\sl $^{31}$ Center for High Energy Physics (TUHEP), Tsinghua University, Beijing, China 100084}

{\sl $^{32}$ Centre de Physique Theorique, CNRS - Luminy, Universiti d'Aix - Marseille II, Campus of Luminy, Case 907, 13288 Marseille Cedex 9, France}

{\sl $^{33}$ Centro de Investigaciones Energ\'eticas, Medioambientales y Technol\'ogicas, CIEMAT, Avenia Complutense 22, E-28040 Madrid, Spain}

{\sl $^{34}$ Centro Nacional de Microelectr\'onica (CNM), Instituto de Microelectr\'onica de Barcelona (IMB), Campus UAB, 08193 Cerdanyola del Vall\`es (Bellaterra), Barcelona, Spain}

{\sl $^{35}$ CERN, CH-1211 Gen\`eve 23, Switzerland}

{\sl $^{36}$ Charles University, Institute of Particle \& Nuclear Physics, Faculty of Mathematics and Physics, V Holesovickach 2, CZ-18000 Praque 8, Czech Republic}

{\sl $^{37}$ Chonbuk National University, Physics Department, Chonju 561-756, Korea}

{\sl $^{38}$ Cockcroft Institute, Daresbury, Warrington WA4 4AD, UK }

{\sl $^{39}$ College of William and Mary, Department of Physics, Williamsburg, VA, 23187, USA}

{\sl $^{40}$ Colorado State University, Department of Physics, Fort Collins, CO 80523, USA}

{\sl $^{41}$ Columbia University, Department of Physics, New York, NY 10027-6902, USA}

{\sl $^{42}$ Concordia University, Department of Physics, 1455 De Maisonneuve Blvd. West, Montreal, Quebec, Canada H3G 1M8}

{\sl $^{43}$ Cornell University, Laboratory for Elementary-Particle Physics (LEPP), Ithaca, NY 14853, USA}

{\sl $^{44}$ Cukurova University, Department of Physics, Fen-Ed. Fakultesi 01330, Balcali, Turkey}

{\sl $^{45}$ D.~V. Efremov Research Institute, SINTEZ, 196641 St. Petersburg, Russia}

{\sl $^{46}$ Dartmouth College, Department of Physics and Astronomy, 6127 Wilder Laboratory, Hanover, NH 03755, USA}

{\sl $^{47}$ DESY-Hamburg site, Deutsches Elektronen-Synchrotoron in der Helmholtz-Gemeinschaft, Notkestrasse 85, 22607 Hamburg, Germany}

{\sl $^{48}$ DESY-Zeuthen site, Deutsches Elektronen-Synchrotoron in der Helmholtz-Gemeinschaft, Platanenallee 6, D-15738 Zeuthen, Germany}

{\sl $^{49}$ Durham University,  Department of Physics, Ogen Center for Fundamental Physics, South Rd., Durham DH1 3LE, UK}

{\sl $^{50}$ Ecole Polytechnique, Laboratoire Leprince-Ringuet (LLR), Route de Saclay, F-91128 Palaiseau Cedex, France}

{\sl $^{51}$ Ege University, Department of Physics, Faculty of Science, 35100 Izmir, Turkey}

{\sl $^{52}$ Enrico Fermi Institute, University of Chicago, 5640 S. Ellis Avenue, RI-183, Chicago, IL 60637, USA}

{\sl $^{53}$ Ewha Womans University, 11-1 Daehyun-Dong, Seodaemun-Gu, Seoul, 120-750, Korea}

{\sl $^{54}$ Fermi National Accelerator Laboratory (FNAL), P.O.Box 500, Batavia, IL 60510-0500, USA}

{\sl $^{55}$ Fujita Gakuen Health University, Department of Physics, Toyoake, Aichi 470-1192, Japan}

{\sl $^{56}$ Fukui University of Technology, 3-6-1 Gakuen, Fukui-shi, Fukui 910-8505, Japan}

{\sl $^{57}$ Fukui University, Department of Physics, 3-9-1 Bunkyo, Fukui-shi, Fukui 910-8507, Japan}

{\sl $^{58}$ Georg-August-Universit{\"a}t G{\"o}ttingen, II. Physikalisches Institut, Friedrich-Hund-Platz 1, 37077 G{\"o}ttingen, Germany}

{\sl $^{59}$ Global Design Effort}

{\sl $^{60}$ Gomel State University, Department of Physics, Ul. Sovietskaya 104, 246699 Gomel, Belarus}

{\sl $^{61}$ Guangxi University, College of Physics science and Engineering Technology, Nanning, China 530004}

{\sl $^{62}$ Hanoi University of Technology, 1 Dai Co Viet road, Hanoi, Vietnam}

{\sl $^{63}$ Hanson Professional Services, Inc., 1525 S. Sixth St., Springfield, IL 62703, USA}

{\sl $^{64}$ Harish-Chandra Research Institute, Chhatnag Road, Jhusi, Allahabad 211019, India}

{\sl $^{65}$ Helsinki Institute of Physics (HIP), P.O. Box 64, FIN-00014 University of Helsinki, Finland}

{\sl $^{66}$ Henan Normal University, College of Physics and Information Engineering, Xinxiang, China 453007}

{\sl $^{67}$ High Energy Accelerator Research Organization, KEK, 1-1 Oho, Tsukuba, Ibaraki 305-0801, Japan}

{\sl $^{68}$ Hiroshima University, Department of Physics, 1-3-1 Kagamiyama, Higashi-Hiroshima, Hiroshima 739-8526, Japan}

{\sl $^{69}$ Humboldt Universit{\"a}t zu Berlin, Fachbereich Physik, Institut f\"ur Elementarteilchenphysik, Newtonstr. 15, D-12489 Berlin, Germany}

{\sl $^{70}$ Hungarian Academy of Sciences, KFKI Research Institute for Particle and Nuclear Physics, P.O. Box 49, H-1525 Budapest, Hungary}

{\sl $^{71}$ Ibaraki University, College of Technology, Department of Physics, Nakanarusawa 4-12-1, Hitachi, Ibaraki 316-8511, Japan}

{\sl $^{72}$ Imperial College, Blackett Laboratory, Department of Physics, Prince Consort Road, London, SW7 2BW, UK}

{\sl $^{73}$ Indian Association for the Cultivation of Science, Department of Theoretical Physics and Centre for Theoretical Sciences, Kolkata 700032, India}

{\sl $^{74}$ Indian Institute of Science, Centre for High Energy Physics, Bangalore 560012, Karnataka, India}

{\sl $^{75}$ Indian Institute of Technology, Bombay, Powai, Mumbai 400076, India}

{\sl $^{76}$ Indian Institute of Technology, Guwahati, Guwahati, Assam 781039, India}

{\sl $^{77}$ Indian Institute of Technology, Kanpur, Department of Physics,  IIT Post Office, Kanpur 208016, India}

{\sl $^{78}$ Indiana University - Purdue University, Indianapolis, Department of Physics, 402 N. Blackford St., LD 154, Indianapolis, IN 46202, USA}

{\sl $^{79}$ Indiana University, Department of Physics, Swain Hall West 117, 727 E. 3rd St., Bloomington, IN 47405-7105, USA}

{\sl $^{80}$ Institucio Catalana de Recerca i Estudis, ICREA,  Passeig Lluis Companys, 23, Barcelona 08010, Spain}

{\sl $^{81}$ Institut de Physique Nucl\'eaire, F-91406 Orsay, France }

{\sl $^{82}$ Institut f\"ur Theorie Elektromagnetischer Felder (TEMF), Technische Universit\"at Darmstadt, Schlo{\ss}gartenstr. 8, D-64289 Darmstadt, Germany}

{\sl $^{83}$ Institut National de Physique Nucleaire et de Physique des Particules, 3, Rue Michel- Ange, 75794 Paris Cedex 16, France}

{\sl $^{84}$ Institut Pluridisciplinaire Hubert Curien, 23 Rue du Loess - BP28, 67037 Strasbourg Cedex 2, France}

{\sl $^{85}$ Institute for Chemical Research, Kyoto University, Gokasho, Uji, Kyoto 611-0011, Japan}

{\sl $^{86}$ Institute for Cosmic Ray Research, University of Tokyo, 5-1-5 Kashiwa-no-Ha, Kashiwa, Chiba 277-8582, Japan}

{\sl $^{87}$ Institute of High Energy Physics - IHEP, Chinese Academy of Sciences, P.O. Box 918, Beijing, China 100049}

{\sl $^{88}$ Institute of Mathematical Sciences, Taramani, C.I.T. Campus, Chennai 600113, India}

{\sl $^{89}$ Institute of Physics and Electronics, Vietnamese Academy of Science and Technology (VAST), 10 Dao-Tan, Ba-Dinh, Hanoi 10000, Vietnam}

{\sl $^{90}$ Institute of Physics, ASCR, Academy of Science of the Czech Republic, Division of Elementary Particle Physics, Na Slovance 2, CS-18221 Prague 8, Czech Republic}

{\sl $^{91}$ Institute of Physics, Pomorska 149/153, PL-90-236 Lodz, Poland}

{\sl $^{92}$ Institute of Theoretical and Experimetal Physics, B. Cheremushkinskawa, 25, RU-117259, Moscow, Russia}

{\sl $^{93}$ Institute of Theoretical Physics, Chinese Academy of Sciences, P.O.Box 2735, Beijing, China 100080}

{\sl $^{94}$ Instituto de Fisica Corpuscular (IFIC), Centro Mixto CSIC-UVEG, Edificio Investigacion Paterna, Apartado 22085, 46071 Valencia, Spain}

{\sl $^{95}$ Instituto de Fisica de Cantabria, (IFCA, CSIC-UC), Facultad de Ciencias, Avda. Los Castros s/n, 39005 Santander, Spain}

{\sl $^{96}$ Instituto Nazionale di Fisica Nucleare (INFN), Laboratorio LASA, Via Fratelli Cervi 201, 20090 Segrate, Italy}

{\sl $^{97}$ Instituto Nazionale di Fisica Nucleare (INFN), Sezione di Ferrara, via Paradiso 12, I-44100 Ferrara, Italy}

{\sl $^{98}$ Instituto Nazionale di Fisica Nucleare (INFN), Sezione di Firenze, Via G. Sansone 1, I-50019 Sesto Fiorentino (Firenze), Italy}

{\sl $^{99}$ Instituto Nazionale di Fisica Nucleare (INFN), Sezione di Lecce, via Arnesano, I-73100 Lecce, Italy}

{\sl $^{100}$ Instituto Nazionale di Fisica Nucleare (INFN), Sezione di Napoli, Complesso Universit{\'a} di Monte Sant'Angelo,via, I-80126 Naples, Italy}

{\sl $^{101}$ Instituto Nazionale di Fisica Nucleare (INFN), Sezione di Pavia, Via Bassi 6, I-27100 Pavia, Italy}

{\sl $^{102}$ Instituto Nazionale di Fisica Nucleare (INFN), Sezione di Pisa, Edificio C - Polo Fibonacci Largo B. Pontecorvo, 3, I-56127 Pisa, Italy}

{\sl $^{103}$ Instituto Nazionale di Fisica Nucleare (INFN), Sezione di Torino, c/o Universit{\'a}' di Torino facolt{\'a}' di Fisica, via P Giuria 1, 10125 Torino, Italy}

{\sl $^{104}$ Instituto Nazionale di Fisica Nucleare (INFN), Sezione di Trieste, Padriciano 99, I-34012 Trieste (Padriciano), Italy}

{\sl $^{105}$ Inter-University Accelerator Centre, Aruna Asaf Ali Marg, Post Box 10502, New Delhi 110067, India}

{\sl $^{106}$ International Center for Elementary Particle Physics, University of Tokyo, Hongo 7-3-1, Bunkyo District, Tokyo 113-0033, Japan}

{\sl $^{107}$ Iowa State University, Department of Physics, High Energy Physics Group, Ames, IA 50011, USA}

{\sl $^{108}$ Jagiellonian University, Institute of Physics, Ul. Reymonta 4, PL-30-059 Cracow, Poland}

{\sl $^{109}$ Jamia Millia Islamia, Centre for Theoretical Physics, Jamia Nagar, New Delhi 110025, India}

{\sl $^{110}$ Jamia Millia Islamia, Department of Physics, Jamia Nagar, New Delhi 110025, India}

{\sl $^{111}$ Japan Aerospace Exploration Agency, Sagamihara Campus, 3-1-1 Yoshinodai, Sagamihara, Kanagawa 220-8510 , Japan}

{\sl $^{112}$ Japan Atomic Energy Agency, 4-49 Muramatsu, Tokai-mura, Naka-gun, Ibaraki 319-1184, Japan}

{\sl $^{113}$ Johannes Gutenberg Universit{\"a}t Mainz, Institut f{\"u}r Physik, 55099 Mainz, Germany}

{\sl $^{114}$ Johns Hopkins University, Applied Physics Laboratory, 11100 Johns Hopkins RD., Laurel, MD 20723-6099, USA}

{\sl $^{115}$ Joint Institute for Nuclear Research (JINR), Joliot-Curie 6, 141980, Dubna, Moscow Region, Russia}

{\sl $^{116}$ Kansas State University, Department of Physics, 116 Cardwell Hall, Manhattan, KS 66506, USA}

{\sl $^{117}$ KCS Corp., 2-7-25 Muramatsukita, Tokai, Ibaraki 319-1108, Japan}

{\sl $^{118}$ Kharkov Institute of Physics and Technology, National Science Center, 1, Akademicheskaya St., Kharkov, 61108, Ukraine}

{\sl $^{119}$ Kinki University, Department of Physics, 3-4-1 Kowakae, Higashi-Osaka, Osaka 577-8502, Japan}

{\sl $^{120}$ Kobe University, Faculty of Science, 1-1 Rokkodai-cho, Nada-ku, Kobe, Hyogo 657-8501, Japan}

{\sl $^{121}$ Kogakuin University, Department of Physics, Shinjuku Campus, 1-24-2 Nishi-Shinjuku, Shinjuku-ku, Tokyo 163-8677, Japan}

{\sl $^{122}$ Konkuk University, 93-1 Mojin-dong, Kwanglin-gu, Seoul 143-701, Korea}

{\sl $^{123}$ Korea Advanced Institute of Science \& Technology, Department of Physics, 373-1 Kusong-dong, Yusong-gu, Taejon 305-701, Korea}

{\sl $^{124}$ Korea Institute for Advanced Study (KIAS), School of Physics, 207-43 Cheongryangri-dong, Dongdaemun-gu, Seoul 130-012, Korea}

{\sl $^{125}$ Korea University, Department of Physics, Seoul 136-701, Korea}

{\sl $^{126}$ Kyoto University, Department of Physics, Kitashirakawa-Oiwakecho, Sakyo-ku, Kyoto 606-8502, Japan}

{\sl $^{127}$ L.P.T.A., UMR 5207 CNRS-UM2, Universit{\'e} Montpellier II, Case Courrier 070, B{\^a}t. 13, place Eug{\`e}ne Bataillon, 34095 Montpellier Cedex 5, France}

{\sl $^{128}$ Laboratoire d'Annecy-le-Vieux de Physique des Particules (LAPP), Chemin du Bellevue, BP 110, F-74941 Annecy-le-Vieux Cedex, France}

{\sl $^{129}$ Laboratoire d'Annecy-le-Vieux de Physique Theorique (LAPTH), Chemin de Bellevue, BP 110, F-74941 Annecy-le-Vieux Cedex, France}

{\sl $^{130}$ Laboratoire de l'Acc\'el\'erateur Lin\'eaire (LAL), Universit\'e Paris-Sud 11, B\^atiment 200, 91898 Orsay, France}

{\sl $^{131}$ Laboratoire de Physique Corpusculaire de Clermont-Ferrand (LPC), Universit\'e Blaise Pascal, I.N.2.P.3./C.N.R.S., 24 avenue des Landais, 63177 Aubi\`ere Cedex, France}

{\sl $^{132}$ Laboratoire de Physique Subatomique et de Cosmologie (LPSC), Universit\'e Joseph Fourier (Grenoble 1), 53, ave. des Marthyrs, F-38026 Grenoble Cedex, France}

{\sl $^{133}$ Laboratoire de Physique Theorique, Universit\'e de Paris-Sud XI, Batiment 210, F-91405 Orsay Cedex, France}

{\sl $^{134}$ Laboratori Nazionali di Frascati, via E. Fermi, 40, C.P. 13, I-00044 Frascati, Italy}

{\sl $^{135}$ Laboratory of High Energy Physics and Cosmology, Department of Physics, Hanoi National University, 334 Nguyen Trai, Hanoi, Vietnam}

{\sl $^{136}$ Lancaster University, Physics Department, Lancaster LA1 4YB, UK}

{\sl $^{137}$ Lawrence Berkeley National Laboratory (LBNL), 1 Cyclotron Rd, Berkeley, CA 94720, USA}

{\sl $^{138}$ Lawrence Livermore National Laboratory (LLNL), Livermore, CA 94551, USA}

{\sl $^{139}$ Lebedev Physical Institute, Leninsky Prospect 53, RU-117924 Moscow, Russia}

{\sl $^{140}$ Liaoning Normal University, Department of Physics, Dalian, China 116029}

{\sl $^{141}$ Lomonosov Moscow State University, Skobeltsyn Institute of Nuclear Physics (MSU SINP), 1(2), Leninskie gory, GSP-1, Moscow 119991, Russia}

{\sl $^{142}$ Los Alamos National Laboratory (LANL), P.O.Box 1663, Los Alamos, NM 87545, USA}

{\sl $^{143}$ Louisiana Technical University, Department of Physics, Ruston, LA 71272, USA}

{\sl $^{144}$ Ludwig-Maximilians-Universit{\"a}t M{\"u}nchen, Department f{\"u}r Physik, Schellingstr. 4, D-80799 Munich, Germany}

{\sl $^{145}$ Lunds Universitet, Fysiska Institutionen, Avdelningen f{\"o}r Experimentell H{\"o}genergifysik, Box 118, 221 00 Lund, Sweden}

{\sl $^{146}$ Massachusetts Institute of Technology, Laboratory for Nuclear Science \& Center for Theoretical Physics, 77 Massachusetts Ave., NW16, Cambridge, MA 02139, USA}

{\sl $^{147}$ Max-Planck-Institut f{\"u}r Physik (Werner-Heisenberg-Institut), F{\"o}hringer Ring 6, 80805 M{\"u}nchen, Germany}

{\sl $^{148}$ McGill University, Department of Physics, Ernest Rutherford Physics Bldg., 3600 University Ave., Montreal, Quebec, H3A 2T8 Canada}

{\sl $^{149}$ Meiji Gakuin University, Department of Physics, 2-37 Shirokanedai 1-chome, Minato-ku, Tokyo 244-8539, Japan}

{\sl $^{150}$ Michigan State University, Department of Physics and Astronomy, East Lansing, MI 48824, USA}

{\sl $^{151}$ Middle East Technical University, Department of Physics, TR-06531 Ankara, Turkey}

{\sl $^{152}$ Mindanao Polytechnic State College, Lapasan, Cagayan de Oro City 9000, Phillipines}

{\sl $^{153}$ MSU-Iligan Institute of Technology, Department of Physics, Andres Bonifacio Avenue, 9200 Iligan City, Phillipines}

{\sl $^{154}$ Nagasaki Institute of Applied Science, 536 Abamachi, Nagasaki-Shi, Nagasaki 851-0193, Japan}

{\sl $^{155}$ Nagoya University, Fundamental Particle Physics Laboratory, Division of Particle and Astrophysical Sciences, Furo-cho, Chikusa-ku, Nagoya, Aichi 464-8602, Japan}

{\sl $^{156}$ Nanchang University, Department of Physics, Nanchang, China 330031}

{\sl $^{157}$ Nanjing University, Department of Physics, Nanjing, China 210093}

{\sl $^{158}$ Nankai University, Department of Physics, Tianjin, China 300071}

{\sl $^{159}$ National Central University, High Energy Group, Department of Physics, Chung-li, Taiwan 32001}

{\sl $^{160}$ National Institute for Nuclear \& High Energy Physics, PO Box 41882, 1009 DB Amsterdam, Netherlands}

{\sl $^{161}$ National Institute of Radiological Sciences, 4-9-1 Anagawa, Inaga, Chiba 263-8555, Japan}

{\sl $^{162}$ National Synchrotron Radiation Laboratory, University of Science and Technology of china, Hefei, Anhui, China 230029}

{\sl $^{163}$ National Synchrotron Research Center, 101 Hsin-Ann Rd., Hsinchu Science Part, Hsinchu, Taiwan 30076}

{\sl $^{164}$ National Taiwan University, Physics Department, Taipei, Taiwan 106}

{\sl $^{165}$ Niels Bohr Institute (NBI), University of Copenhagen, Blegdamsvej 17, DK-2100 Copenhagen, Denmark}

{\sl $^{166}$ Niigata University, Department of Physics, Ikarashi, Niigata 950-218, Japan}

{\sl $^{167}$ Nikken Sekkai Ltd., 2-18-3 Iidabashi, Chiyoda-Ku, Tokyo 102-8117, Japan}

{\sl $^{168}$ Nippon Dental University, 1-9-20 Fujimi, Chiyoda-Ku, Tokyo 102-8159, Japan}

{\sl $^{169}$ North Asia University, Akita 010-8515, Japan}

{\sl $^{170}$ North Eastern Hill University, Department of Physics, Shillong 793022, India}

{\sl $^{171}$ Northern Illinois University, Department of Physics, DeKalb, Illinois 60115-2825, USA}

{\sl $^{172}$ Northwestern University, Department of Physics and Astronomy, 2145 Sheridan Road., Evanston, IL 60208, USA}

{\sl $^{173}$ Novosibirsk State University (NGU), Department of Physics, Pirogov st. 2, 630090 Novosibirsk, Russia}

{\sl $^{174}$ Obninsk State Technical University for Nuclear Engineering (IATE), Obninsk, Russia}

{\sl $^{175}$ Ochanomizu University, Department of Physics, Faculty of Science, 1-1 Otsuka 2, Bunkyo-ku, Tokyo 112-8610, Japan}

{\sl $^{176}$ Osaka University, Laboratory of Nuclear Studies, 1-1 Machikaneyama, Toyonaka, Osaka 560-0043, Japan}

{\sl $^{177}$ {\"O}sterreichische Akademie der Wissenschaften, Institut f{\"u}r Hochenergiephysik, Nikolsdorfergasse 18, A-1050 Vienna, Austria}

{\sl $^{178}$ Panjab University, Chandigarh 160014, India}

{\sl $^{179}$ Pavel Sukhoi Gomel State Technical University, ICTP Affiliated Centre \& Laboratory for Physical Studies, October Avenue, 48, 246746, Gomel, Belarus}

{\sl $^{180}$ Pavel Sukhoi Gomel State Technical University, Physics Department, October Ave. 48, 246746 Gomel, Belarus}

{\sl $^{181}$ Physical Research Laboratory, Navrangpura, Ahmedabad 380 009, Gujarat, India}

{\sl $^{182}$ Pohang Accelerator Laboratory (PAL), San-31 Hyoja-dong, Nam-gu, Pohang, Gyeongbuk 790-784, Korea}

{\sl $^{183}$ Polish Academy of Sciences (PAS), Institute of Physics, Al. Lotnikow 32/46, PL-02-668 Warsaw, Poland}

{\sl $^{184}$ Primera Engineers Ltd., 100 S Wacker Drive, Suite 700, Chicago, IL 60606, USA}

{\sl $^{185}$ Princeton University, Department of Physics, P.O. Box 708, Princeton, NJ 08542-0708, USA}

{\sl $^{186}$ Purdue University, Department of Physics, West Lafayette, IN 47907, USA}

{\sl $^{187}$ Pusan National University, Department of Physics, Busan 609-735, Korea}

{\sl $^{188}$ R. W. Downing Inc., 6590 W. Box Canyon Dr., Tucson, AZ 85745, USA}

{\sl $^{189}$ Raja Ramanna Center for Advanced Technology, Indore 452013, India}

{\sl $^{190}$ Rheinisch-Westf{\"a}lische Technische Hochschule (RWTH), Physikalisches Institut, Physikzentrum, Sommerfeldstrasse 14, D-52056 Aachen, Germany}

{\sl $^{191}$ RIKEN, 2-1 Hirosawa, Wako, Saitama 351-0198, Japan}

{\sl $^{192}$ Royal Holloway, University of London (RHUL), Department of Physics, Egham, Surrey TW20 0EX, UK }

{\sl $^{193}$ Saga University, Department of Physics, 1 Honjo-machi, Saga-shi, Saga 840-8502, Japan}

{\sl $^{194}$ Saha Institute of Nuclear Physics, 1/AF Bidhan Nagar, Kolkata 700064, India}

{\sl $^{195}$ Salalah College of Technology (SCOT), Engineering Department, Post Box No. 608, Postal Code 211, Salalah, Sultanate of Oman}

{\sl $^{196}$ Saube Co., Hanabatake, Tsukuba, Ibaraki 300-3261, Japan}

{\sl $^{197}$ Seoul National University, San 56-1, Shinrim-dong, Kwanak-gu, Seoul 151-742, Korea}

{\sl $^{198}$ Shandong University, 27 Shanda Nanlu, Jinan, China 250100}

{\sl $^{199}$ Shanghai Institute of Applied Physics, Chinese Academy of Sciences, 2019 Jiaruo Rd., Jiading, Shanghai, China 201800}

{\sl $^{200}$ Shinshu University, 3-1-1, Asahi, Matsumoto, Nagano 390-8621, Japan}

{\sl $^{201}$ Sobolev Institute of Mathematics, Siberian Branch of the Russian Academy of Sciences, 4 Acad. Koptyug Avenue, 630090 Novosibirsk, Russia}

{\sl $^{202}$ Sokendai, The Graduate University for Advanced Studies, Shonan Village, Hayama, Kanagawa 240-0193, Japan}

{\sl $^{203}$ Stanford Linear Accelerator Center (SLAC), 2575 Sand Hill Road, Menlo Park, CA 94025, USA}

{\sl $^{204}$ State University of New York at Binghamton, Department of Physics, PO Box 6016, Binghamton, NY 13902, USA}

{\sl $^{205}$ State University of New York at Buffalo, Department of Physics \& Astronomy, 239 Franczak Hall, Buffalo, NY 14260, USA}

{\sl $^{206}$ State University of New York at Stony Brook, Department of Physics and Astronomy, Stony Brook, NY 11794-3800, USA}

{\sl $^{207}$ Sumitomo Heavy Industries, Ltd., Natsushima-cho, Yokosuka, Kanagawa 237-8555, Japan}

{\sl $^{208}$ Sungkyunkwan University (SKKU), Natural Science Campus 300, Physics Research Division, Chunchun-dong, Jangan-gu, Suwon, Kyunggi-do 440-746, Korea}

{\sl $^{209}$ Swiss Light Source (SLS), Paul Scherrer Institut (PSI), PSI West, CH-5232 Villigen PSI, Switzerland}

{\sl $^{210}$ Syracuse University, Department of Physics, 201 Physics Building, Syracuse, NY 13244-1130, USA}

{\sl $^{211}$ Tata Institute of Fundamental Research, School of Natural Sciences, Homi Bhabha Rd., Mumbai 400005, India}

{\sl $^{212}$ Technical Institute of Physics and Chemistry, Chinese Academy of Sciences, 2 North 1st St., Zhongguancun, Beijing, China 100080}

{\sl $^{213}$ Technical University of Lodz, Department of Microelectronics and Computer Science, al. Politechniki 11, 90-924 Lodz, Poland}

{\sl $^{214}$ Technische Universit{\"a}t Dresden, Institut f{\"u}r Kern- und Teilchenphysik, D-01069 Dresden, Germany}

{\sl $^{215}$ Technische Universit{\"a}t Dresden, Institut f{\"u}r Theoretische Physik,D-01062 Dresden, Germany}

{\sl $^{216}$ Tel-Aviv University, School of Physics and Astronomy, Ramat Aviv, Tel Aviv 69978, Israel}

{\sl $^{217}$ Texas A\&M University, Physics Department, College Station, 77843-4242 TX, USA}

{\sl $^{218}$ Texas Tech University, Department of Physics, Campus Box 41051, Lubbock, TX 79409-1051, USA}

{\sl $^{219}$ The Henryk Niewodniczanski Institute of Nuclear Physics (NINP), High Energy Physics Lab, ul. Radzikowskiego 152, PL-31342 Cracow, Poland}

{\sl $^{220}$ Thomas Jefferson National Accelerator Facility (TJNAF), 12000 Jefferson Avenue, Newport News, VA 23606, USA}

{\sl $^{221}$ Tohoku Gakuin University, Faculty of Technology, 1-13-1 Chuo, Tagajo, Miyagi 985-8537, Japan}

{\sl $^{222}$ Tohoku University, Department of Physics, Aoba District, Sendai, Miyagi 980-8578, Japan}

{\sl $^{223}$ Tokyo Management College, Computer Science Lab, Ichikawa, Chiba 272-0001, Japan}

{\sl $^{224}$ Tokyo University of Agriculture Technology, Department of Applied Physics, Naka-machi, Koganei, Tokyo 183-8488, Japan}

{\sl $^{225}$ Toyama University, Department of Physics, 3190 Gofuku, Toyama-shi 930-8588, Japan}

{\sl $^{226}$ TRIUMF, 4004 Wesbrook Mall, Vancouver, BC V6T 2A3, Canada}

{\sl $^{227}$ Tufts University, Department of Physics and Astronomy, Robinson Hall, Medford, MA 02155, USA}

{\sl $^{228}$ Universidad Aut\`onoma de Madrid (UAM), Facultad de Ciencias C-XI, Departamento de Fisica Teorica, Cantoblanco, Madrid 28049, Spain}

{\sl $^{229}$ Universitat Aut\`onoma de Barcelona, Institut de Fisica d'Altes Energies (IFAE), Campus UAB, Edifici Cn, E-08193 Bellaterra, Barcelona, Spain}

{\sl $^{230}$ University College of London (UCL), High Energy Physics Group, Physics and Astronomy Department, Gower Street, London WC1E 6BT, UK }

{\sl $^{231}$ University College, National University of Ireland (Dublin), Department of Experimental Physics, Science Buildings, Belfield, Dublin 4, Ireland}

{\sl $^{232}$ University de Barcelona, Facultat de F\'isica, Av. Diagonal, 647, Barcelona 08028, Spain}

{\sl $^{233}$ University of Abertay Dundee, Department of Physics, Bell St, Dundee, DD1 1HG, UK}

{\sl $^{234}$ University of Auckland, Department of Physics, Private Bag, Auckland 1, New Zealand}

{\sl $^{235}$ University of Bergen, Institute of Physics, Allegaten 55, N-5007 Bergen, Norway}

{\sl $^{236}$ University of Birmingham, School of Physics and Astronomy, Particle Physics Group, Edgbaston, Birmingham B15 2TT, UK}

{\sl $^{237}$ University of Bristol, H. H. Wills Physics Lab, Tyndall Ave., Bristol BS8 1TL, UK}

{\sl $^{238}$ University of British Columbia, Department of Physics and Astronomy, 6224 Agricultural Rd., Vancouver, BC V6T 1Z1, Canada}

{\sl $^{239}$ University of California Berkeley, Department of Physics, 366 Le Conte Hall, \#7300, Berkeley, CA 94720, USA}

{\sl $^{240}$ University of California Davis, Department of Physics, One Shields Avenue, Davis, CA 95616-8677, USA}

{\sl $^{241}$ University of California Irvine, Department of Physics and Astronomy, High Energy Group, 4129 Frederick Reines Hall, Irvine, CA 92697-4575 USA}

{\sl $^{242}$ University of California Riverside, Department of Physics, Riverside, CA 92521, USA}

{\sl $^{243}$ University of California Santa Barbara, Department of Physics, Broida Hall, Mail Code 9530, Santa Barbara, CA 93106-9530, USA}

{\sl $^{244}$ University of California Santa Cruz, Department of Astronomy and Astrophysics, 1156 High Street, Santa Cruz, CA 05060, USA}

{\sl $^{245}$ University of California Santa Cruz, Institute for Particle Physics, 1156 High Street, Santa Cruz, CA 95064, USA}

{\sl $^{246}$ University of Cambridge, Cavendish Laboratory, J J Thomson Avenue, Cambridge CB3 0HE, UK}

{\sl $^{247}$ University of Colorado at Boulder, Department of Physics, 390 UCB, University of Colorado, Boulder, CO 80309-0390, USA}

{\sl $^{248}$ University of Delhi, Department of Physics and Astrophysics, Delhi 110007, India}

{\sl $^{249}$ University of Delhi, S.G.T.B. Khalsa College, Delhi 110007, India}

{\sl $^{250}$ University of Dundee, Department of Physics, Nethergate, Dundee, DD1 4HN,  Scotland, UK}

{\sl $^{251}$ University of Edinburgh, School of Physics, James Clerk Maxwell Building, The King's Buildings, Mayfield Road, Edinburgh EH9 3JZ, UK}

{\sl $^{252}$ University of Essex, Department of Physics, Wivenhoe Park, Colchester CO4 3SQ, UK}

{\sl $^{253}$ University of Florida, Department of Physics, Gainesville, FL 32611, USA}

{\sl $^{254}$ University of Glasgow, Department of Physics \& Astronomy, University Avenue, Glasgow G12 8QQ, Scotland, UK}

{\sl $^{255}$ University of Hamburg, Physics Department, Institut f{\"u}r Experimentalphysik, Luruper Chaussee 149, 22761 Hamburg, Germany}

{\sl $^{256}$ University of Hawaii, Department of Physics and Astronomy, HEP, 2505 Correa Rd., WAT 232, Honolulu, HI 96822-2219, USA}

{\sl $^{257}$ University of Heidelberg, Kirchhoff Institute of Physics, Albert {\"U}berle Strasse 3-5, DE-69120 Heidelberg, Germany}

{\sl $^{258}$ University of Helsinki, Department of Physical Sciences, P.O. Box 64 (Vaino Auerin katu 11), FIN-00014, Helsinki, Finland}

{\sl $^{259}$ University of Hyogo, School of Science, Kouto 3-2-1, Kamigori, Ako, Hyogo 678-1297,  Japan}

{\sl $^{260}$ University of Illinois at Urbana-Champaign, Department of Phys., High Energy Physics, 441 Loomis Lab. of Physics1110 W. Green St., Urbana, IL 61801-3080, USA}

{\sl $^{261}$ University of Iowa, Department of Physics and Astronomy, 203 Van Allen Hall, Iowa City, IA 52242-1479, USA}

{\sl $^{262}$ University of Kansas, Department of Physics and Astronomy, Malott Hall, 1251 Wescoe Hall Drive, Room 1082, Lawrence, KS 66045-7582, USA}

{\sl $^{263}$ University of Liverpool, Department of Physics, Oliver Lodge Lab, Oxford St., Liverpool L69 7ZE, UK}

{\sl $^{264}$ University of Louisville, Department of Physics, Louisville, KY 40292, USA}

{\sl $^{265}$ University of Manchester, School of Physics and Astronomy, Schuster Lab, Manchester M13 9PL, UK}

{\sl $^{266}$ University of Maryland, Department of Physics and Astronomy, Physics Building (Bldg. 082), College Park, MD 20742, USA}

{\sl $^{267}$ University of Melbourne, School of Physics, Victoria 3010, Australia}

{\sl $^{268}$ University of Michigan, Department of Physics, 500 E. University Ave., Ann Arbor, MI 48109-1120, USA}

{\sl $^{269}$ University of Minnesota, 148 Tate Laboratory Of Physics, 116 Church St. S.E., Minneapolis, MN 55455, USA}

{\sl $^{270}$ University of Mississippi, Department of Physics and Astronomy, 108 Lewis Hall, PO Box 1848, Oxford, Mississippi 38677-1848, USA}

{\sl $^{271}$ University of Montenegro, Faculty of Sciences and Math., Department of Phys., P.O. Box 211, 81001 Podgorica, Serbia and Montenegro}

{\sl $^{272}$ University of New Mexico, New Mexico Center for Particle Physics, Department of Physics and Astronomy, 800 Yale Boulevard N.E., Albuquerque, NM 87131, USA}

{\sl $^{273}$ University of Notre Dame, Department of Physics, 225 Nieuwland Science Hall, Notre Dame, IN 46556, USA}

{\sl $^{274}$ University of Oklahoma, Department of Physics and Astronomy, Norman, OK 73071, USA}

{\sl $^{275}$ University of Oregon, Department of Physics, 1371 E. 13th Ave., Eugene, OR 97403, USA}

{\sl $^{276}$ University of Oxford, Particle Physics Department, Denys Wilkinson Bldg., Keble Road, Oxford OX1 3RH England, UK }

{\sl $^{277}$ University of Patras, Department of Physics, GR-26100 Patras, Greece}

{\sl $^{278}$ University of Pavia, Department of Nuclear and Theoretical Physics, via Bassi 6, I-27100 Pavia, Italy}

{\sl $^{279}$ University of Pennsylvania, Department of Physics and Astronomy, 209 South 33rd Street, Philadelphia, PA 19104-6396, USA}

{\sl $^{280}$ University of Puerto Rico at Mayaguez, Department of Physics, P.O. Box 9016, Mayaguez, 00681-9016 Puerto Rico}

{\sl $^{281}$ University of Regina, Department of Physics, Regina, Saskatchewan, S4S 0A2 Canada}

{\sl $^{282}$ University of Rochester, Department of Physics and Astronomy, Bausch \& Lomb Hall, P.O. Box 270171, 600 Wilson Boulevard, Rochester, NY 14627-0171 USA}

{\sl $^{283}$ University of Science and Technology of China, Department of Modern Physics (DMP), Jin Zhai Road 96, Hefei, China 230026}

{\sl $^{284}$ University of Silesia, Institute of Physics, Ul. Uniwersytecka 4, PL-40007 Katowice, Poland}

{\sl $^{285}$ University of Southampton, School of Physics and Astronomy, Highfield, Southampton S017 1BJ, England, UK}

{\sl $^{286}$ University of Strathclyde, Physics Department, John Anderson Building, 107 Rottenrow, Glasgow, G4 0NG, Scotland, UK}

{\sl $^{287}$ University of Sydney, Falkiner High Energy Physics Group, School of Physics, A28, Sydney, NSW 2006, Australia}

{\sl $^{288}$ University of Texas, Center for Accelerator Science and Technology, Arlington, TX 76019, USA}

{\sl $^{289}$ University of Tokushima, Institute of Theoretical Physics, Tokushima-shi 770-8502, Japan}

{\sl $^{290}$ University of Tokyo, Department of Physics, 7-3-1 Hongo, Bunkyo District, Tokyo 113-0033, Japan}

{\sl $^{291}$ University of Toronto, Department of Physics, 60 St. George St., Toronto M5S 1A7, Ontario, Canada}

{\sl $^{292}$ University of Tsukuba, Institute of Physics, 1-1-1 Ten'nodai, Tsukuba, Ibaraki 305-8571, Japan}

{\sl $^{293}$ University of Victoria, Department of Physics and Astronomy, P.O.Box 3055 Stn Csc, Victoria, BC V8W 3P6, Canada}

{\sl $^{294}$ University of Warsaw, Institute of Physics, Ul. Hoza 69, PL-00 681 Warsaw, Poland}

{\sl $^{295}$ University of Warsaw, Institute of Theoretical Physics, Ul. Hoza 69, PL-00 681 Warsaw, Poland}

{\sl $^{296}$ University of Washington, Department of Physics, PO Box 351560, Seattle, WA 98195-1560, USA}

{\sl $^{297}$ University of Wisconsin, Physics Department, Madison, WI 53706-1390, USA}

{\sl $^{298}$ University of Wuppertal, Gau{\ss}stra{\ss}e 20, D-42119 Wuppertal, Germany}

{\sl $^{299}$ Universit\'e Claude Bernard Lyon-I, Institut de Physique Nucl\'eaire de Lyon (IPNL), 4, rue Enrico Fermi, F-69622 Villeurbanne Cedex, France}

{\sl $^{300}$ Universit\'e de Gen\`eve, Section de Physique, 24, quai E. Ansermet, 1211 Gen\`eve 4, Switzerland}

{\sl $^{301}$ Universit\'e Louis Pasteur (Strasbourg I), UFR de Sciences Physiques, 3-5 Rue de l'Universit\'e, F-67084 Strasbourg Cedex, France}

{\sl $^{302}$ Universit\'e Pierre et Marie Curie (Paris VI-VII) (6-7) (UPMC), Laboratoire de Physique Nucl\'eaire et de Hautes Energies (LPNHE), 4 place Jussieu, Tour 33, Rez de chausse, 75252 Paris Cedex 05, France}

{\sl $^{303}$ Universit{\"a}t Bonn, Physikalisches Institut, Nu{\ss}allee 12, 53115 Bonn, Germany}

{\sl $^{304}$ Universit{\"a}t Karlsruhe, Institut f{\"u}r Physik, Postfach 6980, Kaiserstrasse 12, D-76128 Karlsruhe, Germany}

{\sl $^{305}$ Universit{\"a}t Rostock, Fachbereich Physik, Universit{\"a}tsplatz 3, D-18051 Rostock, Germany}

{\sl $^{306}$ Universit{\"a}t Siegen, Fachbereich f{\"u}r Physik, Emmy Noether Campus, Walter-Flex-Str.3, D-57068 Siegen, Germany}

{\sl $^{307}$ Universit{\`a} de Bergamo, Dipartimento di Fisica, via Salvecchio, 19, I-24100 Bergamo, Italy}

{\sl $^{308}$ Universit{\`a} degli Studi di Roma La Sapienza, Dipartimento di Fisica, Istituto Nazionale di Fisica Nucleare, Piazzale Aldo Moro 2, I-00185 Rome, Italy}

{\sl $^{309}$ Universit{\`a} degli Studi di Trieste, Dipartimento di Fisica, via A. Valerio 2, I-34127 Trieste, Italy}

{\sl $^{310}$ Universit{\`a} degli Studi di ``Roma Tre'', Dipartimento di Fisica ``Edoardo Amaldi'', Istituto Nazionale di Fisica Nucleare, Via della Vasca Navale 84, 00146 Roma, Italy}

{\sl $^{311}$ Universit{\`a} dell'Insubria in Como, Dipartimento di Scienze CC.FF.MM., via Vallegio 11, I-22100 Como, Italy}

{\sl $^{312}$ Universit{\`a} di Pisa, Departimento di Fisica 'Enrico Fermi', Largo Bruno Pontecorvo 3, I-56127 Pisa, Italy}

{\sl $^{313}$ Universit{\`a} di Salento, Dipartimento di Fisica, via Arnesano, C.P. 193, I-73100 Lecce, Italy}

{\sl $^{314}$ Universit{\`a} di Udine, Dipartimento di Fisica, via delle Scienze, 208, I-33100 Udine, Italy}

{\sl $^{315}$ Variable Energy Cyclotron Centre, 1/AF, Bidhan Nagar, Kolkata 700064, India}

{\sl $^{316}$ VINCA Institute of Nuclear Sciences, Laboratory of Physics, PO Box 522, YU-11001 Belgrade, Serbia and Montenegro}

{\sl $^{317}$ Vinh University, 182 Le Duan, Vinh City, Nghe An Province, Vietnam}

{\sl $^{318}$ Virginia Polytechnic Institute and State University, Physics Department, Blacksburg, VA 2406, USA}

{\sl $^{319}$ Visva-Bharati University, Department of Physics, Santiniketan 731235, India}

{\sl $^{320}$ Waseda University, Advanced Research Institute for Science and Engineering, Shinjuku, Tokyo 169-8555, Japan}

{\sl $^{321}$ Wayne State University, Department of Physics, Detroit, MI 48202, USA}

{\sl $^{322}$ Weizmann Institute of Science, Department of Particle Physics, P.O. Box 26, Rehovot 76100, Israel}

{\sl $^{323}$ Yale University, Department of Physics, New Haven, CT 06520, USA}

{\sl $^{324}$ Yonsei University, Department of Physics, 134 Sinchon-dong, Sudaemoon-gu, Seoul 120-749, Korea}

{\sl $^{325}$ Zhejiang University, College of Science, Department of Physics, Hangzhou, China 310027}

{\sl * deceased } 

\end{center}

\end{center}

\chapter*{Acknowledgements} 
We would like to acknowledge the support and guidance of the International Committee on Future Accelerators (ICFA), chaired by A. Wagner of DESY, and the International Linear Collider Steering Committee (ILCSC), chaired by S. Kurokawa of KEK, who established the ILC Global Design Effort, as well as the World Wide Study of the Physics and Detectors.
      
\medskip
We are grateful to the ILC Machine Advisory Committee (MAC), chaired by F. Willeke of DESY and the International ILC Cost Review Committee, chaired by L. Evans of CERN, for their advice on the ILC Reference Design. We also thank the consultants who particpated in the Conventional Facilities Review at CalTech and in the RDR Cost Review at SLAC. 

\medskip 
We would like to thank the directors of the institutions who have hosted ILC meetings: KEK, ANL/FNAL/SLAC/U. Colorado (Snowmass), INFN/Frascati, IIT/Bangalore, TRIUMF/U. British Columbia, U. Valencia, IHEP/Beijing and DESY.

\medskip 
We are grateful for the support of the Funding Agencies for Large Colliders (FALC), chaired by R. Petronzio of INFN, and we thank all of the international, regional and national funding agencies whose generous support has made the ILC 
Reference Design possible.

\medskip 
Each of the GDE regional teams in the Americas, Asia and Europe are grateful for the support of their local scientific societies, industrial forums, advisory committees and reviewers.

\cleardoublepage
\tableofcontents %

\setcounter{secnumdepth}{0}





\mainmatter
\setcounter{page}{1} \setcounter{chapter}{0}


\newcommand{\picturefolder}{}

\clearpage
\setcounter{figure}{0}
\setcounter{table}{0}
\setcounter{secnumdepth}{4}
\setcounter{subsection}{0} 
\setcounter{subsubsection}{0}
\usecounter{subsubsection}
\chapter{Introduction}
\label{detector_introduction}

The physics potential of the ILC, discussed in Volume 2 of this document, has captured
the imagination of the world high energy physics community. Understanding the mechanism
behind mass generation and electroweak symmetry breaking, searching for and perhaps 
discovering supersymmetric particles and confirming their supersymmetric nature,
and hunting for signs of extra space-time dimensions and quantum gravity, constitute
some of the major physics goals of the ILC. In addition, making precision measurements of
standard model processes will open windows on physics at energy scales beyond our direct reach. 
The unexpected is our fondest hope.

The detectors which will record and measure the charged and neutral particles produced 
in the ILC's high energy $e^+e^-$ collisions, are the subject of this Volume 4 of the International
Linear Collider Reference Design Report, which is also called Detector Concept Report (DCR). Experimental conditions at the ILC provide an ideal environment for
the precision study of particle production and decay, and offer the unparalleled 
cleanliness and well-defined initial conditions conducive to recognizing new phenomena.
Compared to hadronic interactions, $e^+e^-$ collisions generate events essentially free 
from backgrounds due to multiple interactions; provide accurate knowledge of the
center-of-mass energy, initial state helicity,
and charge; and produce all particle species democratically. In fact, $e^+e^-$ collisions afford
full control of the initial state helicity by appropriately selecting electron and positron polarizations,
providing a unique and powerful tool for measuring asymmetries, boosting desired signals, and reducing unwanted backgrounds.
ILC Detectors need not contend
with extreme data rates or high radiation fields. They can in fact record events
without need for electronic preselection and without the biases such selection may introduce.
The detectors, however, need to achieve unprecedented precision to reach the performance 
required by the physics.
The physics does pose significant challenges for detector performance, and pushes the
limits of jet energy resolution, tracker momentum resolution, and vertex impact parameter
resolution, to name a few. Multi-jet final states and SUSY searches put a premium on hermeticity 
and full solid angle coverage. The ILC environment, although benign by LHC standards, 
does pose some interesting challenges of its own. 

The world-wide linear collider physics and detector community has wrestled with these challenges for
more than a decade, and has made impressive progress in developing the new sensor technologies
an ILC detector will need. Concepts for the detectors have evolved throughout the world, with early 
designs recorded in several reports \cite{ref-TESLA:2001,ref-ACFA:2001,ref-SM01:2001}. Rapid progress on the machine
side, first with the ITRP's choice of superconducting RF in 2004, then with the formation of the ILC's Global
Design Effort in early 2005, and most recently
with the design and costing exercise recorded in the Reference Design Report, have spurred the
experimental community to keep in step. With this in mind, the World Wide Study of Physics and 
Detectors for Future Linear Colliders charged the high energy physics community to prepare 
Detector Outline Documents, to capture both the thinking behind and the present status of 
the existing detector designs.  Four reports from the concept teams, GLD~\cite{ref-GLD}, LDC~\cite{ref-LDC},
SiD~\cite{ref-SiD}, and 4th~\cite{ref-4th}, were presented in Spring of 2006. These documents discuss design 
philosophy, conceptual designs, R\&D readiness and plans, subsystem performance, and overall 
physics performance for each of the concepts; and they form the basis for the present report. 

Development of the
concepts goes hand in hand with sub-detector research and development, which is occurring both on the
somewhat orthogonal axis of the R\&D collaborations, and in some cases, within the concepts themselves. 
Ideally, advances in the detector arts benefit all four concepts.

Progress to date indicates that current designs will deliver the performance
ILC physics demands, and that they are buildable with technologies that are within
reach. Not all is demonstrated, but a growing community is involved in refining
and optimizing designs, and advancing the technologies. Continued and expanded support
of detector R\&D and concept designs can lead to full engineering designs, and proof of
principle technology demonstrations on the timetable being proposed for the ILC Engineering
Design Report.

In the following chapters, this report will make the case for ILC detectors. Challenges from the 
physics and ILC environment that drive the detector design and technology are outlined in 
chapter~\ref{Challenges}. An overview of the four detector concepts is given in 
chapter~\ref{detector_concepts}. Chapter~\ref{detector_MDI} reviews the machine detector interface, 
including interaction region design, evaluation of backgrounds, and properties of the bunch train timing.
It delineates and details the ILC environment. The status of subsystem designs and technologies, 
sampled across the concepts, is given thorough review in Chapter~\ref{detector_technologies}. 
Sub-detector performance is the subject of increasingly sophisticated and realistic simulations; 
it is reviewed in Chapter~\ref{subdetector_performance}. The integrated performance of the detectors 
has long been approximated only with fast Monte Carlo. New studies, based on full Monte Carlo, are 
given in Chapter~\ref{physics_performance}. They evidence the growing maturity of ILC physics studies, 
and promise more believable results. Chapter~\ref{detector_2detectors} argues the need for two, 
complementary detectors at the ILC. Chapter~\ref{detector_cost} gives the ballpark cost of the present 
concepts, derived from comparisons of the individual cost accounts. The present detector concepts are 
designed to study $e^+e^-$ interactions over the full range accessible to the ILC, from 500~GeV in the 
first phase of the machine to 1000~GeV in the machine's second phase. The physics may lead us to detours 
at the Z pole, or an exploration of gamma gamma collisions or other options. These options are discussed 
in Chapter~\ref{detector_options}. The report concludes in Chapter~\ref{detector_conclusions} with
a look at the next steps before the ILC detector community.

\cleardoublepage
\chapter{Challenges for Detector Design and Technology}
\label{Challenges}

The physics of the ILC and the ILC machine environment present real challenges to ILC 
detector designers. ILC physics puts a premium on high resolution
jet mass reconstruction, which pushes calorimetry well beyond the current state of the art.
Particle Flow calorimetry promises the high performance needed, but demands that new detector
technologies and new reconstruction algorithms be developed.
Higgs studies need charged track momentum resolution well beyond what was achieved at LEP/SLD
and even substantially beyond that developed for LHC. High field magnets and high precision/low mass
trackers are under development to reach this goal. Flavor and quark charge tagging at the ILC, 
needed for precision measurements of Higgs branching fractions and quark asymmetries, demand
development of a new generation of vertex detectors. 

The ILC environment is 
benign by LHC standards, and so admits designs and technologies which have not been considered 
in the context of LHC detector R\&D. However, it still poses 
fundamental challenges for many of the detector subsystems. The vertex detector and the very forward
calorimetry, in particular, must contend with very high backgrounds primarily from the soft $e^+e^-$ pairs produced
by beamstrahlung when the beams collide. High occupancies require fast vertex readouts; fast readouts
require extra power; and both must be accommodated with very low mass detectors and supports. This
is a significant challenge. High radiation loads and bunch crossings every 300~ns complicate the design of the very forward
calorimeters. The need for sensitivity to single, tell-tale electrons in the haystack of pairs adds to
the challenge.

Table~\ref{tab-performance-requirement} summarizes several selected 
benchmark physics processes and fundamental measurements
that make particular demands on one subsystem or another, and set the requirements for detector performance.

\begin{table}[hbt]
\begin{center}
\caption{Sub-Detector Performance Needed for Key ILC Physics Measurements.
\label{tab-performance-requirement}}
\begin{scriptsize}
\begin{tabular}{|l@{\hspace{-8pt}}|l@{\hspace{-8pt}}|c@{\hspace{-8pt}}|c@{\hspace{-8pt}}|c|}
\hline
Physics Process & Measured Quantity 
 & \begin{tabular}{@{\hspace{-6pt}}c} Critical \\ System  \end{tabular}
 & \begin{tabular}{@{\hspace{-6pt}}c} Critical Detector \\ Characterstic \end{tabular}
 & \begin{tabular}{@{\hspace{-6pt}}c} Required \\ Performance  \end{tabular} \\ \hline

 \begin{tabular}{@{\hspace{-6pt}}l} $ZHH$ \\ $HZ \rightarrow q\bar{q}b\bar{b}$ 
  \\ $ZH \rightarrow ZWW^*$  \\ $\nu\bar{\nu} W^+W^-$ \end{tabular}
& \begin{tabular}{@{\hspace{-6pt}}l} Triple Higgs Coupling \\ Higgs Mass 
  \\  B($H\rightarrow WW^*$) \\  $\sigma({e^+e^- \rightarrow \nu\bar{\nu} W^+W^-})$ \end{tabular} 
& \begin{tabular}{@{\hspace{-6pt}}c} Tracker \\ and \\ Calorimeter \end{tabular} 
& \begin{tabular}{@{\hspace{-6pt}}c} Jet Energy \\ Resolution, \\ $\Delta E/E$ \end{tabular} 
& $3 $to$ 4 \%$ 
\\ \hline

\begin{tabular}{@{\hspace{-6pt}}l} $ZH \rightarrow \ell^+\ell^-X$ \\ $\mu^+ \mu^- (\gamma)$ 
  \\ $ZH + H \nu\nu \rightarrow \mu^+\mu^- X$ \end{tabular} 
& \begin{tabular}{@{\hspace{-6pt}}l} Higgs Recoil Mass \\ Luminosity Weighted $\rm E_{cm} $ 
  \\ B($H\rightarrow \mu^+ \mu^-$) \end{tabular} 
& Tracker\hspace*{24pt}
& \begin{tabular}{@{\hspace{-6pt}}c} Charged Particle \\ Momentum Res.,
  \\ $\Delta p_t/p_t^2$  \end{tabular} 
& $5\times 10^{-5}$ 
\\ \hline

 \begin{tabular}{@{\hspace{-6pt}}l} $HZ, H \rightarrow b\bar{b},c\bar{c},gg$ 
  \\ $b\bar{b}$ \end{tabular} 
& \begin{tabular}{@{\hspace{-6pt}}l} Higgs Branching Fractions \\ $b$ quark charge asymmetry \end{tabular} 
& \begin{tabular}{@{\hspace{-6pt}}c} Vertex \\ Detector \end{tabular} 
& \begin{tabular}{@{\hspace{-6pt}}c} Impact \\ Parameter, $\delta_b$ \end{tabular} 
& \begin{tabular}{@{\hspace{-6pt}}c} $ 5 \mu \mathrm{m} ~\oplus $ \\ 
    $10 \mu \mathrm{m} / p(\mathrm{GeV/c})\sin^{3/2}\theta$ \end{tabular} 
\\ \hline

 \begin{tabular}{@{\hspace{-6pt}}l} SUSY, eg. $\tilde \mu$ decay \end{tabular} 
& \begin{tabular}{@{\hspace{-6pt}}l} $\tilde \mu$ mass  \end{tabular} 
& \begin{tabular}{@{\hspace{-6pt}}l} Tracker,   \\ Calorimeter \end{tabular} 
& \begin{tabular}{@{\hspace{-6pt}}l} Momentum Res.,   \\ hermeticity \end{tabular} 
& \begin{tabular}{@{\hspace{-6pt}}l}  \ \end{tabular}

\\ \hline

\end{tabular}
\end{scriptsize}
\end{center}
\end{table}

\section{Jet Energy Resolution Requirements}

Many of the interesting
physics processes at the ILC appear in multi-jet final states, often accompanied by charged leptons
or missing energy. The reconstruction of the invariant mass of two or more jets will 
provide an essential tool for identifying and distinguishing $W$'s, $Z$'s, $H$'s, and top, and
discovering new states or decay modes. Ideally, the di-jet mass resolution should be comparable 
to the natural decay widths of the parent particles, around a few GeV or less. Improving the jet 
energy resolution to $\sigma_{E}/E  < 3 \sim 4\%$ ($30\%/\sqrt{E}$ for jet energies below approx. $100$~GeV),
which is about a factor of two better than that achieved at LEP, will provide such di-jet mass
resolution. But achieving such resolution represents a considerable technical
challenge for ILC detectors. 

It appears possible to reach such jet mass resolutions with the combination of an excellent, highly efficient and nearly 
hermetic tracking system and a calorimeter with very fine transverse and longitudinal segmentation. The energy from 
charged particles is first measured in the tracker, and then isolated in the calorimeter. By excluding the energy deposited by charged 
particles in the calorimeter, but including that from neutral hadrons and photons, a significant improvement in the overall jet resolution is possible.
This so-called "particle flow" concept is undergoing extensive
study in simulation, and has motivated the development of high granularity electromagnetic 
and hadronic calorimeters, and highly efficient tracking systems.


The jet energy resolution challenge has inspired another approach as well. A transversely segmented, 
dual readout 
calorimeter also promises excellent jet energy resolution, and its performance in an ILC detector is under study. 



Several fast simulation physics studies document the importance of achieving very high jet energy resolution
in ILC detectors, by plotting how the errors in key physics measurements depend on the resolution parameter $\alpha$,
given implicitly by the relation $\Delta E/E = \alpha/\sqrt{E}$. Reduced errors are 
equivalent to a luminosity bonus, and the added effective luminosity is often considerable.
Precision studies of the Higgs boson will be central part of the ILC physics program.
The measurement of the Higgs self coupling, via the reaction
$e^+e^- \rightarrow ZHH \rightarrow qqbbbb $, is extremely interesting, and probably unique to the
ILC. The low cross-section,
multi-jet signature, and high backgrounds make this measurement very challenging as well.
Excellent jet energy resolution might make the 
difference between being able to measure this reaction at the ILC, or not.
Measurements of the mass of the Higgs in the four jet channel, $e^+e^- \rightarrow ZH \rightarrow qqbb$, 
can utilize momentum-energy constraints and large statistics, and will benefit significantly from improved jet energy resolution.
Measuring the $WW$* branching fraction via the reaction $e^+e^- \rightarrow ZH \rightarrow {ZWW}$* 
$\rightarrow qqqql\nu$, is more challenging, since momentum and energy constraints have limited utility
in this multi-jet, missing-energy final state. Boosting the jet energy resolution significantly
reduces the error on this measurement as well. Finally, in a scenario where the Higgs does not 
appear at the ILC, and studies of $WW$ scattering move to the fore, improving the jet energy resolution will
improve the discrimination of $WW\nu\nu, WZ\nu e$, and $ZZee$ final states, thereby increase the effective 
integrated luminosity, and thus increase the reach of the ILC for new physics beyond its kinematic range.

The ability to distinguish $W$ and $Z$ decays cleanly will pay benefits in SUSY studies as well.
For example, to distinguish chargino from neutralino pair production, one must distinguish final states
with two $W$'s and missing energy, from those with two $Z$'s and missing energy. In addition, improved jet energy resolution
will sharpen the determination of the endpoints of the energy spectrum of the $W$ which results from chargino
decay, and so improve the measurement of the chargino mass.

\subsection {Higgs Self-Coupling Measurement}
The measurement of the cross-section for $e^+e^- \rightarrow ZHH$ will allow the determination of
the trilinear Higgs self-coupling, $\lambda_{hhh}$, which provides a determination of the
shape of the Higgs potential, independent of that inferred from the Higgs mass. This will
constitute a fundamental test of the Higgs mechanism. The Higgs self-coupling measurement at $\sqrt{s}=500$~GeV 
using the reaction $e^+e^- \rightarrow ZHH \rightarrow qqbbbb$ is a challenging measurement that 
requires excellent $W$, $Z$, and $H$ boson identification in a high track multiplicity environment with
6 jets. The total cross-section for $e^+e^- \rightarrow ZHH$, before factoring in $Z$ and $H$ 
branching ratios, is only 0.18 fb.  Major backgrounds include $e^+e^- \rightarrow t\bar{t} \rightarrow bbWW \rightarrow bbcscs$
and $e^+e^- \rightarrow ZZZ, ZZH \rightarrow qqbbbb$.

\begin{figure}[htb]
\begin{center}
\includegraphics[height=8cm]{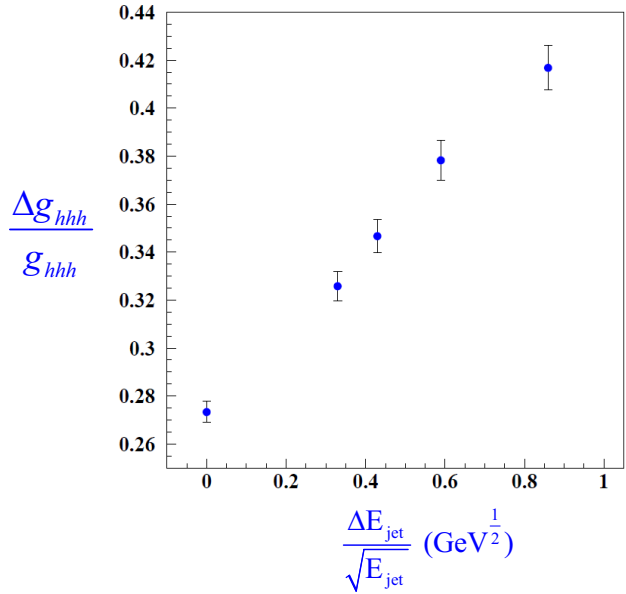}
\caption{Error in the Triple Higgs Coupling vs Jet Energy Resolution}
\label{fig-dellamhhhvsjetres}
\end{center}
\end{figure}

How the Higgs self coupling measurement depends on jet energy
resolution \cite{Castanier:2001sf,ref-Barklow:2006hhz} is shown in Figure~\ref{fig-dellamhhhvsjetres},
where an integrated luminosity of 2000 fb$^{-1}$ is assumed. Gluon radiation is fully suppressed in this study.
Neutrinos in the decay of bottom hadrons limit the Higgs
mass resolution, while neutrinos in the decay of charm and bottom degrade the Z boson mass 
resolution relative to what is obtained assuming Z decays to u,d,s, quarks only. Future
analyses which correct for the missing neutrino energies should improve the Higgs mass
resolution and reduce backgrounds, and so reduce the errors. 
The error in the coupling shows a significant reduction as the jet energy resolution
changes from $ \Delta E/ \sqrt E = 60 \%$ to $30 \%$, corresponding to an equivalent 
40\% luminosity gain and a marked reduction in the error of a critical quantity.

\subsection[Higgs mass in the four jet final state]
	{\bf Higgs Mass in the 4-jet Channel $\bf e^+e^- \rightarrow ZH \rightarrow \bf q \bar q b \bar b$}

The measurement of the Higgs mass through the recoil mass technique is limited
statistically by the relatively small branching ratio for Z boson decays to charged lepton
pairs. The much larger statistics associated with hadronic Z boson decay can be utilized by
measuring the Higgs mass in the 4-jet channel, $e^+e^- \rightarrow HZ \rightarrow qqbb$, so long as the 
Higgs mass is small enough that the branching ratio to b-quarks pairs is large enough. The dependence of
the accuracy of the mass measurement on the jet energy resolution has been explored \cite{ref-Barklow:2005lcws}
assuming a Standard Model Higgs with a mass of 120~GeV, as favoured by current electroweak precision measurements, a branching ratio to b-quark pairs
of 68\%, $\sqrt{s}$=350 GeV, and an integrated luminosity of 500 fb$^{-1}$. 

\begin{figure}[htb]
\begin{center}
\includegraphics[width=13.0cm,bbllx=0,bblly=100,bburx=780,bbury=550]{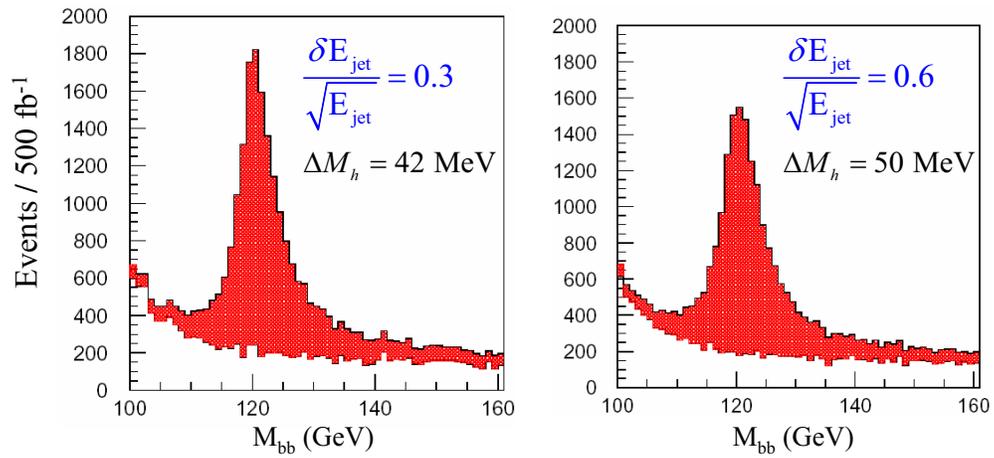}
\caption[Higgs di-jet invariant mass for different jet energy resolutions]{Reconstructed Higgs di-jet invariant mass for different jet energy resolutions. The analysis has been performed for a center of mass energy of 350~GeV and a total integrated luminosity of 500 fb$^{-1}$.}
\label{fig-delmhsvsjetres}
\end{center}
\end{figure}

The analysis selects hadronic final states with large visible energy, and forces the charged
and neutral tracks into a 4-jet topology. If one jet-pair has a mass consistent with
the Z boson and the other is consistent with having two b-quark jets, the b quark jet-pair
is considered the Higgs candidate. By imposing beam energy-momentum constraints, the
resolution of the Higgs mass and the signal/ background ratio in the signal region are improved significantly. The result is given in Figure~\ref{fig-delmhsvsjetres}
which shows the invariant mass of the two b-quark jets for different jet energy resolutions. The error
in the Higgs mass improves by a factor 1.2 in going from $ \Delta E/ \sqrt{E} = 60 \%$ to $30 \%$,
corresponding again to an equivalent 40 \% luminosity gain.

\subsection[Branching fraction for $\bf H\rightarrow WW$*]{\bf Branching fraction for $\bf H\rightarrow WW$*}

\begin{figure}[htb]
\begin{center}
\includegraphics[height=8cm,width=7cm]{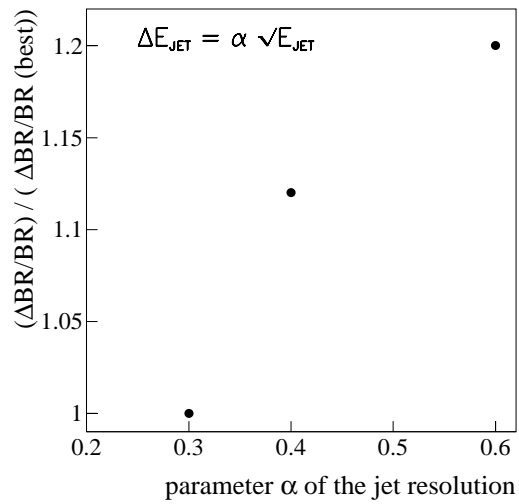}
\caption[Error in H$\rightarrow WW^*$ vs jet energy resolution.]{Relative error in the Higgs branching ratio to $WW$* vs jet energy resolution}
\label{fig-delhbrvsjetres}
\end{center}
\end{figure}

One of the principal motivations for building a detector with excellent jet energy
resolution is the need to distinguish hadronically decaying $W$ bosons from $Z$ bosons in 
events where beam energy-momentum constraints either cannot be imposed or have
limited utility, such as events with 6 or 8 fermions in the final state. A test of this kind of
$W$/$Z$ separation is provided by the measurement of the $H \rightarrow  WW$* branching ratio in 
the reaction $e^+e^- \rightarrow ZH \rightarrow ZWW$*$\rightarrow qqqql\nu$. The
dependence of the $H \rightarrow WW$* branching ratio error on jet 
energy resolution \cite{ref-Brient:2001ww} is summarized in Figure~\ref{fig-delhbrvsjetres}. 
There is a factor of 1.2 improvement
in the branching fraction error in going from $ \Delta E/ \sqrt E = 60 \%$ to $30 \%$,
corresponding again to an equivalent 40\% luminosity gain.

\subsection[Cross Section for $e^+e^- \rightarrow \nu\nu WW$]{\bf Cross Section for $\bf e^+e^- \rightarrow \bf\nu\bar\nu W^+W^-$}  

This process was originally considered in scenarios with no elementary Higgs, as a way of
probing the strong WW scattering that could moderate the resulting divergences in the WW scattering amplitude.
It is a fundamental electroweak process, and by virtue of the missing neutrinos in the
final state, a suitable benchmark process for distinguishing W's and Z's. Study \cite{ref-Chierici:2001ar} has shown,
like the studies above, the improvement in going from $ \Delta E/ \sqrt E = 60 \%$ to $30 \%$, 
is equivalent to an increase of $30\%$ to $40\%$ in luminosity.


\section{Additional Challenges for the Calorimetry}
As discussed above, the concept of particle flow requires that a highly granular calorimeter be developed with 
good segmentation both transversely and longitudinally. Within this concept, high granularity
becomes much more important than very good energy resolution. Cell sizes around $1 \times 1~cm^2$ or less 
seem appropriate for the electromagnetic and possibly even the hadronic sections, while energy resolutions of 
$\approx 15 \%$ for the electromagnetic part and $>40\%$ for the hadronic part are considered good enough.
This is certainly the principal challenge for Calorimetry. 

ILC physics demands more than good jet energy resolution of the calorimeter. Searches for SUSY will utilize missing
energy signatures, requiring both good resolution in missing energy, and the hermeticity to ensure that
energy losses down the beamline are minimal. Lepton identification is very important in ILC physics.
Efficient electron and muon ID and accurate momentum measurements over the 
largest possible solid angle will be required for detailed studies of leptons from $W$ and $Z$ decays. 
Identifying electrons and muons within jets is of course 
more difficult, but is important for flagging the presence of neutrinos from heavy quark decays 
and identifying jet flavor and quark charge. 
In some models the precise reconstruction of photons which are not pointing back at the 
origin is crucial, stressing the importance of the calorimeter's spatial and angular resolution.

The identification and measurement of tau lepton decays is a particularly difficult and important case, 
critical in analyzing tau polarization states, and it will require differentiating tau decays 
to $\pi$, $\rho$, $A_1$, and $\rho'$. This may impose the most severe requirements on segmentation
in the electromagnetic calorimeter.

Overall, lepton ID requires a lot of the calorimetry: high granularity, 
excellent hermiticity, sensitivity to minimum ionizing particles, compact electromagnetic shower development,
and good electromagnetic energy resolution.

\section{Challenges for the Tracking}

Tracking at the ILC poses multiple challenges. The momentum resolution specification is well beyond the current state of the art. 
Full solid angle coverage for tracks with energies ranging from the beam energy to very low momenta 
is required for particle flow calorimetry and missing energy measurements.
The pattern recognition algorithm must be robust and highly efficient in the presence of backgrounds.  
All the while the tracker must be built with minimal
material to preserve lepton id and high performance calorimetry. Here we consider the impact of the tracker
momentum resolution on key physics measurements.   

\subsection{Higgs Mass from Dilepton Recoil}
Studies of the Higgs Boson are expected to take center stage at the ILC. The production of the Higgs through
"Higgs-strahlung" in association with a $Z$, will allow a precision 
Higgs mass determination, precision studies of the Higgs branching fractions, measurement of the production 
cross section and accompanying tests of SM couplings, and searches 
for invisible Higgs decays.  When the associated $Z$ boson decays leptonically, it is possible to reconstruct 
the mass of the object recoiling against it with high precision, and without any assumptions on the nature 
of the recoiling particle or its decays. The 
resolution in the recoil mass, which translates into how sharply the Higgs signal rises above the $Z Z$ background, 
depends on the accuracy with which the beam energy can be measured, 
the initial beam energy spread, which at ILC is about $0.1\%$, and the precision with which the lepton 
momenta are measured.  

\begin{figure}[htb]
\begin{center}
\includegraphics[height=10cm,bbllx=12mm,bblly=0mm,bburx=260mm,bbury=231mm]{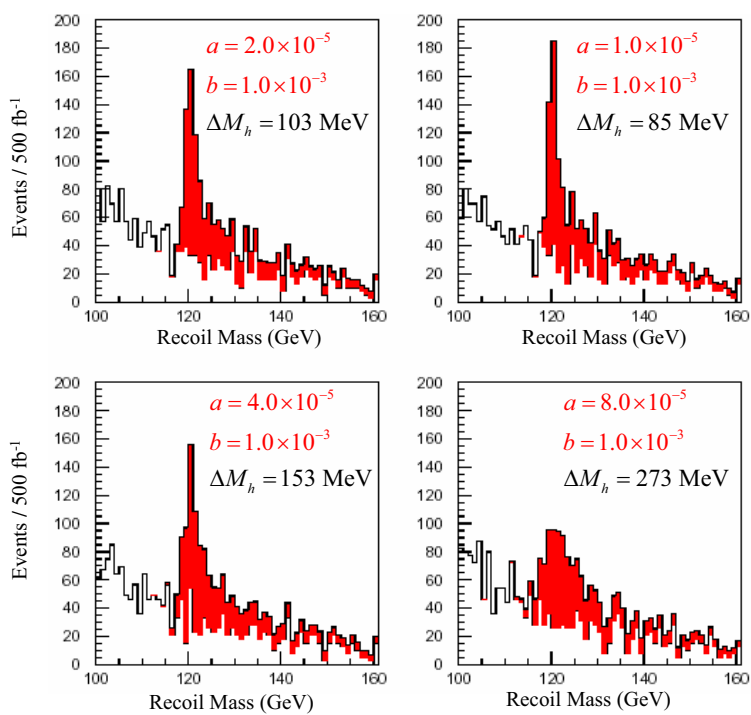} 
\caption[Higgs boson recoil mass spectra]{Higgs recoil mass spectra for several values of parameters characterizing the 
tracker momentum resolution, which is parameterized as $\delta p_t/p_t^2 = a\oplus   b/(p_t\sin\theta)$.}
\label{fig-delmhvsmrecoil}
\end{center}
\end{figure}

It is interesting to see how the precision of the mass measurement depends on the momentum resolution of the 
tracker \cite{ref-schreibermomentum, ref-Barklow:2005lcws, ref-Yang:2005hk}.  
Figure~\ref{fig-delmhvsmrecoil} shows the recoil mass distribution 
opposite the Z for four different values of tracker momentum resolution, characterized by the parameters $a$ and $b$, 
assuming the Higgs mass is 120~GeV,  $\sqrt{s} = 350$~GeV, and the integrated luminosity is 500~fb$^{-1}$.  
Here the momentum resolution is written  $\delta p_t/p_t^2 = a\oplus   b/(p_t\sin\theta)$ .   For example, we find 
that the Higgs mass can be determined with a precision of 150 MeV for Z decays to muon pairs assuming $a=4 \times 10^{-5}$ 
and  $b=1 \times 10^{-3}$. Accuracy in the mass measurement improves significantly as the tracker momentum 
resolution improves.

%

\subsection{Slepton Mass Measurement from Lepton Energy Spectrum Endpoints}

The ILC offers the possibility of determining the masses of the sleptons to high precision,
if they are kinematically accessible. Studies of the production of smuon and 
selectron pairs, and their subsequent decays to charged leptons and neutralinos \cite{ref-martynsleptons}, provide 
another example where the measurement sensitivity can depend on the tracker's momentum resolution.

In a recent study the impact of the momentum resolution \cite{ref-Barklow:2005lcws}on measuring the mass of a smuon
in the co-annihilation region is studied. The smuon mass is taken to be 224~GeV, and the neutralino mass, 212~GeV. 
The study assumed running at  $\sqrt{s}= 500$~GeV with an integrated luminosity of 500~fb$^{-1}$.
The smuon mass is determined by looking at the endpoints of the momentum spectrum of the decay
muons. The error on the mass is studied as a function of the parameters a and b which
characterize the momentum resolution of the tracker, as above. 

The accuracy of the smuon mass, in fits 
where the neutralino mass is assumed to be held fixed at some predetermined value, is 
independent of variations of $a$ in the range $1.0 \times 10^{-5}$ to $8.0\times 10^{-5}$, 
and independent of those in $b$ in the range $0.5 \times 10^{-3}$ to $4.0 \times  10^{-3}$. 
This is not surprising, since the muon momentum spectrum is relatively soft and the 
tracker's momentum resolution in this region is especially good. The beam energy spread and
radiative tail are reflected in the low and high ends of the muon energy spectrum, respectively, 
and dominate the effects of tracker momentum resolution in the observed spectral shape.

The tracker momentum resolution plays a much more important role if the slepton and neutralino
are not nearly degenerate in mass. Two studies have been performed assuming the SPS1A SUSY parameter space point
with slepton and neutralino masses of 143~GeV and 96~GeV, respectively.
In the first \cite{ref-Barklow:2006smu}, measuring the masses of both the smuon
and the neutralino is considered. The study 
assumed running at  $\sqrt{s}= 500$~GeV, and an integrated luminosity of 500~fb$^{-1}$.
The two masses are simultaneously determined by looking at the endpoints of the momentum spectrum of the
muon produced in the smuon decay, and are studied as a function of the parameter $a$, as above. 
The accuracy of the smuon (neutralino) mass varies from 98 (86) MeV to 115 (97) MeV as
the parameter $a$ is varied from 0 to $2.0\times 10^{-5}$, and degrades  to 139 (113) MeV when the
parameter $a$ is increased to $8.0\times 10^{-5}$. The improvement in the 
smuon (neutralino) resolution as the parameter $a$ is decreased from $8.0\times 10^{-5}$ to  $2.0\times 10^{-5}$
is equivalent to a ~45\% (~35\%) gain in luminosity. In the other study \cite{ref-Gerbode:2005ke}, 
the authors study selectron pair-production at  $\sqrt{s} = 1000$~GeV. 
The higher energy leads predictably to a very much higher lepton energy endpoint, 
225~GeV in contrast to the 125~GeV above, and consequently an even greater dependence on the momentum resolution of 
the tracker. Very good momentum resolution, especially in the forward direction, will allow even high 
energy data sets to be usefulin measuring masses.

\subsection{${\bf \rm E_{\rm cm}}$ Determination from $\bf e^+e^-  \to \mu^+ \mu^- (\gamma)$}

\begin{figure}[htb]
\begin{center}
\includegraphics[height=8cm]{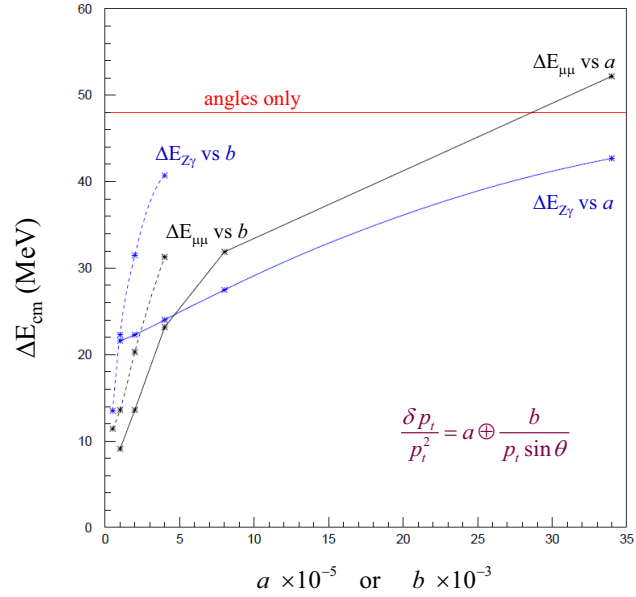} 
\caption[Error in $E_{\rm cm}$ as a function of momentum resolution]{Error in ${\rm E_{\rm cm}}$ as a function of the 
parameters describing the tracker momentum resolution. Results from simulated measurements of lepton angles, muon pair 
production, and radiative returns to the Z are shown.}
\label{fig-ecm_from_mumu}
\end{center}
\end{figure}

Accurately determining the center of mass energy at the ILC is prerequisite for many physics studies, 
and major efforts are being devoted to measuring the beam energy before and after the interaction point. 
Because the ${\rm E_{\rm cm}}$ measured upstream and downstream of the interaction point can differ 
from the luminosity-weighted ${\rm E_{\rm cm}}$ by as much as 250 ppm, it is important to be able to 
compare such measurements with a direct detector measurement of the center-of-mass energy based on physics events.  
The latter measurement directly measures the luminosity-weighted center-of-mass energy.  As shown 
in \cite{ref-Barklow:2005lcws}, the excellent momentum resolution of the tracker is particularly advantageous in this measurement,
which can be done by studying muon pair production and radiative returns to the Z, where the Z subsequently 
decays to muon pairs. ${\rm E_{\rm cm}}$ measurements at LEP using $e^+e^-  \to \mu^+ \mu^- \gamma$ relied 
solely on lepton angle measurements because little additional information could be gleaned from a direct 
muon momentum measurement. The resolution was inadequate. However, with the trackers being considered for 
ILC detectors, the momentum measurement can significantly improve the ${\rm E_{\rm cm}}$ measurement over 
what can be achieved with angles alone.  Figure~\ref{fig-ecm_from_mumu} shows the accuracy with which 
${\rm E_{\rm cm}}$ can be determined with a data sample of 100 fb$^{-1}$ by utilizing radiative returns($Z\gamma$)
or full energy muon pairs ($\mu\mu$) as a function of the parameters which describe the momentum resolution. 
For variations of the curvature error parameter, a, the multiple scattering parameter b is set to $ 1.0\times10^{-3}$; for
variations in b, a is set to $2.0\times10^{-5}$. For 
comparison, the accuracy obtained by using an angles-only measurement is also shown. For full energy mu pair 
production there is a strong dependence on the curvature error, and for both methods there is a strong dependence
on the multiple scattering term. In any 
case, excellent tracker momentum resolution will allow the determination of ${\rm E_{\rm cm}}$ to about 20 MeV. 
At the same time this analysis makes a strong case for excellent forward tracking, with a minimal material
budget to minimize the multiple scattering term.

\subsection{$\bf {\rm BR}(H\to \mu^+ \mu^-)$ Measurement at $\bf {\rm E_{\rm cm}=1\ TeV}$}

At high energies close to 1~TeV even rare Higgs decays modes like $H \to \mu^+ \mu^-$ become accessible, thanks
to growing t channel cross sections, increased machine luminosity, polarization enhancements, and improved signal
to background. At an 
energy of $\sqrt{s}=800$~GeV the 
${\rm BR}(H\to \mu^+ \mu^-)$ could be measured with a relative
accuracy of 30\% assuming a 120~GeV Higgs mass, 1000~fb$^{-1}$ luminosity and no initial state polarization \cite{ref-Battaglia:2001vf}.  
The signal is visible in the di-muon invariant mass distribution as a Higgs resonance above the background
from $e^+e^- \rightarrow W^+W^- \rightarrow \mu^+\mu^-\nu_{\mu}\bar{\nu}_{\mu}$. Improving the tracker
momentum resolution sharpens the peak, improving the signal to background ratio, and lowering
the error in the branching fraction. 
The branching fraction error can be reduced $15\%$ if the momentum resolution term a is reduced from $4\times 10^{-5}$ to
$2\times 10^{-5}$



\section{Challenges for Vertexing}
 
\begin{figure}
\begin{center}
\includegraphics[height=8.0cm]{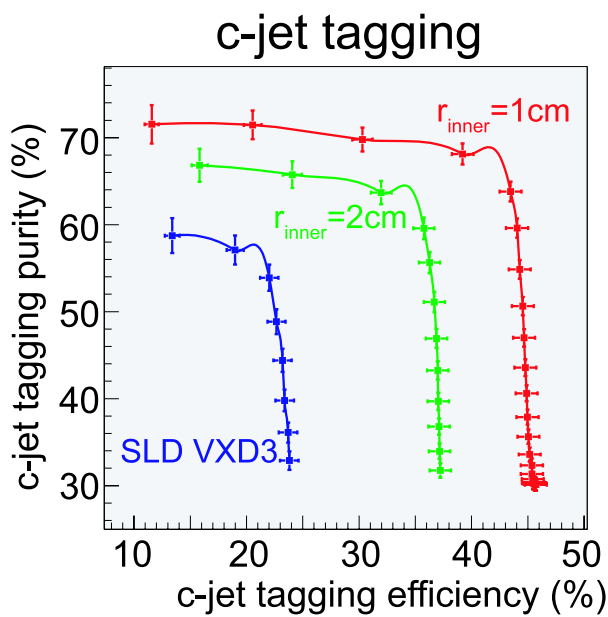} 
\caption[Charm tagging efficiency vs purity for different beam pipe radii]{Charm Tagging Efficiency and Purity for Vertex Detectors with inner radius of 1~cm, 2~cm, and 2.6~cm}
\label{fig-Abe:2001ctag}
\end{center}
\end{figure}

Ideally, vertex detection allows the full vertex topology of heavy particle production and decay to be determined. 
Vertex detection is critical at the ILC. Identifying heavy particle decay vertices, and 
measuring the invariant mass of their charged decay products, tags their flavor.
Heavy flavor identification is the key to measuring the Higgs branching fractions with high precision. 
Charm identification, in conjunction with the 
observation of missing energy, could provide startling evidence for new physics, e.g. when  
stop decays to charm and a neutralino. Maximizing the efficiency and purity of heavy flavor tags pushes
vertex detector efficiency, angular coverage, and impact parameter resolution beyond the 
current state of the art. Improving the point resolution per measurement, 
minimizing the beam pipe radius, and reducing the thickness of the detector sensors and supports can result 
in significant enhancements to the flavor tagging efficiency \cite{ref-Abe:2001pe}, as shown for charm quark tagging in 
Figure~\ref{fig-Abe:2001ctag}.

Measuring the net charge associated with secondary and tertiary
heavy quark decays can provide a determination of quark charge, which makes it possible to measure asymmetries and polarizations.  
For example, $b\bar{b}$ forward backward asymmetries are most sensitively probed with quark charge measurements. 
If these asymmetries are anomalous, as measurements at the Z have suggested, sensitive quark charge measurements
could measure the effect, and even reveal evidence for 
extra dimensions or other new physics signatures. Quark flavor and charge determinations also permit analyzing top quark polarizations,
and thereby test for anomalous couplings in $t \bar{t}$ production and decay and access 
SUSY parameters if stop or sbottom decay to top.

Vertex detection also plays an important role in tracking generally. Multi-layer vertex detectors provide 
efficient stand-alone pattern recognition and even momentum measurement, which may well be essential in 
measuring soft tracks. Since pixel detectors provide excellent pattern recognition capability, vertex detectors 
may also provide the seeds for recognizing tracks in forward and central trackers.
  

The ILC environment also poses significant challenges to vertex detector design.  While ILC requirements for rate 
capability and radiation load are dwarfed in comparison to those for the LHC, 
the production of prodigious numbers of $e^+e^-$ pairs, the inevitable consequence of colliding nanometer 
sized beams at high luminosity, results in severe backgrounds in the ILC vertex detector. These pairs produce 
of order 100 hits/mm$^2$/train in the innermost layer of the vertex detector, more than an order of magnitude 
more than pattern reconstruction can comfortably handle. Consequently, it is essential to time-slice the bunch train into 
more manageable pieces, and integrate over $<$ 150 bunch crossings as opposed to the full 3000. 
To do so requires a readout technology much faster than that of traditional pixellated vertex detectors.
This fact has led to the active development of many new technologies. The simultaneous challenges
of rapid readout, constrained power budgets, transparency, and high resolution, have made this a challenging undertaking.
However, the ILC data rates, which are significantly lower than at LHC, admit designs which use much less power 
per channel, and hence can 
be thinner and more highly pixellated than their LHC counterparts, with consequently better resolution. The low ILC radiation load admits 
a much wider selection of technologies than are possible at the LHC. 

In the following the impact of vertex detector performance on several key physics measurements is discussed.

\subsection{Measuring the Higgs Branching Fractions}

The measurement of the Higgs Branching Fractions, and their dependence on vertex detector resolution, is by 
far the best studied of the suite of possible vertex physics topics. If the Higgs has relatively low mass,
as the precision electroweak fits prefer, precision measurements of the branching fractions will establish
how the Higgs couplings depend on mass and will differentiate Standard Model behavior from that of other
Higgs models. While measuring the Higgs Branching Fractions is not the measurement most
demanding of vertex detector performance, it is certainly one benchmark worth noting.
Several groups have undertaken the study of the process \cite {ref-Battaglia:2000jb, ref-Battaglia:2002av, 
ref-Aguilar-Saavedra:2001rg, ref-Abe:2001np, ref-Kuhl:2004ri}. There is reasonable agreement as 
to how well the branching fractions can be measured, with the fractional errors in BR($H \rightarrow b\bar{b}$),
BR($H \rightarrow c\bar{c}$), and BR($H \rightarrow gg$) around 1\%, 12\%, and 8\% respectively for a 120~GeV Higgs,
at $\sqrt{s} = 350$~GeV, and an integrated luminosity of 500~fb$^{-1}$~\cite {ref-Battaglia2000,ref-Barklow:2004th}. 
Other final states, like $WW$, 
$ZZ$, or $tt$, are measured with intermediate precisions. By including running at 1~TeV for 1~ab$^{-1}$, most of 
the relative branching fractions are determined to the level of 2-5\% \cite {ref-Barklow:2004th}.

Several authors \cite{ref-Brau:2000dq,ref-Abe:2002qs,ref-Ciborowski:2005sq}
have studied the impact of the choice of detector parameters on the measured 
accuracies of the Branching Fractions.  They all evaluated the impact of improving the spatial resolution 
and varying the radius of the innermost layer (and hence the beam pipe radius). The studies are in rough agreement, 
finding that halving the inner radius from about 2.4~cm to 1.2~cm reduces the errors in the 
Charm Branching Fraction by of order 10\%. Similar, but smaller effects, 
are seen when the resolution is halved, or the material budget significantly reduced. In sum, modest but not insignificant
reductions in the Higgs Branching Fraction errors are seen as critical detector parameters, especially the inner radius,
are optimized.

\subsection{Measuring Quark Charge}
\begin{figure}
\begin{center}
\includegraphics[width=10.0cm]{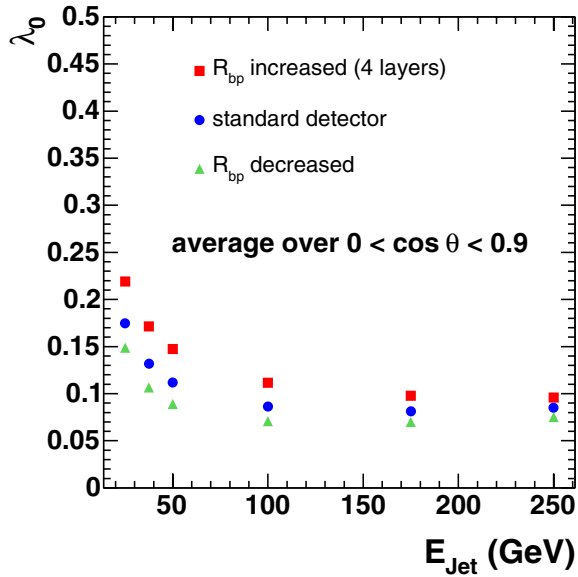} 
\caption[Probability to misreconstruct the neutral B vertex as charged]{Probability of misreconstructing a neutral B vertex as charged, vs B jet energy, for beam pipe radii
of 0.8 cm, 1.4 cm, and 2.5 cm}
\label{fig-qerrvsjetenergy}
\end{center}
\end{figure}

     Determination of the quark charge may be more demanding of vertex detector performance than the Higgs 
Branching Fraction measurements discussed above, because it demands correct association of even low momentum 
tracks to the correct decay vertex. These tracks of course suffer the most from multiple Coulomb scattering in the beam 
pipe and inner detector layers. A preliminary study~\cite{ref-Hillert:2005ss}
of how the quark charge 
determination depends on the radius of the innermost vertex layer has indicated that the probability 
of misreconstructing neutral vertices as charged decreases rather significantly as the beam pipe radius is 
reduced, as shown in Figure~\ref{fig-qerrvsjetenergy}. Further study must evaluate the full impact on 
measurements of  $b\bar{b}$ asymmetries at the ILC, but a significant advantage for detectors with the smallest 
inner radii seems an inescapable conclusion.

\section{Challenges for Very Forward Calorimeters}

\begin{figure}
\begin{center}
\includegraphics[height=8cm]{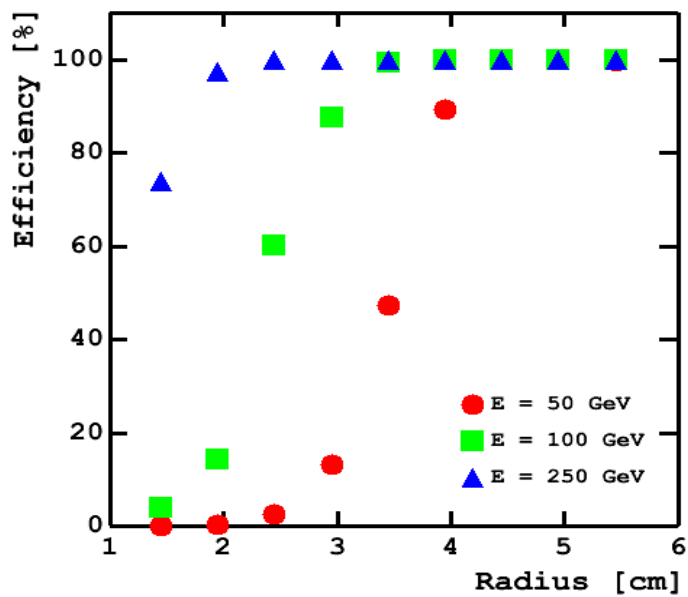} 
\caption[Electron detection efficiency in the very forward calorimeter]{Efficiency to detect an electron of 50, 100, or 250~GeV energy
in the BeamCal as a function the impact distance from the beamline at center of mass energy of 500~GeV. The
beamstrahlung background was simulated using Guinea Pig. }
\label{fig-beamcalvetoeff}
\end{center}
\end{figure}

The very forward region of the ILC detector will be instrumented with two electromagnetic calorimeters, the Lumical and Beamcal.
The Lumical extends
the hermeticity of the detector's calorimetry to very small polar angles, provides a fast
luminosity measurement. 
The Beamcal, which is located even below the Lumical, in front of the final 
focussing magnets, primarily is to be used to monitor the beam parameters at the interaction point.
It 
must survive in the very high radiation environment generated by e$^+$e$^-$ pairs and beamstrahlung and
be independently read out each bunch crossing, to provide bunch-by-bunch machine diagnostics.
It must be capable of vetoing electrons at small polar angles with high efficiency, in order to suppress 
backgrounds when searching for new particles whose signatures involve large missing 
energy and momentum in the final state. This is e.g. the case in supersymmetric models where the mass difference
between the primary produced sleptons and the LSP is small.
Backgrounds arise from two-photon events and radiative Bhabha events, which are
characterized by electrons or positrons scattered at
small angles.  Beamstrahlung leads to the production of a very large number of relatively low
energy e$^+$e$^-$ pairs hitting the BeamCal, amounting to several TeV of energy deposited per bunch crossing.
To identify a single high energy electron on top of this
broadly distributed background, the BeamCal must be dense and finely segmented, 
both transversely and longitudinally \cite{ref-TESLA:2001,ref-Lohrmann:2005}. The efficiency for 
identifying electrons is
shown in Figure~\ref{fig-beamcalvetoeff} for a diamond-tungsten sampling calorimeter
with a Moliere radius and a transverse segmentation of about 1 cm.

Measurements of the energy deposited by beamstrahlung pairs in the BeamCal, together with measurement
of the beamstrahlung photons in another, downstream calorimeter, the GamCal, allow a bunch-by-bunch luminosity measurement
and  intra-train luminosity optimization, by providing information to the beam delivery 
feedback system. In addition, beam parameters can be determined by analyzing the shapes 
of the observed energy depositions. Since both calorimeters must be read out after each bunch-crossing, the
development of a fast readout electronics with adequate resolution is necessary. The high 
occupancy requires a high bandwidth data transmission and processing system.  The
absorbed radiation dose is up to 10 MGy per year for the sensors near the beampipe
and changes rapidly, depending on the position, the beam parameters, and the magnetic
field in the detector. Novel sensors have to be developed whose
response is independent of the absorbed dose.

\section{Other Detector Challenges}

There are many other detector subsystems which challenge experimental ingenuity and require R\&D before final 
detector designs can be put in place. Systems at the machine-detector interface, 
like polarimeters
and beam energy spectrometers, need continued development to reach the new levels of
systematic understanding required for precision measurements at the ILC.  Beam energy must 
be measured to the 100 ppm level to achieve the desired accuracy in threshold energy scans and
mass measurements.  Polarization must be measured to the 1000 ppm level for precision 
measurements at the Z, and the 2000 ppm level for measurements at higher energies.  
These requirements are beyond today's state of the art.  

The detectors planned for the ILC all use
high field superconducting solenoids, with designs based loosely on the recent success
of the 4 Tesla CMS coil.  Additional research and development is needed to refine the
designs, develop new conductors, and accommodate the requirements of field uniformity imposed
by the tracking. Preserving emittance for the beams as they pass through the solenoidal field
at finite crossing angles, and minimizing pair disruption for the exiting beams, requires
the addition of a small dipole component to the solenoidal field, called the Dipole in Detector
(or DID, or anti-DID). The DID needs design, as does the compensation solenoid. 

More traditional systems, like that used for muon identification, must be
adapted to the particular problems posed by the ILC, e.g., handling the flux of
background muons produced in upstream beam collimation, and providing
tail catching for hadron showers which originate in the HCAL and Solenoid. Providing robust,
reliable, and economical muon tracking coverage over very large areas remains
a significant challenge. 

Particle Identification, other than that for electrons and muons, has received
modest attention in the context of ILC detectors. If there is appreciable running
at the $Z$, $\pi/K/p$ identification will be important for flavor tagging and 
charm and B reconstructions. With the exception of the dE/dx capabilities of TPCs,
and some discussion of time-of-flight measurements, specialized PID detectors
have been largely ignored. The challenge here is to understand
the physics motivation for PID in the high energy ILC environment.

Detector system integration presents significant challenges, and demands serious engineering and design.
ILC detectors must support the final quadrupoles and the fragile beampipe with its
massive masking, stably, adjustably, and without vibration. The detectors themselves may be required to move
on and off beamline rapidly and reproducibly, and maintain or monitor inter-system alignments
at the few micron level. The various components of the detector must be integrated in a
way that allows assembly, access, repair, calibration, and alignment, and that doesn't compromise
performance or solid angle coverage. These very real challenges lie ahead.

\section{Conclusions}
To fully exploit the physics opportunities presented at the ILC requires a detector
with capabilities far beyond the detectors at LEP or LHC. The ILC machine environment, although not without its
own challenges, admits detector designs of much higher performance than the 
detectors planned for the LHC, with much better jet energy resolution, tracker momentum resolution, 
and vertex detector impact parameter resolution.  This increased performance
is needed at the ILC, to make precision measurements of 
masses and branching fractions, distinguish final state quanta,
extract low cross-section signals, see new phenomena, and exploit the 
delivered luminosity as well as possible. Detector research and
development is needed to realize these advances. Activities have 
been ongoing for a couple of years, and big advances have 
already been achieved; it must be
expanded and accelerated in order to prepare believable ILC detector designs
and costs in time for the ILC machine.

\cleardoublepage
\chapter{Detector Concepts}
\label{detector_concepts}
 
Four ILC detector concepts have emerged in the last few years.
All four designs strive to provide highly efficient tracking, charged particle momentum resolution 
$\delta p/p^{2} \approx 5 \times 10^{-5}$, 
dijet mass resolution at the 3\% level, excellent heavy quark identification capability, and full and
hermetic solid angle coverage. Three of the concepts use traditional solenoidal magnet designs and 
adopt the particle flow calorimetry strategy, where highly segmented electromagnetic and hadronic calorimeters 
allow separation of the energy deposited by charged tracks, photons, and neutral hadrons. 
The technical realizations of these three concepts differ, however, and utilize complementary subdetector technologies. 
High granularity and excellent spatial shower reconstruction are at the center of 
attention for these concepts. 
The fourth concept stresses excellent energy resolution, relies 
less on spatial resolution, and utilizes a novel
dual readout scheme to allow efficient software compensation. 


In this chapter brief reviews of the rationales and main characteristics of each of the different concepts
are given. More details may be found in the respective Detector Outline Documents~\cite{ref-GLD, ref-LDC, ref-SiD, ref-4th}. 
Few technical details are discussed in this chapter; for these the reader is referred to chapter
\ref{detector_technologies} on subdetector technologies.

The need to extract the maximum information from ILC events dictates a few design characteristics 
which are shared by all the detector concepts. All four concepts utilize similar pixellated 
vertex detectors, which provide high precision vertex reconstruction and  
serve as powerful tracking detectors in their own right.  All four concepts have sophisticated tracking systems,
which have been optimized for high track reconstruction efficiency and excellent momentum resolution. Since much 
of the physics relies on high quality calorimetry, all the concepts have chosen to arrange the calorimeters inside the
coil. All the concepts have also incorporated high field solenoids, ranging between 3 and 5 Tesla, to insure
excellent momentum resolution and help disperse charged energy in the calorimeters, and all have
relied on the recent success of the CMS solenoid to give their magnet designs credibility.






\section{The SiD Concept}

     The SiD concept utilizes silicon tracking and a silicon-tungsten sampling 
calorimeter, complemented by a powerful pixel vertex detector, finely segmented hadronic 
calorimeter, and a muon system. Silicon detectors are fast and robust, and they can be finely segmented. 
Since silicon sensors are fast, most SiD 
systems will only record backgrounds from the single bunch crossing accompanying a physics event,
maximizing event cleanliness. Since silicon detectors are tolerant of background 
mishaps from the machine, the vertex detector, the tracker, and the calorimeter 
can all absorb significant radiation bursts without `tripping' or sustaining damage,
thereby maximizing running efficiency.

     The SiD concept recognizes the fundamental importance of calorimetry for ILC physics, and adopts
a strategy based on Particle Flow Calorimetry. This leads naturally to the 
choice of a highly pixellated silicon-tungsten electromagnetic calorimeter, and a multi-layered, highly segmented hadron 
calorimeter. Achieving excellent jet energy resolution requires both the calorimeters to be 
located within the solenoid.  Since a high granularity silicon-tungsten calorimeter is 
expensive, as is a large solenoid, cost considerations push the design to be as compact as possible, 
with the minimum possible radius and length.
The use of a high field solenoid (5 Tesla) compensates for the reduced radius by improving
the separation of charged and neutral particles in the calorimeters. Given the high field, an 
all-silicon tracker, with its high intrinsic resolution, can provide superb charged particle 
momentum resolution, despite the limited real estate.  The high field also constrains e$^+$e$^-$-pair
backgrounds to minimal radius, and so allows a beam-pipe of minimal radius for high performance vertex detection. 

\begin{figure}
	\centering
		\includegraphics[height=10cm,bbllx=0,bblly=0,bburx=174mm,bbury=165mm]{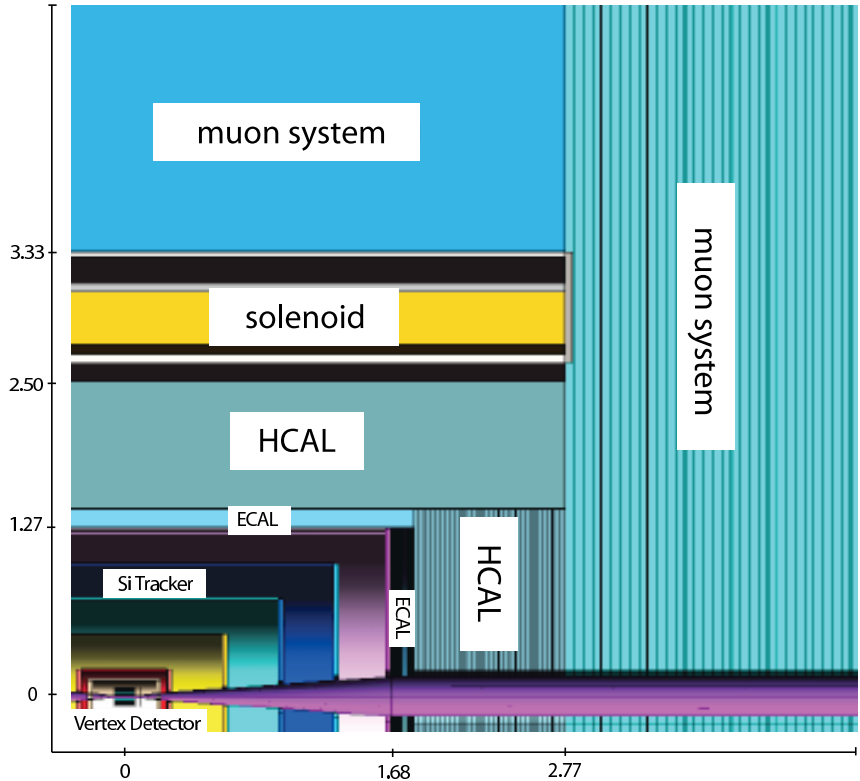}
	\caption[Illustration of a quadrant of SiD]{Illustration of a quadrant of SiD. The scale shown is in meters.}
	\label{fig-SiD-layout}
\end{figure}

    The SiD Starting Point is illustrated in Figure~\ref{fig-SiD-layout}. The overall SiD design, its performance, and
technology options are described in more detail in Ref.~\cite{ref-SiD}.

     The SiD baseline detector has the following components, moving from small to large radii:

\begin{itemize}
\item The vertex tracker has five  barrel layers of pixel detectors augmented with four endcap layers
on each side, beginning at a radius of 1.4~cm and extending to 6.1~cm. The endcap design
insures excellent pattern recognition capability and impact parameter resolution over the full solid angle.

\item The main tracking system consists of five layers of silicon microstrip sensors, which tile
low mass carbon fiber/rohrcell cylinders and endcap planes. The baseline design calls for
axial-only measurements in the barrels, and stereo measurements on each endcap. Individual layers are 
only 0.8 \% $X_{0}$ thick, including sensors, readout ASICs, and cables.

\item The electromagnetic calorimeter (ECAL) begins at a radius of 1.27~m and consists of 30 alternating layers
of silicon pixel sensors and tungsten absorber. The pixel area is about 14~mm$^{2}$; roughly 1000 pixels
on each sensor are read out  with an ASIC chip. Care is taken to minimize the gap between absorbers
in order to preserve a small Moliere radius. The device is 29 $X_{0}$ deep.

\item The hadronic calorimeter (HCAL) follows the ECAL, beginning at a radius of 1.41 m. The SiD baseline calls
for roughly 40 layers of Fe absorber, with highly pixellated (1 cm$^{2}$) RPCs for readout between absorber layers. Scintillator,
GEM, and Micromegas detectors are also being considered. The device is 4 interaction lengths deep.

\item The 5 Tesla solenoid is based on the CMS design, with inner (outer) radius of 2.50 (3.30) m. The high
field helps disperse particles entering the calorimeters, provides high momentum resolution in the tracker, and
constrains the pair background produced to small radii.

\item The flux return and muon system begins at a radius of 3.33~m, and extends to 6.45~m. Iron plates
about 10 cm thick make up the flux return, and not all the gaps are instrumented. Both RPCs and scintillator
strips are being considered as technology options. 

\item Forward systems aren't shown in the diagram, but consist of a luminosity calorimeter, and a beamcal, to 
catch very forward produced pairs. A gamcal, which helps with the instantaneous luminosity measurement, is 
designed to measure beamstrahlung photons downstream of the detector.

\end{itemize}

			
			
	

\subsection{Integrated Tracking}

     The tracking system in SiD is to be regarded as an integrated system, incorporating the 
vertex detector, the central tracker, and the electromagnetic calorimeter.

    The vertex detector plays a key role in track pattern recognition. Most tracks are first
found in the vertex detector and then extrapolated 
into the central tracker, where they pick up the additional hits needed 
to measure their curvature accurately. This procedure misses roughly 5 \% of tracks, because they result 
from neutral decays outside the vertex 
detector proper. Those originating from within the second layer of the central tracker, are reconstructed by a 
stand-alone central tracking algorithm. Tracks produced by decays beyond the second layer of the central tracker, 
but within the ECAL, are captured with a 
calorimeter-assisted tracking algorithm. This algorithm uses the track entry points and directions
as measured in the electromagnetic calorimeter to provide seeds for extrapolation back into the tracker.
Altogether, the track pattern recognition efficiency is very high, even in the presence of backgrounds. 
Realistic simulation studies are
underway to aid in optimizing the final tracking system design. Studies of 
calorimeter-assisted tracking are not yet fully developed, but have already demonstrated that 
~60\% of pions from $K_s^0$ decays at the Z can be reconstructed using a barrel-only algorithm. 
The momentum resolution of the combined system is excellent, with $\sigma_{p}/p^2 < 2 \times 10^{-5}$ GeV$^{-1}$
at high momentum.   
 

\begin{figure}
	\centering
		\includegraphics[height=7cm
		]{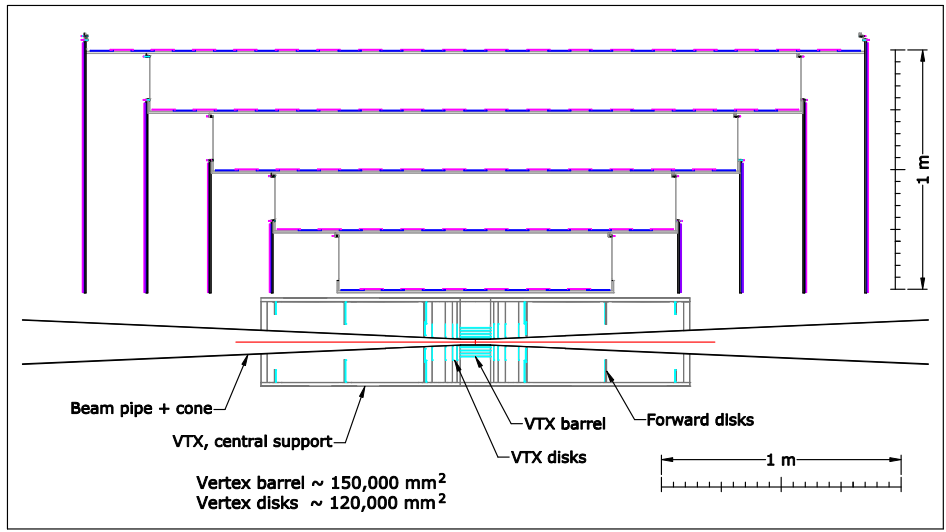}
	\caption[Mechanical concept for the SiD vertex detector]{Mechanical concept for supporting the SiD vertex detector barrel and endcaps, tracker 
forward disks, and the beam pipe}
	\label{fig-SiD_vertex}
\end{figure} 

     Mechanical designs for the vertex detector and the central tracker have been developed.
Figure~\ref{fig-SiD_vertex} shows the vertex detector and forward tracker, and half the
central tracker.  A double-walled carbon fiber cylinder supports the vertex detector, 
forward disks of the central tracker, and the beampipe.  Services are located at the ends of the barrels, and 
cooling is provided by forced convection with dry air. This is adequate because individual sensor modules are readout with
a bump-bonded ASIC chip (described below), which is power-pulsed ``on'' only during the bunch train.

\subsection{Electromagnetic Calorimetry}

     The SiD ECAL consists of alternating layers of tungsten radiator and large-area silicon 
diode detectors. The design minimizes the 
effective Moliere radius by packing 300 $\mu$ m thick silicon sensors
into 1 mm gaps between tungsten plates.  Longitudinally, the ECAL 
consists of 30 alternating layers of tungsten and silicon. The first 20 layers of tungsten 
each have a thickness of 2.7 mm; the last 10 layers have 
double this thickness, making a total depth of about 29 radiation lengths at normal 
incidence. This results in an energy resolution of $17\%/\sqrt{E(\mathrm{GeV})}$. The inner radius (length) of the 
barrel is kept relatively small 127 (359) cm, to minimize the required area of silicon 
needed. The endcaps are located inside the barrel and start at a distance of 168 cm from 
the interaction point.

\begin{figure}
	\centering
		\includegraphics[height=8cm]{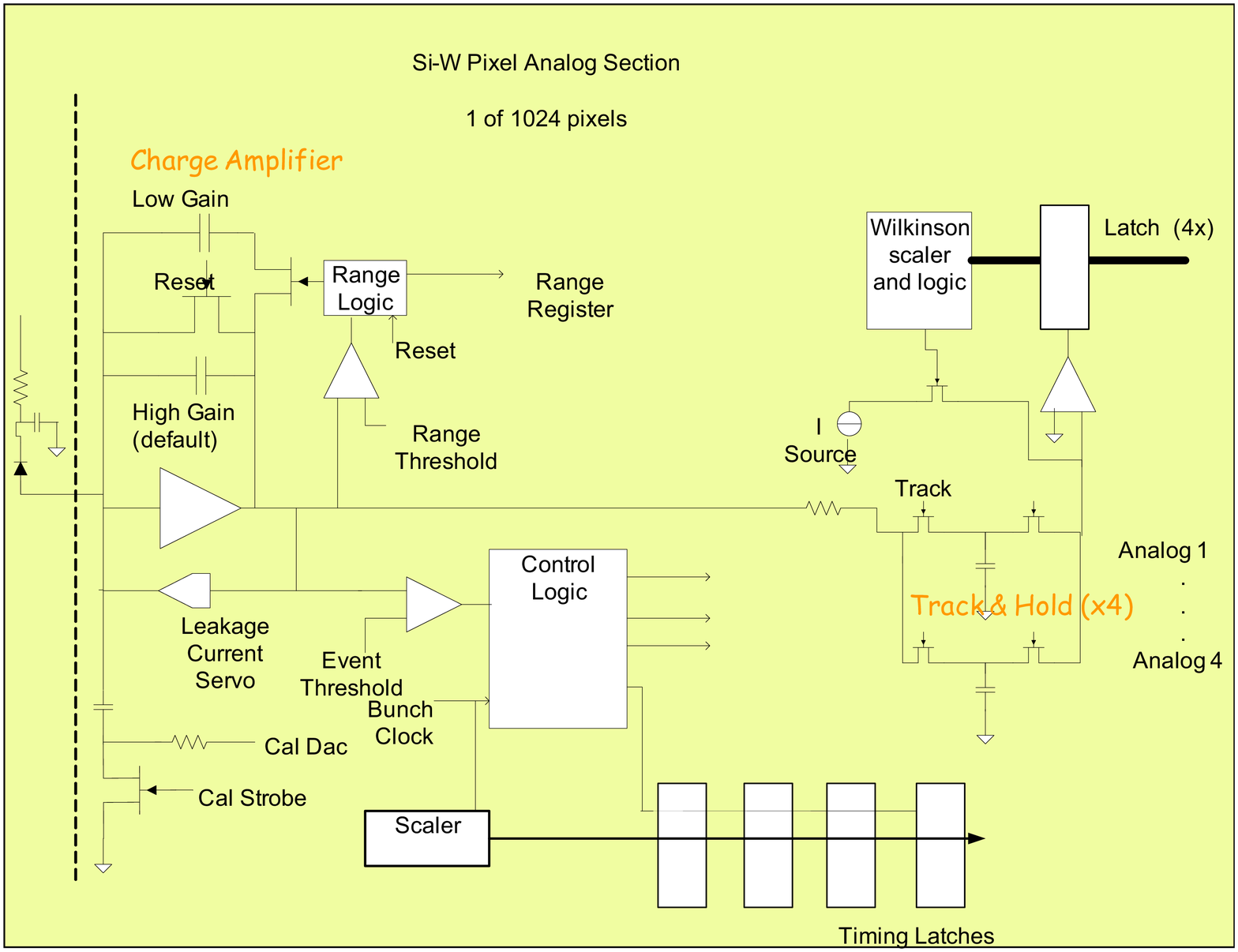}
	\caption[Finctional diagram of the KPIX chip]{Functional diagram of one channel (of 1024) of the KPiX chip.  The silicon 
detector pixel is indicated by the diode and capacitor at left.}
	\label{fig-SiD_channel}
\end{figure}

     Figure ~\ref{fig-SiD_channel} is a diagram of a single channel of the 1024-channel ASIC readout chip,
called KPiX, indicating its functional features. KPiX has a 1:2500 dynamic range to accommodate the tremendous 
range in energy densities between MIPs and the cores of very high energy EM showers. 
The calculated noise level is about 1000 e's, to be compared with the MIP signal charge 
25 times larger. The chip can store four hits (times and pulse heights) per bunch train  for each 
pixel. The chip, a modification of which is adapted to reading out the tracker microstrip sensors,
is power-pulsed. The chip has been prototyped, and is in the final debug stages prior to a full submission.

     The HCAL is a sandwich of absorber plates and detector elements.  The SiD starting point
uses steel for the absorber and resistive plate 
chambers (RPCs) as the detector.   One of the 
criteria for the HCAL is to minimize the gaps between absorber plates, because an increase in the gap size has a 
large impact on the overall detector cost.  The current gap size is 12mm. To satisfy the 
stringent imaging requirements of the PFA algorithm, the transverse segmentation is 
required to be as small as 1 to a few cm$^2$, and every layer is read out separately. The absorber 
consists of steel plates with a thickness of 20 mm (approximately 1.1 X$_{0}$). The cell 
structure, which is the same for the barrel and the endcaps, is repeated 34 times, 
leading to an overall depth of the HCAL corresponding to four interaction lengths. 
Tungsten is also being considered for the absorber. 
     Several detector options are under consideration. Glass RPCs have been shown to be 
reliable and highly efficient. The development of economical large area GEM foils and Micromegas
are making these approaches viable as well. Scintillating tiles, readout with silicon photomultipliers, are 
another option.

\subsection{Solenoid and Flux Return}

     The SiD concept incorporates a large 5 Tesla 
superconducting solenoid which provides a clear bore 5.0~m in diameter by 5.6~m long.  
An iron flux-return system limits the fringe field of the solenoid and provides absorber for muon 
identification and tracking. 
     The CMS solenoid, now operational, provides a substantial proof-of-concept for the SiD solenoid. 
Although providing 20\% lower field than the SiD solenoid, the CMS solenoid is 
physically larger and stores 2.6 Giga-Joules (GJ) magnetic energy vs.~1.4 GJ stored by 
the SiD solenoid. The ratio of operating current to critical current is comparable to that in CMS, and the 
ratio of stored energy to cold mass is lower in SiD than CMS. A detailed finite-element model study 
has indicated that the realization of the SiD solenoid is not less credible than that of CMS.
A detailed design is still required.
 
    The conceptual design for the flux return includes an iron yoke, consisting of an octagonal central barrel 
and endcaps of steel plates 10 cm thick, with 5 cm gaps for muon chambers.  A total of 
23 layers of steel was chosen for both the barrel and the endcaps to adequately shield the 
region external to the detector from stray magnetic field.

\subsection{Muon System}

     The SiD muon system is designed to identify muons from the interaction point with 
high efficiency and to reject almost all hadrons (primarily pions and kaons). 
The muon detectors will be inserted in the gaps between the iron plates which comprise 
the flux return. It is unlikely that all gaps will be instrumented.

     Present studies indicate that a resolution of ~1-2~cm is more than adequate. This can 
be achieved by both of the technologies under consideration (see below). Simulations show that the muon 
identification efficiency is greater than 96\% above a momentum of 4 GeV/c. Muon 
purity approaches 90\%. Muons perpendicular to the $e^+e^-$ beamline reach the SiD muon system 
when their momentum exceeds $\approx 3 $GeV/c.    

     Two technologies are under consideration for the muon system: scintillator strips
and RPCs. As the MINOS experiment has already proved,  a strip-scintillator detector works well for identifying 
muons and for measuring hadronic energy in neutrino interactions.   RPCs have often been used as muon 
detectors (BaBar and BELLE) and will be used in both LHC experiments. The major concern with RPCs are 
their aging characteristics (BaBar was forced to replace its original RPCs and BELLE 
had startup problems).  However, significant progress has been made in recent years in 
understanding aging mechanisms, and recent RPC installations have performed reliably.  

\subsection{Forward Calorimeters}
     
      The forward region is defined as polar angles $\cos\theta > 0.99$  ($\theta < 140~\mathrm{mrad}$), i.e.
angles below the coverage of the SiD Endcap ECAL. The physics missions in this region are the precision measurement of 
the luminosity using forward Bhabha pairs (LumCal), the measurement of the bunch-by-bunch luminosity and
bunch diagnostics using the beamstrahlung gammas and pairs (GamCal and BeamCal, respectively), and 
the extension of calorimeter coverage into the very forward region to provide full hermeticity for physics searches.
 
       The BeamCal is a highly segmented Si-W electromagnetic calorimeter located on the front face of the final 
focus quadrupole magnets which covers the region 3 mrad to 20 mrad. These calorimeters will have to be supported 
by the forward support tube, and employ detectors with exceptional radiation hardness. 


\subsection{Machine Detector Interface}

     SiD has explored IR Hall design, and developed a detector footprint, preliminary 
assembly procedures, and access strategies. The total pit area required for SiD's on-beamline 
configuration is 20~m transverse to the beam, and 2~m along it. This footprint allows for
detector access, which is accomplished by moving the endcap away from the central 
detector along the beam line, and  self-shielding, with the use of a beamline absorber plug. 
Assembly off-beamline in an underground pit requires an IR hall 48~m transverse to the 
beam and 28~m along the beam. A Hall height of 33m accommodates the needed 
assembly space and room for the crane and lifting fixtures. 
   
 


     

\subsection{Conclusions and Future Plans}
     The principal goal of the SiD Design Study has been, and remains, to design a detector optimized 
for studies of 0.5-1.0 TeV $e^{+}e^{-}$ collisions, which is rationally constrained by costs, and which 
utilizes Si/W electromagnetic calorimetry and all silicon tracking. So far, the conceptual mechanical 
design of the SiD Starting Point has been developed and captured 
in a full Geant4 description of SiD.  SiD's physics performance has been simulated, costs estimated, and work begun
on the needed detector technologies. 

     The next step involves moving beyond the SiD Starting Point, evolving toward an optimized detector design.
The simulation and costing tools needed for this process are now largely in place.
SiD will study integrated physics performance and cost vs variations in 
B field strength, ECAL inner radius, ECAL length, and HCAL depth. SiD also plans to optimize subsystem parameters,
proceed with conceptual engineering designs for all subsystems, benchmark integrated detector performance
on key physics measurements, and select favored subsystem technologies.

\section{The LDC (Large Detector) Concept}
\label{sec-LDC}
\begin{figure}
	\centering
		\includegraphics[height=8cm,bbllx=10mm,bblly=80mm,bburx=200mm,bbury=200mm]{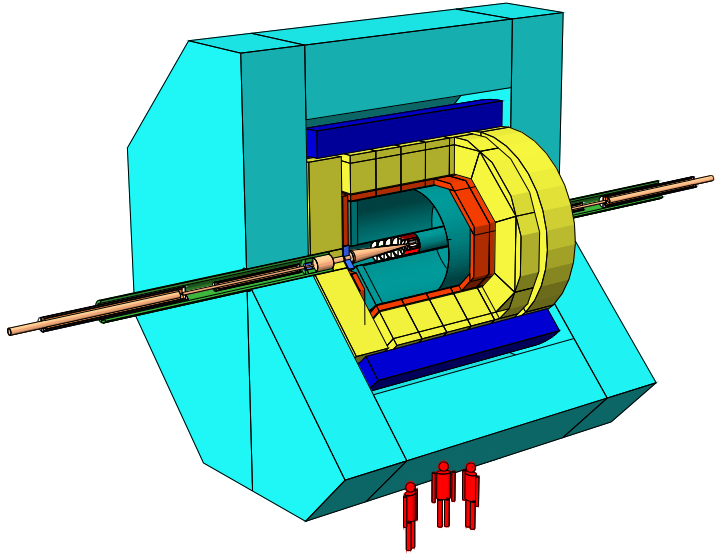}
	\caption[3D-view of the LDC detector]{View of the LDC detector concept, as simulated with the MOKKA simulation package.}
	\label{fig:ldc01}
\end{figure}

The LDC detector starts from two basic assumptions on how the physics at the ILC should be dealt with: a precision, 
highly redundant and reliable tracking system, and particle flow as a means to do complete event reconstruction. 
This sets the stage for the overall layout of the detector, which consists of a large volume tracker, and  
highly granular electromagnetic and hadronic calorimeters, all inside a large volume magnetic field of up to 4 Tesla, 
completed by a precision muon system which covers nearly the complete solid angle outside the coil. A detailed 
description of the LDC detector may be found in Ref.~\cite{ref-LDC}. A view of the simulated detector is shown 
in Figure~\ref{fig:ldc01}.

The tracker has as its central component a Time Projection Chamber (TPC) which provides up to 200 precise 
measurements along the track of a charged particle. This is supplemented by a sophisticated system of Si-based 
tracking detectors, which provide additional points inside and outside of the TPC, and which significantly extend 
the angular coverage of the TPC to very small angles. A vertex detector, also realized in Si technology, gives 
unprecedented precision in the reconstruction of long lived particles like $b-$ or $c-$hadrons.

Tracking in the high multiplicity environment at the ILC poses significant challenges, if the requirement of close 
to 100\% efficiency over a large momentum range and large solid angle is to be reached. For a number of physics 
channels, excellent momentum resolution is of utmost importance, and has large impact on the overall design of 
the system. 
The combination of a precision TPC with a small number of Si-detector layers has been chosen because of its 
potential for excellent performance and great robustness of the detector. 

Over the past few years the concept of particle flow has become widely accepted as the best method to reconstruct 
events at the ILC. Particle flow aims at reconstructing every particle in the event, both charged and neutral ones. 
This pushes the detector design in a direction where the separation of particles is more important than the precise measurement 
of its parameters. In particular in the calorimeter, the spatial reconstruction of individual particles takes precedence 
over the measurement of their energy with great precision. Because of this the proposed calorimeters - 
both electromagnetic 
and hadronic - are characterised by very fine granularity, both transversely and longitudinally while sacrificing 
somewhat the energy resolution. The concept of particle flow in addition requires a detection of charged particles with 
high efficiency in the tracker. Thus the overall design of the detector needs to be optimised in the direction of efficient 
detection of charged particles, and a good measurement of the neutral particles through the calorimeters.  

In more detail the proposed LDC detector has the following components:
\begin{itemize}
\item	A five layer pixel-vertex detector (VTX). To minimise the occupancy of the innermost layer, it is only half 
as long as the outer four. The detector, the technology of which has not yet been decided, is optimised for excellent point 
resolution and minimum material thickness;
\item	a system of Si strip and pixel detectors beyond the VTX detector. In the barrel region two layers of Si strip 
detectors (SIT) are arranged to bridge the gap between the VTX and the TPC. In the forward region a system (FTD) of Si pixels 
and Si strip detectors cover disks to provide tracking coverage to small polar angles; 
\item	a large volume time projection chamber (TPC) with up to 200 points per track. The TPC has been optimized 
for excellent 3D point resolution and minimum material in the field cage and in the endplate;
\item	a system of "linking" detectors behind the endplate of the TPC (ETD) and in between the TPC outer radius and 
the ECAL inner radius (SET). Silicon strip technology is investigated as a prime candidate for both detectors, but other 
technologies are explored as well. A solution where the detector is closely integrated with the ECAL is favoured, especially in
the barrel; 
\item	a granular Si-W electromagnetic calorimeter (ECAL) providing up to 30 samples radially, with a transverse 
segmentation of $0.55 \times 0.55~\mathrm{cm}^{2}$ throughout;
\item	a granular Fe-scintillator hadronic calorimeter (HCAL) with up to 40 samples longitudinally, and a cell size 
of $3 \times 3~\mathrm{cm}^{2}$ for the inner layers. The option of a gas hadronic calorimeter which would allow much finer segmentation, but 
would use only binary readout of each cell, is also being considered;
\item	a system of high precision extremely radiation hard calorimetric detectors in the very forward region, to measure 
luminosity and to monitor the quality of the collision (LumiCAL, BCAL, LHCAL);
\item	a large volume superconducting coil, creating a longitudinal B-field of nominally 4 Tesla;
\item	an instrumented iron return yoke, which returns the magnetic flux of the magnet, and at the same time serves as a 
muon detector by interspersing a number of layers of tracking detectors among the iron plates;
\item	a sophisticated data acquisition system which enables the monitoring of the electron-positron collisions without 
an external trigger, to maximize the sensitivity to physics signals and possible discoveries.
\end{itemize}

\begin{figure}[htb]
	\centering
\begin{tabular}{cc}
  \includegraphics[height=4cm]{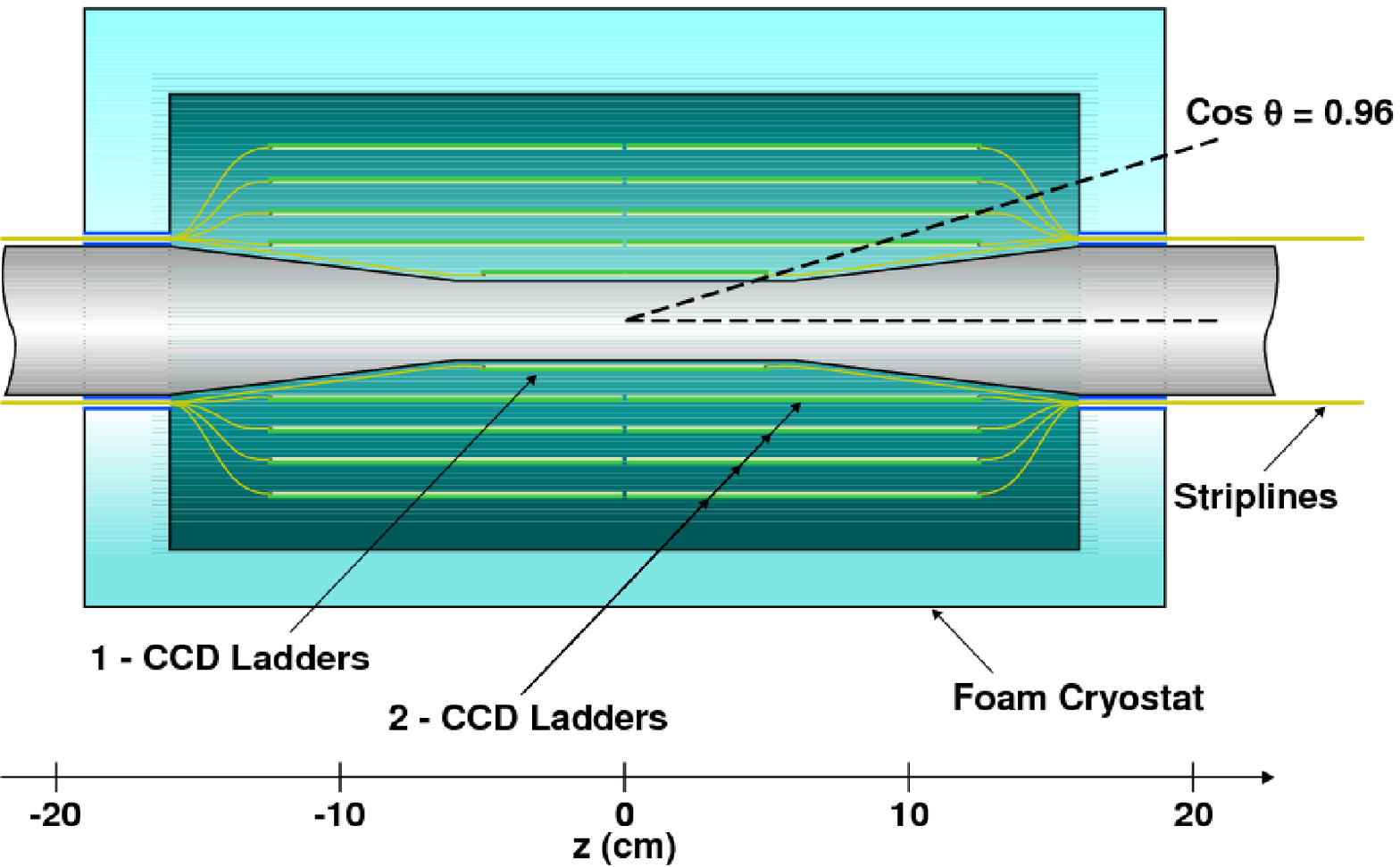} & 
		\includegraphics[width=8cm]{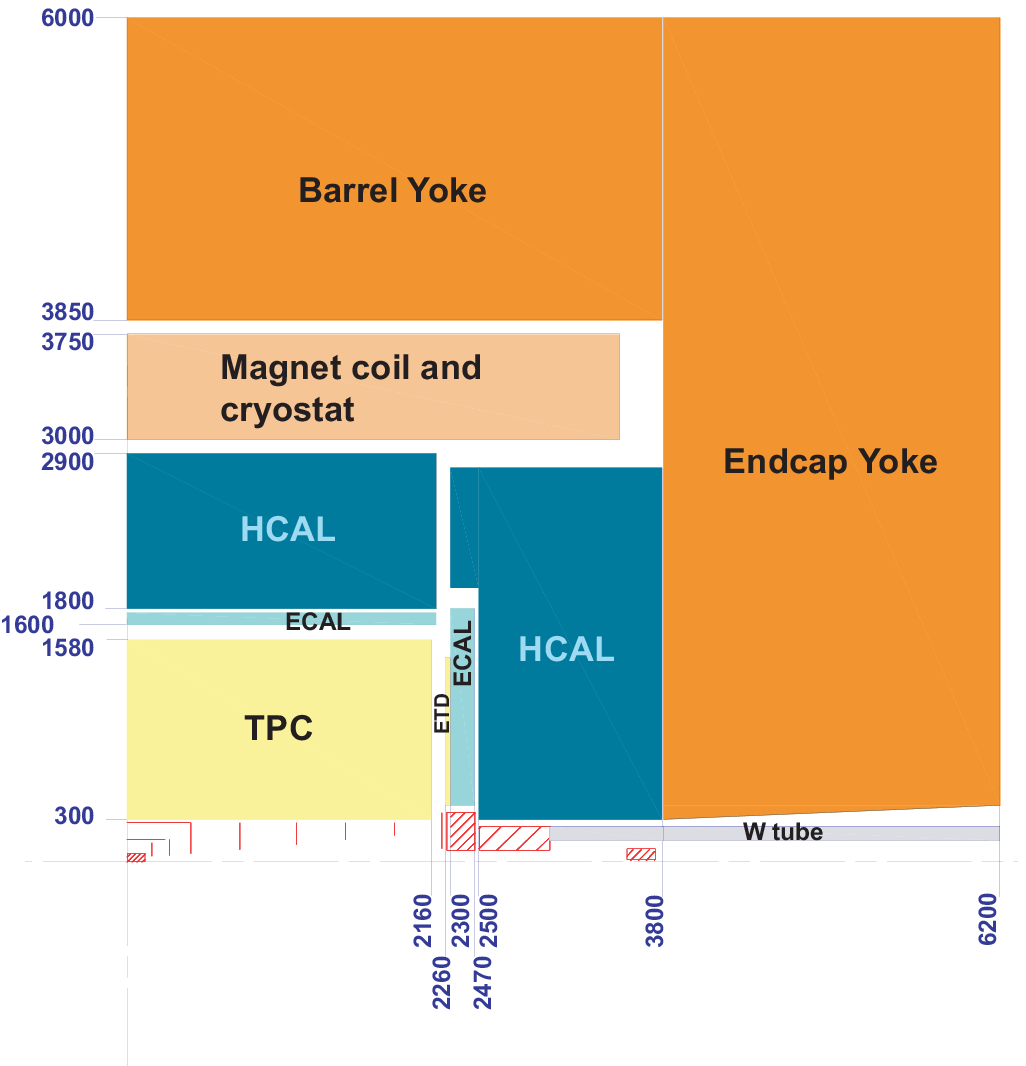}
\end{tabular}
	\caption[Sideview of the LDC detector]{left: enlarged side-view of the vertex detector; right: 1/4 view of the LDC detector concept.}
	\label{fig:LDC280806}
\end{figure}

Altogether the detector has a total height of around 14~m and a length of 20~m. It will feature around $10^9$ electronic 
channels, needed to record every detail of the collision. It is expected that a collaboration similar in size to the one currently 
constructing the LHC detectors will be needed to build and later operate this detector. A schematic view of one quarter of 
this detector is shown in Figure~\ref{fig:LDC280806}. 

\subsection{The Tracking System}
The tracking system for LDC is designed to provide redundant pattern recognition capability and excellent momentum resolution
over the full solid angle, including sufficient momentum resolution in the far forward direction to distinguish the charges of high momentum
particles. The design seeks to minimize material, so as to minimize interference with the measurement of electrons and photons 
in the calorimeters. It incorporates a high precision vertex detector to provide very good bottom and charm tagging capabilities, and to
measure the full vertex topology in physics events.
 
The complete tracking system is immersed in a 4~T solenoidal magnetic field, aligned with the z-axis.  For the proposed 14~mrad crossing
angle, a small dipole field (Dipole in Detector, or DID, or anti-DID) may be superimposed on the main B-field to maintain the beam's emittance
as it passes through the detector field to the interaction point. This is discussed in more detail in Chapter~\ref{detector_MDI}, ``The Machine Detector Interface''.
 
The tracking system includes several distinct sub-detector systems. The high precision pixel vertex detector surrounds the interaction
point. It 
consists of five concentric layers with radii between 1.55~cm for the innermost layer and 6.00~cm for the outermost layer. This detector provides 
excellent point resolution. Its material budget has been minimized in order to optimize the impact parameter resolution over the widest
possible solid angle.

Moving outward, there is a system of Si-strip and pixel detectors which provide excellent linkage between tracks measured in the TPC and 
those in the vertex detector, and which extend coverage to very forward angles. Two concentric Si strip detector layers are arranged outside the 
vertex detector, in the barrel region, and six disks, the first two of which are implemented as pixel detectors, cover the forward region.

A large TPC provides robust pattern recognition, even in complicated, background-laden events, and excellent momentum resolution.
Up to 200 three dimensional space points are measured per track, with point resolutions in the $r-\phi$ plane of 100 $\mu$m or better. The chamber is readout 
with GEMs or Micromegas, which reduce positive ion feedback, reduce E~$\times$~B effects, and improve position resolution compared to
traditional wire chamber readouts. The endplate thickness is minimized to reduce its impact on the calorimetric measurements which follow. 
An additional chamber is located behind the TPC endplate, to provide a space point between the TPC and the ECAL endcap system.
This can serve to improve the polar angle definition of forward going tracks, and improve linkage between the TPC and the endcap. Several
technology options for the chamber are being considered. A layer of Si strip detectors in the barrel region outside of the TPC is being considered as an upgrade option. Such a layer would provide additional calibration points,
improve the overall momentum resolution, and help improve linkage between the TPC and the barrel ECAL system.

\subsection{The Calorimeter System }
\label{sec-LDC-calo}
Proper identification on an event-by-event basis of the hadronic decays of W, Z, and possibly of the H bosons, is required to maximize 
the physics output of the Linear Collider. Precision measurements at the 3\% level of the mass of pairs of hadronic jets ($30\%/\sqrt(E) $ at jet energies up to approx. 100 GeV, $3\%$ for higher energy jets) are needed to fully 
exploit the physics potential of the machine. The key to this unprecedented mass resolution is the individual measurement of the energy of all particles 
in a jet. The contribution of charged particles, which on average make up 65\% of a jet's energy, is measured with the tracking system. 
Neutral energy, i.e.~that from photons and neutral hadrons, is measured with the calorimeters. The energy deposited in the calorimeters by the charged particles
must be distinguished and isolated from the neutral energy depositions. This approach, often referred to as the ``particle flow'' method (PFA), drives the concept of calorimetry for the LDC.

The calorimeter is divided in depth into an electromagnetic section, optimized for the measurement of photons and electrons, and a hadronic 
section for measurement of hadronic showers. The two parts are installed inside the coil to avoid energy losses  in any inactive material in front of the calorimeters. 

To optimize the separation of showers from photons and hadrons the electromagnetic part uses tungsten (or lead) as absorber 
material, providing a large ratio of interaction to radiation lengths and a small Moli\`ere radius, interleaved with layers of Si detectors. 
In order to maintain the smallest 
possible Moli\`ere radius, the detector and readout must be accommodated in a small gap between tungsten layers. Gap sizes between 2 and 3~mm
are being considered. Silicon diodes, with $5.5 \times 5.5~\mathrm{mm}^{2}$ 
readout cells, roughly one third of the Moli\`ere radius, provide very fine readout segmentation. 
To reach an adequate energy resolution (which impacts also position and angular resolution) with an acceptable polar angle dependence, the following 
sampling is under study: 12 radiation lengths are filled with 20 layers of $0.6 X_0$ thick tungsten absorbers (2.1~mm) 
and another 11 radiation lengths are made out of 9 layers of tungsten $1.2 X_0$ thick. The calorimeter starts with an active layer.
Overall, the electromagnetic calorimeter is divided into a cylindrical barrel and two endcaps. 

For the hadronic part, the emphasis is as well on small readout cells, to provide the best possible separation of energy deposits from 
neutral and charged hadrons. The single particle resolution needs to be adequate, but is not the driving design criterion. Two options 
are currently under study and a technology choice will be based on the results of extensive beam tests. The first uses 
scintillator cells with roughly $3 \times 3\mathrm{cm}^{2}$ granularity and multi-bit (or analogue) readout. The second is based on gaseous 
detectors and uses even 
finer granularity, perhaps $1 \times 1\mathrm{cm}^{2}$.  Due to the large number of cells, in the second case single-bit (or digital) readout is 
sufficient. In both cases the absorber material is iron (stainless steel). However, the use of tungsten or brass in the hadronic section is also under study.  





The HCAL is arranged in 2 cylindrical half barrels  and two end caps. 
The barrel HCAL fills the magnetic field volume between the ECAL and the cryostat within $180 < r < 290 \mathrm{cm}$.
In the magnetic field direction the barrel extends from $-220<~z~<~220~\mathrm{cm}$. The end caps close the barrel on either side in order to fully 
cover the solid angle. The gap between the barrel and the end cap is needed for support and for cables from the inner detectors, and for 
the readout data concentration electronics of the barrel HCAL itself. Care has been taken to maximize the absorber material in the space 
available, so that the probability for leakage is minimized. Even though the muon system will act as a tail catcher, the uninstrumented 
material associated with the cryostat and the coil (1.6 $\lambda$ thick) 
severely limits its energy resolution. Each HCAL half barrel is subdivided into 16 modules, 
each of the end caps into 4 modules. Two HCAL modules together form an octant, and support the ECAL modules in this azimuthal range.

Three calorimeters are planned in the very forward region of the detector: The BeamCal, which is  adjacent to the beam pipe, and the LumCal,
which covers larger 
polar angles, are electromagnetic calorimeters. The LHCAL is a hadron calorimeter covering almost the same angular range as the LumCal. 
These calorimeters will have several functions. All of them improve the hermeticity of the detector, which is important for new particle searches and 
jet energy resolution, and they help to shield the central detectors from backscattered particles. 

Due to their large charge and small size, the crossing bunches generate a significant amount of beamstrahlung.  Beamstrahlung photons  which
interact with photons or electrons or positrons from the opposing beam, can convert to electron-positron pairs. These pairs, in turn,  are deflected 
by the electromagnetic fields of the passing bunch, and deposit much of their energy on the BeamCal. The pattern of this energy deposition
provides information to the beam delivery feedback system which is useful in optimizing the luminosity, bunch by bunch.

The LumiCal is the luminometer of the detector. From the physics program an accuracy of the luminosity measurement of better that $10^{-3}$ is required. 
Small angle Bhabha scattering will be used for this measurement. 

\subsection{The Solenoidal Magnet}
\label{sec-LDC-coil}
The tracker and the complete calorimeter in the LDC are contained within a solenoidal coil, which produces a field of up to 4 Tesla. 
Except for the its length and required field homogeneity, the LDC magnet parameters are very similar to those for the CMS magnet, 
which has now operated successfully.

The magnet system consists of the superconducting coil, a solenoid made of five modules which includes correction coils, and an iron yoke, composed of the central barrel
yoke and two end cap yokes.
Preliminary calculations show that a total coil length of about 7 m and an iron thickness of about 2.5 m were good compromises to obtain the requested field parameters.
The required field homogeneity can only be obtained if special correction devices are introduced. They are incorporated into the main windings 
of the coil, by adding an extra current in appropriate locations of the windings. 

\subsection{The Muon System}
\label{sec-LDC-muon}
Lepton identification is one of the prerequisites for ILC experimentation: identifying leptons and their charge will be used, for instance, to tag flavour and decay chains of 
heavy quarks, to charge-tag gauge bosons, and to tag various SUSY particle decays. Lepton tagging can be 
also used to flag the presence of neutrinos in the underlying event, thus signaling missing energy.
The energy range one has to cover for lepton identification is quite large, spanning from a few GeV up to hundreds of GeV. The electromagnetic calorimeter, in conjunction
with the charged particle tracking and dE/dx, will identify electrons. Particles which have penetrated the calorimeters, solenoidal coil, and iron
flux return without interacting are identified as muons.

The muon identification system, the outermost device of the experimental apparatus, uses the iron of the flux return as absorber, with the gaps between 
the iron slabs instrumented with detectors. Several detector technologies are being considered. 
 Resistive plate chambers have been successfully used in previous experiments and are also 
considered for the LDC detector. The choice is driven by the need for 
reliable, sturdy and inexpensive devices, as the area to cover is quite big and once installed, replacing detectors would be both time-consuming and difficult. 
An active layer is placed right behind the coil, so that the system can also be used as a tail 
catcher for highly energetic showers which leak out of the back of the calorimeter. 

\subsection{Data Acquisition}
The LDC detector has been designed without a traditional trigger system. Each bunch crossing of the 
accelerator is recorded. A selection of events is only performed by a software trigger. This ensures a very 
high efficiency and sensitivity to any type of new physics but at the same time puts fairly 
stringent requirements on the frontend electronics of each sub-detector. However the rather clean events, low levels of 
background, and relatively low event rate allow one to pursue such a design.

The data acquisition system would provide a dead time free pipeline of 1 ms, the time required for one pulse train from the ILC, and
be ready for another train within 200 ms, the nominal time between trains. Event selection would proceed in software.
The high granularity of the detector and the 2820 collisions in 1 ms still require a substantial bandwidth to read the data in time before 
the next bunch train. To achieve this, the detector front end readout will provide zero suppression and data compression as much as 
possible. Due to the high granularity it is mandatory to multiplex many channels into a few optic fibers to avoid a large number 
of readout cables. Such multiplexing will also reduce dead material and gaps in the detector as much as possible.

The data of the full detector will be read out via an event building network for all bunch crossings in one train. After the readout the data 
of a complete train will be situated in a single PC. The event selection will be performed on this PC based on the full event information 
and bunches of interest will be defined. The data of these bunches of interest will then be stored for further physics analysis as well as 
for calibration, cross checks and detector monitoring.

The machine operation parameters and beam conditions are vital input for the high precision physics analysis and will therefore be needed 
alongside the detector data. Since the amount of data and time structure of this data is similar to that from the detector, a common data acquisition system and 
data storage model is envisaged.

To ensure the smooth functioning of this concept a well-calibrated detector is important. Strategies for a fast 
online calibration of key detector elements will be needed, and will have to be developed over the next 
few years. 

The hardware for the data acquisition should be defined as late as possible, to profit from the latest industrial 
developments. It will rely heavily on commodity hardware, and avoid custom developments 
wherever possible. Even so the development, the building and the commissioning of the 
data acquisition system will present a significant challenge for LDC. 

\subsection{Conclusion}
The LDC design is an example of a detector optimized for the particular physics at the ILC. Particle flow, with the 
rather unique requirements it puts on detector design, has been one of the driving forces of the conceptual 
layout. The conceptual design has reached a rather mature state, and has not changed significantly since 
first published in 2001~\cite{ref-TESLA:2001}. Over the past few years significant progress however has been made in 
the transfer of the conceptional design state into a real, technically understood design. 

%
%
\section{The GLD Concept}
\label{SectionConcept}

The physics to be studied at the International Linear Collider (ILC)
encompasses a wide variety of processes over
the energy region from $\sqrt{s} \approx M_Z$ to 1~TeV\cite{ref-TESLA:2001, ref-ACFA:2001, ref-SM01:2001}.
 Key ILC physics processes include production of 
gauge bosons ($W$ or $Z$), heavy flavor quarks ($b$ and $c$),
and/or leptons ($e, \mu, \tau$), either as direct products
of $e^+ e^-$ collisions or
as decay daughters of heavy particles (SUSY particles,
Higgs boson, top quark, etc.). For these studies, it is essential to reconstruct
events at the level of the fundamental quanta, the quarks, leptons, and gauge bosons. The detectors at the ILC
must identify them efficiently, and measure their four-momenta precisely. 
In order to satisfy these requirements, the detector
must have superb jet energy resolution ($\Delta E_j/E_j = 30\%/\sqrt{E_j \ \rm{(GeV)}}$),
efficient jet flavor identification, excellent charged-particle momentum resolution ($\Delta p_t/p_t^2 \leq 5 \times 
10^{-5} (\rm{GeV/c})^{-1}$), and hermetic coverage to veto 2-photon background processes. 

The GLD detector concept has been developed in order to meet these requirements. It is 
based on a large gaseous tracker 
and highly segmented calorimeter placed within
a large bore solenoidal field with a 3 Tesla  magnetic field. A detailed description of the design of GLD can be 
found in Ref.~\cite{ref-GLD}.

The basic design of  GLD incorporates a calorimeter
with fine segmentation and large inner radius to
optimize it for particle flow.
Charged tracks are measured in a large gaseous
tracker, a Time Projection Chamber (TPC),
with excellent momentum resolution.
The TPC reconstructs tracks with high efficiency, even those from decaying particles, 
such as $K^0$, $\Lambda$, and
new unknown long-lived particles, and allows
efficient matching between tracks and hit clusters in the calorimeter.
The solenoid magnet is located outside the
calorimeter, so as not to degrade energy resolution. Because the detector volume is huge,
a moderate magnetic field of 3 Tesla has been chosen.

\begin{figure}
\begin{center}
\includegraphics[height=8cm]{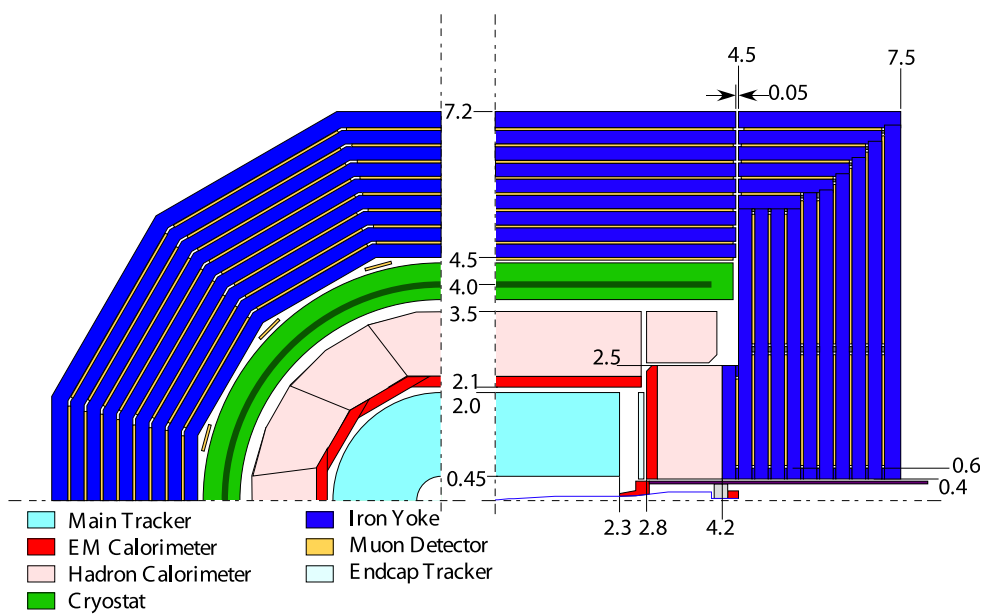}
\end{center}
\caption[Schematic view of the GLD detector concept]{Schematic view of two different quadrants 
of the GLD Detector. The left figure shows the $r\phi$ view 
and the right shows the $rz$ view. Dimensions are given in meters. The vertex detector
and the silicon inner tracker are not shown here.}
\label{fig-GLD}
\end{figure}

\begin{figure}
\begin{center}
\includegraphics[width=10cm]{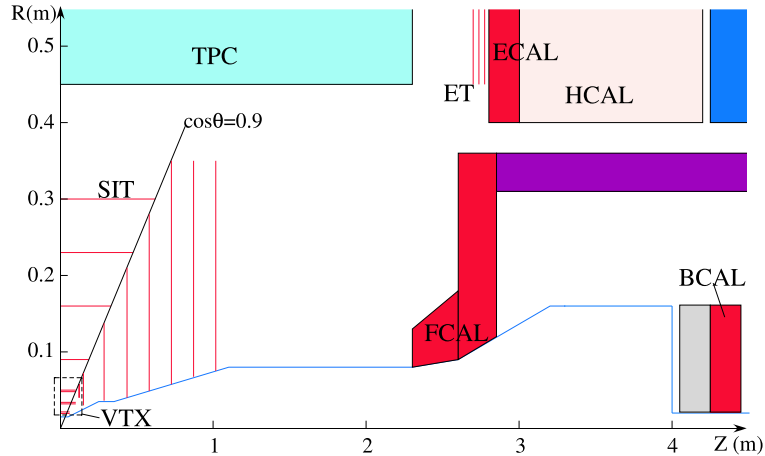}
\end{center}
\caption[Schematic view of the inner and forward detector of GLD]{Schematic view of the inner and forward
detectors of GLD. The horizontal scale and the vertical scales are 
not the same, as indicated in the middle of the figure.}
\label{fig-GLDIR}
\end{figure}

Figure~\ref{fig-GLD} shows a schematic
view of two different quadrants of the baseline design of GLD.
The inner and forward detectors are schematically
shown in Figure~\ref{fig-GLDIR}.
The baseline design has the following sub-detectors:
\begin{itemize}
\item A Time Projection Chamber (TPC) as a large gaseous central tracker;
\item A highly segmented electromagnetic calorimeter (ECAL) placed at large radius and based on a
tungsten-scintillator sandwich structure;
\item A highly segmented hadron calorimeter (HCAL)
with a lead-scintillator sandwich structure and radial thickness of 
$\sim 6\lambda$;
\item Forward electromagnetic calorimeters (FCAL and BCAL) which extend solid angle 
coverage down to very forward angles; 
\item A precision silicon micro-vertex detector(VTX);
\item Silicon inner (SIT) and endcap(ET) trackers;
\item A beam profile monitor (BCAL) in front of the final quadrupoles;
\item A muon detector interleaved with the iron plates of the return yoke; and
\item A solenoidal magnet to generate the 3 Tesla magnetic field.
\end{itemize}
The iron return yoke and barrel calorimeters 
are 12-sided polygons,  and the outer edge of the HCAL is a 24-sided polygon
in order to reduce  any unnecessary gaps between
the muon system and the solenoid, the 
HCAL and the solenoid, and the  TPC and ECAL.

In addition to the baseline configuration, GLD is considering adding 
silicon tracking between the TPC and the ECAL in the barrel  
region to improve the momentum resolution further, and
TOF counters in front of the ECAL to improve the particle identification capability.
%


%
\subsection{Vertex Detector}
The inner radius of the vertex detector is 20 mm and the outer radius is 50 mm.
It consists of three doublet layers, where each doublet
comprises two sensor layers separated by  2 mm. 

In the baseline design, fine pixel CCDs (FPCCDs) serve as the sensors
for the vertex detector. 
The FPCCD is a fully depleted CCD with a pixel size 
of order $5 \times 5~\mu\mathrm{m}^2$.  Signals integrated during the entire ILC beam train of 
about 1 msec are stored in the pixels and read out 
during the 200 ms between trains.
             
In the FPCCD option, the pixel occupancy is expected to be less than 0.5\%
for the inner most layer (R=20 mm, assuming B=3 T and the
ILC nominal machine parameters~\cite{ref-raubenheimer}).
The hit density is, however, as high as 40/$\rm{mm}^2$.
Coincidences between the two sensors in a doublet layer help to determine the track 
vectors locally and discriminate against these background hits. Signal shape is also 
taken into account to reduce the number of background hits during the track reconstruction.

Track reconstruction errors are minimized by using very thin sensors (much less than 100~$\mu$m). 
This puts special importance on the R\&D effort on wafer thinning. The fabrication
of the small pixel sensors is also being pursued.

\subsection{Silicon Trackers}
The silicon trackers of GLD consist of the silicon inner tracker,
the silicon forward tracker and the silicon endcap tracker.

The silicon inner tracker is located between the vertex detector and the TPC.
It consists of the barrel inner tracker (BIT)
and the forward inner tracker (FIT). The BIT consists of four layers 
of silicon strip detector located between the radii of 9 cm
and 30 cm. It is used to improve the efficiency for linking tracks between the main tracker
and the vertex detector, to boost reconstruction efficiency for low 
$p_t$ tracks, and to improve momentum resolution.
Time stamping capability is crucial for this device in order
to identify the bunch corresponding to the track measured
in the main tracker.
                                                                               
%

The forward silicon tracker (FIT) consists of seven layers 
of disks which 
cover the angular range
down to about 150 mrad, matching the
coverage of the endcap calorimeter.
The technologies to be used for the FIT 
depend on the track density of jets and
the background level from beam backgrounds
and 2-photon backgrounds. Detailed 
simulation studies are underway to determine the technology choice. 
  Pixel sensors for the first 
three layers and silicon strip sensors for the outer four layers
are assumed at the moment.

%

Several layers of silicon strip detectors are placed
in the gap between the TPC
and the endcap ECAL. The endcap silicon tracker (ET)
will improve momentum resolution for charged particles
which have a small number of TPC hits, and
improve track matching between the TPC and shower clusters
in the ECAL.

\subsection{Main Tracker}
A TPC (Time Projection Chamber) with 40 cm inner radius and 200 cm 
outer radius is used as the main tracker of GLD.
The number of radial samples is 200 and 
the maximum drift length in the z-direction is 230 cm.
For signal readout, a micro pattern gaseous detector (MPGD)
is utilized, which achieves better point resolution and 
two-track separation than the usual wire chamber readout. 
Technologies under study are Micromegas~\cite{micromegas}
and GEMs (Gas Electron Multiplier) foils~\cite{GEM}.
Both devices are gas chambers, in which the drift electrons 
are amplified in high electric fields produced by
microscopic structures (with size of 
the order of 50 $\mu$m) within the MPGD.  

Depending on the drift length, point resolutions between
50 to 150 $\mu$m in the $r\phi$ plane and 
500 $\mu$m  in the z direction are expected.
In combination with the silicon inner tracker and the vertex detector, the fractional momentum resolution 
is expected to be less than $5 \times 10^{-5} p_t$~(GeV/c)$^{-1}$ in the high $p_t$ limit.
                                                                              
\subsection{Calorimeters}
The calorimeter system of GLD consists of the electro-magnetic calorimeter,
the hadron calorimeter, and the forward calorimeters.

The electro-magnetic calorimeter (ECAL)
consists of 30 layers of tungsten and scintillator sandwich,
with
thicknesses of 3~mm and 2~mm, respectively, and with an additional 
1mm gap for the readout.  Lead absorber is considered as an option.
The scintillator has a rectangular shape with dimension $1 \times 4~\mathrm{cm}^2$. With adjacent layers
at right angles, it achieves an effective 
cell size of $1 \times 1~\mathrm{cm}^2$ while reducing the
number of readout channels. A tile structure with dimension $2 \times 2~\mathrm{cm}^2$ is considered as an option.
The light emitted in the scintillator
is detected by  a Multi Pixel Photon Counter (MPPC), which is now under 
development.


The hadron calorimeter (HCAL),
consists of  46 layers
of lead and scintillator
sandwich with 20 mm and 5 mm thicknesses, respectively,  
and a 1 mm gap for readout.
This configuration is expected to be compensating, i.e. to provide relatively equal responses to hadrons
and electrons, thus giving the best energy resolution for individual hadron showers.
Installing strips of $1 \times 20~\mathrm{cm}^2$, interleaved with $4 \times 4~\mathrm{cm}^2$
tiles, the effective cell size of HCAL could be $1 \times 1~\mathrm{cm}^2$. 
As in the ECAL, MPPC will be used as the photon sensor 
to read scintillation light via a wavelength
shifting fiber.
A digital hadron calorimeter option is also being considered for the HCAL , which would reduce the cost of the 
read out electronics.
For the digital HCAL, the base line design
using scintillator strips may
have shower overlap problems. This is being studied with a realistic PFA model,
so as to determine the optimal width and length of the strips.

%

The forward calorimeter of GLD consists of two parts:
FCAL and BCAL. The z-position of the FCAL is close to
that of the endcap ECAL, and radially outside of
the dense core of the beam-induced pair background. 
The BCAL is located just in front of the final
quadrupole magnet at 4.5 m from the interaction point. 
For the case of a 14 mrad beam crossing angle, the BCAL has holes of radius 1.0 cm and 1.8 cm 
for the incoming and outgoing beams.  These holes are the
only regions not covered by GLD for particle detection, although large energy depositions
in parts of the BCAL may render it insensitive close to the beams.

A mask made of low-Z material with the same inner radius as the BCAL is positioned 
in front of it to absorb
low energy  backscattered $e^\pm$. The z-position of the FCAL
is chosen so that it works as a mask for
photons backscattered from the
BCAL, so they cannot hit the TPC directly.
The FCAL(BCAL) will consist of 55 (33) layers of tungsten and Si 
sandwich.  For the BCAL, radiation hard sensors, such as
diamond, might be necessary.
\subsection{Muon System}
The muon detectors of the GLD are placed between the iron blocks 
which comprise the magnet flux return yoke.  The iron return yoke must be roughly 2.5 m thick
to keep the leakage of the  magnetic field along the beamline acceptable for  machine operation.
9 layers of muon detectors are placed between the iron return yoke blocks, 
each layer consisting of a two-dimensional array of
scintillator strips with wavelength-shifter fiber
readout by MPPCs.
\subsection{Detector Magnet and Structure}
The detector magnetic field is generated by a
superconducting solenoid with correction windings
at both ends. The radius of the coil is
4.0 m and the length is 8.9 m. 
Additional serpentine
windings for the detector integrated dipole (DID)
might be necessary to correct for effects
arising from the finite crossing angle of the beams.                      
The total size
of the iron structure has a height of 15.3 m and
a length of 16 m. Its thickness is determined by the requirement that the leakage field be sufficiently low. 



\newcommand {\eeWW} {e^+e^- \rightarrow W^+W^- }
\newcommand {\eeZZ} {e^+e^- \rightarrow Z^0Z^0 }

\newcommand {\Wqq} {$W \rightarrow q \bar{q}$ }
\newcommand {\Zqq} {$Z \rightarrow q \bar{q}$ }

\newcommand {\Wjj} {$W \rightarrow j j$ }
\newcommand {\Zjj} {$Z \rightarrow j j$ }

\newcommand {\Zee} {$Z \rightarrow e e$ }
\newcommand {\Wev} {$W \rightarrow e \nu$ }
\newcommand {\Zvv} {$Z \rightarrow \nu \nu$ }

\newcommand {\Zuu} {$Z \rightarrow \mu \mu$ }
\newcommand {\Wuv} {$W \rightarrow \mu \nu$ }

\newcommand {\Ztt} {$Z \rightarrow \tau \tau$ }
\newcommand {\Wtv} {$W \rightarrow \tau \nu$ }

\newcommand {\We}  {$W \rightarrow e$ }
\newcommand {\Wv}  {$W \rightarrow \nu$ }

\newcommand {\WWjjjj} {$WW \rightarrow jjjj$ }
\newcommand {\dream} {{\sc dream }}
\newcommand {\spacal} {{\sc spacal }}

\newcommand {\fem}    {\ensuremath{f_{em}}}    
\newcommand {\geant}   {{\sc geant} }

\newcommand{\Cv} {\v{C} }
\newcommand{\pu} {$\pi\rightarrow\mu$}
\newcommand{\uu} {$\mu\rightarrow\mu$}
\newcommand{\x}   {\cdot 10^}
\newcommand{\Ppunch} {P$_{\rm punch}$}
\newcommand{\Pdream} {P$_{\rm dream}$}
\newcommand{\Pinter} {P$_{\pi \rm int}$}
\newcommand{\Pmatch} {P$_{\rm Ematch}$}

\newcommand {\qcd}  {{\sc qcd} }

\newcommand {\inpp}     {in $p\bar{p}$ Collisions at $\sqrt{s}$ = 1.8 TeV}
\newcommand {\pbarp}    {$p\bar{p}$ }
\newcommand {\prl}      {{\em Phys. Rev. Letts.} }
\newcommand {\prd}      {{\em Phys. Rev. D} }
\newcommand {\prdrc}    {{\em Phys. Rev. D Rapid Comm.} }
\newcommand {\PRD}      {{\em PRD} }
\newcommand {\PRL}      {{\em PRL} }
\newcommand {\pl}       {{\em Phys. Letts.} }
\newcommand {\plb}      {{\em Phys. Letts. B} }
\newcommand {\nim}      {{\em Nucl. Instrum. and Methods} }
\newcommand {\abachi}   {Aba\-chi, S., {\em et al.,} }
\newcommand {\abbott}   {Abbott, B., {\em et al.,} }
\newcommand {\abazov}   {Abazov, V.M., {\em et al.,} }
\newcommand {\RECO}     {{\sc reco} }
\newcommand {\PAPOOSE}  {{\sc papoose} }
\newcommand {\GEANT}    {{\sc geant} }
\newcommand {\ISAJET}   {{\sc isajet} }
\newcommand {\HERWIG}   {{\sc herwig} }
\newcommand {\PYTHIA}   {{\sc pythia} }
\newcommand {\POMPYT}   {{\sc pompyt} }
\newcommand {\aH}       {{\sc hep}}
\newcommand {\hep}      {{\sc hep}}
\newcommand {\C}        {{\v{C}er\-enk\-ov} }
\newcommand {\Ce}       {{\v{C}erenkov} }
\newcommand {\Sc}       {{scintillation} } 
\newcommand {\etal}     {{\it et al.}}
\newcommand {\mET}      {$\not\!\!{E_T}$}
\def\D0{D\O}
\newcommand {\qq} {$q \bar{q}$}
\newcommand {\tautau} {$\tau \bar{\tau}$}
\newcommand {\bb} {$b \bar{b}$}
\newcommand {\bbmu} {$b \bar{b}_{\mu}$}
\newcommand {\mutau} {$\mu \bar{\tau}$}

\newcommand{\hapub} {''Hadron and Jet Detection with a Dual-Readout
  Calorimeter'', {\em Nucl. Instrum. and Methods}
   {\bf A537} (2005)  537-561.}

\newcommand{\epub} {''Electron Detection with a Dual-Readout
  Calorimeter'', {\em Nucl. Instrum. and Methods}
    {\bf A536} (2005) 29-51.}

\newcommand{\mupub} {''Muon Detection with a Dual-Readout
  Calorimeter'', {\em Nucl. Instrum. and Methods}
    {\bf A533} (2004) 305-321.}

\newcommand{\profilepub} {''Comparison of Scintillation and \C Light in
  High-energy Electromagnetic Profiles'', 
  {\em Nucl. Instrum. and Methods} {\bf A548} ( 2005) 336-354.}

\newcommand{\separpub} {''Separation of Scintillation and \C Light
  in an Optical Calorimeter'',  
  {\em Nucl. Instrum. and Methods} {\bf A 550} (2005) 185-200.}

\newcommand {\ee}       {$\em{e^+e^-}$ }
\newcommand {\eePEP}    {$\em{e^+e^-}$ annihilation at 29 GeV~}
\newcommand {\eea}      {$\em{e^+e^-}$ annihilation~}
\newcommand {\eeA}      {$\em{e^+e^-}$ Annihilation~}
\newcommand {\Bu}   {B$_{\Large u}$~}
\newcommand {\Bs}   {B$_{\Large s}$~}
\newcommand {\Ds}   {D$_{\Large s}$~}
\newcommand {\DtoK}     {$D^0 \rightarrow K^-$~}
\newcommand {\Dmix}     {$D^0-\bar{D^0}$ mixing~}
\newcommand {\Bmix}     {$B^0-\bar{B^0}$ mixing~}
\newcommand {\ET}   {E$_{\rm T}$}

\newcommand {\aihara} {Aihara, H., {\em et al.,} }
\newcommand {\BATRUN} {{\sc batrun} }
\newcommand {\dumand} {{\sc dumand} }

\newcommand {\bdm} {\begin{displaymath}}
\newcommand {\edm} {\end{displaymath}}

\def\ss{\makebox[16pt][l]{$^1\hspace*{-2pt} S_0\;$}}
\def\ts{\makebox[16pt][l]{$^3S_1\;$}}
\def\1p1{\makebox[16pt][l]{$^1\hspace*{-2pt} P_1\;$}}
\def\etal{{\sl et al.}}
\def\pbarp{$\overline{p}$-$p~$} \def\pbar{$\overline{p}$}
\def\D0{D\O}  \def\d0{D\O}

\def \TPt {\rm P_{\rm T}}
\def \TEt {\rm E_{\rm T}}
\def \pb {\rm pb}
\def \nb {\rm nb}
\def \GeV {\rm GeV}
\def \Gevtwo {\rm GeV_{\rm 2}}
\def \pt {P_{T}}
\def \et {E_{T}}
\newcommand {\Et}{E_{T}}
\newcommand {\Pt}{P_{T}}

\def \ETmiss {{E}_{T}^{miss}}


\def  \met {\not\!\!\et }
\newcommand  {\vmet}{\vec{\met}}
\def  \abseta {\mid\eta\mid}
\def \radlen {X_{0}}
\def \etaphi {\eta,\phi}
\newcommand {\pizero}{\pi^{0}}
\newcommand {\DR}{\Delta R}

\section{Fourth Concept  (``4th'') Detector  }


The Fourth Concept detector differs from the other three concepts in several respects.
In contrast to the particle flow calorimetry adopted in the other concepts, the 4th concept utilizes
a novel implementation of compensating calorimetry, which balances the 
response to hadrons and electrons and so is insensitive to fluctuations in 
the fraction of electromagnetic energy in showers.
The demonstrated performance of the \dream dual-readout calorimeter lends
credibility to this concept. 4th is innovative in other respects as well, incorporating
dual solenoids and endcap coils to manage magnetic flux return and identify muons.

The key elements in the design are as follows:

\begin{itemize}
\item  The 4th concept uses projective towers of dual-readout fiber sampling calorimeters
to measure separately the hadronic and electromagnetic components of a shower, and so provide
``software compensation'' and excellent hadronic energy resolution. The towers have good transverse
segmentation, no longitudinal segmentation, a depth of 10 $\lambda$, and are read out with photo 
detectors at their outer radius.  
\item  The electromagnetic calorimeter is based on a crystal calorimeter, with readout of both 
Cerenkov and scintillation light to provide compensation, placed directly before the fiber
towers.
\item  Central tracking is provided by a large Time Projection Chamber (TPC). The excellent pattern
recognition capability of a TPC, its ability to measure $dE/dx$, and the high momentum resolution
possible with very high precision individual measurements, are a natural match to the new
ILC physics regime.   4th includes the option of a low mass, cluster-counting KLOE-style drift chamber 
which can be readout at each bunch crossing.
\item  A pixel vertex detector will be used for $b$ and $c$ quark tagging and accurate vertex reconstruction.
\item  The tracking chambers and calorimeter will be inside a 3.5 T axial field provided by a large
radius solenoid. A second, larger radius and lower field solenoid, with its B field opposite to that
of the inner solenoid, will provide flux return and a region where high spatial resolution drift
tubes can measure muon momenta to high precision.
\item  An endcap "wall of coils" confines the flux of the two solenoids in z, and eliminates the 
need for massive iron flux return system.
\end{itemize}

\begin{figure}[htb!]
 \epsfysize=7.0cm
  \centerline{\epsffile{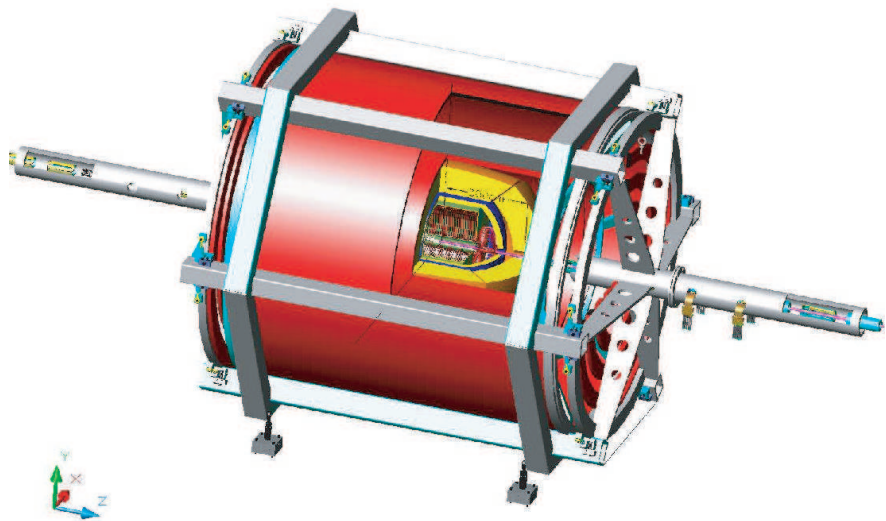}}
  \caption[3D view of the 4th concept detector]{Cut-away view of pixel vertex (blue), TPC (green),
     calorimeter (yellow), dual solenoids (red) and supports for
       muon spectrometer tubes, inside a frame, and the common
       support for beam line elements (purple).}
  \label{fig:AM-3}
\end{figure}

The 4th Concept detector is shown in Figure~\ref{fig:AM-3}. The detector and its performance is 
described in more detail in Ref.~\cite{ref-4th}.
  
\subsection{Tracking}

The pixel vertex detector is the same design as the {\normalsize SiD}
detector, and utilizes a 50 $\mu$m thick depletion region
with 15 $\mu$m $\times$ 15 $\mu$m pixels and sophisticated  front end 
processing and zero-suppression.  Its inner and outer radii are about 1.5~cm
and 8~cm, respectively,  in a  3.5~T field.  This  high
precision pixel vertex detector is essential for the tagging of $b$ and $c$
quarks and $\tau$ leptons, and the suppression of hit occupancies  so
near  to the beam.

The Time Projection Chamber (TPC) is very similar to those being developed 
by the {\sc gld} and {\sc ldc} concepts, in collaboration with the TPC R\&D
groups. It has sophisticated readout in a 3.5 T magnetic field and uses low
diffusion gas at moderate electron drift velocity.
In the new experimental physics regime of a TeV $e^+e^-$ collider, a
three-\-dimensional imaging tracking detector such as a TPC could well be
a significant advantage.
It presents very little material to passing particles; its two-track 
discrimination and spatial precision are ideal for observing long-lived
($\gamma \beta c \tau \approx$ 1-100~cm) decaying states; it offers
essentially complete solid angular coverage for
physics events; its measurement of ionization allows searches for free
quarks at $(1/3)^2$ or $(2/3)^2$ ionization, for magnetic monopoles, 
and for any
other exotically ionizing tracks. In addition, the multiple measurements
of the $z$-coordinates along the trajectory of a track yield a 
measurement of magnetic charge ($m$) by {\bf F} $= m${\bf B} bending. 
Finally, the $dE/dx$ ionization measurement of a TPC will assist physics
analyses involving electron identification, discrimination of singly
ionizing $e^-$ from a doubly ionizing $\gamma \rightarrow e^+ e^-$
conversion for aligned tracks, and other track  backgrounds.

With sufficiently high precision in the TPC,
{\it e.g.}, single-electron digital capabilities in a low diffusion
gas, it will not be necessary to incorporate auxilliary detectors (such 
as silicon strips surrounding the TPC on all its boundaries) in order to meet the momentum
resolution goal of $\delta(1/p_T) \approx 3 \times 10^{-5}$ (GeV/c)$^{-1}$.

An option for a gaseous central tracker is a cluster-counting drift chamber
modelled on the successful KLOE main tracking chamber.  This drift chamber 
(CluCou) maintains very low multiple scattering due to a He-based gas and aluminum 
wires in the tracking volume and utilizes carbon fiber end plates. Forward tracks 
(beyond $\cos \theta \approx 0.7$) which 
penetrate the wire support frame and electronics pass through only about 15-20\% $X_{0}$ of material.
The low mass of the tracker directly improves momentum resolution in the multiple scattering dominated region 
below 50 GeV/c.  The He gas has a low drift velocity which allows a new cluster 
counting technique\cite{ref-TrRev} that clocks in individual ionization clusters on every 
wire, providing an estimated 50 micron spatial resolution per point, a $dE/dx$ resolution 
near 3\%, and $z$-coordinate information on each track segment through an effective dip 
angle measurement.  The maximum drift time in each cell is less than the 300 ns beam crossing 
interval, so this chamber sees only one crossing per readout.  

The critical issues of occupancy and two-track resolution are being simulated for ILC 
events and expected machine and event backgrounds, and direct GHz cluster counting 
experiments are being performed.  This chamber has timing and pattern recognition
capabilities midway between the faster, higher precision silicon tracker and the slower, full imaging 
TPC, and is superior to both with respect to its low multiple scattering.

\subsection{Calorimetry}

The calorimeter is a spatially fine-grained dual-readout fiber
sampling calorimeter augmented with the ability to measure the 
neutron content of a shower.  The dual fibers are sensitive to scintillation and 
Cerenkov radiation, for separation of the hadronic and electromagnetic components of 
hadronic showers\cite{dream-ha}.  The energy
resolution of the tested \dream calorimeter should be surpassed with finer spatial 
sampling, neutron detection for the measurement of fluctuations
in binding energy losses, and use of a larger test module, to reduce
leakage fluctuations.  The calorimeter modules
will have fibers up to their edges, and will be constructed for sub-millimeter
close packing, with signal extraction done at the outer radius so that the
calorimeter system will approach full coverage without cracks.
A separate {\sc em} section is planned. It would be located in front of the
dual-readout calorimeter and consist of a crystal calorimeter
with  readout of both scintillation and \C light. This would provide 
better photoelectron statistics and therefore achieve better 
energy and spatial resolution for photons and electrons than is 
possible in the fiber calorimeter modules.  The dual
readout of these crystals is essential: over one-half of all hadrons
interact in the so-called {\sc em} section, depositing widely 
fluctuating fractions of {\sc em} and hadronic energy losses.

\begin{figure}[htb!]
 \epsfysize=12.0cm
  \centerline{\epsffile{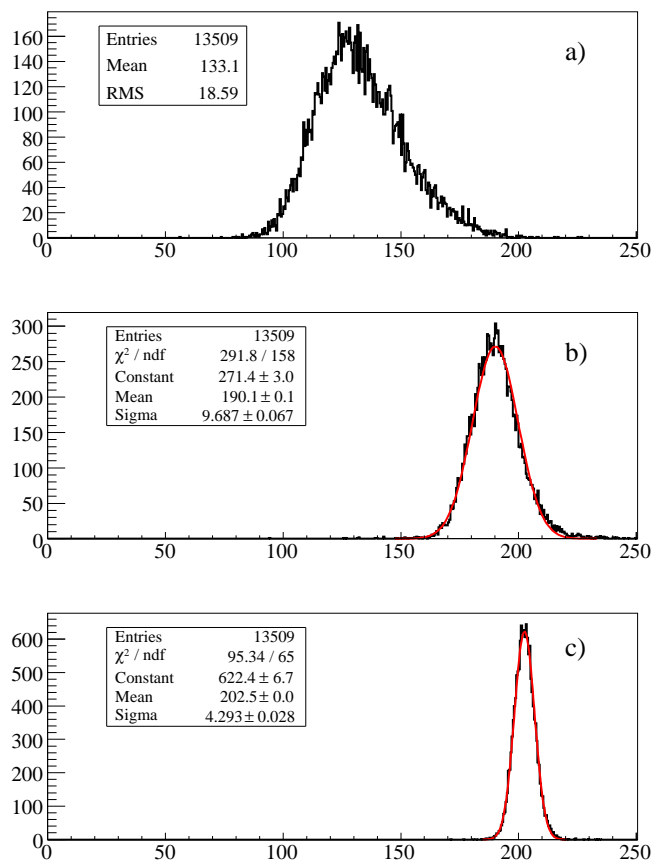}}
  \caption[Scintillator signal and energy distribution for 200 GeV $\pi^-$]{(a) The distribution of the scintillator ($S$) signal for 200 
   GeV $\pi^-$.  This is the raw resolution that a typical scintillating 
   sampling  calorimeter would achieve; (b) the leakage-dominated 
   energy distribution using only the 
   $S$ and $C$ (\C) signals  for each event.  
   (c) The energy distribution with leakage fluctuations suppressed 
   using the known beam energy (=200 GeV)  to make a better estimate
   of \fem~ each event.  The actual energy resolution of a fiber dual-readout
   calorimeter lies between Figures (b) and (c).} 
  \label{fig:3frames}
\end{figure}

The energy resolution achieved in the \dream calorimeter for incident 
200 GeV $\pi^{-}$ is shown in Fig. \ref{fig:3frames} for both 
leakage-dominated (Fig. \ref{fig:3frames}(b)) and leakage-suppressed
(Fig. \ref{fig:3frames}(c)) analyses.  The true resolution for a simple 
dual readout calorimeter is between  these two cases.
 
\begin{figure}[htb!]
 \epsfysize=9.0cm
  \centerline{\includegraphics[height=8cm,bbllx=0,bblly=0,bburx=183mm,bbury=135mm]{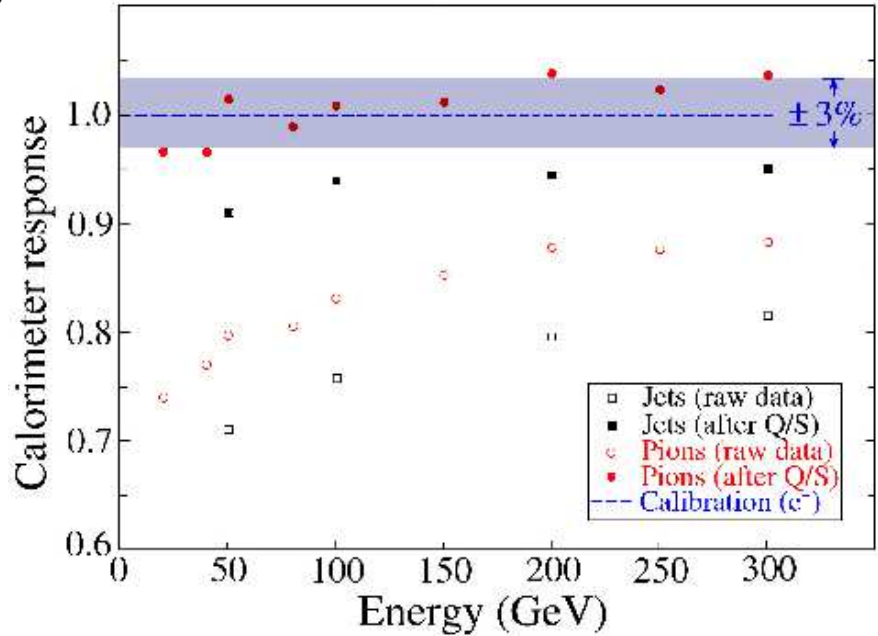}}
  \caption[Measured response of the \dream calorimeter to 20 and 300 GeV hadrons]{Measured response of the dual readout calorimeter for
  hadrons from 20 to 300 GeV.  The \dream module was calibrated
  only with 40 GeV electrons. }
  \label{fig:linearity}
\end{figure}

Finally, and very importantly, the hadronic response of this 
dual-readout calorimeter is demonstrated to be linear in hadronic
energy from 20 to 300 GeV having been {\it calibrated only with 
40 GeV electrons}. See Fig. \ref{fig:linearity}. This is a critical advantage at the ILC where
calibration with 45 GeV electrons from $Z$ decay could suffice to calibrate
the device up to 10 times this energy for physics.

\subsection{Magnetic field, muons and machine-detector interface}

\begin{figure}[htb!]
 \epsfysize=7.0cm
 \centerline{\epsffile{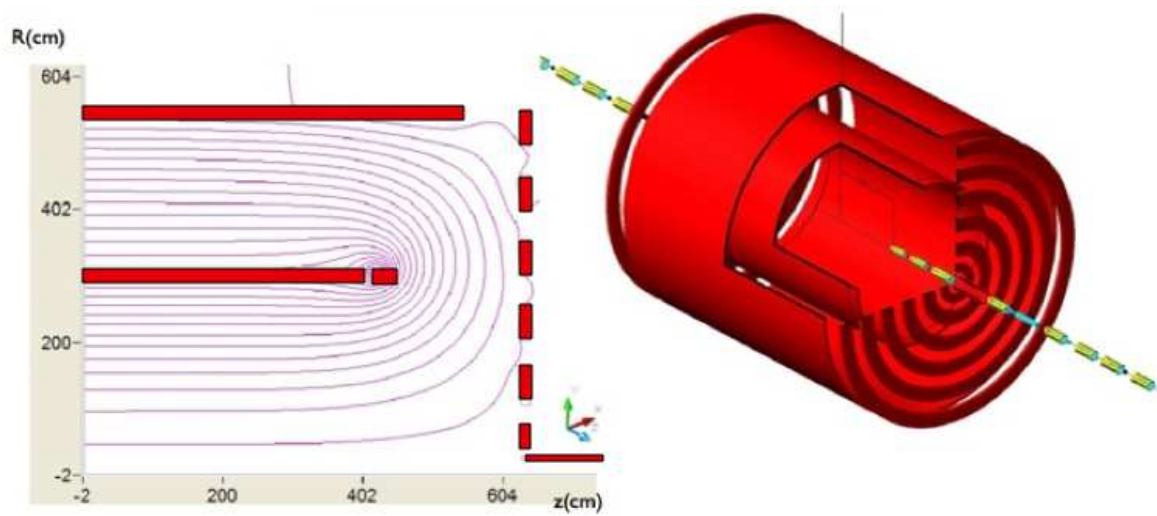}}
 \caption[The coils of the 4th concept and the field lines]{Drawings showing the two solenoids 
   and the ``wall of coils''
   that redirects the field out radially, and the resulting field lines in an
     $r-z$ view.  This field is uniform to 1\% at $3.5$ T in the TPC tracking
     region, and also uniform and smooth at $-1.5$ T in the muon
     tracking annulus between the solenoids. }
 \label{fig:B+coils}
\end{figure}

The muon system utilizes a dual-solenoid magnetic field configuration 
in which the flux from the inner solenoid is returned
through the annulus between this inner solenoid and an 
outer solenoid oppositely driven with a smaller turn density.  
The magnetic field in the volume between 
the two solenoids will back-bend muons which have penetrated the calorimeter 
and allow, with the addition of tracking chambers, a second momentum measurement.
This will achieve high precision without the limitation of 
multiple scattering in $Fe$, a limitation that fundamentally limits
momentum resolution in conventional muon systems to 10\%.    
High spatial precision drift tubes with cluster counting
electronics are used to measure tracks in this volume.  
The dual-solenoid field is terminated by a novel ``wall of coils''
that provides muon bending down to small angles 
($\cos \theta \approx 0.975$) and also
allows good control of the magnetic environment 
on and near the beam line. The design is illustrated in Fig~\ref{fig:B+coils}

The path integral of the field in the annulus for  a muon from 
the origin is about 3 T$\cdot$m
over $0 < \cos \theta < 0.85$ and remains larger than 0.5 T$\cdot$m
out to $\cos \theta = 0.975$, allowing both good momentum resolution
and low-momentum acceptance over almost all of $4\pi$.

For isolated tracks, the dual readout calorimeter independently provides a unique 
identification of muons relative to pions with a background track
rejection of $10^4$, or better, through its separate measurements
of ionization and radiative energy losses.

The detector's magnetic field is confined essentially to a cylinder with
negligible fringe fields, without the use of iron flux return. This scheme
offers flexibility in controlling the fields 
along the beam axis.  The twist compensation solenoid
just outside the wall of coils is shown in the above figure, along
with the beam line elements close to the IP.  The iron-free 
configuration \cite{Mik}
allows us to mount all beam line elements on a single support and
drastically reduce the effect of vibrations at the final focus (FF),
essentially because the beams will coherently move up and down together.
In addition, the FF elements can be brought close to the vertex chamber
for better control of the beam crossing.   

The open magnetic geometry of the 4th Concept also allows for future 
physics flexibility for asymmetric energy collisions, the installation
of specialized detectors outside the inner solenoid, and
magnetic flexibility for non-zero dispersion FF optics at the IP, 
adiabatic focussing at the IP, and monochromatization of the
collisions to achieve a minimum energy spread \cite{Mik}.
Finally, this flexibility and openness does not prevent additions
in later years  to the detector or to the beam line, and therefore no
physics  \cite{polar} is precluded by this detector concept. 

\subsection{4th Conclusions}

The four sub-detectors are integrated, at least at this concepts
stage, to achieve high precision
measurements of all the partons of the standard model,
including \Wjj and \Zjj decays and $\nu$'s by their
missing momentum vector.  The ability to use precision
calorimeter calibrations at the $Z$ for even very high energy
jet energy measurements will be a significant advantage.

\section{Concepts Summary}
The four detector concepts described above have been developed in response to the
physics and environmental challenges posed by the ILC. All deliver levels of performance
beyond the current state of the art, and all employ new detector technologies currently
under development. As shown in Chapter~\ref{subdetector_performance} on subdetector performance and in Chapter~\ref{physics_performance} on 
integrated physics performance, these detector concepts will do justice to the ILC physics
program, and they will do so with technologies that are within reach. Table~\ref{Detectors:concepts}
 presents, for comparison, some of their key parameters. The four concepts use complementary approaches
and a variety of technology choices. Three of the concepts
choose TPCs combined with silicon detectors for charged particle tracking; one chooses a pure silicon tracker. Three of the
concepts rely on particle flow calorimetry, although their implementations vary: SiD chooses
a compact design with high magnetic field; GLD pushes the calorimeters out in radius and
along z, with a comparatively lower field; and LDC is intermediate in size and B field. 
The Fourth concept follows a different philosophy altogether, with
compensating calorimetry and a (comparatively) moderate field. Other aspects of the designs
differ too. SiD, with its higher B field, can move its vertex detector to smaller radii. Fourth,
with its novel dual-solenoid design, eliminates the mass of a traditional iron flux return.
GLD adopts a common readout technology, using Multi Pixel Photon Counters, to readout scintillator
in the electromagnetic and hadronic calorimeters, and in the muon system as well. LDC and
SiD employ the extremely fine segmentation possible with Si/W electromagnetic calorimetry.

\begin{table}[hbt]
\begin{center}
\caption{Some key parameter of the four ILC detector concepts.
\label{Detectors:concepts}}
\begin{footnotesize}
\begin{tabular}{lcccc}
\hline
           & GLD   & LDC    & SiD   & 4th \\
\hline
VTX       & pixel  & pixel  & pixel & pixel \\
\# of layers  & 6  & 5      & 5     & 5     \\
\# of disks   & 2  & 0      & 4     & 4     \\
inner radius (cm) & 2.0 & 1.6 & 1.4 & 1.5\\
outer radius (cm) & 5.0 & 6.0 & 6.1 & 6.1  \\
\hline
Main tracker & TPC/ Si & TPC/ Si & Si & TPC/ drift \\
inner radius (TPC/ Si)(cm) &45 & 30 (16) & 20 &  20 \\
outer radius (TPC/Si)(cm) & 200& 158 (27) & 127 & 140 \\
half length (TPC/Si)(cm)  & 230& 208 (140) &  168& 150 \\
\# of TPC points & 200 & 200 & - &  200/ 120 \\
\# of Si points (barrel) & 4 & 2& 5 &  \\
\# of Si points (endcap) & 7 & 7& 4 & \\ 
\hline
ECAL & Scint.-W & Si-W & Si-W & Cystal\\
inner radius (cm) & 210 & 160 & 127 & 150 \\
outer radius (cm) & 229.8 & 177 & 140 & 180    \\
half length (barrel,cm ) & 280 & 230 & 180 & 240 \\
\# $X_0$ & 27 & 23 & 29 & 27 \\
\hline
HCAL & Scint-Fe & Scint - Fe & RPC/ GEM - W & fiber Dream \\
inner radius (cm)        & 229.8 & 180 & 141 & 180    \\
outer radius (cm)        & 349.4 & 280 & 250&  2.80 \\
half length (barrel, cm) & 280 & 230   & 277.2& 2.8    \\
\# of $\lambda$          & 5.8 & 4.6   & 4.0   & 9 \\
\hline
Magnet & & & \\
type & main & main & main & inner/ outer \\
field strength (T) & 3 & 4 & 5 & 3.5 / -1.5\\
radius (cm) & 400 & 300 & 250 & 300 / 550 \\
half length (cm) & 475 & 330 & 275 & 400/ 600 \\
\hline
Overall Detector & & & \\
radius (cm) & 720 & 600 & 645 & 550 \\
half-length (cm) & 750 & 620 & 589 & 650 \\
\hline 
\end{tabular}
\end{footnotesize}
\end{center}
\end{table}

The next step for each of the concepts involves moving beyond their present baselines,
and developing optimized detector designs. Recent advances with Particle Flow Algorithms
and further developments in realistic detector simulations are making this possible.
Each concept must study integrated physics performance and cost vs. variations in 
B field strength, ECAL inner radius, ECAL length, and HCAL depth. Once global parameters are
refined, subsystem parameters must be optimized, subsystem conceptual engineering designs
developed, integrated physics performance benchmarked, and favored subsystem technologies
selected. 



\cleardoublepage

\chapter{Machine Detector Interface}
\label{detector_MDI}
Even more so than at previous colliders, the final part of the accelerator, 
the beam delivery and the final focus, are closely coupled with the experiment 
at the interaction region. The design and management of this 
machine-detector-interface (MDI) is therefore a very important part 
of the design of the detector and the machine, and has consequences 
for both. 

The machine detector interface is concerned with the consequences of the 
beam delivery system to the experiment, and all design aspects of the 
interaction region (crossing angle, final focusing elements, 
etc.) and the interfacing of the detector with this interaction region. 
Of particular importance is the optimization of the interaction region 
in view of beam induced backgrounds. 

Closely related though strictly speaking not part of the MDI are the 
measurement of the luminosity, the measurement of the beam energy and 
the determination of the beam polarization. 

Since the infrastructure needed to assemble and operate the detectors 
has repercussions on the design and the layout of the machine in this 
region as well, a brief discussion of these aspects is included in 
this part as well. 

\section{Interaction Regions}
\label{sec-MDI-IR}
The interaction region is meant to include the design of the machine and 
of relevant parts of the detector between the final focusing elements 
and the interaction point. The design of the interaction region seen from 
a MDI point of view has to serve a number of different functions: The beam
has to be delivered through the largest possible aperture to the interaction 
point. A series of detectors record the interaction, in particular the 
remnants from the interaction in the very forward direction. The interaction 
region also has to shield the rest of the detector efficiently from 
backgrounds produced in the collision and from sources upstream and downstream of the detector. 

A particular challenge to the design of the interaction region is that 
it has to accommodate the wide range of parameter sets discussed in the 
Reference Design Report (RDR), and for a wide range of beam energies, from 90~GeV to 1~TeV CM. 

To serve these needs the interaction region designs of the different 
detector concepts all include a masking scheme, often realized as 
tungsten masks which shield the incoming and the outgoing beam, and 
a set of detectors, to detect the background particles. 

The baseline of the ILC foresees one interaction region with 
a crossing angle between the electron and the positron beam of 14~mrad. 
Alternatives are interaction regions with much smaller (2~mrad) crossing 
angles, and interaction regions with 20~mrad and more. Two detectors that share occupancy of the interaction point in ``push-pull'' mode are planned. 

\subsection{Beam Induced Backgrounds Sources}
\label{sec-MDI-bgd-src}
A number of different processes create backgrounds related to the beam 
which are potentially problematic for the detector. The main 
sources of such backgrounds are: 
\begin{itemize}
\item ``Beamstrahlung'' created in the interaction of the tightly focused 
electron and positron beams. Beamstrahlung generates: 
  \begin{itemize}
    \item Disrupted beam
    \item Photons, radiated into a very narrow cone in the forward direction;
    \item Electron-positron pairs, radiated into the forward direction, 
      and steered by the 
      collective field of the opposing beam and the central magnetic field of the detector solenoid. 
  \end{itemize}
\item Synchrotron Radiation, created upstream in the beam delivery system, in particular, by
  the non-Gaussian tail of the beam interacting with
  the final focusing elements near the interaction point. 
\item Muons, created by interaction between collimators that define the maximum aperture and tails in 
  the electron or positron bunches, and transported through the tunnel 
  into the detector. 
\item Neutrons created from off energy e+e- pairs and disrupted beam that strike beam line components before the beam dumps, and neutrons created in the beam dumps that are backscattered into the detector.
\item Hadrons and muon pairs created by $\gamma\gamma$ interactions.
\end{itemize}
Although particles from the beamstrahlung go primarily into the very forward 
directions, and mostly exit the detector together with the outgoing 
beam, a small but still significant fraction have sufficient 
transverse momentum to hit detector or beam-line components, and interact
with them. Particles created or backscattered in these interactions 
are a major source of background in the detector. 

\subsection{Interaction Region Layout}
\label{sec-MDI-IR-Layout}
\begin{figure}
	\centering
		\includegraphics[height=6cm]{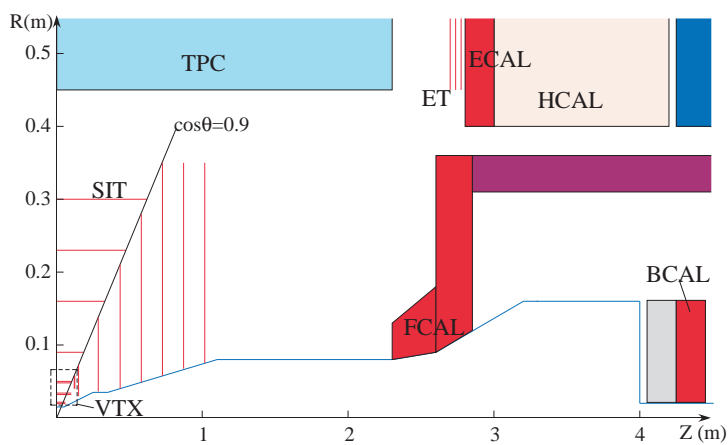}
	\caption[Sideview of an interaction region.]{Side view of the interaction region and very forward region of a typical ILC detector.}
	\label{fig:gldfwd060320}
\end{figure}

A typical layout of an interaction region is shown in figure~\ref{fig:gldfwd060320}. The beam pipe has its smallest 
radius right at the interaction point, at the left edge of the figure. It flares to larger radii as z increases, to 
give room to the charged background particles, which are channeled by the magnetic field into the forward directions. 
Two components of the detector are of particular importance to background studies: the BeamCal and the FCAL/Lumical (different names have been chosen by different concepts). Both are small, compact calorimeters located close to the beams, and both are 
subject to significant background radiation. 

The different background sources have a significant impact on the design of the 
magnetic field in the interaction region. To guide the charged background 
particles out of the detector, the direction of the field should 
point in the direction of the outgoing beam which is passing through the solenoid off-axis. This can be achieved through 
the superposition of the conventional solenoidal field from the detector 
with a dipole field, produced by adding some dedicated dipole windings 
to the detector solenoid. This so-called Detector Integrated Dipole (DID)
becomes effective once the crossing-angle increases beyond a few mrad. For 
historical reasons, such a situation whereby the field is aligned with the outgoing 
beam, is called an 'Anti-DID', and is the preferred solution. 

\subsection{Background Estimation}
\label{sec-MDI-bgd}
The consequences of the different background sources discussed in 
section~\ref{sec-MDI-bgd-src} have been studied in simulation for all 
detector concepts. To simulate the beam-beam interaction the 
Guinea Pig \cite{ref-Guinea-Pig} and the CAIN \cite{ref-CAIN} programs 
have been used. Background simulations dealing with muons and neutrons are made using MUCARLO \cite{ref-MUCARLO}, MARS\cite{ref-MARS}, and FLUKA\cite{ref-FLUKA}. 
Synchrotron radiation (SR) was simulated by tracking the scattering of beam halo particles, and the consequent radiated photons, using GEANT.

The output from these programs is input to a 
complete and detailed material simulation of the interaction region elements and the detectors 
\cite{ref-BRAHMS,ref-MOKKA,ref-SLIC, ref-JUPITER}, which are based on GEANT4 \cite{ref-GEANT3, ref-GEANT4}. The simulations have been 
done, if not otherwise noted, for the nominal parameter set \cite{ref-ILCparameters}, but some studies have been performed for a range of parameters as well. 
All studies include an Anti -DID field and are based on a 14~mrad 
Crossing-angle scenario. In a few cases the variations expected 
for different crossing angles are shown for comparison. Backgrounds due to primaries, as well as secondary and tertiary particles produced in interactions of primaries with the IR materials, are tracked and evaluated at different critical detector subsystem locations.

\subsubsection{Pair Background}
\label{sec-MDI-bgd-pair}
Electron-positron pairs are created in great number in the interaction 
of the primary electron  and positron bunches. They travel mostly in the 
direction of the outgoing beams. The magnetic field will tend to 
focus one charge of particle, and tend to defocus slightly the other, 
depending on the direction of travel. A small number of pairs are produced with
large enough transverse momenta to enter directly into detector components. 
An important source of backgrounds are secondary particles, 
created in the interaction of pairs with detector or machine elements. 
Some of these secondary particles may travel back into the detector, and create 
background hits. 

\begin{figure}
	\centering
		\includegraphics[height=8cm]{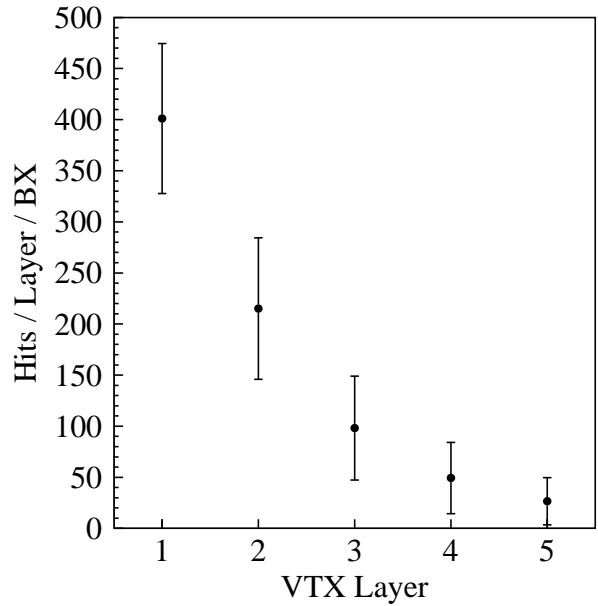}
	\caption{Background induced hits in the VTX detector per bunch crossing.}
	\label{fig-MDI-pairs1}
\end{figure}

\begin{figure}
	\centering
		\includegraphics[height=8cm]{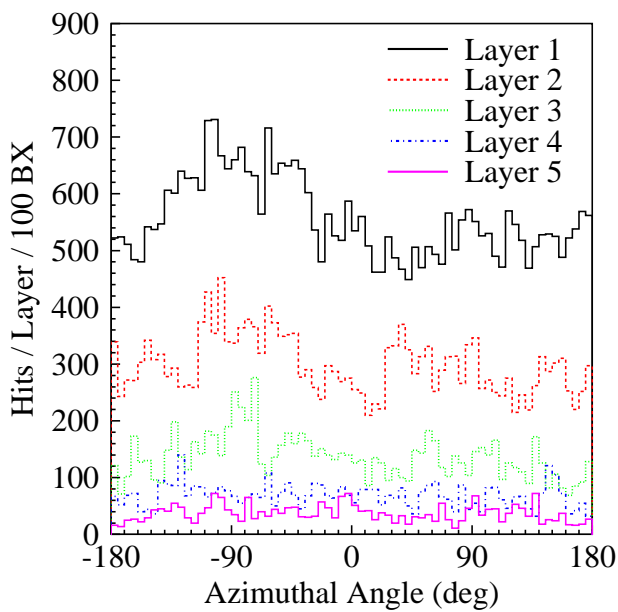}
	\caption[Background hits in the VTX detector vs. azimuthal angle.]{Background induced hits in the VTX detector, as a function of the azimuthal angle, $\Phi$}
	\label{fig-MDI-pairs2}
\end{figure}

The detector most sensitive to this is the vertex detector. Depending on detector concept, its innermost 
layer sits at a radius that varies from $1.3$ to $2.0$ cm from the 
interaction point. The total number of hits as a function of the layer number 
in the Vertex detector is shown in Figure~\ref{fig-MDI-pairs1}. While the majority of these hits are caused by the $e^+e^-$ pairs directly reaching the vertex detector layers, some hits are also caused by secondary $e^+e^-$ produced in the far forward detector. The azimuthal 
distribution of these hits is shown in Figure~\ref{fig-MDI-pairs2}. The clear 
non-uniformity observed in backscattered hits is an effect of the Anti-DID field, and is basically an image of the hole for the outgoing beam. The backscattering rate is, however, highly dependent on the fringe field of the detector solenoid, the Anti-DID field, and the far forward detector geometry, and further optimization is possible.

The VTX hits shown in Figures~\ref{fig-MDI-pairs1} are for one bunch crossing, 
while Figure~ \ref{fig-MDI-pairs2} shows the results from 100 superimposed 
bunchcrossings. The innermost layer hits will reach 1\% occupancy in about 150 bunches, imposing a constraint on the vertex detector readout speed. This constraint has been long appreciated and is motivating the development of several new technologies for pixilated vertex detectors. 

\begin{figure}
	\centering
		\includegraphics[height=10cm]{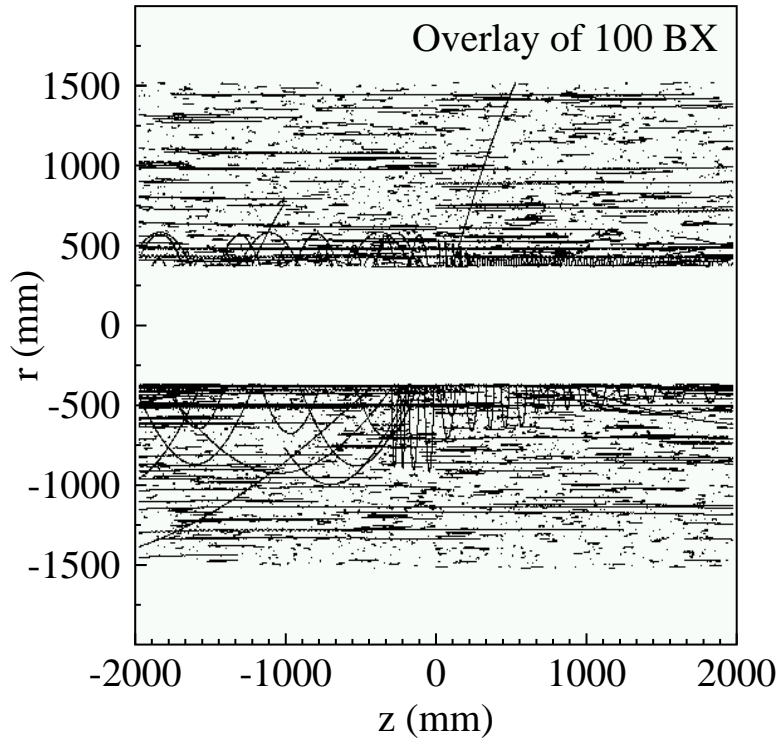}
	\caption{Hits produced in the TPC from pairs}
	\label{fig-MDI-TPCpair}
\end{figure}

The number of hits at radii outside the Vertex detector, i.e. $>  10 $cm, 
falls off very rapidly. For a Silicon-based tracking system they are not a 
real concern. For a TPC-based tracker, where a large (O(100)) bunches is 
integrated into one readout frame, they are potentially more 
important. In Figure~\ref{fig-MDI-TPCpair} the distribution of hits in the 
TPC is shown, integrated for 100 bunch crossings (though hits from different 
bunch crossings are not displayed in time in this picture). The total occupancy of the 
TPC in this case is far below one percent, and does not present a problem. 

In a few rare cases, pair-induced background creates photons of 
high enough energy to actually create tracks in the detector. The 
tracks expected from 100 bunch crossings are also visible in Figure~\ref{fig-MDI-TPCpair}. Their number is small and does not present a problem. 

\begin{figure}
	\centering
	\begin{tabular}{cc}
		\includegraphics[height=7cm]{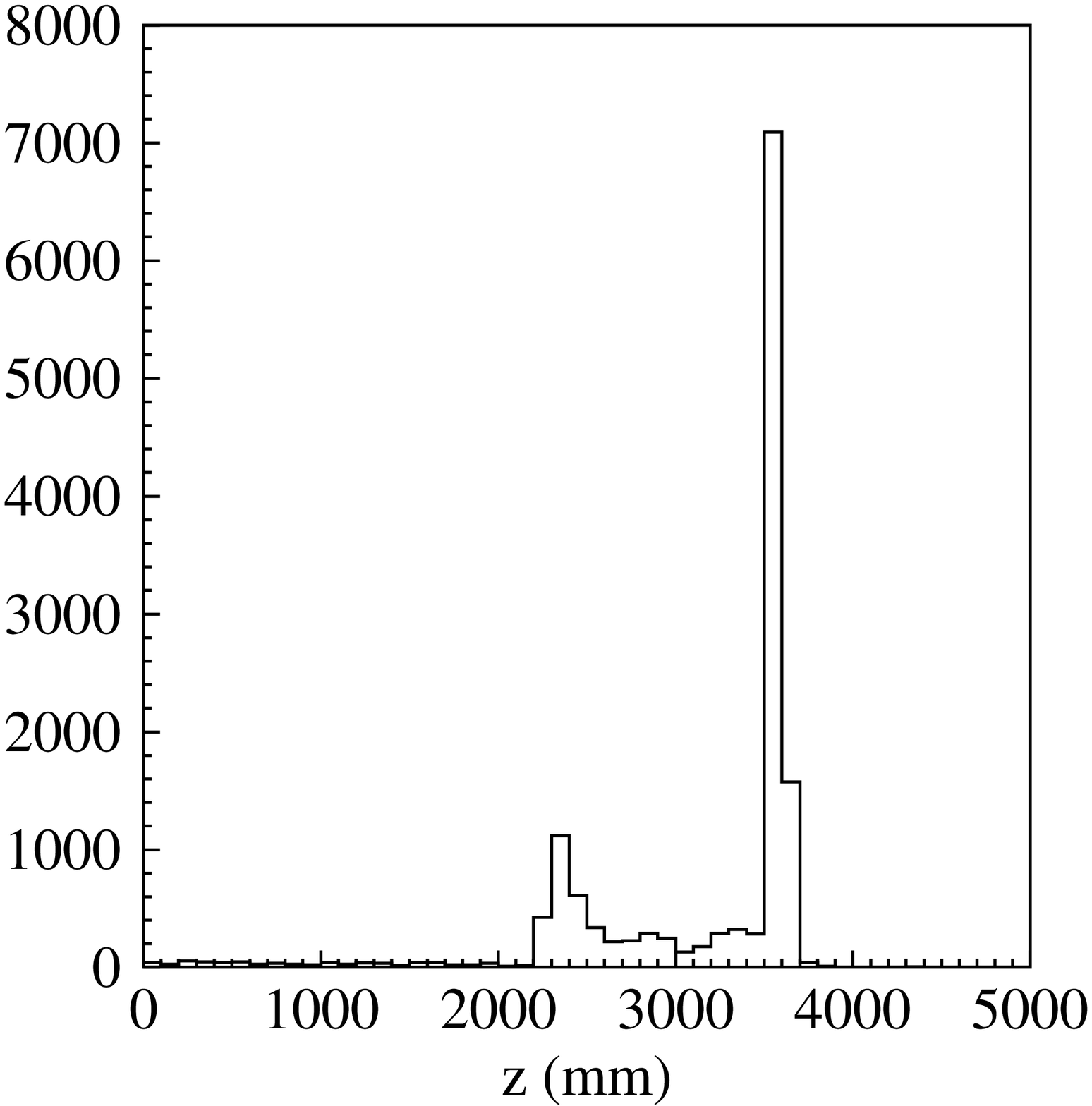} & 
		\includegraphics[height=7cm]{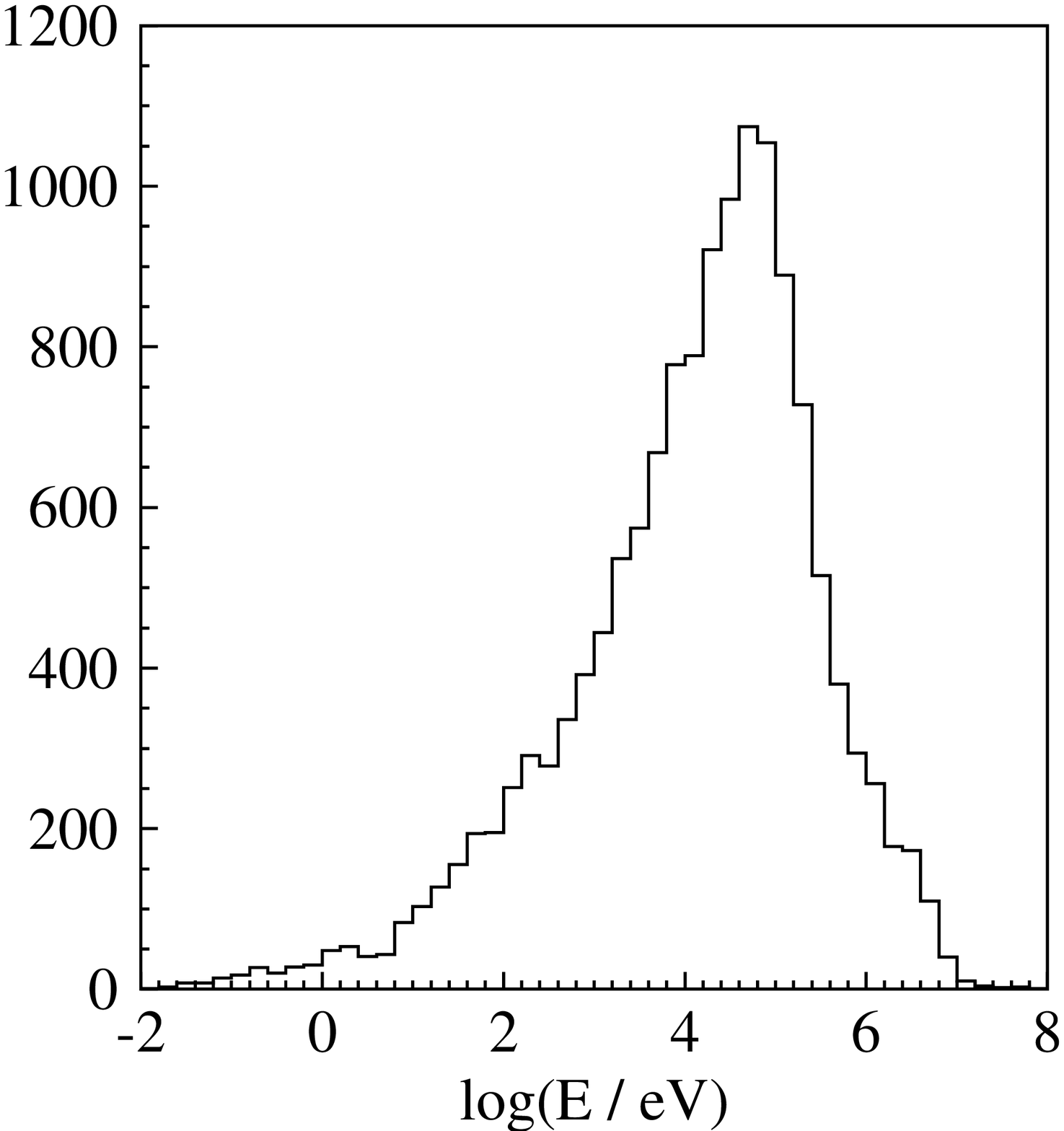}
  \end{tabular}
	\caption[Position of major sources for neutron background in the detector]{Position of the major sources for neutron background in the detector, as a function of the position along the beam (left), and the energy distribution of the neutrons which reach the TPC (right).}
	\label{fig-MDI-neutrons}
\end{figure}

The pairs background also produces a significant neutron background in the 
detector. Most of these neutrons are created in electromagnetic 
showers in hot regions of the innermost calorimeter, as well as 
the closest beam elements. The origin of neutrons is illustrated in 
Figure~\ref{fig-MDI-neutrons}, together with their energy spectrum. These 
neutrons are important for a number of reasons: 
The Si-based vertex detector and trackers are sensitive to bulk damage by 
neutrons. The total dose of neutrons collected should stay below a 
flux of $10^{10}$/cm$^{2}$/year. In the detectors equipped with a hydrogenous gas-filled TPC 
the neutrons can create spurious hits in the gas. The possibly most 
affected detector however is the end cap of the calorimeter, in 
particular the hadronic calorimeter, where the neutrons create 
spurious hits, and contribute to the confusion term in the 
particle flow measurement. 

\subsubsection{Photon Background}
\label{sec-MDI-photons}
A by-product of the beam-beam interaction is a large number of photons, which are radiated primarily in the forward
direction. These photons carry a significant amount of energy. They follow within a very narrow 
cone the direction of the incoming beam, and are thus mostly exiting the detector through the 
outgoing beam hole. Nevertheless there are tails in the distribution of these photons to larger transverse 
momenta, so that some photons hit the different elements in the beam line in the very forward 
direction. Similar to the case with pairs, these photons initiate showers in the forward detectors, 
and some particles from these showers make it back into the detector. After  pair-related backgrounds, particles 
created from beamstrahlung photons are the most important background in the detector.

\subsubsection{Synchrotron Radiation Background}
\label{sec-MDI-sync}

Another potentially important source of background in the detector are synchrotron radiation photons. These can be produced in wakefield-induced beam scattering from the jaws of the upstream collimation system, as well as in the final focusing elements of the beam delivery system. The collimation system is being designed to ensure that no direct or single-scattered photons can reach any sensitive detector parts. Detailed studies of this are still ongoing. The impact on the SR flux in the IR due to variation in (non-ideal) beam conditions, eg. beam position jitter near the collimator jaws and in the strong fields of the final focus magnets, is a source of concern; this will require extensive study for an ensemble of realistic machine conditions.

\subsubsection{Beam Halo Muon Background}
\label{sec-MDI-muon}

Muons are a major source of background as they can be produced in
abundance in the collimation section upstream of the interaction
point. The beam halo, whose population is difficult to predict, is
scraped away by the collimators producing electromagnetic showers a
few hundred meter upstream of the detector. The muons produced in these
reactions can travel through the beam delivery system (BDS) tunnel towards the detector and
can eventually result in unwanted mostly horizontal tracks in the
tracking systems.

Simulations \cite{ref:MDI-schreiber}, \cite{ref:MDI-keller}, \cite{ref:MDI-SiD} have been performed with simulation software tools based on GEANT, MUCARLO, and MARS,
which model the collimation system, the
beam line elements and the full tunnel up to the detector hall. The
baseline of the ILC BDS foresees a 5m long magnetised iron spoiler
inside the tunnel which should help to reduce the muon flux from the
collimation system. The simulations predict $\approx 12$ ($\approx 1.7$) muons per
bunch crossing passing through the detector for a 500 (250) ~GeV beam.
This yields a load per bunch crossing of less than 3 (0.5) muons passing through the central 
tracking device
(i.e. at a radius lower than 2.5 m around the nominal beamline). 
These muons are potentially a more serious problem for those detector concepts which foresee 
a TPC as a central tracker, since a TPC integrates over around 150 bunches. Thus for a TPC we 
expect less than 400 (60) muons, or about the same number of horizontal tracks overlaid in
one TPC image. The simulations assumed a halo fraction in the beam of 0.1\%,
meaning 0.1\% of each bunch is scraped in the collimation system. This
estimate of the beam halo fraction is considered to be conservative. Studies of the 
impact of this background on the tracking detectors are still ongoing, but first 
results indicate that this level of background tracks does not 
present a problem for a TPC.

\subsubsection{Background Rates: Summary}
\label{sec-MDI-rates}
In Table~\ref{tab-bgd} the expected occupancies for different sub-detectors 
are summarized. In each case a range of expected occupancies is given, which 
covers the range of numbers reported by the different detector concepts. 
Also given is an estimate of the occupancy considered critical, i.e. where 
reconstruction starts to suffer because of the background hits. 

\begin{table}[thb]
\begin{center}
\caption[Estimated detector occupancy for different background sources]{\label{tab-bgd} Estimated detector occupancy from different background sources. Given is the occupancy from the particular background, and the 
value of the critical occupancy, where problems in the reconstruction are 
expected. The expected occupancy is quoted as a range to allow for the 
different detector concepts discussed. }
\begin{tabular}{llll}
\hline
Vertex Detector &&& \\
\hline
Background Source & expected occupancy & critical occupancy & remark
\\
\hline
Pairs: direct & $\leq$1\% & 1\%& r=1.5 cm \\
Pairs: backscatter & $<<$1\%& & \\
\hline
Beam Halo Muons & & & \\
\hline
\hline
Tracking (TPC) &&& \\
\hline
Background Source & expected occupancy & critical occupancy & remark
\\
\hline
Pairs: direct & $<<$0.02\% & 1\% & \\
Pairs: backscatter & $\leq$0.2\%& & \\
\hline
Beam Halo Muons & $\leq$0.15\% (384 $\mu$/200 BX)& under study& ass.\ 0.1\%
loss\\
&&& in coll. sys.\\
\hline
\hline
Tracking (Silicon) &&& \\
\hline
Background Source & expected occupancy & critical occupancy & remark
\\
\hline
Pairs: direct & $\leq$0.2 cm$^{-2}$BX$^{-1}$&0.2 cm$^{-2}$BX$^{-1}$&
forward\\
Pairs: backscatter & $<<$0.2\%& & region\\
\hline
Beam Halo Muons & under study& under study& \\
\hline
\end{tabular}
\end{center}
\end{table}

\section{Detector Integration}
\label{ref-MDI-integration}
The baseline design of the ILC foresees one interaction region, 
equipped with two detectors. The two detectors and the infrastructure serving them 
are laid out in such a way that each can be moved quickly into and out of the 
interaction region (push-pull operation) thus allowing the sharing of 
luminosity between both detectors. Details such as switchover time, 
switchover frequency etc. are still under discussion. Similarly since 
no detailed engineering study has yet demonstrated the feasibility of such a push-pull 
scheme, an option with two beam delivery systems remains under investigation.

To minimise the underground hall size and the interference between 
detector construction and machine construction, most of the detector 
assembly will take place on the surface. Once assembled and 
in part commissioned, sub systems of the detector will be 
brought into the hall for final assembly. 

The hall itself then only has enough space to allow the assembly of the
different major parts into the full detector, and to do detector service 
during shutdown periods. The hall will be designed in such a way that 
one detector can be serviced while the other one is running, to minimise 
the downtime of the accelerator. 

A typical detector underground hall is between 45 and 60 m long, 
and has a transverse dimension around 30 m. 
Installation of each detector requires an access shaft into the hall, 
equipped with a large crane. Depending on the concept, and on the 
maximum size of components to be lowered into the hall, this crane 
might need a load capacity of up to 2000 tons. Inside the hall 
a system of medium sized cranes and air pads will be used to maneuver 
and integrate the different components. 

A major challenge is to design the detector in 
a way which will allow access to its inner parts, in 
particular the vertex detector, in a short time and within the 
space available in the detector hall. The currently favored 
concept followed by SiD, LDC and GLD, foresees a longitudinal 
opening of the detector in the beam position, which will 
provide access to the vertex detector. 

Another major challenge is to engineer the mechanical detector concept for 
push-pull capability. Apart from issues such as maintaining the internal detector alignment and avoiding recalibration, a design must be developed for servicing of the different superconducting 
parts on the detector during and after a move. This will require careful engineering 
to ensure a smooth switchover from one detector to the other.

The elements of the beam delivery system in close proximity to the detector require careful integration and engineering.  These include the final quadrupole doublet (QD0 and QF1) with their integrated sextupole and octupole correction elements, the beam position monitors and kickers that keep the beams in collision, the crab cavities that rotate the beam bunches into head-on collisions, and the extraction line quadrupole magnets that direct the beams cleanly to high power beam dumps. 

The magnets closest to the interaction point will be housed in a common cryostat running with liquid He-II.  The compact winding technology developed by BNL will allow QD0 with its sextupole and octupole elements, the first extraction quad, and a dual solenoid winding to cancel the detector's residual solenoid field on the axis of the incoming beam, to be housed in a common 20-25cm radius cryostat.  This cryostat will be an integral part of the detector, moving with it when the detector is pushed onto or pulled out of the beamline. If a rapid exchange of detectors is to be made possible, each detector will need to house a source of cryogenic fluids for this system that moves with the detector.  If the detector is to be serviced while it is on the beamline, both the support system for the cryostat and the cryogen feed system must accommodate the motion of those parts of the detector (door, endcap segments, etc.) that must move to provide access.

 It is thought that the longitudinal position of the magnet, defining the IP-QD0 drift space, $L^*$, can be optimized for each detector concept.  It will be an engineering challenge to support the magnet in a manner that minimizes vibration transmission from the detector to QD0, allows for alignment and feedback systems to correct its position against slow (diurnal) drifts, and resists any net residual magnetic forces.  

A similar set of engineering challenges exist for the forward calorimetry, forward tracking elements, vertex detector package and beam pipe.  These all occupy the critical 20-25cm inner radial volume of the detector.  Support schemes that work while the detector is closed for data taking or open for minor repairs must be provided for each detector while minimizing materials and allowing cables and tubes to power, readout and cool the detectors.  Both the delicate nature of the thin Be beampipe in the vicinity of the IP and the massive W/Si calorimeters and masks must be taken into account while not jeopardizing the vibration free support of the final quadrupole.

The second magnet cryostat, housing the QF1 quadrupole with its sextupole and octupole correctors and the next elements of the extraction line, will begin about 9m from the interaction point.  Given the 14~mrad crossing angle, this package will require a larger radius.  As it will stay fixed in the hall, its major impact on the detector will be to limit the maximum amount a detector might be opened while it is in its beamline position.

Between the two cryostats a warm section of beamline is foreseen to house the electrostatic kicker that, in conjunction with a BPM just behind the BCAL, measures the beam-beam kick and minimizes it to keep the nanometer size beams in optimal collisions in the face of natural occurring ground vibrations or residual equipment vibrations that might be transmitted to QD0.  The potential of electromotive interference (EMI) to sensitive detector electronics from the feedback kicker, the pulsed crab cavity and the beam itself must be mitigated by careful design and testing.

It is essential that experimenters have full access to the detector in the off-beamline position, whether or not the detector on the IP is taking data.  Radiation safety considerations imply that the personnel servicing the detector be protected by sufficiently thick external shielding walls or that each detector be constructed in such a manner (free of cracks and using sufficient high Z material) so as to be self-shielding.  If the self-shielded-detector model is adopted, devoted shielding around the beamlines outboard of the detector endcaps, moveable to provide access when the detector is opened for quick, on-beamline repairs, will be required.  If external shielding can be avoided, the push-pull switchover time will be shorter and the size of the cavern reduced; moreover, with self-shielding, detector systems of the on-IP detector requiring human access during data taking will not be required to be located behind a second external shielding system, further simplifying the interchange of detectors.

\section{Luminosity, Energy, and Polarisation}
\label{sec-MDI-LEP}
The precise knowledge of the beam parameters are of great 
importance for the success of the physics program at the ILC. 
The main parameters measured by the detectors or instrumentation 
very close to the detectors are the luminosity, the energy 
of the beam, and the polarisation. 

\subsection{Luminosity}

Precision extraction of cross sections depends on accurate knowledge of the luminosity.
For many measurements, such as those based on threshold scans, one needs to know not
only the energy-integrated luminosity, but also the luminosity as a function of energy,
$dL/dE$.
Low-angle Bhabha scattering detected by dedicated calorimeters can provide the
necessary precision for the integrated luminosity. Options include secondary emission
(A) and fast gas Cerenkov (B) calorimetry in the polar angle region from 40-120~mrad.
Acollinearity and energy measurements of Bhabha, $e^+e^-\rightarrow e^+e^-$ , events in the polar
angle region from 120-400~mrad can be used to extract $dL/dE$ and are under study.
Additional input from measurements of the beam energy spread and beam parameters
that control the beamstrahlung spectrum will improve this determination of $dL/dE$.
Techniques include measuring the angular distributions of $e^+e^-$ pairs (C) in the polar
angle region from 5-40~mrad, and measuring the polarization of visible beamstrahlung in
the polar angle region from 1-2~mrad (D).
All the proposed detectors may also be used for real time luminosity monitoring and
tuning.

\subsection{Energy}

Beam energy measurements with an accuracy of (100-200) parts per million are needed
for the determination of particle masses, including m$_{\rm top}$ and m$_{\rm Higgs}$. Energy measurements
both upstream and downstream of the collision point are foreseen by two different
techniques to provide redundancy and reliability of the results. Upstream, a beam
position monitor-based spectrometer is envisioned to measure the deflection of the beam
through a dipole field. Downstream of the IP, an SLC-style spectrometer is planned to
detect stripes of synchrotron radiation (SR) produced as the beam passes through a string
of dipole magnets. The downstream SR spectrometer also has the capability to measure
the beam energy spread and the energy distribution of the disrupted (from beam-beam
effects) beams.

\subsection{Polarization}

Precise measurements of parity-violating asymmetries in the Standard Model require polarization measurements with a precision of 0.25\% or better. High statistics Giga-Z running requires polarimetry at the 0.1\% level. The primary polarization measurement will come from dedicated Compton polarimeters detecting the backscattered electrons.  
To achieve the best accuracy for polarimetry and to aid in the alignment of the spin vector, it is necessary to implement polarimeters both upstream and downstream of the IR. The upstream Compton polarimeter measures the undisturbed beam before collisions. The relatively clean environment allows a laser system that measures every single bunch in the train and a large lever arm in analyzing power for a multi-channel polarimeter, which facilitates internal systematic checks. The downstream Compton polarimeter measures the polarization of the outgoing beam after the collision. The extraction line optics is chosen to be focused at the Compton IP such that its polarization is very similar to the luminosity-weighted polarization at the interaction point. The polarization of the undisturbed beam can be measured as well with non-colliding beams.  Backgrounds in the extraction line require a high power laser that probes a few bunches per train.

The precise measurement of the scattered electrons require high-precision 
detectors. Several technologies are under investigation. The most promising technique 
to-date appears to be Cerenkov detectors. The current baseline design for the Cherenkov detectors consists of  gas tubes read out by photomultipliers. Alternative or additional possibilities are under study.



\section{Summary and Outlook}
\label{sec-MDI-summary}
The understanding of the interaction region of the ILC and its impact 
on the detector performance has matured remarkably over the last few years. 
Good simulation tools are available and serious studies have been done
to understand the background situation. 

In general designs of the interaction exist now which seem to 
control the backgrounds at a level acceptable for the detectors. 
A particular emphasis of the recent past has been the implementation 
and the consequences of an anti-DID field, beneficial for the 
operation of the accelerator at large crossing angles. It appears
that with an anti-DID field backgrounds are controllable and 
not significantly worse than at small crossing angles. 

A recent and rather major change has been the adoption of the ``push-pull'' 
scheme to accommodate two detectors in one interaction hall. The implications of 
this decision are under study, and will need careful evaluation. 

A note of caution though is in place: all conclusions rely on 
simulations, which in many cases have not been tested experimentally. Therefore 
a significant safety factor should be assumed in the 
design of the detectors, maybe as large as a factor of 10, 
for all background rates.

\cleardoublepage

\chapter{Subsystem Design and Technologies}
\label{detector_technologies}
In this section a brief technologically oriented description of the different sub-systems of a 
detector at the ILC is given. The goal of
this section is to describe the different developments, present their state of development, 
identify needed R\&D, and discuss the program of R\&D for the next few years. This chapter 
thus complements the one on the detector concepts, and fills in the missing technical 
details. 
\section{Vertex Detector}

The design of the vertex detector (VTX) needs to be matched to some very challenging physics processes of importance at the ILC, namely multi-jet processes in which the flavors and sign of the quark charge of some of the low energy b and c-jets needs to be determined.  Polar angle coverage needs to be as hermetic as possible, since for some processes the ends of the angular range are most sensitive to new physics.  The measurement of quark charge, based on the procedure of vertex charge determination, imposes the most stringent requirements, since a single low momentum track that is ambiguous between the primary vertex (PV) and the decay chain formed by the secondary and tertiary vertices (SV/TV), invalidates the charge determination.  In practice, studies \cite{ref-VTX-sonja} have shown that efficient discrimination between IP tracks and decay chain tracks is important down to $p_t$ values as low as 100 MeV/c.  

The most decisive parameter in determining the potential physics capability of the vertex detector is the beam-pipe radius $R_{bp}$.  Collimating and controlling backgrounds at the ILC, which is necessary to achieve the minimum beam pipe radius, has been a key feature of the machine design.  Controlling beam-beam disruption, using ``flat'' beams at the IR, much larger in x than in y, is crucial.  The quantitative study of some other background sources has hardly begun, and there will always be trade-offs between boosting the luminosity by applying more aggressive bunch crossing conditions and enhancing the tolerance of vertex detectors to the resulting increased backgrounds. 

Once the final focus conditions have been settled, the value of $R_{bp}$ is determined by the field in the detector solenoid, since higher field is more effective for radial containment of the $e^+e^-$ pair background which dominates the hit density on the VTX inner layer.  The inner section of beam-pipe of length $\approx 14 cm$ is most critical, since this covers the practical polar angle range for high precision tracking.  Beyond $|z| \approx 7 cm$, the beam-pipe radius can expand conically, in order to stay safely beyond the envelope of pair background.  Thus it is the hard edge of this background at $|z| \approx 7 cm$, with appropriate stay-clear, which determines the minimal beam-pipe radius.  

One would of course wish to locate the first layer of the vertex detector just outside the beam-pipe, since the impact parameter resolution for intermediate and low momentum particles is driven by the combined thickness of the beam-pipe and layer-1, together with their distance from the interaction point (IP).  However, this layer may need to be pushed out further, if the background hit rate is excessive.  For a given design of the final focus (FF) and solenoid field, the minimal radius depends on the duration of the sensitive window (SW) of the chosen VTX technology.  There are currently approximately ten technologies being studied for the ILC vertex detector; all use silicon pixels, but the target SW varies from single bunch (ie. $< 300 ns$), through $\approx 50 \mu s$, hence 20 time slices per train, to integration over the entire bunch train of duration $1 ms$.  So once $R_{bp}$ is settled, if backgrounds are expected to be high, the radial position of layer-1 will depend on which technologies can be made to work.  However, if the backgrounds correspond to the calculated  $e^+e^-$ pairs with the nominal FF conditions, all options will work with layer-1 just beyond the beam-pipe.  Since the options with the shorter sensitive windows may have associated disadvantages, the selection of the preferred technology is far from clear.

This issue will depend on numerous factors, such as:
\begin{itemize}
  \item measurement precision, including freedom from induced mechanical oscillations and long term drift in internal alignment
  \item layer thickness (including cooling requirements)
  \item pixel size (needs to be small enough to resolve hits from tracks in the core of high energy jets)
  \item additional material required for end-of-ladder services, cooling, cables and fibers
  \item duration of sensitive window
  \item adequate radiation hardness
  \item preservation of internal alignment in the face of powering the detector and operations such as opening and closing detector end-doors, and push-pull cycles between two detectors
  \item  resistance to electromagnetic interference (EMI) at levels to be encountered at ILC
\end{itemize}

\begin{figure}[htbp]
	\centering
		\includegraphics[bbllx=0, bblly=0,bburx= 210mm, bbury=155mm,height=8cm]{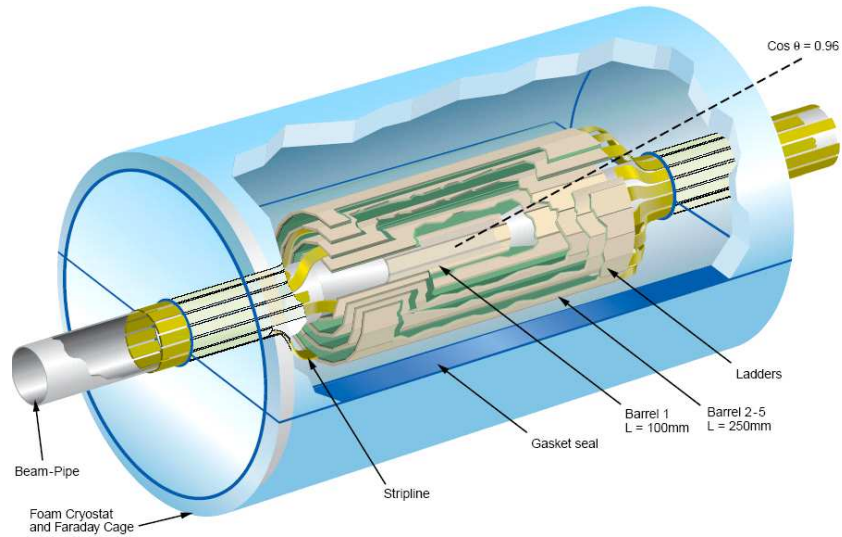}
	\caption['Long Barrel' option for the vertex detector]{ 'Long-barrel' option for the ILC vertex detector.  The cryostat is an almost massless foam construction, and has a negligible impact on physics performance.}
	\label{fig:VTX_VTX1}
\end{figure}

\begin{figure}
	\centering
		\includegraphics[bbllx=0,bblly=0,bburx=159mm,bbury=121mm,height=8cm]{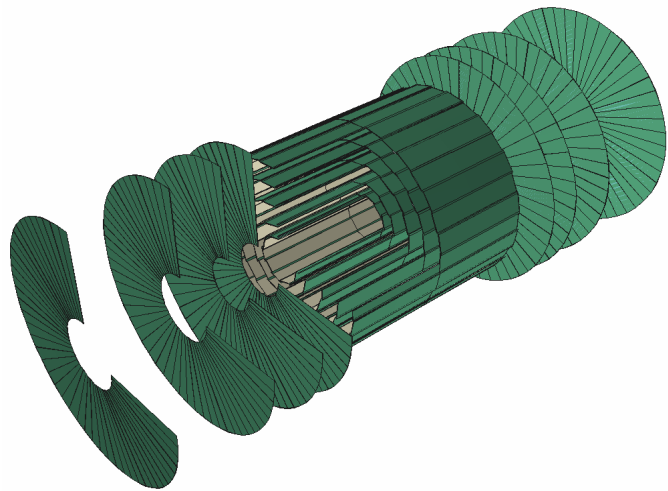}
  \caption['Short barrel' option for the vertex detector]{'Short-barrel plus forward disks' option for the ILC vertex detector }
	\label{fig:VTX_VTX2}
\end{figure}

The ideal vertex detector would provide precision coverage over the full solid angle.  In practice, ILC tracking systems will be cut off by masking below $\theta \approx 7 \deg$, and vertex-quality tracking may cut off around $15 \deg (\cos{\theta} \approx 0.96)$.  At first sight, the optimal performance would be expected from a combination of short barrels and forward disks.  The alternative of long barrels would seem to be less attractive due to the loss of precision resulting from the increased obliquity of tracks at small polar angles.  However, one cannot ignore the fact that the ends of the barrels will inevitably contain extra material from mechanical supports and additional ``baggage'' such as storage capacitors, readout chips, driver chips, electrical connectors and so on. Mechanical supports may need to be relatively robust in order to stabilize structures against considerable Lorentz forces induced by high currents flowing during the bunch train.  While ILC vertex detectors have been sketched with both options (Figures~\ref{fig:VTX_VTX1} and \ref{fig:VTX_VTX2}) the choice will depend on the measured performance of real prototypes, fully tested for operation under realistic conditions. 

Another open question is whether the ``ladders'' that comprise barrel staves should be mechanically linked along their length, perhaps by mounting the sensors on cylindrical support shells, or supported only at their ends.  Again, this will depend on what assembly procedures are practicable for a specific technology.  If thin sensors need to be mounted on substrates in order to handle them for bump-bond attachment of readout chips, mechanically independent barrel staves appear to be natural.  If self-contained sensors need only to be mounted on a substrate and connected by wire bonds, assembly onto a cylindrical support shell may be feasible.  In either case, the cylinders are trapped by the bi-conical beam-pipe, so the detector needs to be constructed as two half-cylinders that are assembled round the beam-pipe and then clamped together.  This requirement also has implications for the preferred scheme of ladder mounting.  

In brief, the requirements for the vertex detector suggested by the physics goals are reasonably well-defined (beam-pipe radius  $\le 15 mm$, $\approx 10^{9}$ pixels of size  $\le 20 {\mu m}^2$ , layer thickness $\approx 0.1\% X_0$).  Given the foundations provided by the SLD vertex detector (307 Mpixels, layer thickness 0.4\% $X_0$), these goals appear reasonable.  Extensive R\&D by many groups round the world over the past 8 years has opened up a number of promising approaches.  The most conservative of these (FPCCDs) could provide a robust solution at least for startup, though they might be pushed to larger radius than is desirable for physics if background levels greatly exceed the current estimates for the baseline FF design. 

In Section \ref{sec-VTX-technologies}, we review the technology options being considered, and in Section \ref{sec-VTX-mechanics} we discuss some mechanical design issues, in both cases noting the accelerated pace of R\&D that will be needed to achieve the goals in time.

\subsection{Technology options}
\label{sec-VTX-technologies}

In contrast to the early days of charm and bottom physics, and the variety of gaseous and silicon technologies used to construct vertex detectors at LEP and SLC, there is now unanimity regarding the basic technology for ILC.  Silicon sensors with small pixels ($\le 20 {\mu m}^2$) are accepted as the only way forward.  Agreement was reached at LCWS 1993, when it was demonstrated \cite{ref-VTX-lcws1993} that this approach was mandated by the hit densities in the core of jets, and by the pair backgrounds.  However, it was equally clear that the CCD approach used for SLD would be far too slow for use at ILC.  Over the past 14 years, a considerable variety of options has been suggested, and some of these are the subject of vigorous international R\&D programmes.  They all have a chance of doing the job, but none is guaranteed to satisfy all requirements.  Some of the most promising may not be ready in time, but may be outstanding candidates for future upgrades.  

\begin{figure}
	\centering
		\includegraphics[height=8cm]{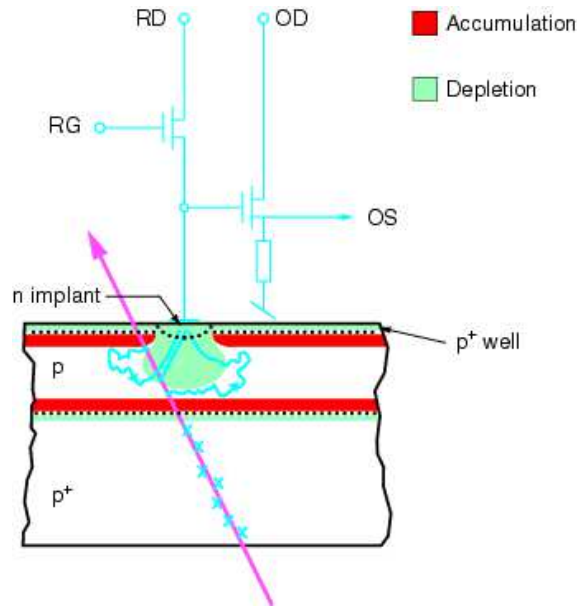}
	\caption[Cross section of a generic sensor architecture]{Cross-section of a generic sensor architecture in which the signal electrons diffuse between the upper and lower reflective layers, until they are captured in the depleted regions associated with the sense nodes ( reverse-biased photogate or diode structures) built into the pixel.}
	\label{fig:VTX_layers}
\end{figure}

It is too early to construct a table of attributes, indicating strengths and weaknesses of the different options - there are too many unknowns.  However, one can attempt a few comments about each, with further details being available in the form of contributions from the detector groups to the ILC Detector R\&D Panel website \cite{ref-VTX-RandDpanel}.  All designs make use of the basic attribute of silicon devices that one can create a buried layer that serves as a sensor, by sandwiching it between appropriately biased neighbouring layers, for example a substrate layer and a readout layer, as sketched in Figure~\ref{fig:VTX_layers}.  The sensor layer may be fully depleted, in which case charge collection to the sense node can be fast (a few ns) or only partly depleted, in which case the signal charge is collected partly by diffusion, which can take $\approx 100 ns$.

A conventional CCD at ILC would collect signal throughout the bunch train, then be read out between trains.  Background hit densities would be excessive.  However, the FPCCD collaboration  \cite{ref-VTX-FPCCD} proposes to solve this problem by using very fine pixels ($\approx 5\times 5 {\mu m}^2$), which not only reduces the percentage of hit pixels, but also permits some measure of background rejection from the shape of the mini-vectors generated by the traversing particles. Due to long signal integration time, FPCCD will need to operate at temperature below room temperature.
The CPCCD design \cite{ref-VTX-CPCCD} achieves background reduction by multiple readouts ($\approx 20$ frames per train), as does the DEPFET sensor \cite{ref-VTX-DEPFET}.  Among the MAPS options \cite{ref-VTX-MAPS}, the CAP \cite{ref-VTX-CAP} or FAPS \cite{ref-VTX-faps} approach considers storage of $\approx 20$ time-sliced signals per train, using in-pixel capacitors after charge-voltage conversion at the readout node.  The ISIS approach \cite{ref-VTX-ISIS} retains the stored signal charge ( $\approx ~20$ samples) in the buried channel of a tiny CCD register within each pixel.  This is considered to be more robust wrt EMI problems such as those observed at SLD.  The chronopixels \cite{ref-VTX-chrono} are altogether more ambitious - they aim to achieve single bunch time stamping of the hit pixels.  They plan to store only binary hit information, since their small pixels $(10\times 10 {\mu m}^2)$ should yield sufficient tracking precision.  This design is ambitious in at least two respects; it needs $45$~nm processing technology, and these sophisticated pixels are likely to be power hungry, so supplying the current during the train could be a challenge.  The SOI-based approach \cite{ref-VTX-SOI} and 3-D pixels \cite{ref-VTX-3D} are even more futuristic.  They aim to interconnect signal sensors with separate readout chips, using closely-spaced metal interconnects, one per pixel.  Implicit in this technology is some degree of wafer thinning, and the ILC application would involve thinning of all silicon layers, each to some tens of microns, so as to satisfy the material budget.  Finally, the SCCD (short-column CCD) is a new idea \cite{ref-VTX-SCCD} to achieve single-bunch timing by alternating the sense in which signals are clocked in adjacent channels.  Only clusters which exhibit a cross-channel match at the time of the bunch crossing are retained, so nearly all out-of-time background is rejected.  Depending on the technology (conventional bump-bonding or the 3-D approach) such structures could be somewhat thicker than desired, or perfectly acceptable.

Whichever technology is considered, one is dealing with at least $10^9$ pixels.  Experience at SLD demonstrated that the LC environment can be challenging as regards beam-related pickup.  This would not be a problem in the case of an uninterrupted metal beam-pipe, but the penetrations for beam-position monitors (BPMs) and other devices in the interaction region (IR) permit high frequency RF power to escape, and this tends to bounce around within and beyond the detector.  There are also other sources of EMI likely to be present during the bunch train.  Strategies to mitigate such effects have been discussed \cite{ref-VTX-EMI}.  One protective measure will be to use correlated double sampling (CDS) for the front-end signal processing.  All technologies say they will do this, but in some cases the time between successive samples is so long that they may be dangerously vulnerable to pickup.  

At this time it is not possible to choose between the different technologies. Development has to 
continue for some time, so that the different options can demonstrate the 
performance they need to operate in the ILC environment. In addition the ILC vertex detector community is discussing evaluation criteria for the different technologies.  

\subsection{Mechanical design issues}
\label{sec-VTX-mechanics}

The two general ideas for vertex detector mechanical design (long barrels vs short barrels and forward disks) are illustrated in Figures~\ref{fig:VTX_VTX1} and \ref{fig:VTX_VTX2}.  How to choose between them?

\begin{figure}
	\centering
		\includegraphics[height=8cm]{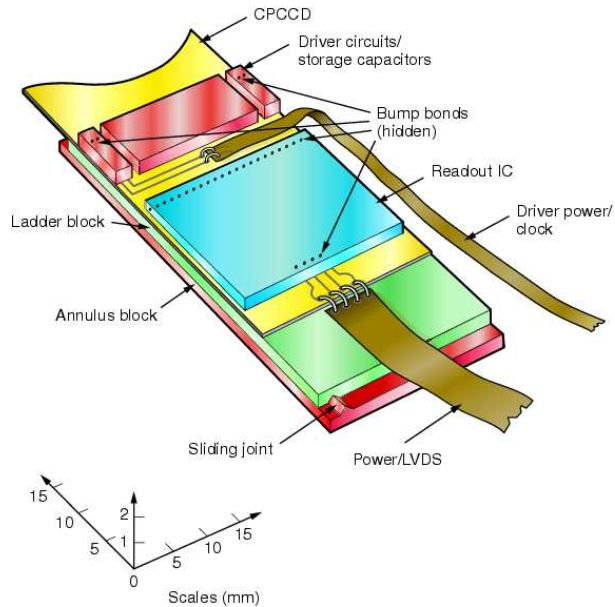}
	\caption[Sketch of CPCCD support structure]{Sketch of mechanical supports and electronics at the end of a ladder for the CPCCD.}
	\label{fig:VTX_CPCCD}
\end{figure}

There are many contributing factors, and it will take several years before they can be resolved.  Firstly, one needs a sensor technology choice, or at least a few compatible options, from the range discussed in the previous section.  Each technology carries with it different ``baggage'', in the form of additional material at the ends of the barrel staves or ladders.  An example is shown in Figure~\ref{fig:VTX_CPCCD}, for the CPCCD.  Even here, while the components can be identified, their actual design and associated material budget are the subject of intense R\&D.  It will be best to wait for working ladders, built with the different technologies, in order to have the necessary input for making this decision.  

As well as the physical differences, the electrical requirements could be decisive.  All options plan to use ``pulsed power'' in order to keep their average power dissipation within limits that will permit gaseous cooling, since liquid cooling (as is obligatory at LHC) would drastically exceed the material budget.  Pulsed power means keeping the detector power switched off, or much reduced, for the $199$~ms between the bunch trains of duration $1$~ms.  In some cases, this means switching on tens of amps of current per ladder during the train.  Given that these ladders are sitting in a magnetic field of 3-5 T, what are the mechanical effects of the associated Lorentz forces?  If we aren't careful, we may have a lot of vibrations exceeding the maximum tolerable limit of about $1 \mu$m.

Apart from vibrations, other mechanical effects (long term creep, distortions, etc) must be held to $1 \mu$m or below.  This tolerance is based on the opportunity for charm tagging efficiencies far above those achieved at LEP or SLD.  The cross-section of the ILC beam-spot (a few nm by $<1 \mu$m), held steady by feedback systems, permits unprecedented discrimination between IP tracks and those in the decay chain.  This is particularly relevant for charm particles, due to the comparatively small impact parameters of their decay products.  The vertex detector will be able to build up knowledge of the IP position in x,y to sub-micron precision, by averaging over a number of events, but this only works if the detector itself is stable to this level.  Issues such as micro-creep of the structure, stiction in sliding joints such as the one shown in Figure~\ref{fig:VTX_CPCCD}, external stresses on the structure, all need to be carefully controlled.  

Regarding external stresses, it is most important that these are effectively eliminated so that the detector retains its shape perfectly between operations such as opening of the end-doors, and push-pull excursions of the detector.  After such operations, there will not be time for re-calibration of the internal geometry by $Z^0$ running or any special calibration runs.    When for example, the end-doors are opened, the beam-pipe to which the vertex detector is attached will inevitably move and flex slightly.  After re-closing, the vertex detector will surely find itself in a different position with respect to other elements of the tracking system (central and forward trackers).  Such overall shifts in position and angle are easily determined by tracking with a small number of events, as long as the internal geometries are not disturbed.  This in turn depends on the mounting systems.  If all the vertex and tracking detectors are attached to their various supports by means of 3-point kinematic mounts (ball, vee and flat, with light springs to maintain contact) no distorting forces can be transmitted to any of the structures.  One still has to be careful about cable design, etc, but the principle is well established. 

Whichever mechanical design is chosen, it remains important to minimize the size of the beam-pipe. Of course a larger beam-pipe leads to reduced impact parameter resolution due to multiple scattering.  A larger beam-pipe necessitates, to some extent, a scaling up of the entire vertex detector, in order to preserve the angular coverage.  Enlarging the detector could have other undesirable consequences such as reduced mechanical stability, forcing an increase in the material budget, and possibly requiring 3 sensors rather than 2 in each ladder of the outer layers.  This would further degrade the performance by requiring more material within the active volume of the vertex detector, in order to service the inner sensor of each ladder.

It should finally be noted that many physics studies so far made regarding the vertex detector design, have used fast simulation programs.  Yet to fully understanding the impact of the material budget, one may need to consider such effects as non-Gaussian tails on distributions of multiple scattering, and secondary interactions in the material of the detector.  For this, full Geant simulations are needed.  The same applies to studies of track-finding efficiencies.  The number of barrel layers, forward disks and external tracking system elements really needed to do the ILC physics is still unclear.  While the current layouts are certainly plausible, they cannot be considered to be at all established at this stage.  Much work has still to be done, both by the detector R\&D groups, and by those doing the simulations.  Ongoing close cooperation between these groups will be essential over the next several years.

\subsection{R\&D leading to technology selection for VTX detectors}
\label{sec-VTX-RandD}

At the moment a number of different collaborations work on developing, testing and understanding the different 
technologies discussed in Section~\ref{sec-VTX-technologies}. The groups have agreed that a major benchmark for 
them will be the production of a detector-scale ladder at or around 2012. Such an achievement would be a 
major benchmark on the road towards developing a vertexing system for the ILC. It will be an important benchmark 
before the community can try to select one or two different technologies to be used in the ILC detectors. 

%
There is an opportunity for coordination of the test facilities to be used in evaluating the different technologies.  
A suite of calibrated test facilities, to be used by everyone for the evaluation of their ladders, with coordinated plans for data collection, analysis and presentation, could make the comparison of technologies more easy and transparent.

\section{Silicon Tracking}
\label{sec-Subs-silicon-tracking}

Both silicon and gaseous tracking technologies are being investigated for tracking charged particles
in the region between vertex detector and the calorimeter.
These systems, working either alone or in combination with the vertex detector and calorimeter,
must provide efficient identification and precise momentum determination of charged particles.
This section is focused on silicon tracking design and technology, while the following section focuses
on gaseous tracking design and technology.

The development of fine-pitch custom readout chips and improvements in the reliability and yield of the strip sensors
has led to the development of ever larger silicon trackers, with the CMS detector having $>200 $~m$^2$ of active silicon.
Silicon strip detectors are now a well established technology.

Silicon strip detectors have a number of attractive features:
\begin{itemize}
\item
Position resolutions of 5-10 $\mu$m are achievable in fine-pitch devices with good 
signal/noise performance,
providing excellent momentum resolution even for very high momentum tracks.
\item
The charge collection time can be made sufficiently fast to identify the beam crossing 
that generated
the hit, minimizing the impact of pileup from beam backgrounds and any detector noise.
\item
The two-hit resolution is superb due to the fine pitch and the small number of strips 
typically associated with
a charged particle.
\item
Silicon strip detectors directly measure points on the charged particle trajectory, 
subject only
to the mechanical alignment precision, and do not require corrections for environmental 
factors or
non-uniform magnetic fields. 
\end{itemize}

While silicon strip detectors have been extensively used in other experiments, 
large detectors typically
have $\approx 2$\% $X_0$ per layer, most of which is attributable to dead material needed 
for support, cooling,
and readout.
This dead material has a number of undesirable effects, including multiple scattering, 
photon conversion,
production of bremsstrahlung photons and delta rays, and hadronic interactions.
In keeping with the ILC goal of making precision measurements, one of the most 
significant R\&D challenges
for silicon tracking at the ILC is to accrue the benefits of silicon strip 
detectors while significantly reducing the amount of material in the tracker.  
This goal, either directly or indirectly, drives much of the silicon tracker
R\&D program that will lead to a new generation of large area silicon trackers.

A detailed description of the silicon tracking R\&D effort can be found in the 
documents prepared for the 2007 Beijing Tracking R\&D Review~\cite{ref-Si-Review} 
by the {\underline Si}licon Tracking for the {\underline L}inear {\underline C}ollider (SiLC)~\cite{ref-Si-SiLC}
and the SiD Tracking Group~\cite{ref-Si-SiDReport}.
In the sections below, the major issues and R\&D efforts underway for silicon tracking at the ILC are summarized.

\subsection{Silicon Sensors}

The baseline silicon tracking technology for the ILC is the silicon microstrip detector.
Made from a thin wafer of high resistivity silicon, a silicon microstrip detector collects ionization
deposited by charged particles onto fine-pitch strips that run the length of the detector.
A typical detector, fabricated with currently established technology, might be made 
from a 150mm diameter, 300$\mu$m thick silicon wafer with 50$\mu$m pitch strips.
When coupled with low-noise readout electronics, such a detector is capable of 
measuring the track coordinate
perpendicular to the strip direction with a precision of $<10\mu$m.  
 
Single-sided detectors excel at measuring a single coordinate, typically the 
azimuthal angle for precise measurement of track curvature.
Where two dimensional hits are required, a number of options have been successfully utilized.
Double-sided detectors with strips on both sides of the silicon wafer can provide a 
2D stereo measurement of the
hit position, although past experience has been that double-sided detectors are 
difficult and costly to fabricate.
A widely used alternative to double-sided detectors is to use back-to-back 
single-sided sensors to provide the 2D stereo measurement.
A third approach that is being studied is to readout both ends of the strip and use
 resistive charge division
to measure the coordinate along the strip direction.
Finally, the use of pixel detectors for the inner region of the tracker, especially in the high background low-angle forward region, is under study, with a possibility to further extend it.

Multiple sensors can be ganged together to effectively create longer strips by gluing sensors end-to-end
and using wire bonds to electrically connect the strips.
The principle advantage of ganging multiple sensors together is to reduce the readout material, power consumed,
and heat that needs to be removed.  The performance, as well as fabrication issues involved in assembly and handling
long sensors, is under study. 

Options based on novel 3D Silicon technology are under active R\&D (SiLC) in collaboration with research centers specialized in these new technologies. It includes new microstrip sensors based on planar 3D Silicon technology and also 3D pixels to be used for fabricating larger area tracking layers.


While silicon microstrip detectors are a well established technology, the ILC community, in collaboration
with industrial partners, is engaged in an active R\&D program to further improve the microstrip detector technology.
The goals of these efforts are to reduce the amount of tracker material and at the same time the detector costs. 
Areas of research include thinning sensors to $\approx 200\mu$m to reduce material thickness,
developing microstrip detectors with larger ($\ge200$mm) wafers to reduce costs,
and fabrication of cost-effective double-sided sensors to provide 2D hit measurements in a single sensor.
The reduced material budget makes it impossible to use pitch adapters and therefore, novel solutions are being investigated to connect the front-end electronics and the readout electronics directly onto the detector. 

\subsection{Readout Electronics}

Readout electronics typically consists of custom front-end integrated circuits that amplify and
detect the strip charge, a hybrid that supports these chips and provides the required
power conditioning and signal termination components, and low-mass cabling that connects the detector
to electronics that interface the detector to the data acquisition, clock distribution, slow control,
and power delivery systems.
 
A key element in the ILC silicon tracking R\&D program is the development of the front-end readout chip.
The front-end readout electronics is designed in a way that preserves at best the intrinsic detector performance and meets the following requirements:

\begin{itemize}
\item Operate within the duty cycle of the ILC machine;
\item Be able to tag individual bunches (BCO electronics);
\item Data sparsification and digitization on sensor;
\item Front-end chips mounted closely onto the sensor;
\item Minimization of power dissipation (typically less than 1~mWatt/channel, all included, without power cycling);
\item Power cycled front-end electronics;
\item Ensure an electronics MIP to noise ratio of order of 25, for detectors from 10 to 60 ~pF capacitance Silicon detector and shaping times between 500 ns and a few $\mu$s;
\item Minimize the on-detector total material to increase the transparency to radiation;
\item Highly multiplexed A/D conversion;
\item Provide a continuous stream of digital data at the end of each bunch train;
\item Ensure the reliability, calibration and monitoring of the whole system over a few millions channels.

\end{itemize}

This is a challenging set of goals, and three different 
chip development efforts are currently in
progress. The KPix chip, which is being developed for use in tracking and calorimeter readout, provides analog
buffering of up to four hits and is structured to provide a high density (1024 channels) bump-bonded interconnect
to the sensor.
The LSTFE chip uses a time-over-threshold technique to determine the charge deposition, allowing digital storage
of hit information. In this scheme, following a low noise pre-amplifier and microsecond-scale shaper, the signal is evaluated by two comparators, one with a high threshold to suppress noise hits, and the second with a lower threshold to provide pulse-integral information in the region surrounding a high-threshold crossing. The gain of the amplification stages is high with pulse height (but not integral) saturating between two and four times minimum ionizing. In this way, the application of the high and low thresholds is made insensitive to irreducible channel-to-channel variations.

\begin{figure}[htbp]
	\centering
		\includegraphics[height=7cm]{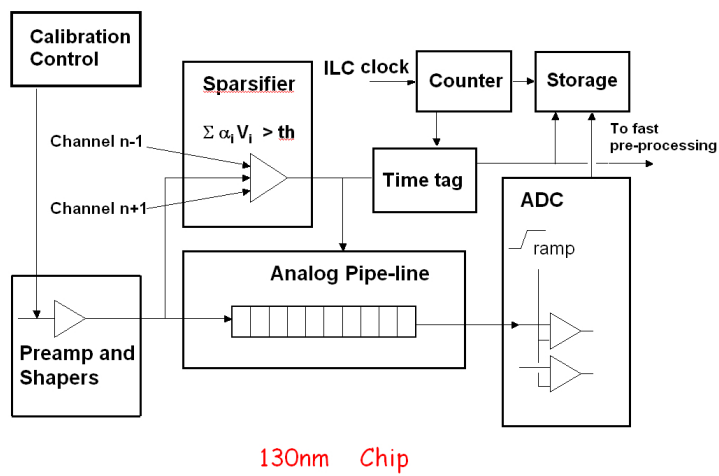}
	\caption[Double metal sensor design]{ Front-end chip architecture using VDSM technology  }
	\label{fig:VDSM-archi}
\end{figure}

\begin{figure}[htbp]
	\centering
		\includegraphics[height=8cm]{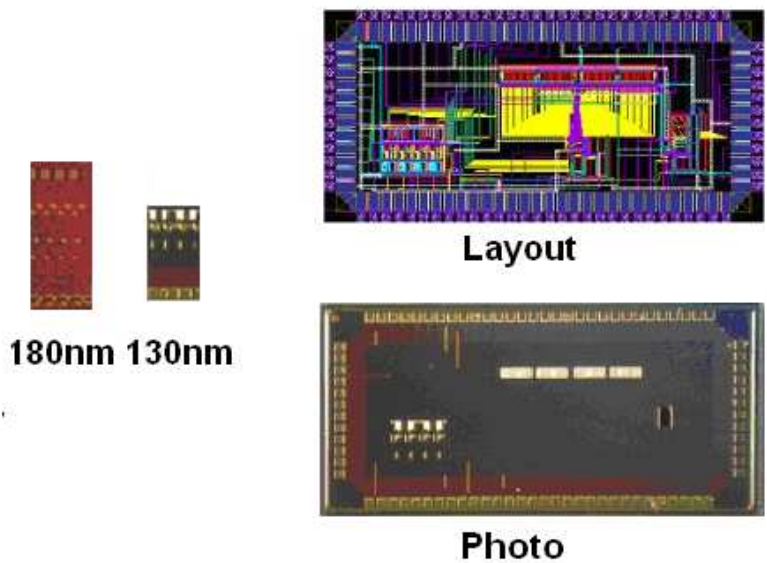}
	\caption[The First Prototype FE Chip]{Layout and photograph of the first prototyped FE chip in 130 nm CMOS technology }
	\label{fig:VDSM-chip_photo-layout}
\end{figure}

The third approach is to develop a fully digitized system, based on 0.25$\mu$m CMOS 
technology to develop readout chips using Very Deep Sub-Micron (VDSM)
CMOS technology. The use of VDSM technology, allows, among other benefits, integrating, 
for instance in 130 nm~CMOS technology, a complete readout channel in less than 50 
$\times$ 2500 $\mu$m$^2$. This front-end readout device allows recording the pulse 
height per cell (Figure~\ref{fig:VDSM-archi}). A resolution transverse to the strip 
of a few micrometers can be achieved using analogue readout and evaluation of centroids. 
One needs a shaping time of typically between 500 ns and 2 $\mu$s (could be even higher 
for very long strips, reaching 2 to 5 $\mu$s) in order to keep a signal to noise ratio 
above 20. Bunch crossing tagging will be achieved for all options. 
The data will be obtained from the detector with pulse sampling allowing accurate amplitude measurement 
and BCO identification. Zero-suppression is to be performed in the front-end electronics, 
using thresholds on analogue sums of adjacent channels 
(Figure~\ref{fig:VDSM-archi}). Calibration will be also integrated into the front-end 
chips using digital to analogue converter and Metal Insulator Metal Capacitors of 
known values as charge reference, together with switched networks. 
A first prototype in 130nm CMOS UMC technology has been successfully produced and 
tested (Figure~\ref{fig:VDSM-chip_photo-layout}). The next versions will also 
include power cycling.

\begin{figure}[htbp]
	\centering
		\includegraphics[height=8cm]{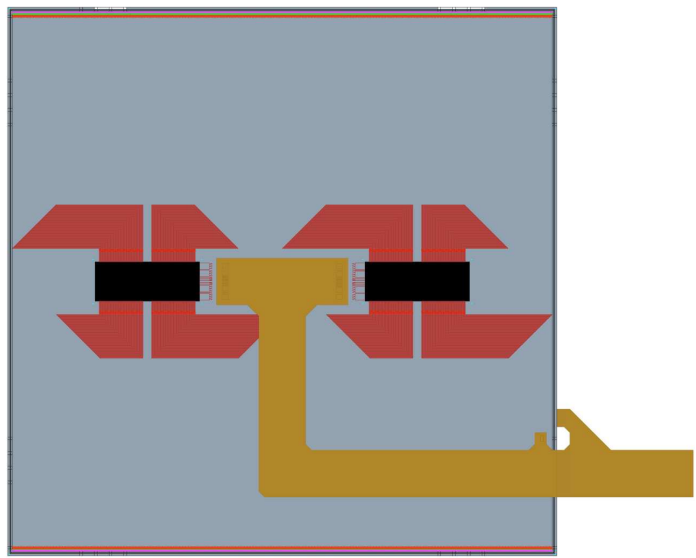}
	\caption[Double metal sensor design]{ Double metal sensor design.  The red double metal traces connect the
black readout chip to the gray readout strips (running vertically) and the tan pig-tail cable.  }
	\label{fig:Si-DoubleMetal}
\end{figure}

Another area of R\&D is to investigate using the silicon microstrip detector itself for signal
routing, simplifying assembly and eliminating the need for the hybrid and further reducing dead material.
The signals from the detector strips are routed to a set of bump bonding pads on the microstrip detector
using a second ``double metal" layer, as shown in Figure~\ref{fig:Si-DoubleMetal}.
The readout chip is then bump bonded directly to the detector.
Additional short traces on the double metal layer provide the necessary interconnect between the readout chip and
the cable for power, clock, and data signals.

Along these lines several techniques for bonding the electronics onto the microstrip detector are available and under investigation in collaboration with industrial partners. They depend on the output pad pitch of the chip. Among them are: the ball solders (for pitch down to 100 $\mu$m), stud bonding (for pitch down to 70 $\mu$m), bump bonding (for pitch down to 30 $\mu$m). Trends in semi-conductor VLSI integration promise for the near future the 
possibility to stack several thin high resitivity Silicon layers to produce optimized 
detectors (3D technology).

Powering the readout electronics, especially the front-end readout chip, is another challenge.
The readout chips require high current at low voltage, whereas minimizing the amount of conductor favors
low current at high voltage.  
Two promising techniques for providing efficient power delivery are serial power and capacitive DC-DC conversion.
While these techniques have been demonstrated to operate for DC loads, R\&D is required to demonstrate that
they can be made to work with power pulsing. Furthermore the need for a quiescent current in a power off mode is currently under investigation.

\subsection{Mechanical Design}

The mechanical design must ensure the stable positioning of the microstrip detectors,
provide cooling to remove the heat dissipated in the readout electronics, incorporate alignment
and position monitoring components, and provide routing and supports for the detectors, cables, 
auxiliary components.  Providing the precision measurements with minimal amounts of material requires
careful design and, in many instances, significant investments in R\&D to demonstrate that the design goals are
achievable in practice.

The mechanical supports must have sufficient rigidity to provide stable support of the detectors, while
also minimizing the material required.
Two approaches have been investigated for supporting the microstrip detectors.  

\begin{figure}[htbp]
	\centering
		\includegraphics[height=6cm]{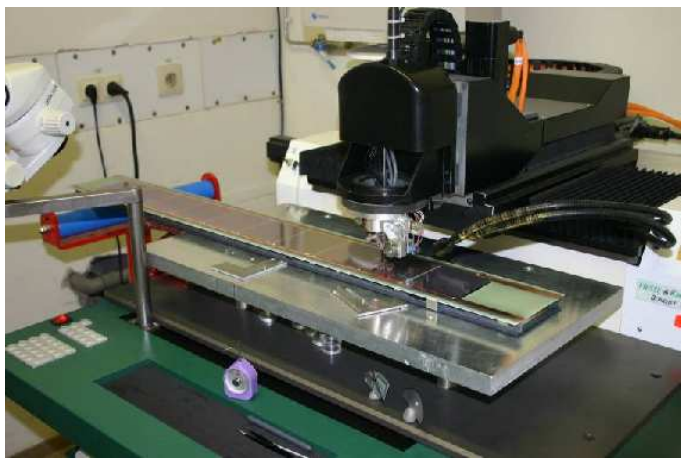}
	\caption[Construction of a Silicon ladder]{Construction of a Silicon ladder made of several Silicon sensors}
	\label{fig:ladder-photo}
\end{figure}

In one approach, several detectors are mounted on a support ``ladder'', which is then attached to a carbon fiber support structure.  This approach allows several sensors to be ganged together, with the goal
of minimizing readout material. A novel approach to construct such elementary modules is under study. The ladder design that is currently investigated includes foam sandwich structures. These are being studied for the ILC vertex detector option developed by the LCFI R\&D group. They have demonstrated that both Silicon Carbide and reticulated Carbon foams can be used to construct stable, extremely low mass ladders. A first step towards this type of ladder support structure is being experienced with the construction of the very first Silicon module prototypes for the present and forthcoming test beams as shown in  Figure~\ref{fig:ladder-photo}.

\begin{figure}[htbp]
	\centering
		\includegraphics[height=8cm,bbllx=80,bblly=70,bburx=690,bbury=550]{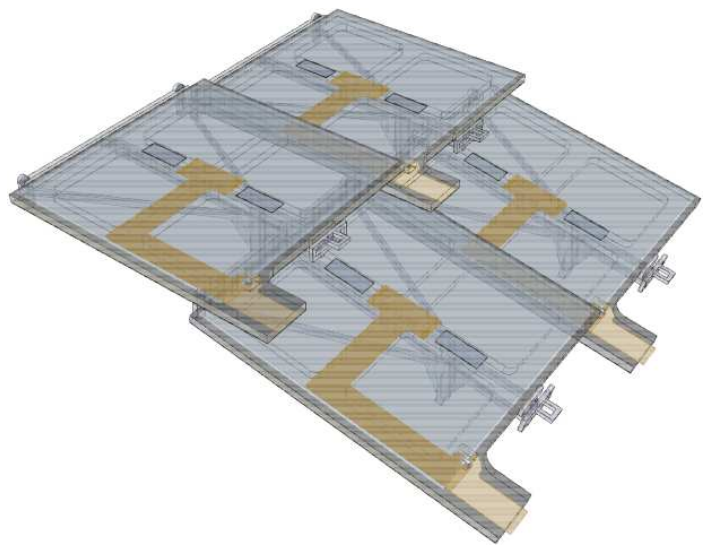}
	\caption[Silicon readout modules]{ Silicon readout modules.}
	\label{fig:Si-ShortModule}
\end{figure}

A second approach is to incorporate individual microstrip detectors into modules that can be directly mounted
onto the support structure, providing a higher degree of segmentation than with ganged sensors, as shown
in Figure~\ref{fig:Si-ShortModule}.

Silicon trackers have traditionally required extensive liquid cooling systems to remove the heat generated 
by the readout electronics.  The small duty cycle of the ILC allows a substantial reduction in the average
power dissipation by power cycling, whereby the current in the input amplifiers is greatly reduced during
the interval between bunch trains.  It is anticipated that the average power can be made sufficiently low
to allow air cooling of the silicon tracker, greatly reducing the amount of material required for cooling.

While power cycling reduces the average current draw, it does not alter the peak current draw, which is typically
of order 1A per readout chip, in the KPix case with 1000 channels per chip. The VDSM chip has proven to give less than 1 mWatt per channel, without power cycling, for the full readout chain (Figure~\ref{fig:VDSM-archi}) as measured on the first prototype.
These peak currents will generate substantial Lorentz forces on the power and ground conductors.
If the forces exerted on these conductors are not well balanced, there
is the danger that impulse forces exerted during the power cycling will induce vibrations in the tracker
and degrade the position resolution that can be achieved.

Precise alignment of the microstrip detectors is required to achieve the desired tracking performance.
Track-based alignment is typically used to provide the most accurate determination of alignment constants.
However, track-based alignment requires a large number of tracks and can only correct for long term alignment changes.
One way to track the relative position of detectors is by shining an infrared laser through several sensors. 
The light generates a signal in the strips illuminated by the laser; since the laser light travels in a straight
line, the relative alignment of several detectors is established. The advantages of this approach are a minimum impact on system integration, the use of the same front-end and readout as for the other sensors. Sensors have to be slightly modified to allow transmission of the beam. R\&D work on Silicon sensors is undertaken in order to still increase the transmittance of these devices.
Another alignment technique under investigation is the use of frequency scanning interferometers to precisely measure
a set of path lengths within the tracker.  In addition to providing measurements of the internal motion of the
silicon tracker, the positioning of the tracker with respect to the vertex detector and calorimeter can be measured.
This may be particularly critical to track alignment changes that occur during the push-pull movements of the detector.

Services have a huge impact in the material budget and are, therefore, an active area of investigation in the ILC.  Power delivery is an important issue, given the restrictions imposed by the given power budget, that will also limit the cooling system. A promising line of investigation is serial powering, where modules are chained in series and are served by a single current source. Analogue and digital voltages are derived by voltage regulators. Serial powering would reduce the number of cables by a factor 2$n$, where $n$ is the number of modules in series. The factor is 2$n$ instead of $n$ since analog voltage is derived from digital power instead of being provided separately. The reduction of cables will lead to a significant reduction of material in the tracking volume. Also, the power efficiency is much higher, reducing the load of the cooling system. Another issue that makes this technology attractive is the fact
that it may reduce the amplitude of the current peaks during the power cycling with the corresponding reduction of the risk of vibrations and the amount of extra conductor required to compensate the IR drop.

\subsection{Detector Prototypes and Tests}
The construction of detector prototypes has started as well as the tests of prototype performances at the Lab test bench and on test beams. The aim is to fully characterize the performances of the new electronics, the new sensors and of the new mechanical structures and designs. Some mechanical effects as those due to power cycling will also be addressed on specific and dedicated test bench. It is of great importance and impact on these detector designs. The Lab test bench activity using radioactive sources have been in progress for a number of years in several laboratories. The test beam activity has started with preliminary prototypes of the detectors and of the readout electronics (SiLC).

   The tests should soon permit to compare the various proposed solutions for the sensors, the mechanical designs and the different readout electronics options. These tests are intended to become combined tests with the different subdetectors, i.e. the vertex detector, the calorimeter and the TPC prototypes. Such combined tests are actively planned and prepared.
   
\subsection{Design of a Silicon Tracking System}
Silicon tracking systems are an integral part of most proposed detectors at the ILC, 
either in combination with a gaseous tracker, or standalone. 
SI detectors will cover the particularly challenging areas in the large angle and the 
very forward direction, which are crucial for the physics program at the ILC, 
but subject to large background problems in the very forward case. 

As part of the R\&D activities studies are under way to investigate 
how to optimally integrate the Silicon detectors into the different 
detector concepts. Questions include the role of material, and distribution 
of the material in the detector due to Silicon, and the granularity 
needed. An open question is where pixel technology is needed, 
and where strip technology is sufficient. Whether or not 2D SI technology 
is needed, and where potentially gains can be realised by using the 
advanced 3D architecture are interesting questions. 

For all these questions, which are adressed in close cooperation with the 
other R\&D groups and the concept groups, powerful and precise 
simulation and reconstruction programs are needed.  

\subsection{Conclusions}

Silicon detectors are unique in their ability to make extremely precise hit position measurement
with a technology that is scalable to large tracking volumes.
They are incorporated, either as a stand-alone tracker or in combination with a gaseous tracker,
in all of the detector concepts except the 4th concept.
Motivated by the goal of making high precision physics measurements at the ILC, silicon tracking
R\&D efforts have a strong focus on developing high precision track measurements while substantially
reducing the minimum amount of material required.
Highlights of this R\&D program include:

\begin{itemize}
\item
Development of power pulsed readout electronics to take advantage of the low duty cycle of the ILC and
significantly reduce the average power consumed.
The goal of this effort is to allow air cooling of the silicon tracker, eliminating the significant amount
of material and complexity required for water cooling.
\item
Development of new detector designs to optimize performance and minimize material.
These efforts include:
\begin{itemize}
  \item Development of long ladders to minimize the number of readout channels.
  \item Development of high density readout chips and bump bonding techniques to minimize the amount of readout material.
  \item Development of thinned silicon wafers to reduce material.
  \item Development of new power delivery components to minimize the material required to power to the tracker.
  \item Development of double-sided detectors and charge division readout to minimize the material required for 3D hit measurements.
  \item R\&D on 3D Silicon technology and on the use of pixels for relatively large Silicon tracking areas.
\end{itemize}
\item
Development of new mechanical designs to provide robust mechanical support.
The goal of this effort is to provide the required mechanical stability while minimizing the amount of material
required.
\item
Development of alignment instrumentation to detect and monitor any movement of the tracker.
These efforts include development of infrared laser and frequency scanning interferometry alignment technologies.
\item
Development and testing of prototype detectors to measure detector performance under realistic conditions.
These efforts are critical to verifying that the expected performance is achieved and that there
are no unexpected problems that would adversely affect construction of silicon trackers for the ILC.
\item
Development of simulation studies on detector performances and on especially important issues such as the design of the large angle and forward region Silicon tracking coverage, and the possibility to build a Silicon tracking system based on pixels. 
\end{itemize}

\section{Gaseous Tracking}
\label{sec-Subs-tracking}

The worldwide effort to develop a gaseous central tracking system for an ILC
 detector is now focused on a design based on the Time Projection Chamber (TPC)  
concept. TPCs have an advantage over other drift chamber designs in that they can
 record a large number of track segments in three dimensions and thereby be more 
robust for tracking particles in high multiplicity jets and in the presence of 
large machine backgrounds. At the same time, the central volume of a TPC has 
very little mass for scattering particles passing through it. The capability 
of the TPC to perform particle identification by measuring the ionization 
energy loss of particles is an additional benefit.

\begin{figure}
	\centering
		\includegraphics[height=8cm]{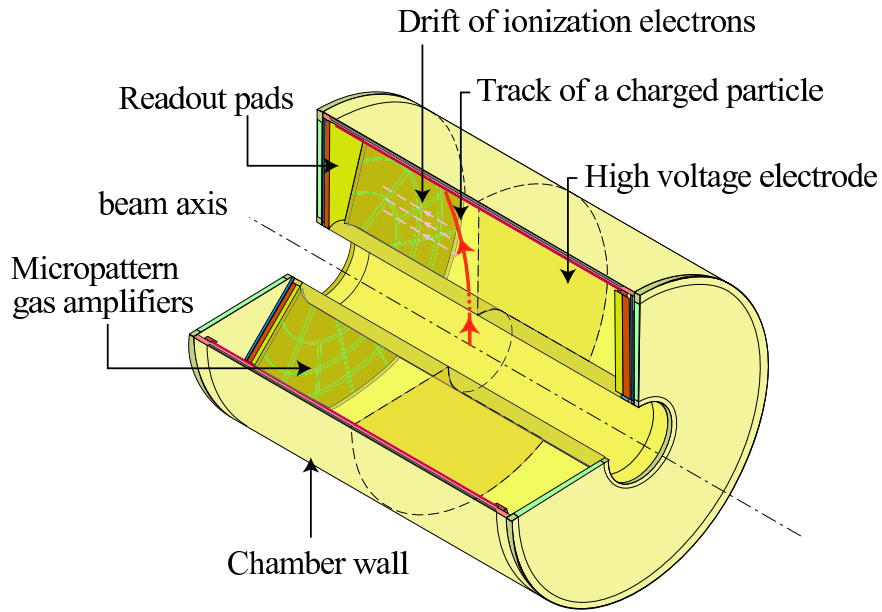}
	\caption[Schematic layout of TPC]{Schematic layout of the TPC}
	\label{fig-tpc-schematic}
\end{figure}

The TPC concept consists of a large container holding a suitable gas in 
which a uniform electric field of a few hundred V/cm is formed parallel 
to the magnetic field of the ILC detector.  Charged particles passing 
through the gas liberate electrons, which then drift towards an endplate 
of the TPC. The electrons undergo gas amplification there and are sampled
in space via a segmented anode (pads), to estimate the coordinates of
track segments in the plane 
parallel to the endplate plane, and in time, to estimate the coordinates 
along the drift direction. A schematic of TPC is shown in Figure~\ref{fig-tpc-schematic}.

TPCs have been used in a number of large particle physics experiments in the past 
with good success. The performance requirements for an ILC TPC, however, greatly 
exceed the achievements of existing TPCs by large factors. In particular, the 
momentum resolution goal of
\begin{equation}
	\sigma (1/p_t) \approx 5 \times 10^{-5} GeV^{-1}
\label{eq-subs-momres}
\end{equation}
(or even less) is a particular challenge. 

\begin{figure}
   \centering
	\includegraphics[height=6cm]{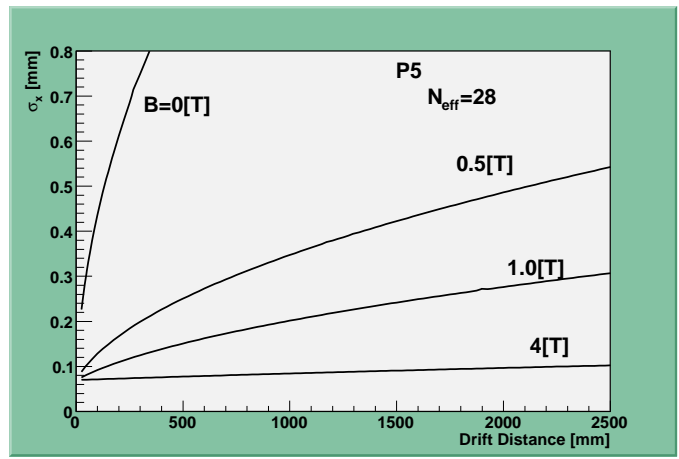}
        \caption[Expected spatial resolution of TPC]{Expected spatial 
resolution 
of TPC with a GEM readout as a function of drift length for several detector 
solenoid magnetic field values.  P5 gas (Ar 95\%, CH$_4$ 5\%) is assumed.}
  \label{fig-tpc-expectedResolution}
\end{figure}

To reach the performance goals, the wire grids used for gas amplification in 
previous TPCs are replaced with micropattern gas detectors, such as Gas 
Electron Multipliers (GEMs) or Micromegas (MMs) and the spatial resolution 
of about 100$\mu$m or less has to be achieved. Due to the minimum wire 
spacing of a few mm, the electric and magnetic fields were not parallel
 near the wire grids, which limited the spatial resolution of wire TPCs.
 The feature sizes of GEMs or MMs are more than an order of magnitude 
smaller, which allows much better precision in determining the spatial
 coordinates of the track samples. An extra benefit of the micropattern 
gas detectors is that the spatial and temporal spread of the signals at 
the endplate are significantly reduced allowing for better two particle 
separation power. The expected spatial resolution of TPC in the case of 
GEM readout is shown in Figure~\ref{fig-tpc-expectedResolution}. As seen
 in the figure, the transverse charge diffusion could be confined by a 
strong detector solenoid magnet of $3 \sim 4$~Tesla within the goal even
 after long drift of more than $2$~m.

Another option using the cluster counting drift chamber is described in section~\ref{ref-gas-cluster}.

\subsection{Basic design concepts of TPC}
\subsubsection{Field cage}
The ILC TPC is foreseen to be a large cylinder of outer diameter 3--4 m 
and total length about 4~m.  Lightweight cylindrical inner and outer 
composite walls hold field forming strips attached to a resistor 
divider network. A central cathode, dividing the TPC into two drift 
volumes, would be held at approximately $50$~kV, with the endplates
 and the other outer surfaces of the TPC at ground potential. The 
composite walls must therefore stand off the large potential of 
the central cathode. Narrow strip pitch with mirror strips are 
being considered in order to keep the field as uniform as possible
 within the active TPC volume.

\subsubsection{Gas system and gas choice}
A conventional recirculating gas system is necessary in order to 
remove atmospheric impurities. Special attention must be given to
 maintain the pressure relative to atmosphere within a tight 
tolerance, in order to limit dynamic field distortions caused
 by flexing of the endplate. 

An argon based gas mixture is the leading candidate but the choice 
of the best quencher to add to the argon is still an open question 
and depends on a number of factors. The electron drift velocity 
should be relatively fast and the transverse diffusion constant 
relatively small in presence of the strong magnetic field.  For 
the GEM readout option, it is beneficial to have 
its larger transverse 
diffusion in the amplification region, in order to ensure the 
charge is spread over more than one pad. To reduce the sensitivity 
to neutron machine backgrounds, it may be important to keep the hydrogen content of the gas 
mixture as small as possible.

\subsubsection{Endplate design}
The endplate should present as little material as possible for 
forward-going particles in order not to compromise the jet energy resolution 
in the forward direction.
High density electronics will allow for
 the possibility of mounting the electronics directly on the back 
of the readout plane, thus reducing the inactive volume. Pulsed 
power operation of the electronics is considered, so that air cooling 
is sufficient to limit the thermal gradient inside the TPC.

\subsection{Amplification and Readout Systems of TPC}
There are a number of different ideas to provide gas gain and sample 
the resulting electron signals. It is important that the system can 
accurately estimate the coordinates for the very narrow distribution
 of electrons that arrive. With moderate size pads, having a width 
of about 1--2 mm, there can be a loss of precision if the charge is 
collected by only one pad within a row. This can be alleviated by
 either having much smaller pads or by spreading the signals after
 the gas gain. The pads should be no larger than about 3 times the
 intrinsic width (standard deviation) of the signal.

GEM foils consist of a thin polyamide film clad on both surfaces 
with copper. Small holes are etched completely through in a fine 
grid pattern with a pitch of about $0.1$~mm. By applying a 
potential difference of about $350$~V between the two copper
 surfaces, large electric fields develop within the holes, 
sufficient to provide gas gain. To reduce the probability 
of sparking to develop, two or three GEM foils are stacked
 up to provide the gas gain in multiple stages.

MM devices have a wire mesh held a very small distance, 
typically less than $0.1$~mm, above the pad plane. A potential 
difference of about $400$~V is sufficient to provide good gas 
gain in the small region between the mesh and pad plane.

With the option of using GEM foils to amplify the drifting electrons, 
the right choice of gas can allow significant diffusion to occur when 
the electrons pass between the GEM foils. This defocussing allows more
 than one pad per row to sample the charge to maintain good spatial resolution.

With the narrow gas amplification region in a MM, it is not possible to use 
gas diffusion to ensure that signals are detected by more than one pad per 
row. An alternative solution to spread the signals over a larger area is to
 affix a resistive foil onto the pad plane. The surface resistance determines 
the spatial extent of the resulting induced signals. The same technique can 
also be applied for GEM gas amplification if the gas diffusion is not sufficient.

Another approach that is being considered is to use CMOS pixel readout, in 
order to measure the charge signals with very fine segmentation. This is 
particularly appropriate for the MM amplification which maintains
 the narrow distribution of the charge signals. The pitch of this readout
 is fine enough that ionization cluster counting may be possible to improve
 the particle identification performance.

\subsection{Challenges for the ILC TPC}
The demand for high precision for the ILC TPC presents significant demands 
on its design and calibration.

\subsubsection{Magnetic field uniformity}
The uniform magnetic field provided by the solenoid may be strongly 
modified by the presence of a detector integrated dipole (DID) or 
possibly anti-DID, which are being considered for helping to guide
 the beams through the detector for an interaction point with a 
large crossing angle (see chapter~\ref{detector_MDI}). Such a 
field affects the track parameter determination in a TPC in two 
ways. Firstly, the helix of a charged particle is distorted. 
Secondly, the paths that the electrons follow towards the endplate
 are no longer straight lines perpendicular to the readout plane. 
It will be important to have magnetic field maps taken under 
different magnetic field configurations to correct the observed 
data. Improved treatment will require good control samples of 
ionization. To produce these a calibration system is foreseen which produces a pattern 
of photo electrons on the cathode. 
Laser induced tracks will also be useful to detect, understand, and correct
track distortions.

\subsubsection{Positive ions in the drift volume}
The drifting electrons can also be affected by the presence of positive 
ions in the TPC. They are produced in the original ionization process, 
but more importantly in the gas gain regions of the TPC. The micropattern
 gas detectors suppress the number of positive ions that reach the drift 
volume from the gas amplification region, but not completely. To eliminate
 this problem, a gating plane, made using a wire grid or possibly an 
additional GEM layer is under consideration.

\subsubsection{Mechanical structure}
It will be challenging to design and build the TPC structure with relatively
 little material, and at the same time be very rigid. The endplate will 
likely be populated by a tiling of removable readout modules. The modules
 will need to be located to high precision with good mechanical stability. 

\subsection{Status of ILC TPC R\&D}
Over the past five years, a number of R\&D efforts around the world have 
been setup to study various aspects of the ILC TPC concept by constructing
 and operating relatively small prototypes. These studies include:
\begin{itemize}
\item Determination of the intrinsic spatial resolution and two particle 
separation power of the different readout options with and without magnetic fields.
In several cases, the spatial resolutions goals for short drift distances
have been achieved;
\item measurements of the charge transfer of electrons and positive ions through the devices;
\item investigation of different field cage designs.
To date these studies have been encouraging. 
\end{itemize}

\begin{figure}
   \centering
	\includegraphics[height=8cm]{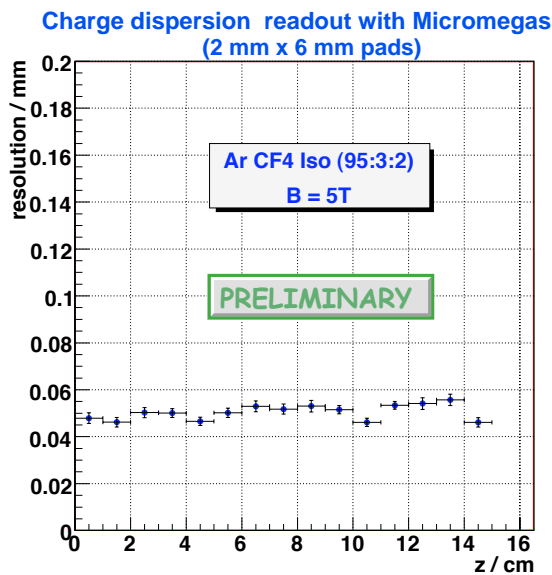}
        \caption[Spatial resolution of TPC with Micromegas and resistive pad readout]{Preliminary 
result of the spatial resolution of Micromegas readout as a function of drift length. 
A resistive pad plane was used to spread the signal charge. }
  \label{fig-tpc-ResolutionRegistiveAnode}
\end{figure}

Figure~\ref{fig-tpc-ResolutionRegistiveAnode} shows a typical result on a 
study of the spatial resolution 
of MM readout in conjunction with  a resistive pad foil used to insure charge spread.
The spatial 
resolution of less than $60\mu$m was reported for a large pad size of $2\times 6$ mm$^2$ and 
Ar-CF4-Isobutane gas using a small TPC.

It is important to verify whether the performance goals can also be reached 
for larger scale TPCs. To that end, the groups have formed a collaboration known as
LCTPC~\cite{ref-LCTPC} to coordinate the R\&D and to build a much larger prototype. 
This work will be carried out in conjunction with the EUDET program~\cite{ref-EUDET}.

\begin{figure}
   \centering
	\includegraphics[height=8cm]{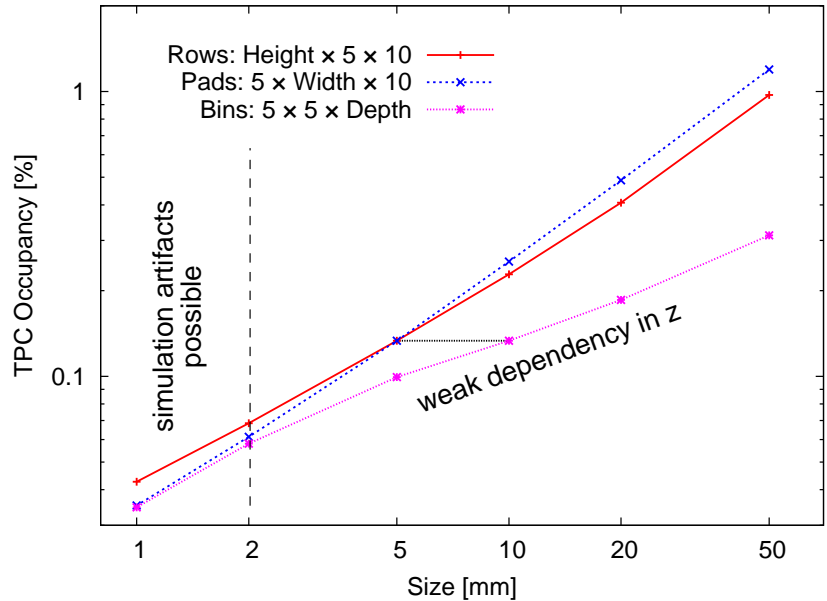}
        \caption[TPC occupancy due to beam-beam backgrounds]{Simulation of TPC occupancy
expected due to background electrons, photons and neutrons, as a function of voxel size. 
A voxel is defined as the space volume which can be resolved. A voxel may contain 
more than one readout channel. The horizontal axis defines the scale of the voxel, 
with the actual spatial extend defined in the picture, for different cases. Even for unrealistically large voxel sizes of a few cm, the occupancy stays below $1\%$.
}
  \label{fig-tpc-Occupancy}
\end{figure}

\begin{figure}
   \centering
	\includegraphics[height=8cm]{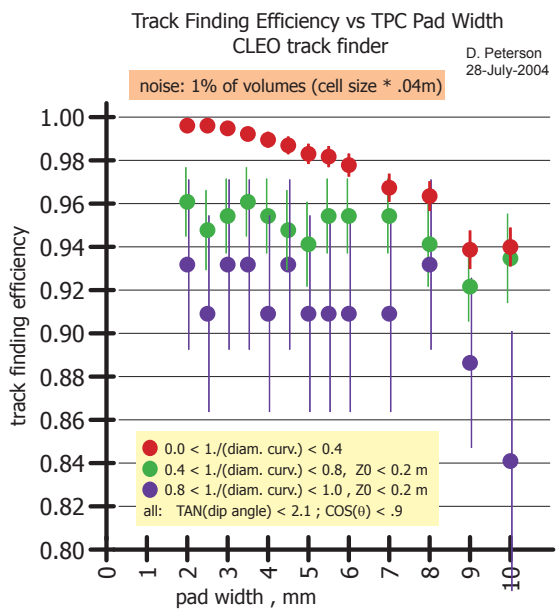}
        \caption[TPC tracking efficiency for $1$\% occupancy]{Study of TPC tracking
efficiency as a function of pad size for 1\% occupancy.}
  \label{fig-tpc-Trackingefficiency}
\end{figure}

A common software framework is under development for studies of the small 
prototype data and will be used for the large prototype. This software 
will also help better define some of the requirements for the full size TPC design.
For example, an important issue is the occupancy due to backgrounds, and 
recent simulation results are shown in Figure~\ref{fig-tpc-Occupancy} where 
it is seen that $< 0.1$\% is expected for pad sizes being considered.
Tracking efficiency remains near 100\% even for 10 times more occupancy, as the study in 
Figure~\ref{fig-tpc-Trackingefficiency} demonstrates.

\subsection{Cluster Counting Drift Chamber}
\label{ref-gas-cluster}

  A second option for a gaseous central tracker is the cluster counting
drift chamber modeled on the successful KLOE\footnote{The KLOE
experiment studies $e^+e^- \rightarrow \phi \rightarrow K^+ K^-$ in
which the slow kaons have both small $p$ and small $\beta$ necessitating
a tracking chamber with a minimum of multiple scattering material.}
main tracking chamber.  This drift chamber (CluCou) maintains very low
multiple scattering due to a He-based gas and aluminum wires in the
tracking volume and, with carbon fiber end planes, forward tracks that
penetrate the wire support frame and the close-in electronics  beyond
$\cos \theta \approx 0.7$ suffer only about 15-20\% Xo of material.  The
KLOE chamber is one of the largest, highest performance and most
transparent tracking chambers ever constructed\cite{ref-TrRev} and has
operated successfully for 10 years.

  The He-based gas reduces substantially the material in the tracking
volume, thereby directly improving momentum resolution in the multiple
scattering dominated region below 50 GeV/c.  This He gas also has a low
drift velocity allowing a new cluster counting technique\cite{ref-TrRev}
that clocks in individual ionization clusters on every wire, providing
an estimated 50 micron spatial resolution per point, a $dE/dx$
resolution near 3\%, and $z$-coordinate information on each track
segment through an effective dip angle measurement.  The drift time in
each cell is less than the 300 ns beam crossing interval, and therefore
this chamber sees only one crossing per readout.  

  The critical issues of occupancy and two-track resolution are being
simulated for ILC events and expected machine and event backgrounds, and
direct GHz cluster counting experiments are being performed.  This
chamber is midway between the faster, higher precision silicon chamber
and the slower 3-d space point information provided by a TPC, and is
orthogonal to both with respect to its low multiple scattering.

\section{Calorimetric Systems}

Calorimeters of the ILC detectors serve for a precise jet energy measurement,
for the precise and fast measurement of the luminosity, and to ensure 
hermeticity down to small polar angles.
To fully exploit the physics potential of the ILC, 
the resolution of the jet energy measurement,
$\sigma_{\rm E}$/E,
is required to be $\approx 3 - 4 \%$, or $30\%/\sqrt(E)$ at energies below 
about 100~GeV.
This resolution, being about a factor of two smaller 
as the best currently operating calorimeters, must be maintained almost over the full polar
angle range. The ability to detect single high energy electrons with nearly 100\% efficiency 
is even required at very small polar angles  
to ensure the potential for new particle searches. 
Special calorimeters 
in the very forward region will make this possible. They will also deliver a fast
and a precise measurement of the delivered luminosity.   

To approach the required jet energy resolution, research is done for two different calorimeter
concepts. The first, followed by the majority of R\&D projects, is the development of extremely 
fine grained and compact calorimeters with single particle shower imaging. 
The particle flow concept is used
to determine jet energy and direction. 
Tracks are matched to their depositions inside the calorimeters. 
Depositions without matched tracks are assumed to originate from neutral particles inside a jet.
The jet energy is then determined from the charged track momenta and the depositions from neutral particles
in the calorimeters. Using Monte Carlo simulations it has been demonstrated that a significant improvement 
of the jet energy resolution is feasible, but substantially more effort is needed to optimize the calorimeter design,
to improve the particle flow algorithms and, most important, to develop
the calorimeter technologies and to verify the Monte Carlo simulations by test-beam
measurements.

The second, followed by one group, exploits the dual readout of scintillation and Cherenkov light of fibers  or crystals (DREAM). 
The
electromagnetic and hadronic component inside a shower can be separated and finally properly recombined
with a gain in resolution due to reduced fluctuations.

\subsection{Electromagnetic Calorimeters for Particle Flow approach}

Electromagnetic calorimeters (ECAL) are designed as 
compact and fine-grained sandwich calorimeters 
optimized for the reconstruction of photons and electrons and for separating them
from depositions of hadrons.
To keep the 
Moliere radius near the minimum possible tungsten or lead are 
used as absorber. Sensor planes are made of silicon pad diodes,
monolithic active pixel sensors (MAPS)
or of scintillator
strips or tiles. Also the combination of silicon 
and scintillator sensor planes was investigated.
The range of energies of electrons and photons suggests a thickness 
of about 24 radiation length for the ECAL.  

\subsubsection{Silicon Tungsten Sandwich Calorimeter}

Tungsten is chosen as a radiator because of its small Moliere radius of 9.5 mm  
minimizing the transversal shower spread.
To reach adequate energy and position resolution over the necessary 
energy range,
the sampling thickness should be finer on the side pointing to the
interaction point than at the rear side, changing e.g. from about 0.6 to 1.2 X$_{\rm o}$.
\begin{figure}[htb]
\begin{minipage}{0.45\textwidth}
\includegraphics[width=0.9\textwidth,height=0.8\textwidth,bbllx=8mm,bblly=5mm,bburx=200mm,bbury=182mm]{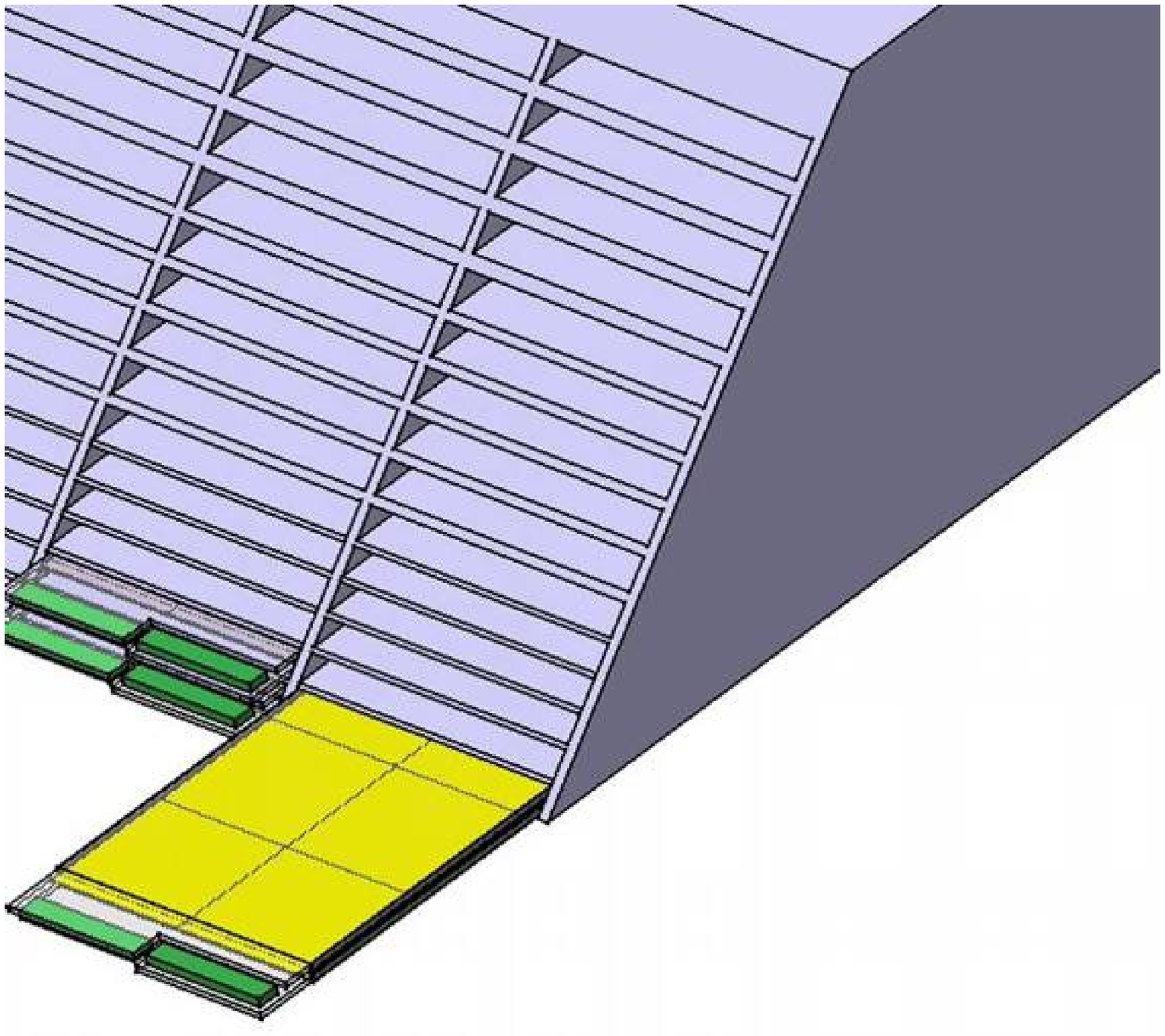}
\caption[Structure of the Si-W calorimeter]{\label{fig:Calice_frame} 
The structure of the silicon tungsten calorimeter, as proposed by the CALICE
collaboration. The slots in the tungsten frame
are equipped with detector slabs.
  }
\end{minipage}
\begin{minipage}{0.02\textwidth}
~~
\end{minipage}
\begin{minipage}{0.45\textwidth}
\includegraphics[width=\textwidth,height=0.8\textwidth]{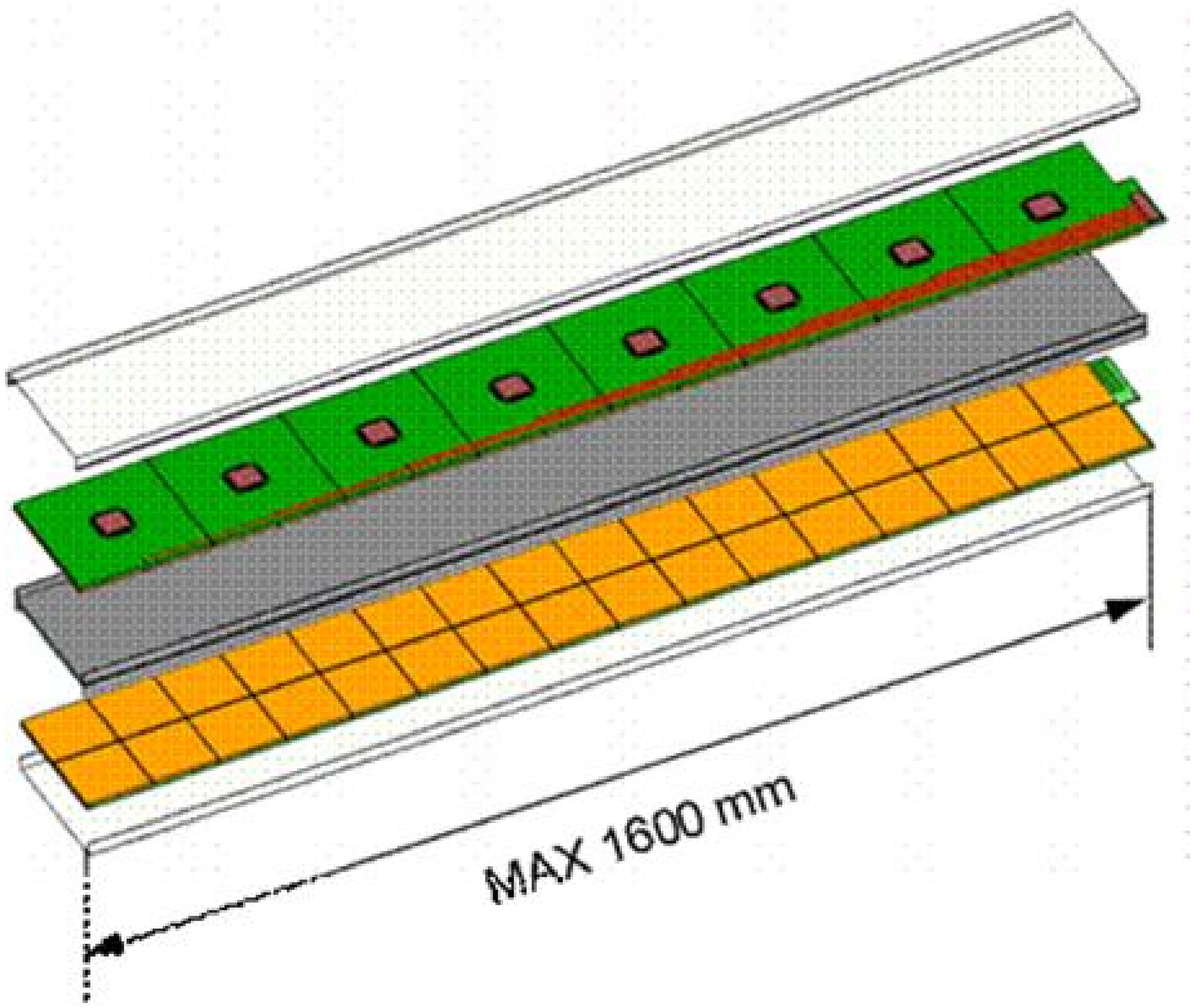}
\caption[Single detector slab of the CALICE proposal]{\label{fig:det_slab} 
A single detector slab which will be inserted into the ECAL support structure. The tungsten 
absorber plate (grey) is attached
at both sides by silicon pad sensors with FE chips on top. 
  }
\end{minipage}
\end{figure}
\begin{figure}[htb]
\begin{minipage}{0.45\textwidth}   
\includegraphics[width=0.9\textwidth,height=0.85\textwidth]{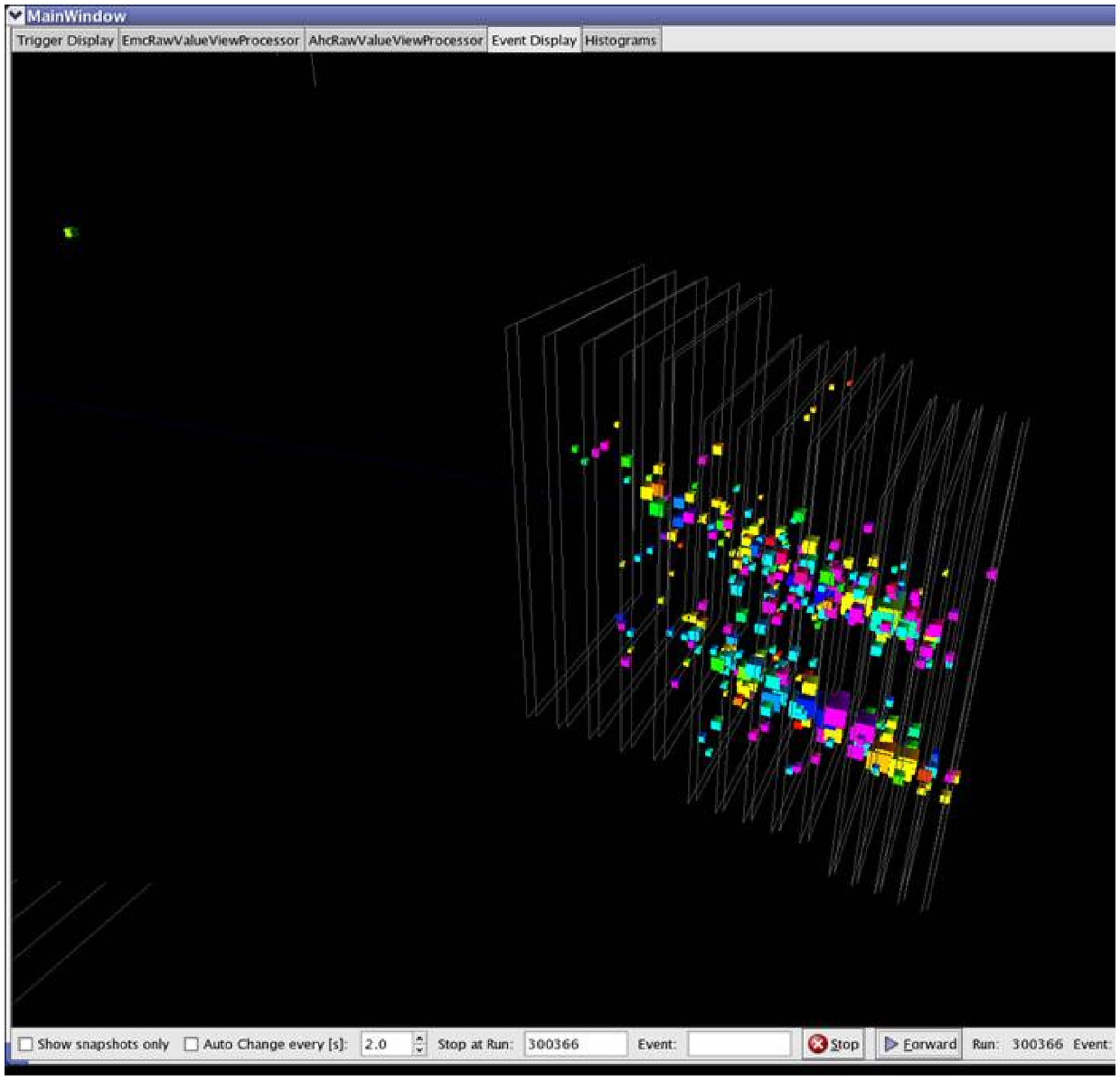}
\caption[Two electron showers in the CALICE prototype detector]{\label{fig:ECAL_test} 
Two electron showers recorded by the CALICE prototype detector
in the test-beam at CERN.
  }
\end{minipage}
\begin{minipage}{0.02\textwidth}
~~
\end{minipage}
\begin{minipage}{0.45\textwidth}    
\includegraphics[width=\textwidth,height=0.8\textwidth]{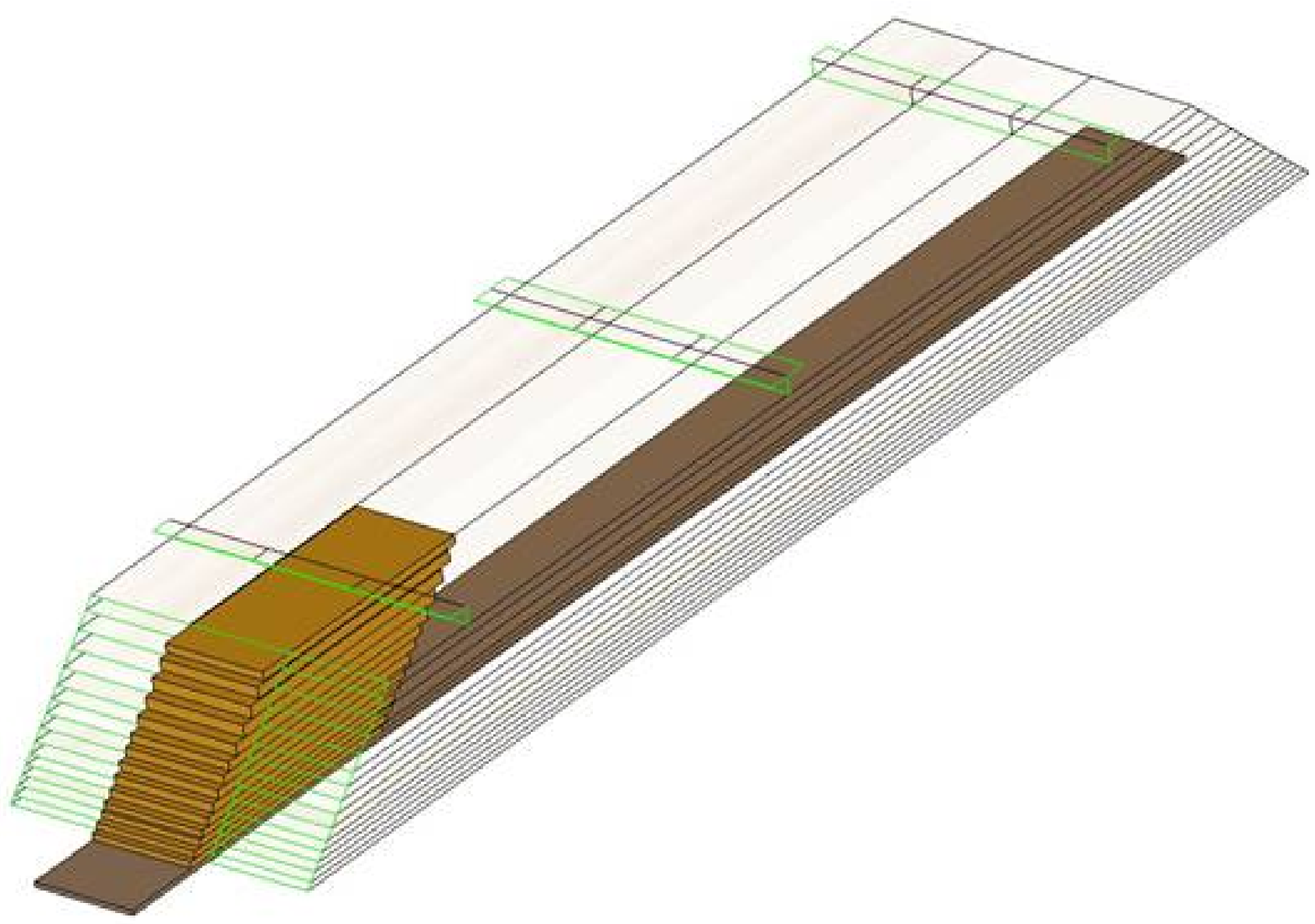}
\caption[The CALICE module]{\label{fig:EU_mod} 
The CALICE module. It will be a full size calorimeter module partly equipped with
500 kg absorber plates and detector slabs. 
  }
\end{minipage}
\end{figure}    
\begin{figure}[htb]
\begin{minipage}{0.45\textwidth}   
\includegraphics[width=1.1\textwidth,height=0.8\textwidth]{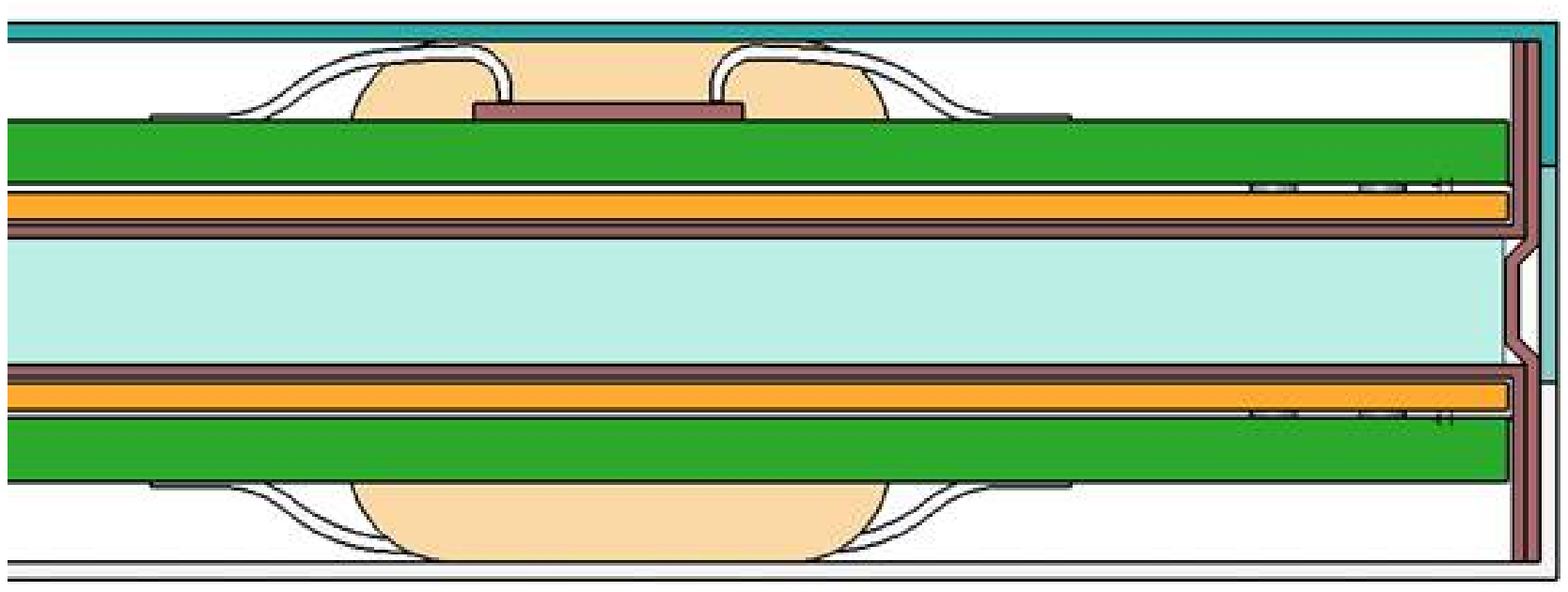}
\caption[Structure of a detector slab]{\label{fig:EU_slab} 
Structure of a detector slab. The tungsten plate in the center (light grey)
is enclosed by silicon sensor planes (light red) and PCBs 
(green) on both sides. The FE chip is integrated in the PCB.
  }  
\end{minipage}
\begin{minipage}{0.02\textwidth}
~~
\end{minipage}
\begin{minipage}{0.45\textwidth}    
\includegraphics[width=\textwidth,height=0.65\textwidth]{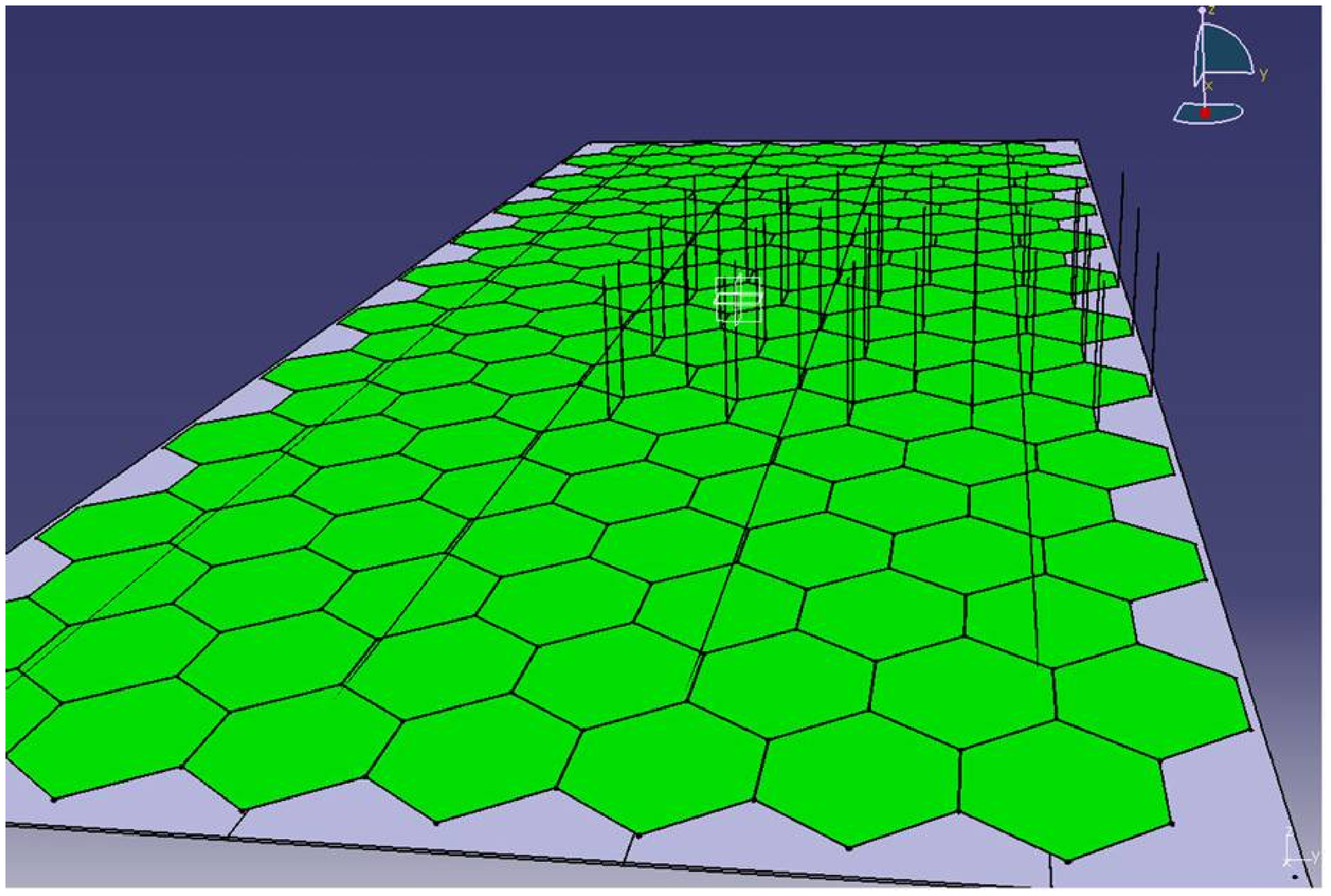}
\caption[Mechanical structure of the SiD Si-W calorimeter]{\label{fig:US_ecal} 
The mechanical structure for the ECAL proposed by the Silicon Detector group. Tungsten planes (grey) of 200 kg
weight are joined to a module by rods (black). Hexagonal sensor planes (green)
 are placed in between the tungsten plates. 
  }
\end{minipage}
\end{figure}  
Two groups study silicon tungsten calorimeters in detail. The first, within the 
CALICE~\cite{calice} 
collaboration,
proposes a mechanical frame made of carbon fiber reinforced epoxy 
with integrated tungsten absorber plates, as shown in Figure~\ref{fig:Calice_frame}. 
Between the absorber plates space is left
for detector slabs, sketched in Figure~\ref{fig:det_slab},
containing silicon sensor planes. 
The silicon sensors are structured 
with quadratic pads of 5x5 mm$^2$ size,
being about a third of the Moliere radius, and about 300 $\mu$m thickness. 
The sensors are glued to a PCB and
to both sides of a tungsten plate wrapped with a H structure made from carbon fiber composite.
The front-end electronics ASICs are soldered on the PCB.
Data are processed
in the front-end ASIC and concentrated by a chip on the edge of the detector slab.
A dynamic range of 15 bit is required. Valid data are shifted to an analog memory,
digitized on chip and stored during the full
bunch train. The concentrator flushes the data after the bunch train and sends them to the DAQ.
In order to avoid active cooling the power dissipation should not exceed 100 $\mu$W per channel.
Power will be pulsed and switched off in between the bunch trains, i.e. 99\% of the time.

CALICE has built a prototype calorimeter with sensors of 1x1 cm$^2$ 
pad size and took data in an
electron beam of about 5~GeV 
at DESY and in a higher energy hadron test-beam at CERN. 
An event display of the showers of two nearby electrons of 
20~GeV recorded when the calorimeter is tilted with respect to the beam axis, is shown in 
Figure~\ref{fig:ECAL_test}. 
The data obtained in the test-beam will be used to determine the performance
of the prototype calorimeter with respect to energy resolution, shower position
resolution and two-shower separation. It will furthermore allow comparison and refinement
of Monte Carlo simulations important for the understanding of the PFA approach.

CALICE prepares in parallel a second prototype, called EUDET module, 
which will be as close as possible to the final design.
This is a full length structure, 
as shown in Figure~\ref{fig:EU_mod}, partly equipped with sensors and absorbers.
The pad size is $5 \times 5$~mm$^2$, resulting in 40k channels to be readout. 
The details of a detector slab are shown in Figure~\ref{fig:EU_slab}. Two 
silicon sensor planes
 are
attached to each side of a tungsten absorber plate. The pads on the sensors are read out
by a FE chip ILC-PHY5 with a 12 bit ADC on chip. The first chip submission occurred at the end of 2006.
Being power pulsed, the total power dissipation will be $25 \mu$W per channel.
The construction and test of the EUDET module will be the proof of the final design
of the CALICE ECAL.
\begin{figure}[htb]
\begin{minipage}{0.35\textwidth}   
\includegraphics[width=0.9\textwidth,height=0.8\textwidth]{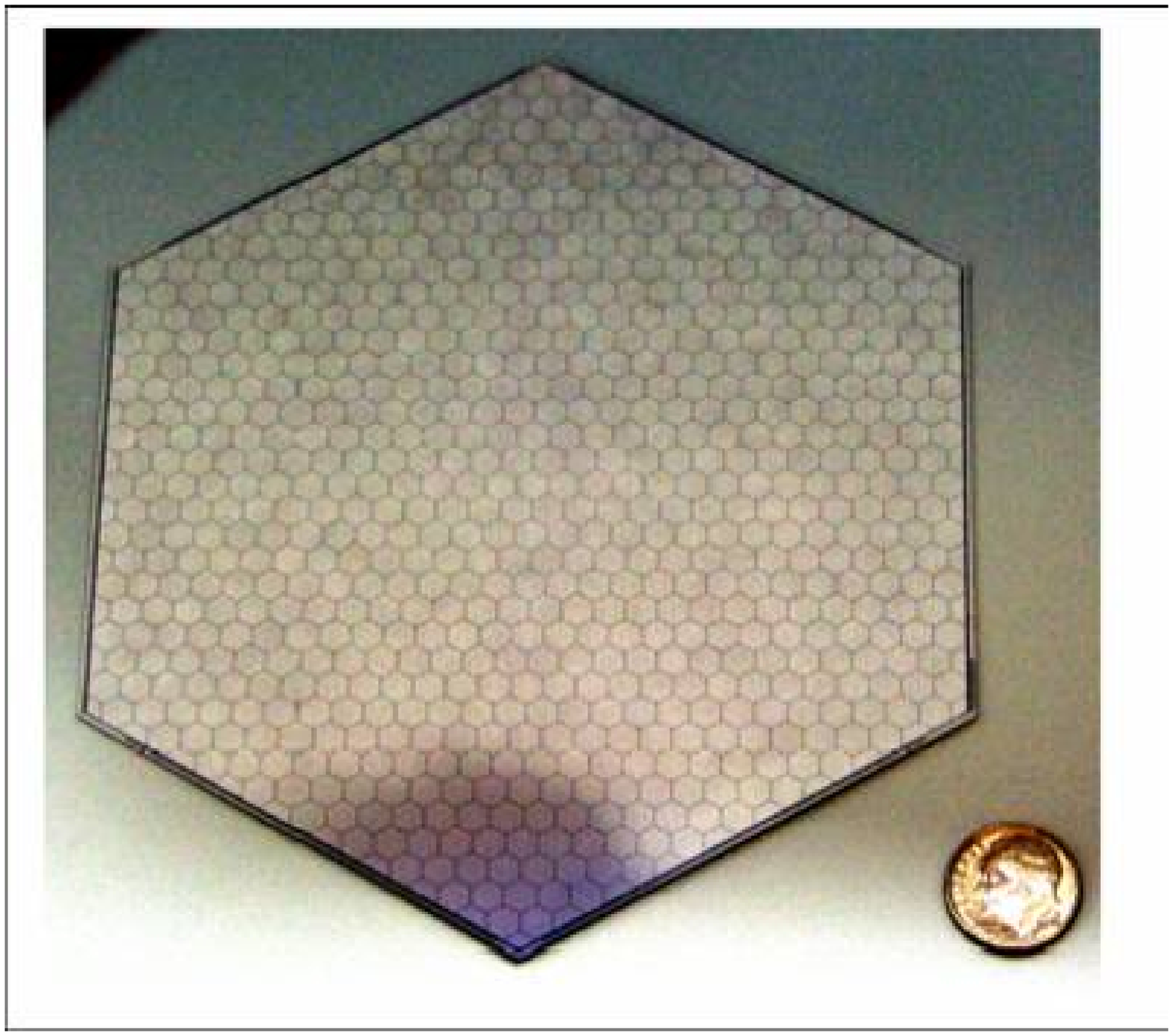}
\caption{\label{fig:sixogon} 
A silicon sensor prototype with hexagon pads.
  }
\end{minipage}
\begin{minipage}{0.02\textwidth}
~~
\end{minipage}
\begin{minipage}{0.63\textwidth}    
\includegraphics[width=0.9\textwidth,height=0.35\textwidth]{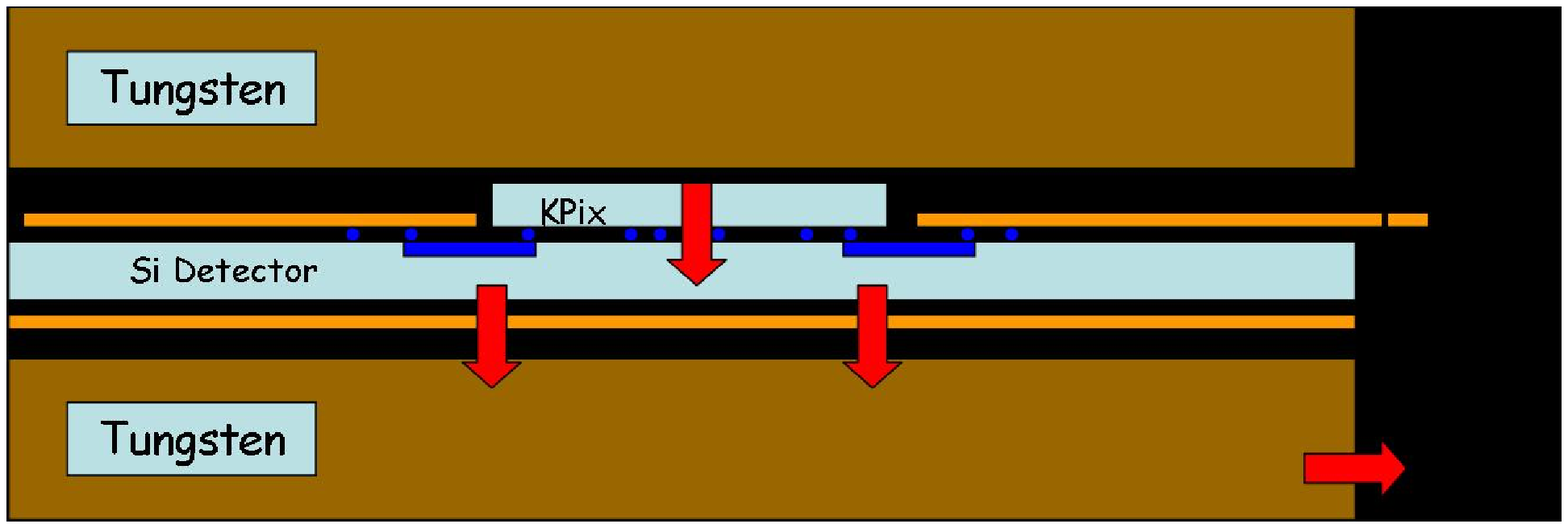}
\caption[Cross section of the ECAL]{\label{fig:sample_struct} 
The cross section of the ECAL. The silicon sensors interspersed between tungsten plates
and readout by the KPiX FE chip bump bonded to metallic traces connected to each sensor cell. 
  }
\end{minipage}
\end{figure}

\begin{figure}[htb]
\begin{minipage}{0.45\textwidth}   
\includegraphics[width=0.9\textwidth,height=0.8\textwidth]{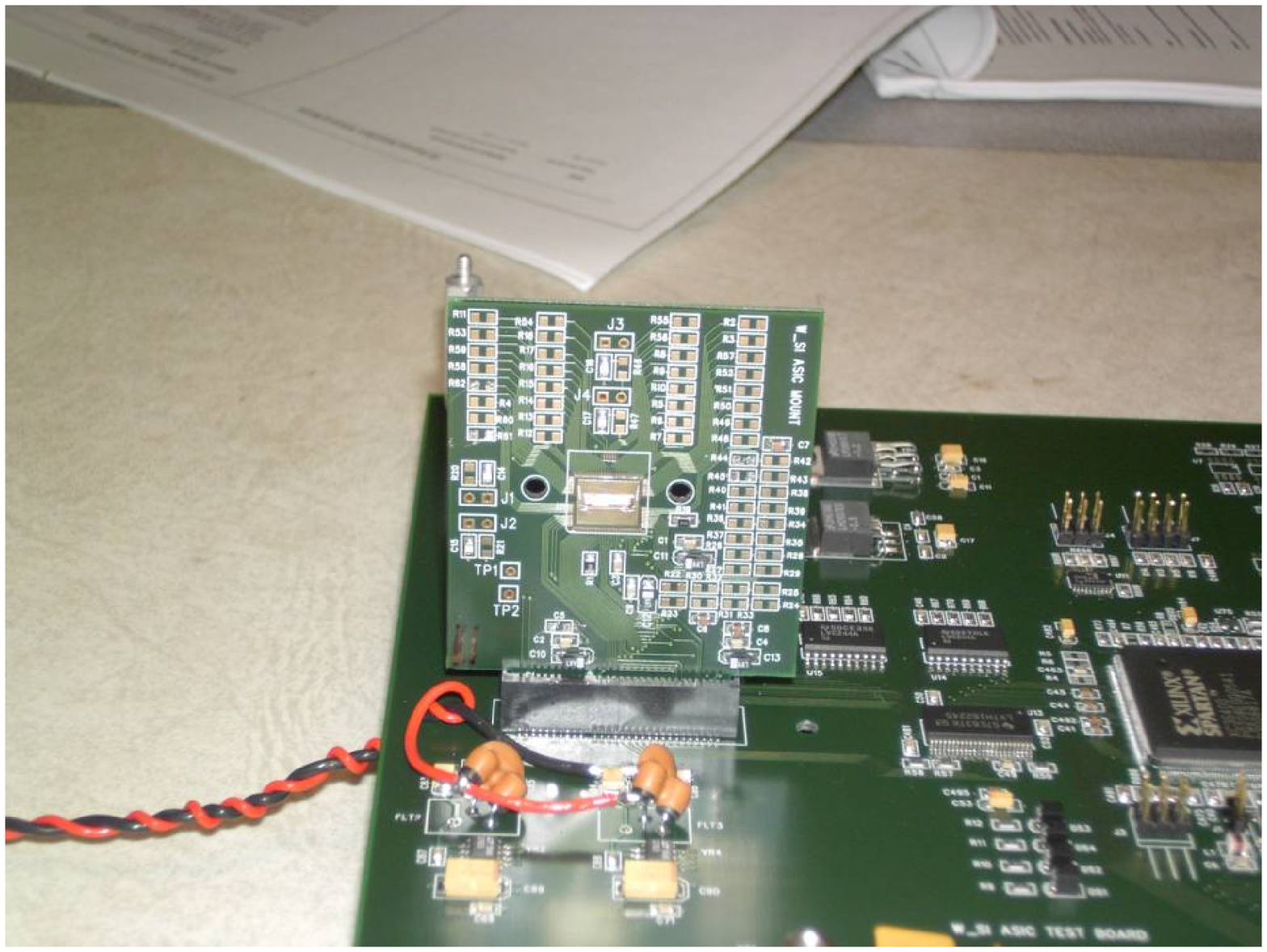}
\caption{\label{fig:kpix} 
The KPiX FE readout chip in a test bench at SLAC.
  }
\end{minipage}
\begin{minipage}{0.02\textwidth}
~~
\end{minipage}
\begin{minipage}{0.45\textwidth}    
\includegraphics[width=\textwidth,height=0.75\textwidth]{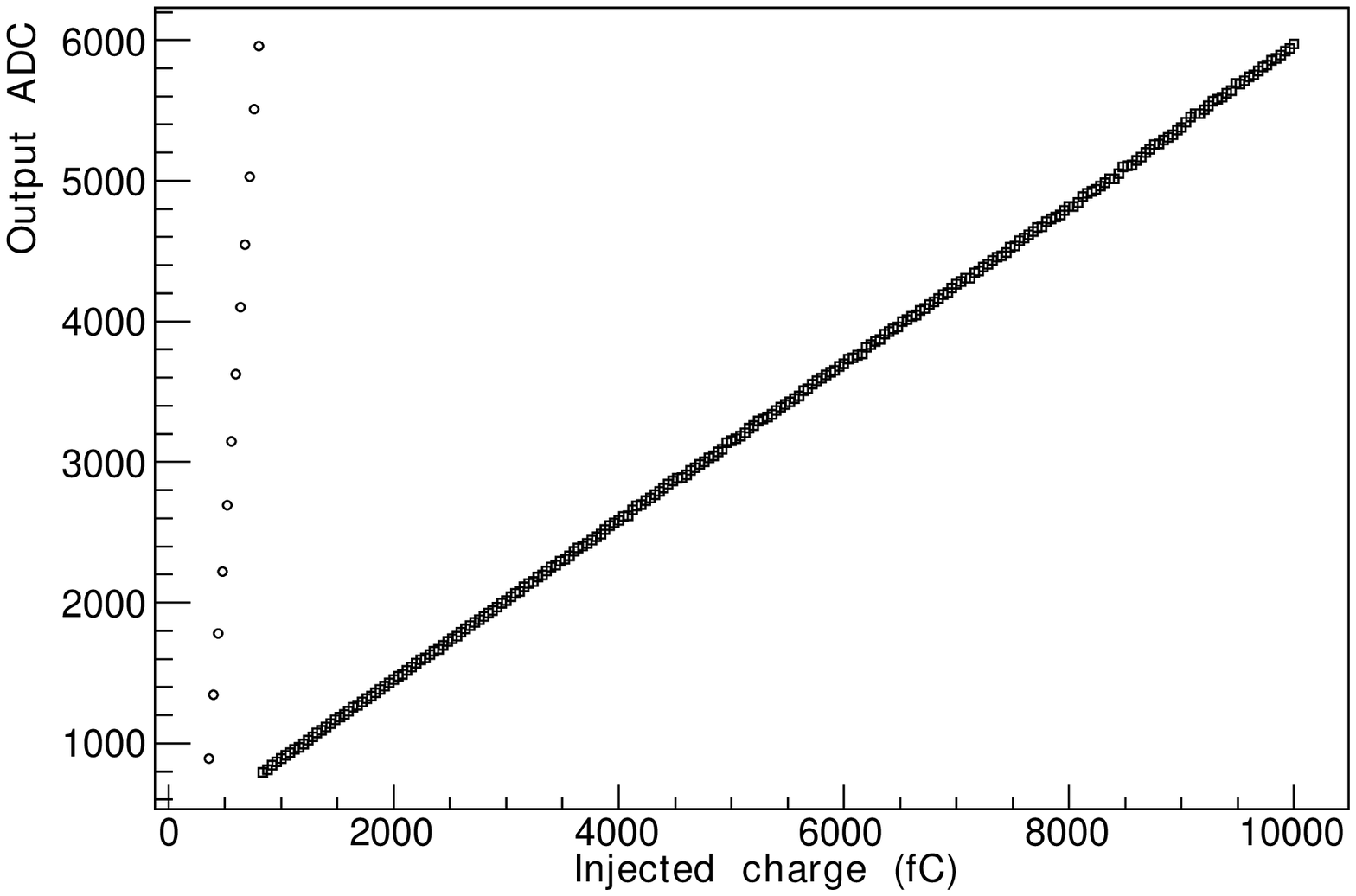}
\caption[Results of linearity test of the KPIX chip]{\label{fig:kpin_lin} 
Linearity test of the KPiX readout chip. The digitized signal is
shown as a function of the injected via a calibration circuit. 
  }
\end{minipage}
\end{figure}

The second project is pursued by groups collaborating on the Silicon Detector Design Study.
~\cite{usuniv}. 
Mechanical stability is obtained by connecting the 
tungsten planes, as shown in Figure~\ref{fig:US_ecal},
 by cylindrical rods. A 1mm gap is left 
for the silicon sensor planes. The total thickness of the tungsten absorber corresponds to 
27 X$_{\rm o}$.
The basic active element consists of hexagonal silicon planes made from 6 inch wafers, maximizing the 
use of sensitive area of a wafer, as shown in Figure~\ref{fig:sixogon}. 
The silicon plane is subdivided in 1024 hexagonal 
diodes of 12 mm$^2$ area each. Each plane will be readout by one 1024 channel ASIC (KPiX).
The chips are bump bonded to
the sensor plane, as can be seen from  Figure~\ref{fig:sample_struct}. 
The KPiX chip performs the analog conditioning and 15-bit digitization for all channels
and tags hits with bunch crossing information, to minimize backgrounds which are out of time.
The data are serialized and transported by transverse data cables to the edge of a calorimeter module.
Several modules are combined to feed the signals to a data concentrator.
It is worth noting that KPiX has been adapted for use with silicon microstrip detectors and RPC and GEM detectors for
the hadronic calorimeter and muon system.
In Figure~\ref{fig:kpix} a KpiX chip is shown in a test bench at SLAC. Measurements done so far
on linearity of the response and timing correspond to the expected performance. 
As an example Figure~\ref{fig:kpin_lin} shows the digitized signal
as a function of the input charge injected via the internal calibration circuit.
A novel feature of the KPiX chip is the dynamic switching, which accommodates the 
large dynamic range required for the ECAL. The charge equivalent of one MIP is 4.1 fC
allowing a good signal/noise for MIP detection.
In Figure~\ref{fig:kpin_lin} the switching occurs around 700 fC.
The upper end of the Figure  corresponds to about 2500 MIPs, roughly the expected 
maximum signal for a 
500 GeV electron incident under 90$^\circ$ at the shower maximum using
12 mm$^2$ pixels.
More tests are needed to understand e.g. channel-by-channel variations. 
 
The goal of the R\&D is to fabricate a full-depth electromagnetic calorimeter prototype module. This will consist of
30 longitudinal layers, each consisting of an about 15 cm 
diameter silicon detector outfitted with a KPiX chip sandwiched between 2.5 mm 
thick tungsten radiator layers. The module will be fully characterized for electromagnetic 
response and resolution in an electron beam, probably at SLAC in 2007.
A first round of 10 silicon detectors, made from a 6 inch wafer, 
has been purchased and tested in the laboratory, and a second round submitted.
Several prototypes of the KPiX chip are successfully tested. 
A second, improved version, is under preparation.  
The light cable for signal transport inside the gap 
is being designed and preparations for bump bonding are underway.

\subsubsection{Monolithic Active Pixel Digital ECAL}

Recently a group from RAL within the CALICE Collaboration has proposed using monolithic active pixel
sensors (MAPS) instead of silicon pad diodes for the ECAL.
MAPS are produced in CMOS technology, widely used in semiconductor industry. To instrument 
an ECAL with MAPS might be of lower cost 
than using the high resistivity silicon needed for the previous designs.
The readout of the pixel will be binary. To ensure that a pixel inside a shower is 
mostly hit only by one  
particle, the pixel size must be about 40x40 $\mu$m$^2$. The total number of pixel for the ECAL will be about 
8 x 10$^{11}$.
The signals on the pixel during a bunch train are stored on the sensor with time stamps and hit pixel numbers
and readout between trains. To avoid a critical amount of noise hits a S/N ratio of larger than 15 is required.
The use of 0.18$\mu$m CMOS technology is planned. 

\subsubsection{Scintillator Tungsten Sandwich Calorimeter}

For a calorimeter with a large radius, a finely segmented scintillator-based sandwich calorimeter 
may have a particle flow performance similar to a compact silicon-tungsten calorimeter,
but might have lower cost. A group of Asian Labs within CALICE~\cite{calice_asia} 
plans a sandwich calorimeter using plastic
scintillator as sensor.
\begin{figure}[htb]
\begin{minipage}{0.55\textwidth}
\begin{center}
\includegraphics[width=0.75\textwidth]{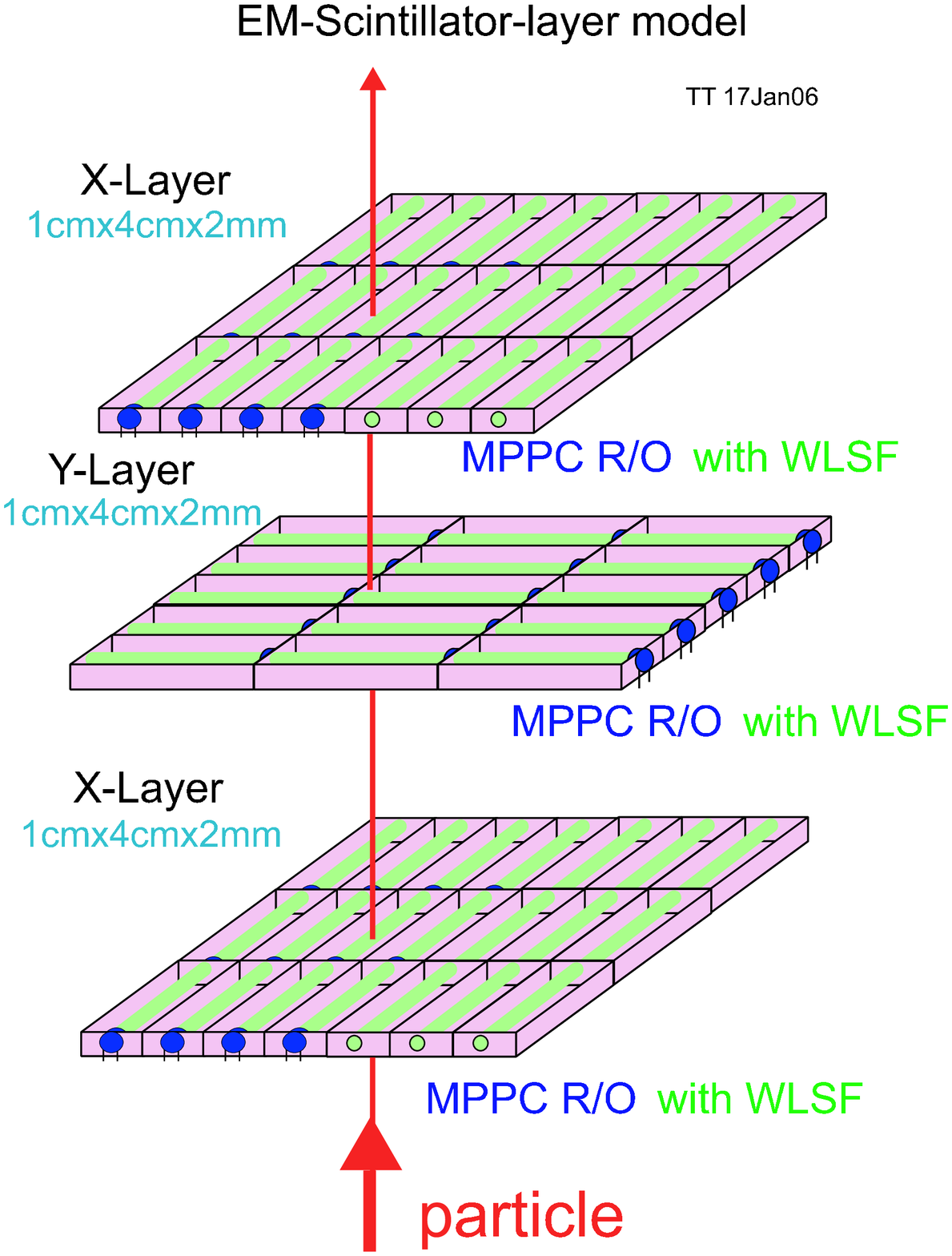}
\caption[A possible strip sequence of the scintillator ECAL]{\label{fig:ECAL_japan} 
A possible strip sequence of the ECAL for GLD. Layers of scintillator strips are oriented
perpendicular to each other. Each strip is equipped with
a wavelength-shifting fiber (green) and readout by a MPPC (blue dots). 
  }
\end{center}
\end{minipage}
\begin{minipage}{0.02\textwidth}
~~
\end{minipage}
\begin{minipage}{0.4\textwidth}    
\includegraphics[width=1.1\textwidth,height=0.7\textwidth]{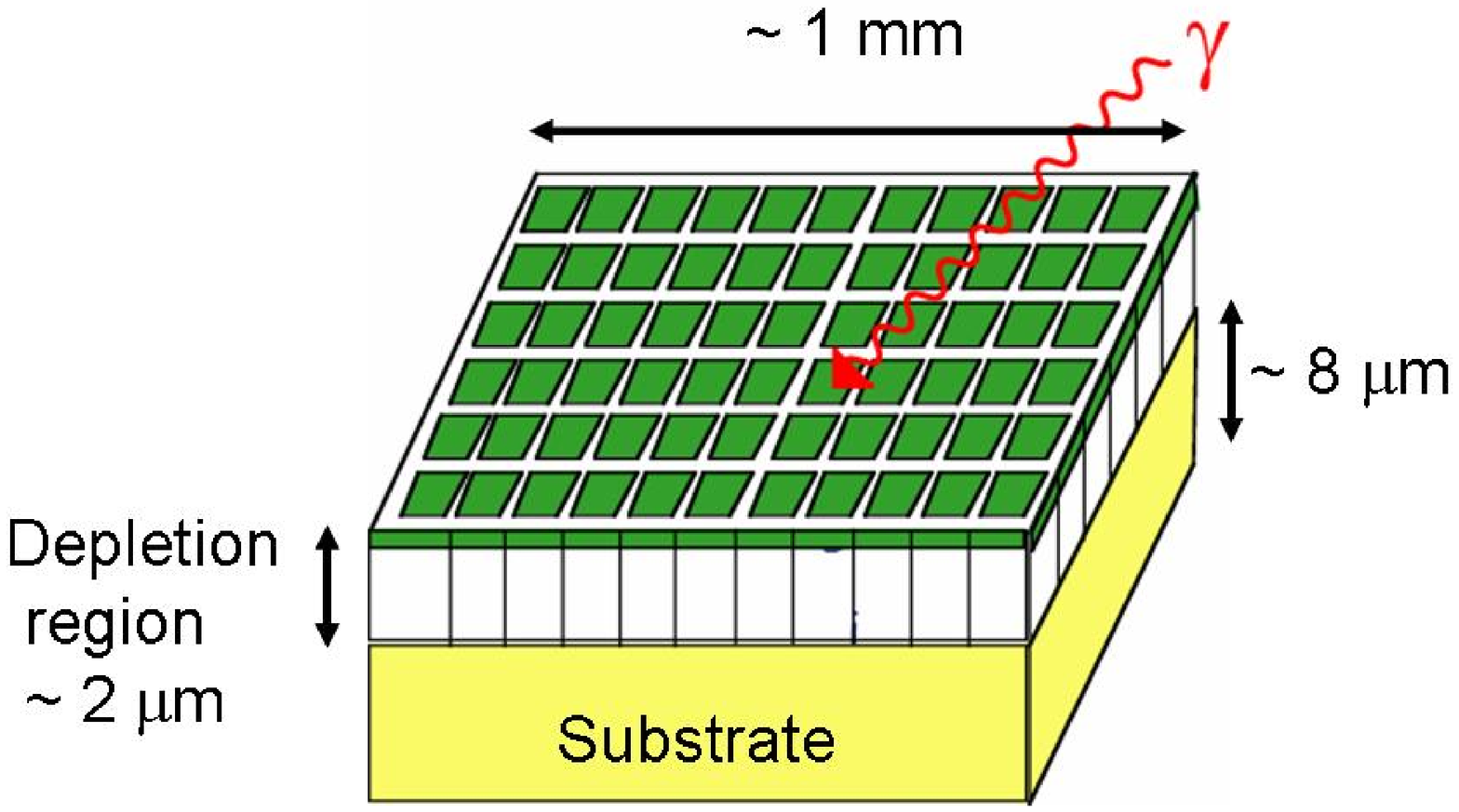}
\caption[Schematic of a MPPC]{\label{fig:MPPC_proto} 
Scheme of a MPPC. A surface of about 1x1 mm$^2$ is structured into pixel diodes. A photon absorbed 
by a pixel induces an avalanche which creates an electrical signal.
  }
\end{minipage}
\end{figure}
Layers of scintillator strips, oriented perpendicular to each other as shown in
Figure~\ref{fig:ECAL_japan},
are placed in between 
tungsten absorber plates. The effective segmentation
given by the strip width is 1x1 cm$^2$.
Each 
strip or tile is equipped  with a wavelength-shifting fiber
 readout by 
novel Geiger mode photo-diodes, called here  multi-pixel photon counter, 
MPPC. 
Figure~\ref{fig:MPPC_proto} illustrates the operation of an MPPC. 
Each pixel is an independent diode with a relatively large
electrical field in the depletion region. A photon absorbed by a pixel
induces an avalanche in the depletion region, inducing a pulse in the
bias voltage circuit. 

\begin{figure}[htb]
\begin{minipage}{0.45\textwidth}
\includegraphics[width=0.9\textwidth,height=0.7\textwidth]{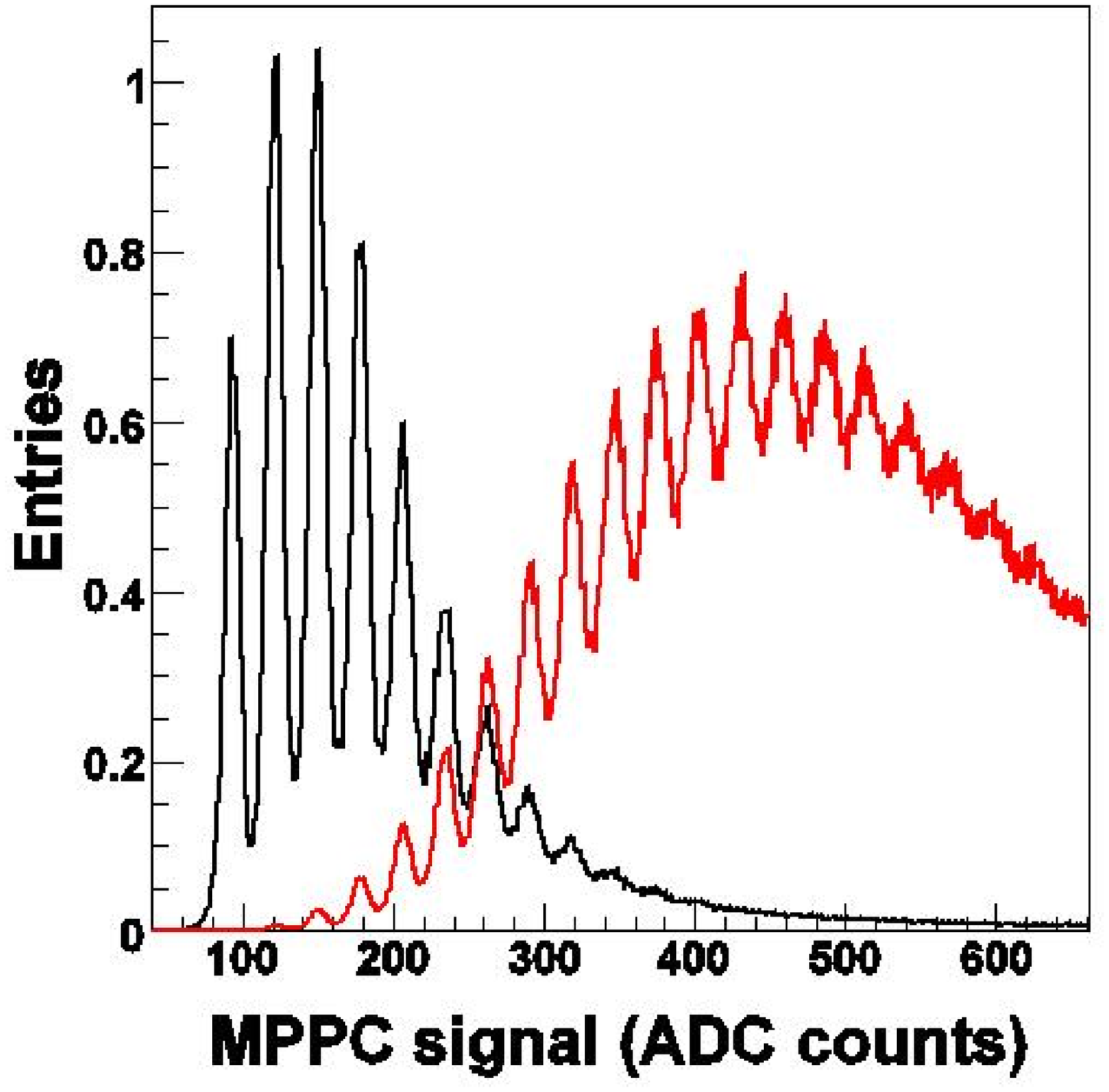}
\caption[Spectrum measured with a MPPC]{\label{fig:MPPC_singleP} 
The spectrum measured with a MPPC of a relatively small (black) and a larger (red) light
signal.  
  }
\end{minipage}
\begin{minipage}{0.02\textwidth}
~~
\end{minipage}
\begin{minipage}{0.45\textwidth}    
\includegraphics[width=0.9\textwidth,height=0.7\textwidth]{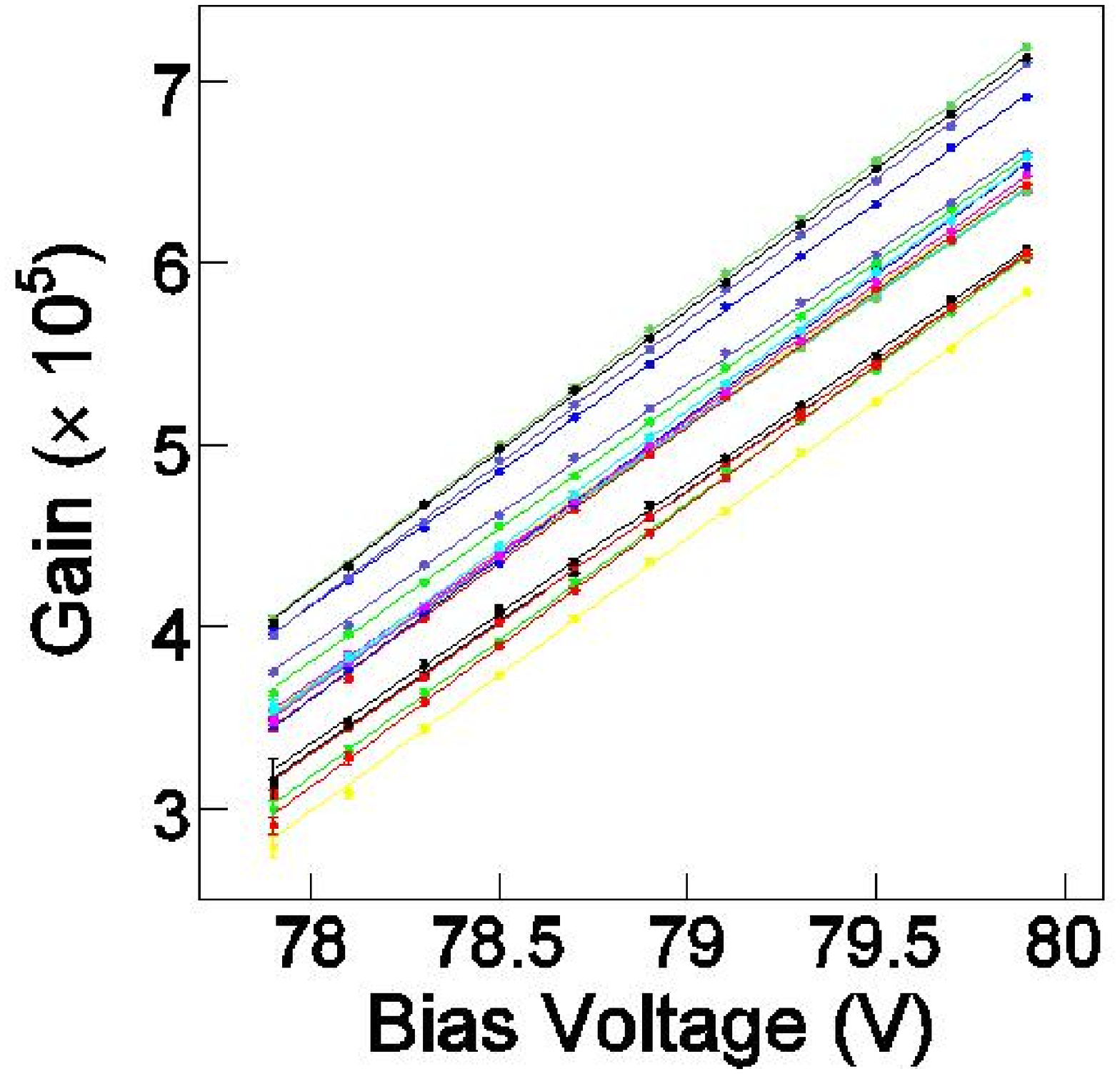}
\caption[The gain of MPPCs]{\label{fig:MPPC_gain} 
The gain of several MPPCs with 1600 pixels as a function of the applied voltage.
}
\end{minipage}
\end{figure}

Prototypes of MPPCs are available with 400 and 1600 pixels. 
From Monte Carlo simulations
it is estimated that for electromagnetic shower reconstruction 
2500 pixels are necessary to match the required performance.
The gain is between 10$^5$ and  10$^6$ for depletion
voltages of 30-70 volts.  The photon detection efficiency is about 25\% 
for devices with 1600 pixels and the time resolution 1 ~ns.   
MPPCs will work in a magnetic field. They show, however, 
a relatively large noise in the range of several MHz.
MPPCs have excellent capability to count 
photoelectrons, as it is shown in the output signal spectra in Figure~\ref{fig:MPPC_singleP}.
Illuminating the MPPCs with faint light pulses (black curve) the peaks
for zero, one and more photoelectrons are nicely visible.

Since the number of pixels on a MPPC is limited, the response as a function of the number of photons is non-linear
for brighter light pulses.
A sample of 20 MPPCs is studied to estimate device-by-device variation. As an example, in  Figure~\ref{fig:MPPC_gain}
the gain as a function of the voltage applied is shown. Above the breakdown voltage a linear 
dependence is observed. For a given voltage the gain variations of the sample are about 30\%. 
In addition, the gain depends on the temperature. The cross-talk between adjacent
pixels is measured to be between 2 and 20\%, depending on the applied voltage. 
Nothing  is known on the long-term  performance stability of MPPCs.
KEK together with Japanese universities launched a Detector Technology Project to develop 
and study MPPCs in collaboration with the Hamamatsu Company~\cite{technology_project}. 

A prototype calorimeter of a structure similar to the GLD design but read-out with classical 
multi-anode
photo-multipliers was tested in an electron beam
of energies from 1 to 4 GeV at KEK. The energy and shower position resolutions of about
$13\%/\sqrt E$ and $4.5/\sqrt E$ mm, respectively, agreed perfectly with Monte Carlo simulations.
In addition, the angle of the shower axis was measured with a resolution
of $4.8/\sqrt E$ degrees, demonstrating the ability to detect photons not originating from the 
interaction point~\cite{KEK_test}.

\begin{figure}[htb]
\begin{minipage}{0.45\textwidth}
\includegraphics[width=0.9\textwidth,height=0.7\textwidth]{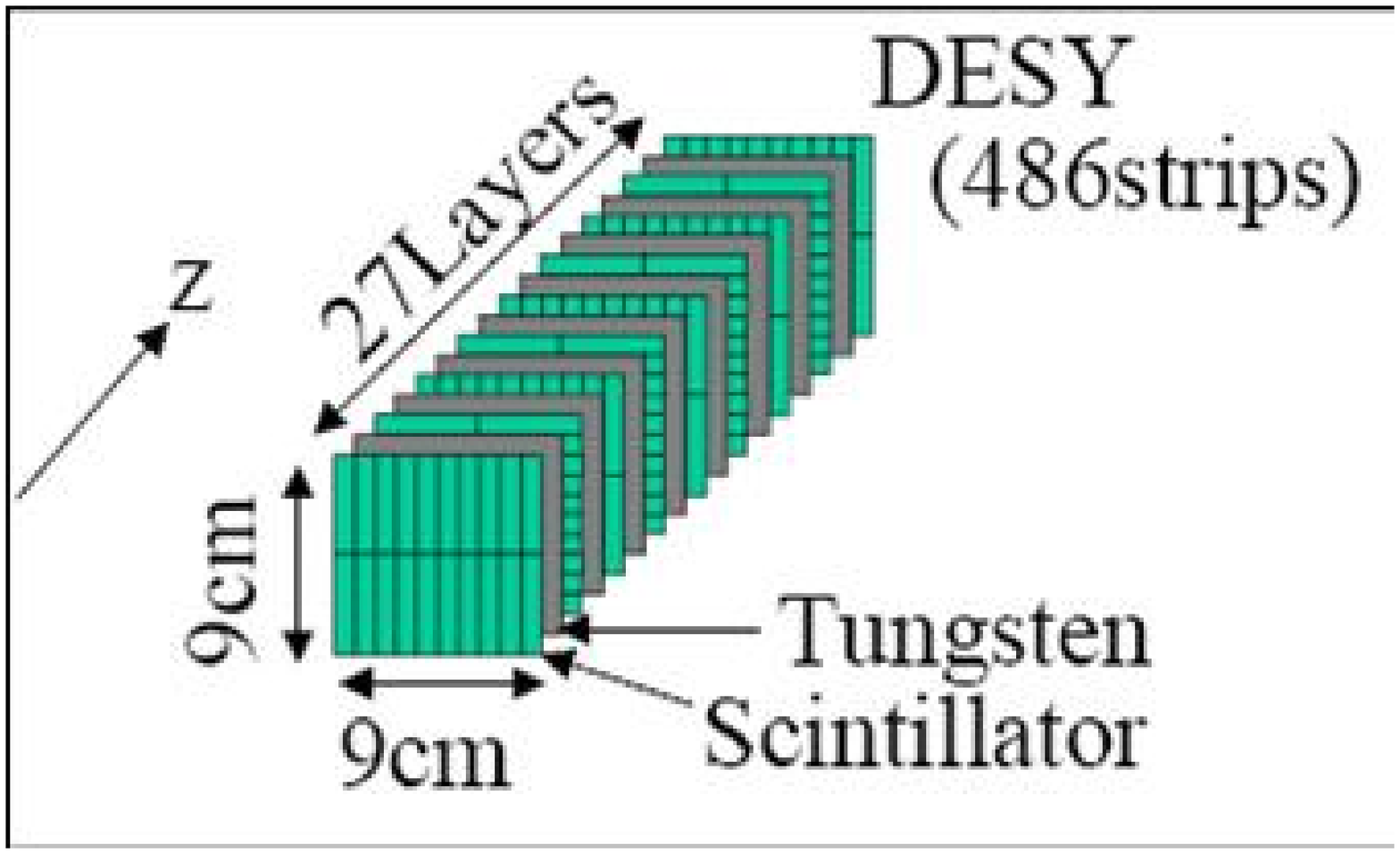}
\caption[Prototype scintillator ECAL]{\label{fig:ECAL_japan_proto} 
A small prototype ECAL instrumented with scintillator strips
readout by MPPCs for test-beam studies. 
  }
\end{minipage}
\begin{minipage}{0.02\textwidth}
~~
\end{minipage}
\begin{minipage}{0.45\textwidth}    
\includegraphics[width=0.9\textwidth,height=0.6\textwidth]{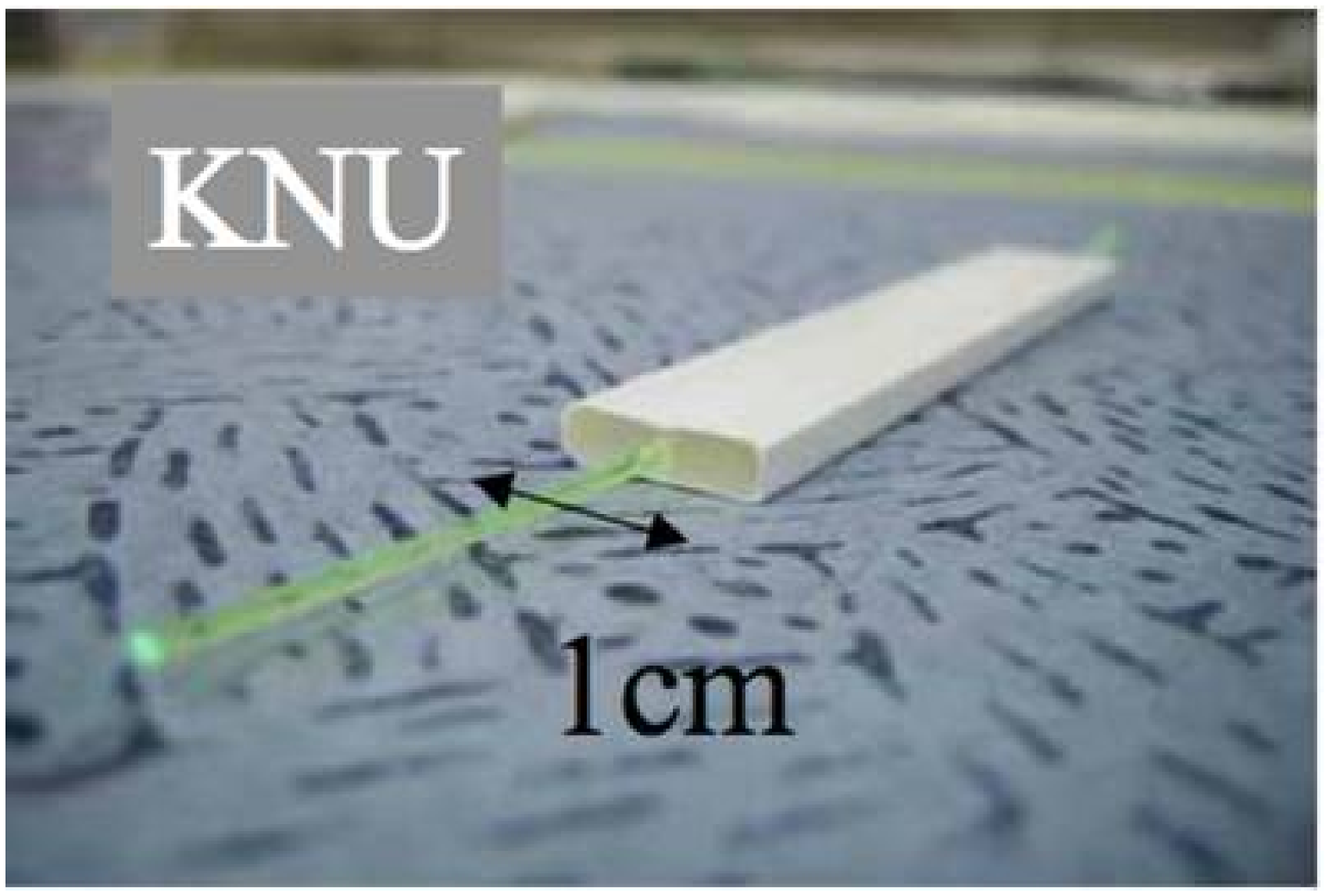}
\caption[A scintillator with an embedded fiber]{\label{fig:scint_korea} 
A scintillator strip with an embedded wavelength shifting fiber produced in Korea. 
}
\end{minipage}
\end{figure}

A new prototype calorimeter has been instrumented with
MPPCs and beam tested at DESY earlier this year.
A sketch is shown in Figure~\ref{fig:ECAL_japan_proto}.  
The thickness of the tungsten absorber plates is 3.5 mm and the thickness of the
scintillator strips is 2 mm. Scintillator strips will be either extruded by Korean partners,
as shown in Figure~\ref{fig:scint_korea},
or made from large planes structured by grooves. About 500 MPPCs will be 
delivered by Hamamatsu. Half of them will be coupled
to the wavelength shifting fibers inside the strips, and half will be attached directly to the scintillator
strip. 
Experience obtained in this test-beam study will then be used for the construction of a larger 
prototype to be tested at FNAL.
\begin{figure}[htb]
\begin{minipage}{0.55\textwidth}    
\includegraphics[width=\textwidth,height=0.5\textwidth]{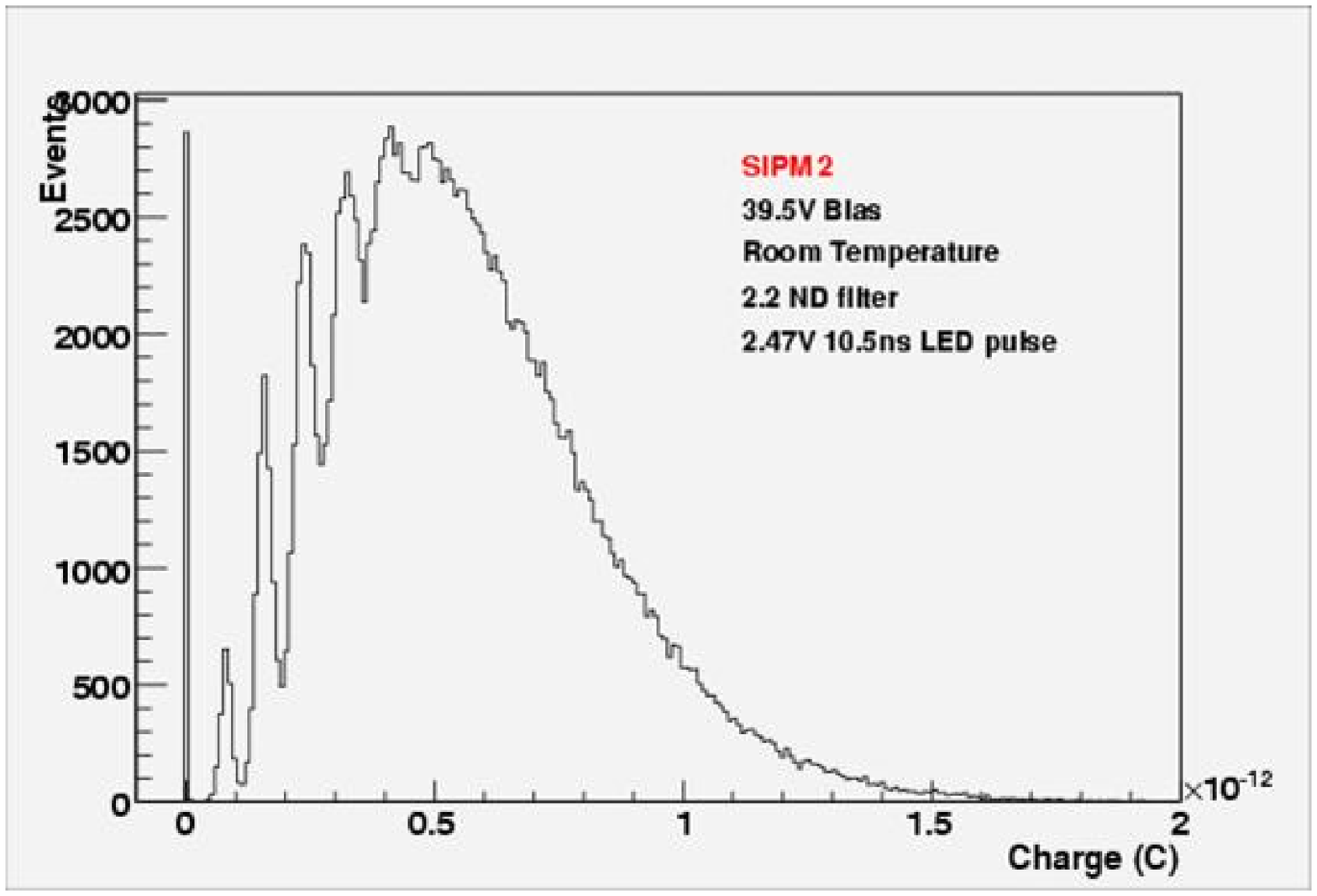}
\caption[A scintillator tile ECAL prototype module]{\label{fig:ECAL_colorado} 
The signal spectrum from cosmic rays crossing a tile module of the University of Colorado group. The tiles consist of 2 mm
thick plastic scintillator tiles readout by a SiPD photo-detector. Nicely seen is the photo-electron counting capability. 
 }
\end{minipage}
\begin{minipage}{0.02\textwidth}
~~
\end{minipage}
\begin{minipage}{0.4\textwidth}
\vspace{-1cm}
\includegraphics[width=\textwidth,height=0.9\textwidth]{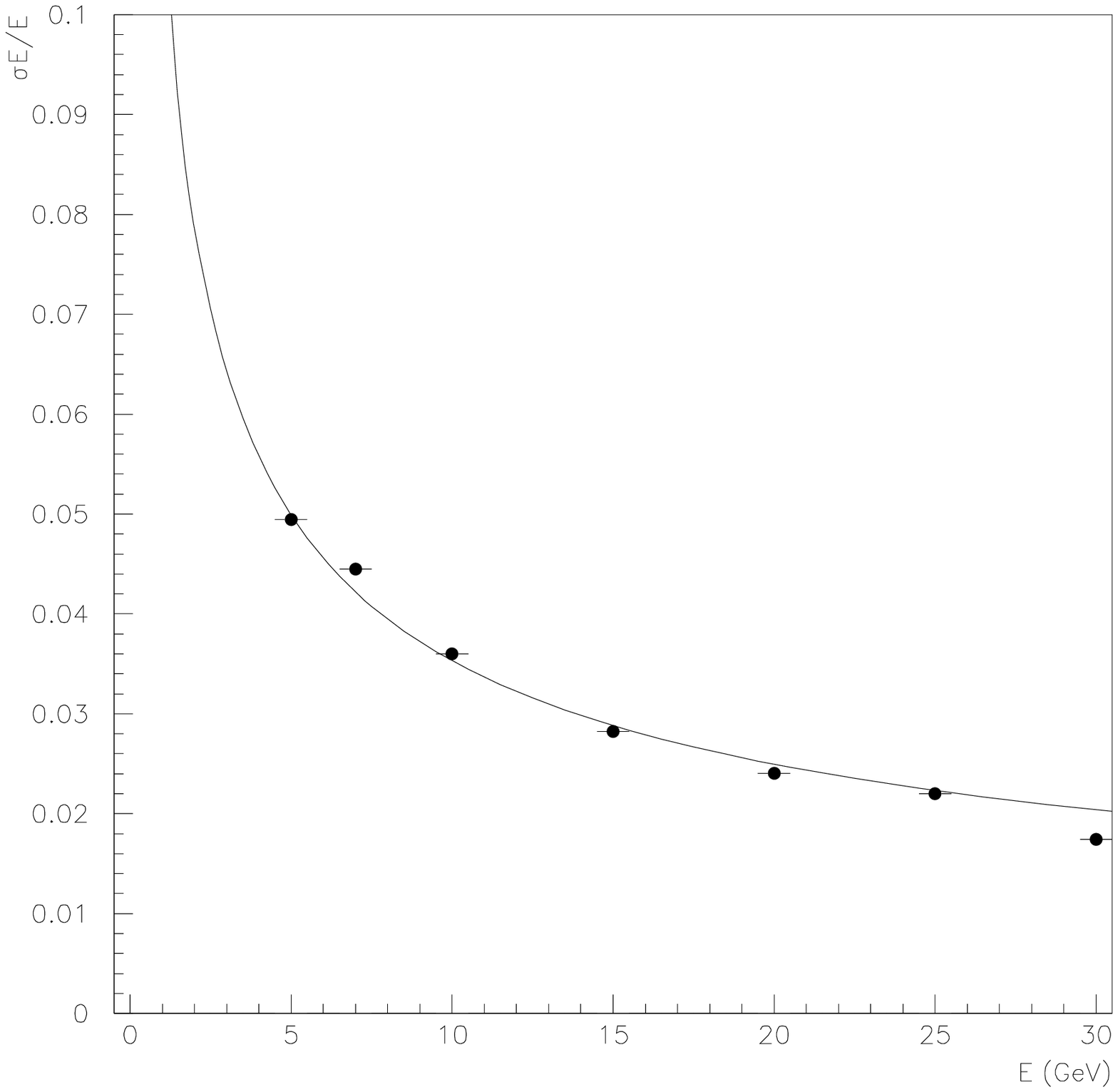}
\caption[Energy resolution in the Si-W prototype calorimeter]{\label{fig:ECAL_E_res} 
The energy resolution measured with a silicon-scintillator lead sandwich calorimeter.
  }
\end{minipage}
\end{figure}   
 
A group from the University of Colorado~\cite{colorado} 
proposes a scintillator-tungsten sandwich calorimeter
using scintillator tiles of $5 \times 5$~cm$^2$ size readout with silicon photo-detectors, 
SiPDs, 
via a wavelength shifting fiber.
In order to improve the shower position resolution the tiles in consecutive layers are offset by $2.5$~cm.
The performance of wavelength shifting fibers with different bend radii has been monitored
over about a year without degradation. 
SiPDs are devices similar to MPPCs described above manufactured by the Photonique Company, Switzerland.
The active area of the devices is 1x1 mm$^2$ and the gain is about 10$^5$.
The 
signal-to-noise ratio is studied as a function 
of the temperature. It improves 
significantly at lower temperatures.  
A small tile module is operated in the lab and performance studies are done 
with cosmic rays. 
 Figure~\ref{fig:ECAL_colorado} shows the signal spectrum from a cosmic ray run using a 2mm thick scintillator tile.
The gate length and position for the integration of the SiPD output is optimized to suppress noise pulses. 

The Colorado group will develop in collaboration with  Photonique a SiPD of larger sensitive area and a gain of about
 10$^6$.

\subsubsection{Mixed Silicon and Scintillator Tungsten}

A prototype of a sandwich calorimeter
consisting of 45 scintillator planes 
and three planes of silicon pads interspersed between lead absorber disks
was built and operated in a test-beam~\cite{checcia_cal}.
The measured energy resolution,
parametrized as $11\%/\sqrt E$, is shown in
Figure~\ref{fig:ECAL_E_res}.
There is no plan for the moment to continue the 
project.  

\subsection{Hadron Calorimeter for Particle Flow approach}
Several technologies 
of fine-segmented sampling calorimeters are under investigation with either 
analog or digital readout. The analog read out calorimeters use scintillator tiles
or scintillator strips
as sensors. Digital calorimeters use GEMs (Gaseous Electron 
Multipliers), Micromegas (Micro mesh gaseous structures) or RPCs (Resistive Plate Chambers) as  
active elements.

\subsubsection{Analog HCAL}
Analog hadron calorimeters use scintillator as detector and steel 
or lead as absorber. The scintillator tiles are readout by novel photo-sensors, e.g. MPPCs
or SiPDs
as described above, or Silicon Photo-multipliers (SiPMs). These photo-sensors
are based on the same working principle but developed in different regions: e.g. MPPCs in Japan and 
SiPMs in Russia~\cite{SiPM}.

Two projects are pursued within  
CALICE. One is based on small area scintillator tiles of a few mm thickness
read out by SiPMs via wavelength shifting fibers. Layers of steel  
serve as absorber and form the mechanical frame.
A small prototype, the MINICAL, was operated successfully in a test-beam. The test
demonstrated that
using SiPMs
as photo-sensors maintains the resolution measured with classical photo-multipliers~\cite{minical}.

\begin{figure}[htb]
\begin{minipage}{0.45\textwidth}
\includegraphics[width=\textwidth,height=0.8\textwidth]{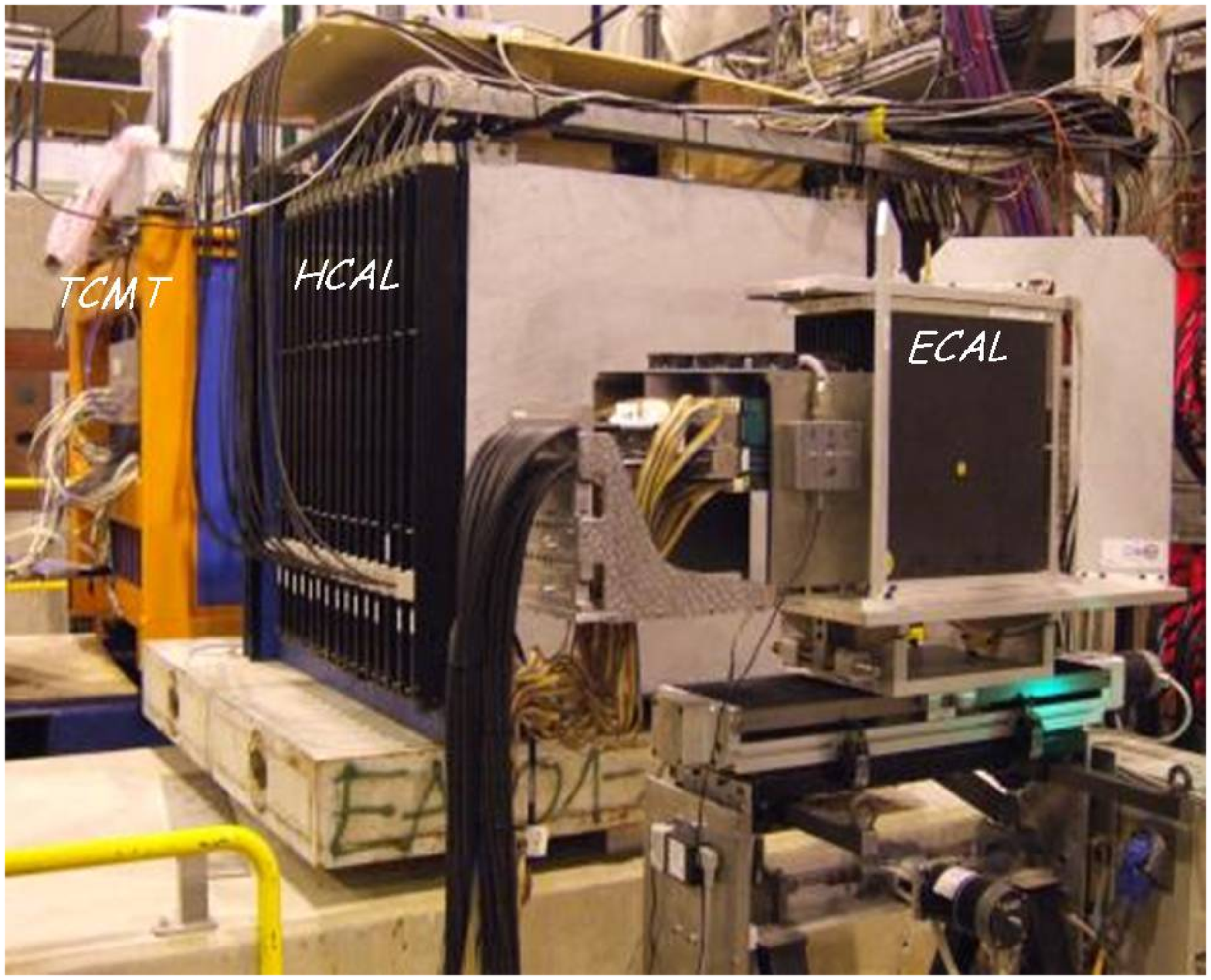}
\caption[Analogue HCAL prototype module]{\label{fig:HCAL_tb} 
The 1 m$^3$ prototype of the CALICE analog HCAL in a test-beam at CERN. Also shown is the prototype of the
ECAL in front of the HCAL and the tail-catcher and muon tracker TCMT.
  }
\end{minipage}
\begin{minipage}{0.02\textwidth}
~~
\end{minipage}
\begin{minipage}{0.45\textwidth}
\includegraphics[width=1.1\textwidth]{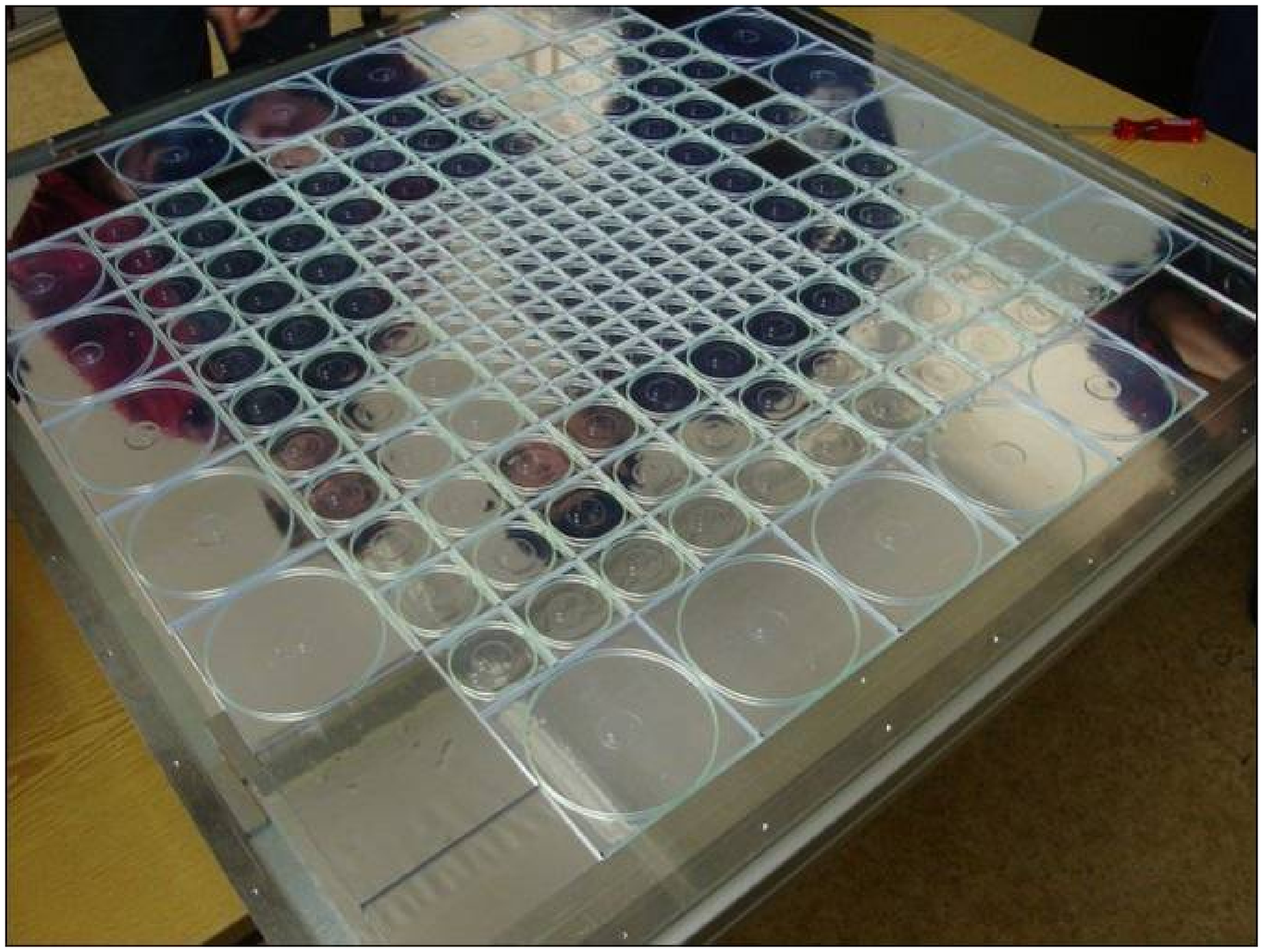}
\caption[Analogue HCAL tile plane]{\label{fig:HCAL_tile_plane} 
A scintillator tile plane for the 1 m$^3$ prototype calorimeter. Each tail is equipped 
with a wavelength shifting fiber read out by a SiPM.  
  }
\end{minipage}
\end{figure}

\begin{figure}[htb]
\begin{center}
\includegraphics[width=0.75\textwidth]{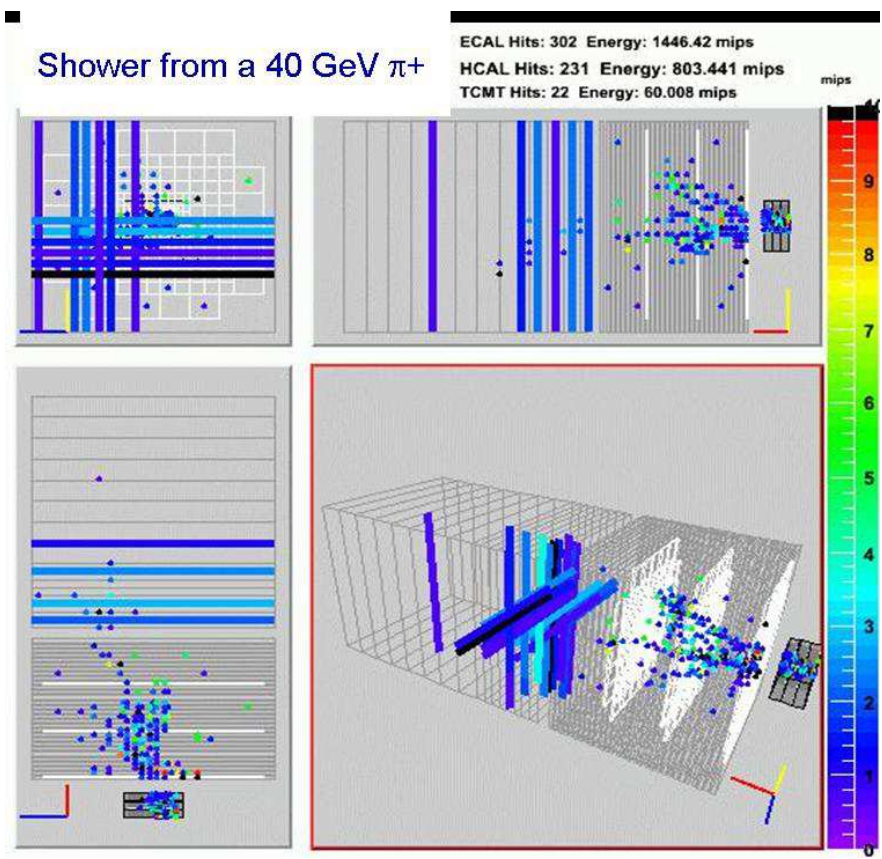}
\caption[Event display of a shower in the test beam]{\label{fig:HCAL_display} 
An event display of the shower of a 40 GeV pion recorded in the CERN
test-beam in several
projections. 
The shower starts 
in the ECAL, continues into the HCAL and ends in the tail-catcher.
  }
\end{center}
\end{figure}

Currently a 1 m$^3$ prototype calorimeter, as shown in Figure~\ref{fig:HCAL_tb}, has been 
partially equipped with 1~m$^2$ sensor layers and tested in a CERN lepton and hadron beam. 
The scintillator
tiles 
are of 3x3 cm$^2$ size
in the core of the calorimeter and 10x10 cm$^2$ in the edge regions,
as shown in Figure~\ref{fig:HCAL_tile_plane}. 
The granularity in the core 
has been chosen to optimize the particle shower separation
power. It is also small enough to test semi-digital (two bit) readout
concepts.
Each tile is equipped with a SiPM. The signals are transported using thin wires to one side of 
the plane and feed in the DAQ electronics which is the same as the one 
being used for the CALICE silicon-tungsten ECAL prototype.
The calorimeter is supplemented by tail catcher and muon tracker to ensure full measurement 
of hadron showers.
The event display of a test-beam hadron in Figure~\ref{fig:HCAL_display} demonstrates
that the full system is working and produces images of the hadron shower structure. 
About 70 million events with electron and hadron beams in the energy
range from 6 to 80 GeV have been collected at CERN in 2006, mostly in
conjunction with the silicon-tungsten ECAL. These data already allow first studies
of particle flow performance with ECAL and HCAL together.
The prototype instrumentation is to be completed in 2007, along with further data
taking at CERN and Fermilab.
A versatile mechanical support structure is under construction, which will
make studies with inclined beam incidence possible. The same structure
shall later also be used with gaseous HCAL modules, as
described below, for a direct
comparison with the purely digital options.
In addition, the use of SiPMs in a large scale prototype
will allow to collect very valuable expertise on the long-term performance 
of these novel photo sensors. 

\begin{figure}[htb]
\begin{center}
\includegraphics[width=0.4\textwidth]{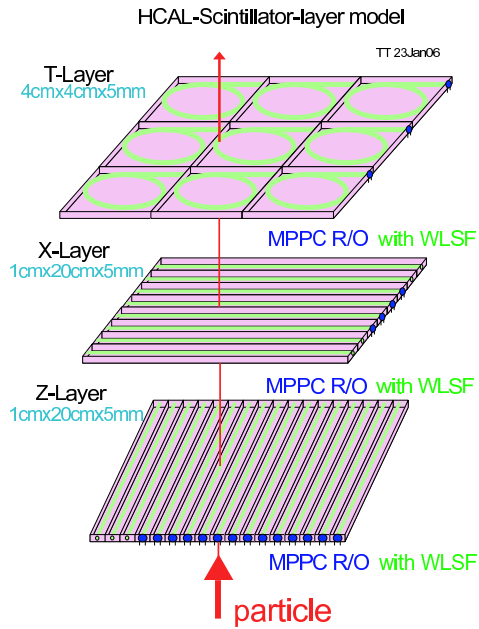}
\caption[Possible tile sequence for the analogue HCAL]{\label{fig:HCAL_strips} 
A possible tile sequence for the analog HCAL. The first two layers consist of scintillator strips oriented
perpendicular to each other, the third layer is made of quadratic tiles. Each strip and tile is equipped with
a wavelength-shifting fiber and readout by a MPPC.
  }
\end{center}
\end{figure}

The second project is focused on a hadron calorimeter design
which uses lead as absorber and scintillator layers as sensors.
Lead is chosen to achieve hardware compensation, i.e. to ensure
almost equal response for the electromagnetic 
and the hadronic shower component and reducing such the fluctuations.
Test-beam measurements have shown that compensation is achieved
by choosing the ratio of lead-to-scintillator thickness to 9.1:2.
In addition, plastic scintillator detects neutrons effectively, improving the energy
measurement of hadrons.
The
scintillators are structured in strips and tiles as shown in Figure~\ref{fig:HCAL_strips}. 
The strip width and length is set to 1 and 20~cm, respectively, and are subject to ongoing optimization.
Wavelength shifting fibers are placed inside grooves along 
the strip center or curled inside the tiles. 
To ensure a sufficient amount of light the scintillator
thickness is set to 5 mm.

The frontend electronics for test-beam studies of this HCAL design
will be based on that developed for the CALICE ECAL.
Test-beam data taken with a prototype calorimeter 
will allow a detailed comparison to the FE/scintillator calorimeter previously.


 \begin{figure}[htb]
\begin{center}
\includegraphics[width=0.8\textwidth,bbllx=80pt,bblly=60pt,bburx=750,bbury=540pt]{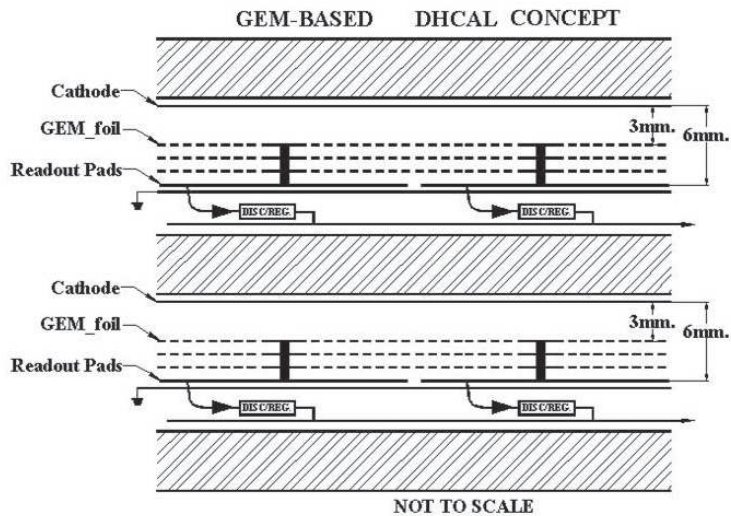}
\caption[Structure of the digital HCAL equipped with GEMs]{\label{fig:HCAL_gem_structure} 
The structure of the digital HCAL equipped with GEMs.
Gas amplification occurs in several layers of GEM foils. The signal is picked up 
from anode pads. The FE electronics unit is placed on the pad. 
  }
\end{center}
\end{figure}
\begin{figure}[htb]
\begin{center}
\includegraphics[width=0.5\textwidth]{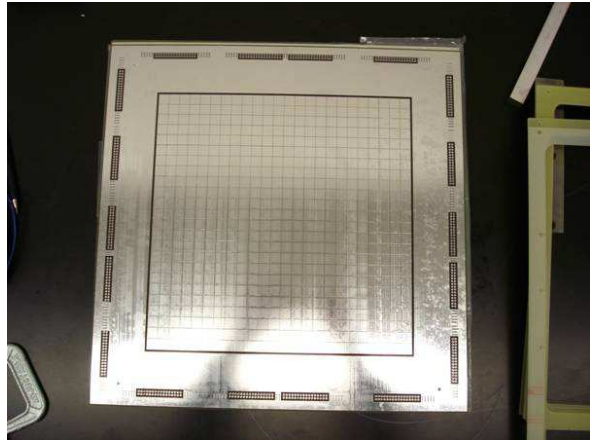}
\caption[GEM chamber for the digital HCAL]{\label{fig:HCAL_gem_TC} 
A 30 x 30 cm$^2$ GEM chamber prepared for test measurements. FE readout electronics is placed on the 
edge of the chamber frame.
  }
\end{center}
\end{figure} 
\begin{figure}[htb]
\begin{minipage}{0.4\textwidth}
\includegraphics[width=\textwidth,height=0.6\textwidth]{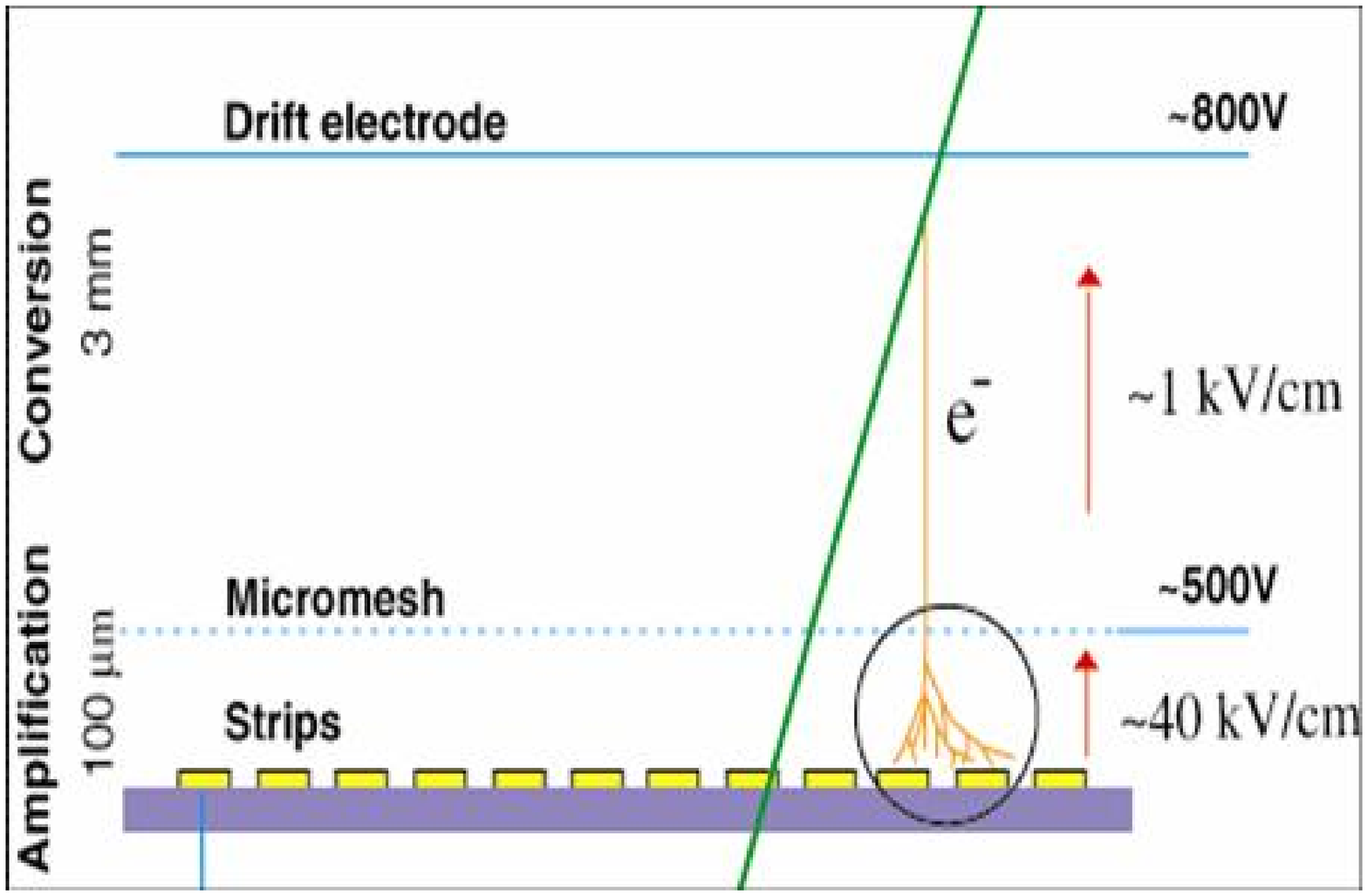}
\caption[Working principle of Micromegas]{\label{fig:micromega1} 
The working principle of Micromegas. Electrons from ionization drift in an electrical field to the mesh
and induce an avalanche when crossing it. Signals can be picked up from anode strips or pads.
  }
\end{minipage}
\begin{minipage}{0.02\textwidth}
~~
\end{minipage}
\begin{minipage}{0.55\textwidth}
\includegraphics[width=1.1\textwidth,height=0.5\textwidth]{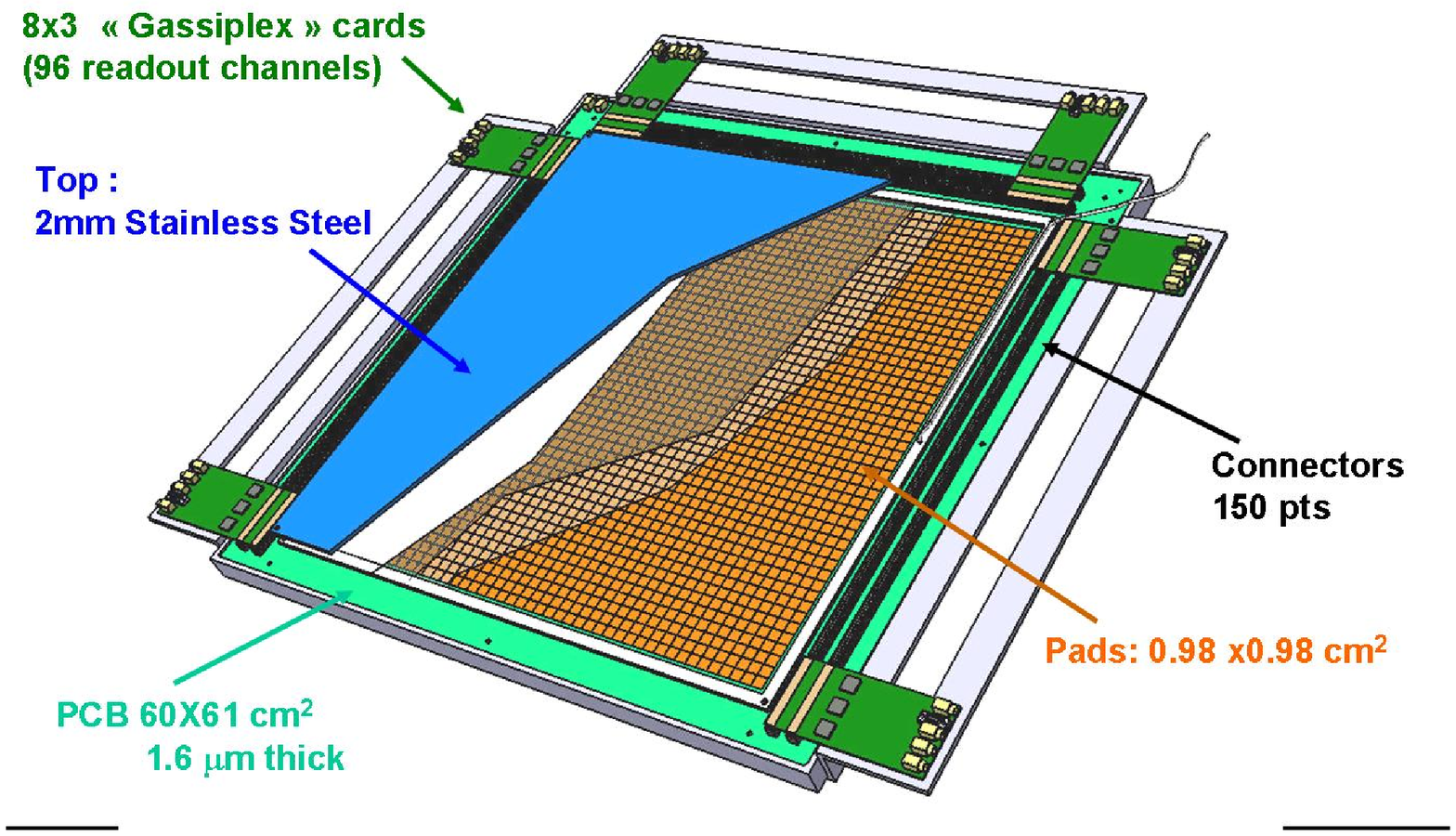}
\caption[A 50 x 50 cm$^2$ chamber using Micromegas for gas amplification]{\label{fig:micromega2} 
A 50 x 50 cm$^2$ chamber using Micromegas for gas amplification. FE readout electronics is placed on the 
edge of the chamber frame.
  }
\end{minipage}
\end{figure} 
\begin{figure}[htb]
\begin{minipage}{0.45\textwidth}
\includegraphics[width=\textwidth]{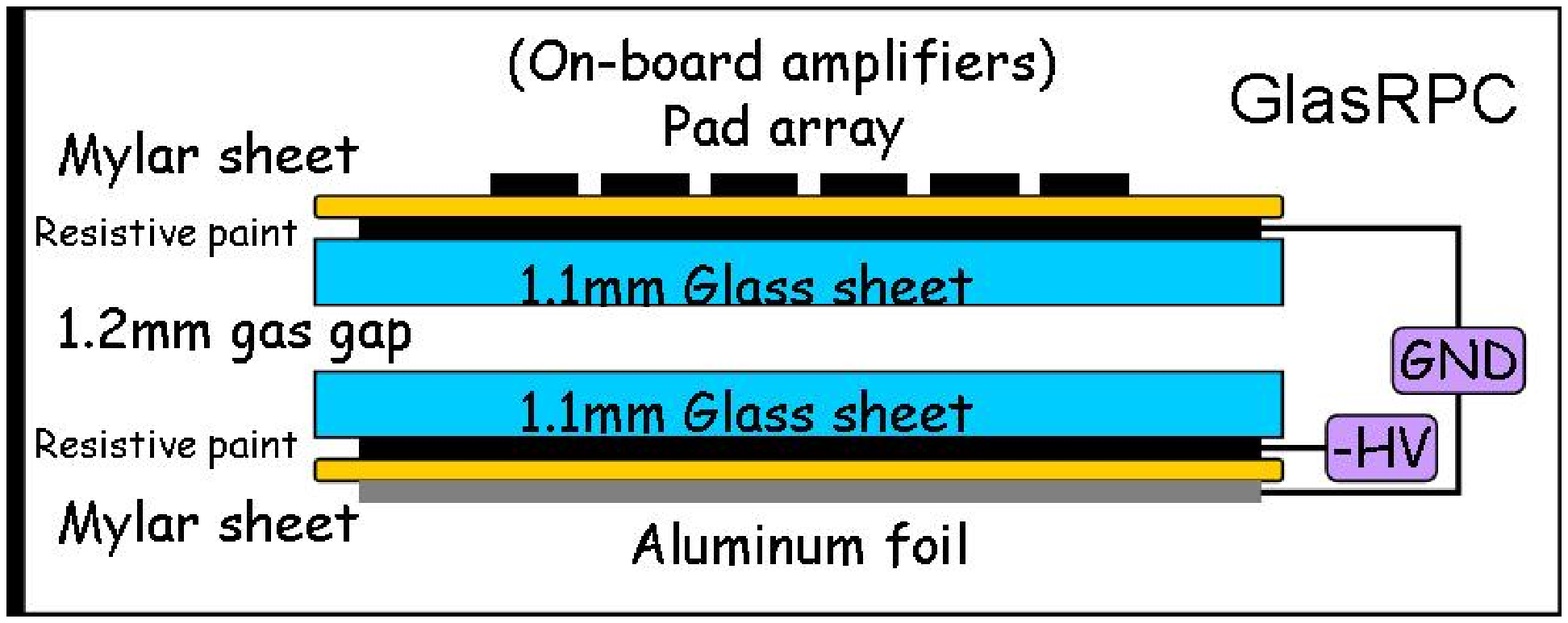}
\caption[Structure of a RPC chamber]{\label{fig:scheme_RPC}
The structure of a RPC chamber. A gap of 2 mm between two glass plates is filled with a working gas mixture. 
Resistive paint on the glass ensures a homogeneous electric field. A pad structure outside allows to detect
the image charge induced by the local discharge induced by a crossing particle.   
  }
\end{minipage}
\begin{minipage}{0.02\textwidth}
~~
\end{minipage}
\begin{minipage}{0.45\textwidth}
\includegraphics[width=\textwidth,height=0.7\textwidth]{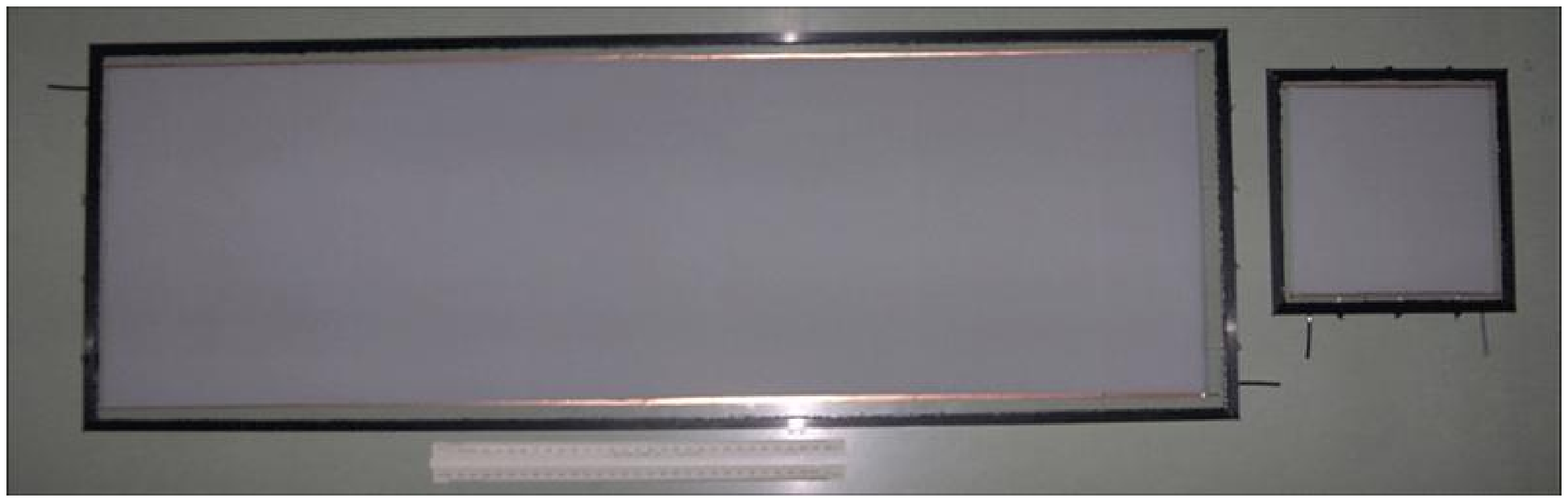}
\caption[Prototypes of RPC chambers]{\label{fig:HCAL_RPC} 
Prototypes of RPCs for performance studies. The larger chamber covers an area of 100x30 cm$^2$.
  }
\end{minipage}
\end{figure} 
\begin{figure}[htb]
\begin{minipage}{0.02\textwidth}
~~
\end{minipage}
\begin{minipage}{0.45\textwidth}
\includegraphics[width=\textwidth]{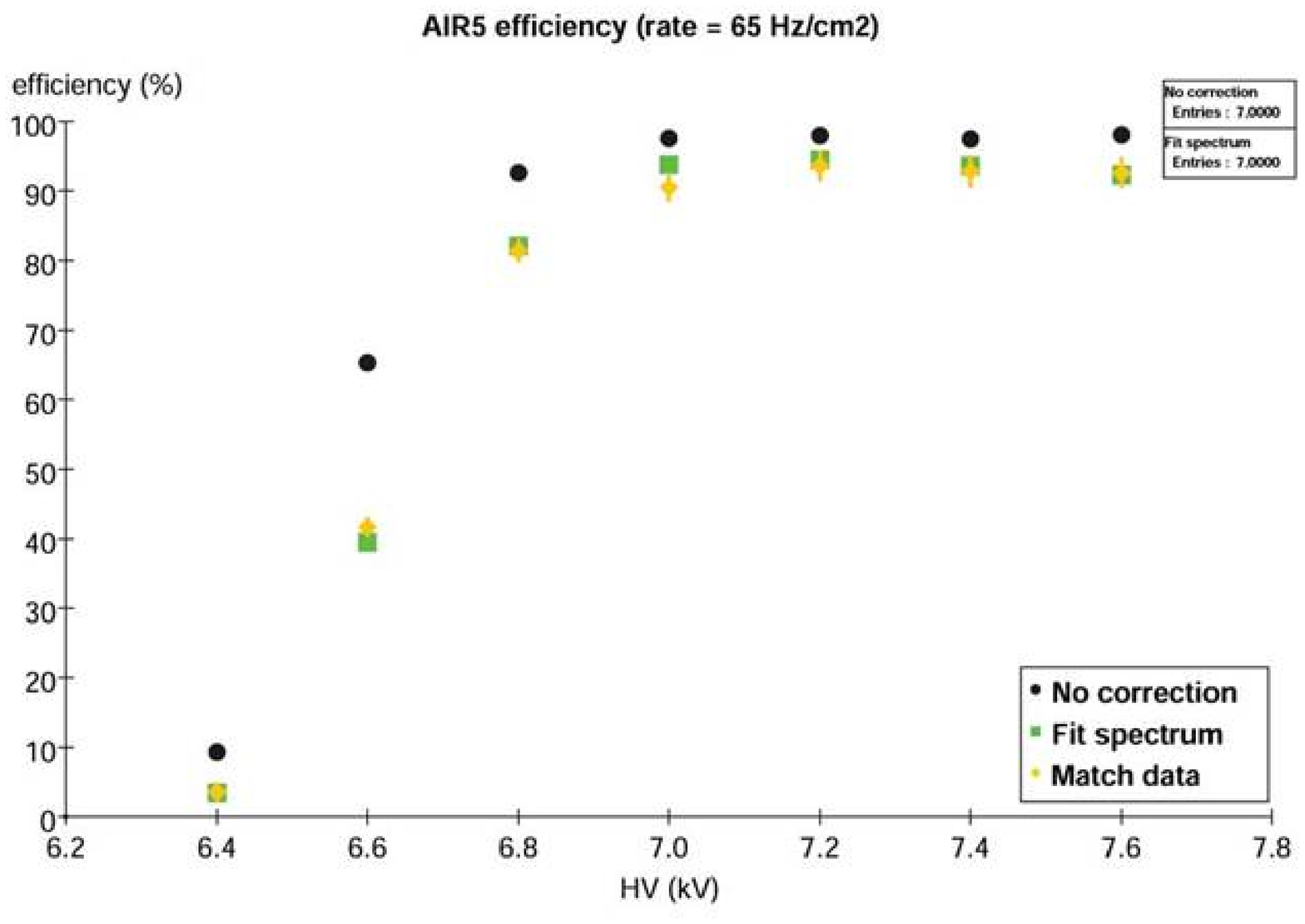}
\caption[Single particle detection efficiency]{\label{fig:HCAL_RPC_test}
The efficiency to detect single particles in a RPC as
a function of the high voltage. A 120 GeV proton beam
at FNAL was used.
}
\end{minipage}
\begin{minipage}{0.02\textwidth}
~~
\end{minipage}
\begin{minipage}{0.45\textwidth}
\includegraphics[width=1.1\textwidth]{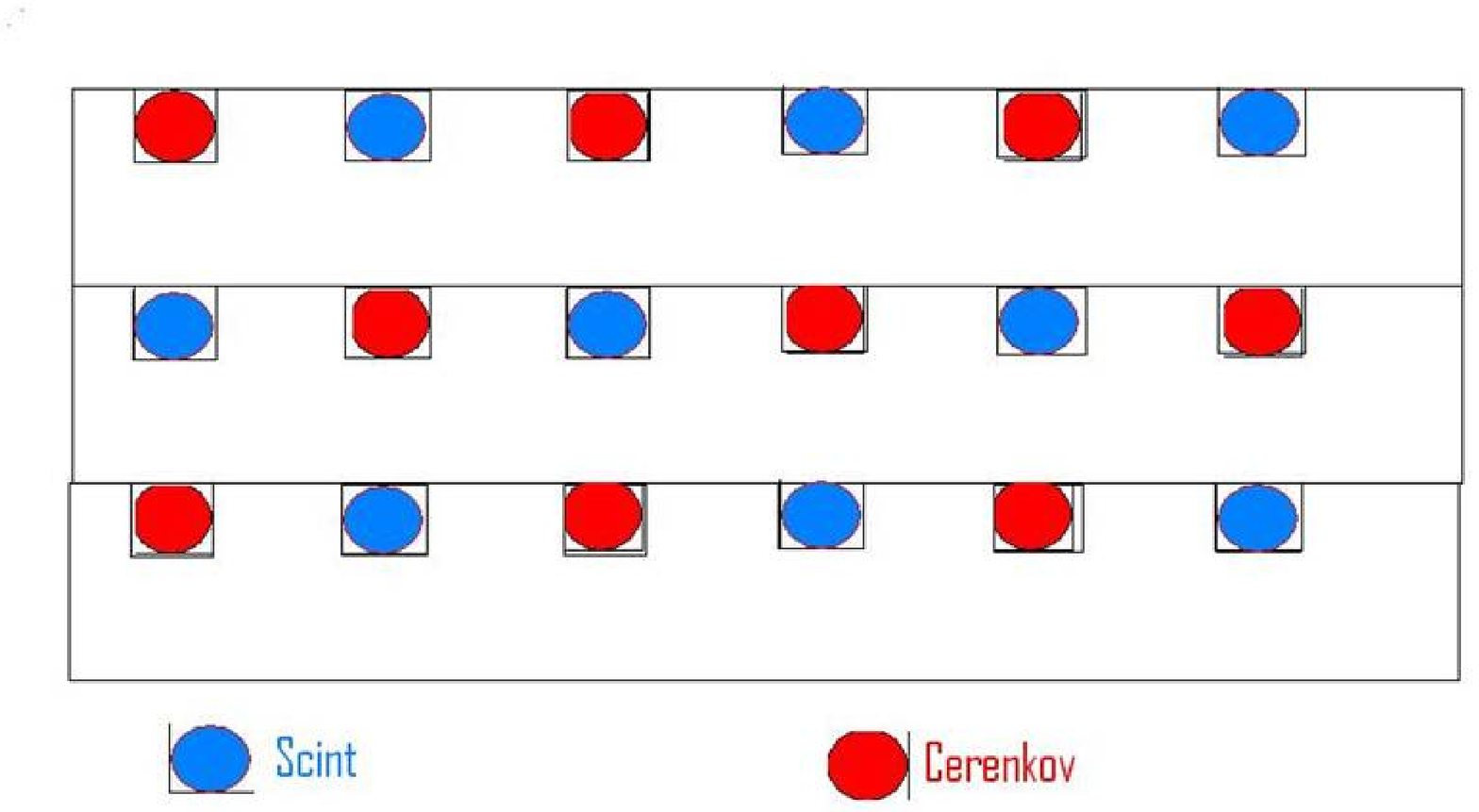}
\caption[Possible structure of the DREAM calorimeter]{\label{fig:DREAM} 
A possible structure of the DREAM calorimeter. Grooves in 2mm thick 
lead or brass absorber plates contain scintillator 
and clear fibers.    
  }
\end{minipage}
\end{figure} 

\begin{figure}[htb]

\begin{tabular}{cc}
\begin{minipage}{0.37\textwidth}
\includegraphics[width=\textwidth,height=0.9\textwidth]{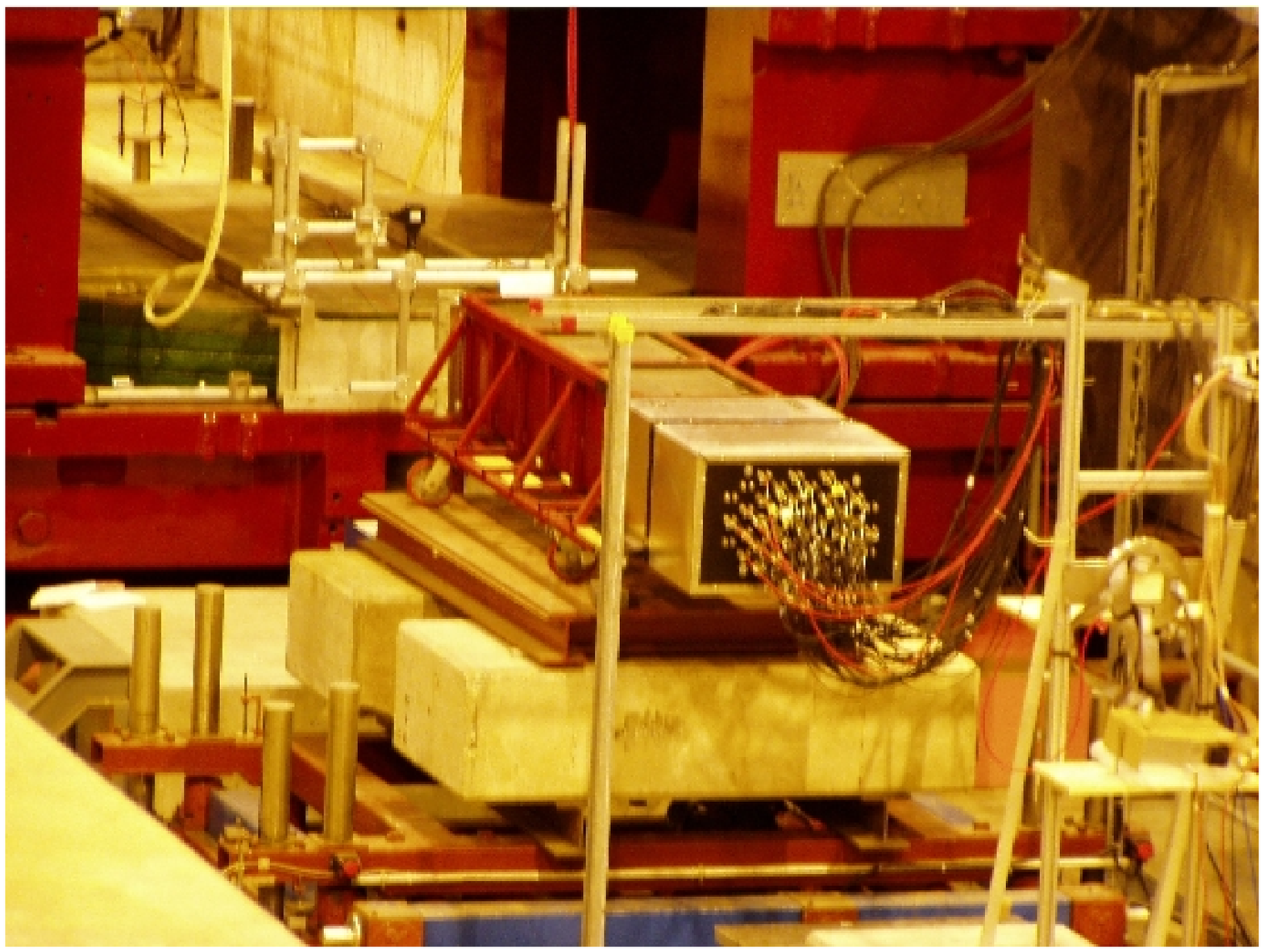}
\caption[A dream module in the CERN test beam]{\label{fig:DREAM1} 
A DREAM module in the test-beam at CERN. The module is about 2~m in depth and 
$32 \times 32$~cm$^2$ in cross section. 
   }
\end{minipage}
&

\begin{minipage}[c]{0.6\textwidth}
\includegraphics[width=0.7\textwidth]{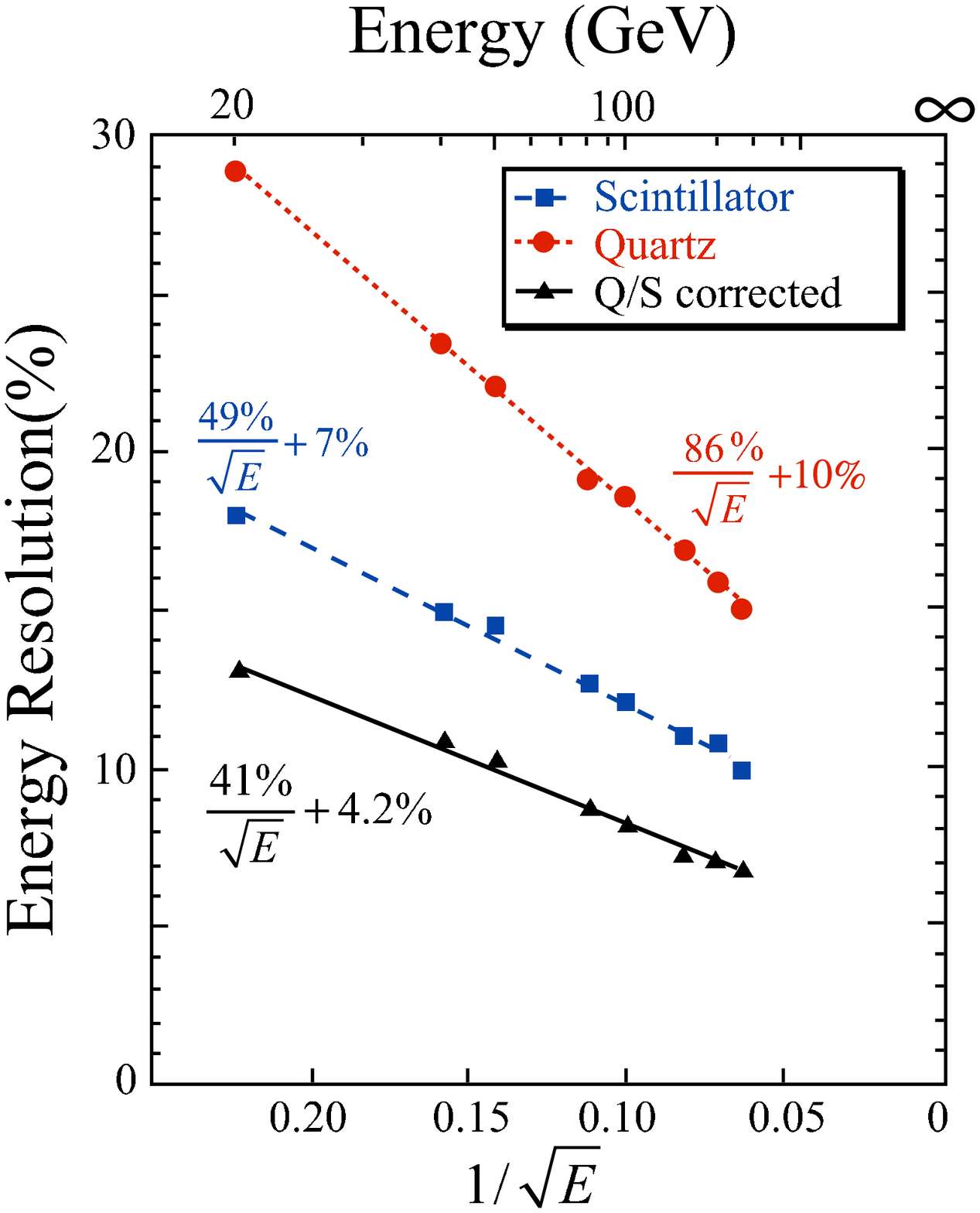}
\caption[The energy resolution of a DREAM module]{\label{fig:DREAM2} 
The energy resolution of a module of the DREAM calorimeter for hadrons. Shown is the energy 
resolution, in percent, as a function of $1/\sqrt E$, increasing towards the left.
  }
\end{minipage} 
\end{tabular}
\end{figure} 
\begin{figure}[htb]
\begin{minipage}{0.45\textwidth}
\includegraphics[width=\textwidth,height=0.7\textwidth]{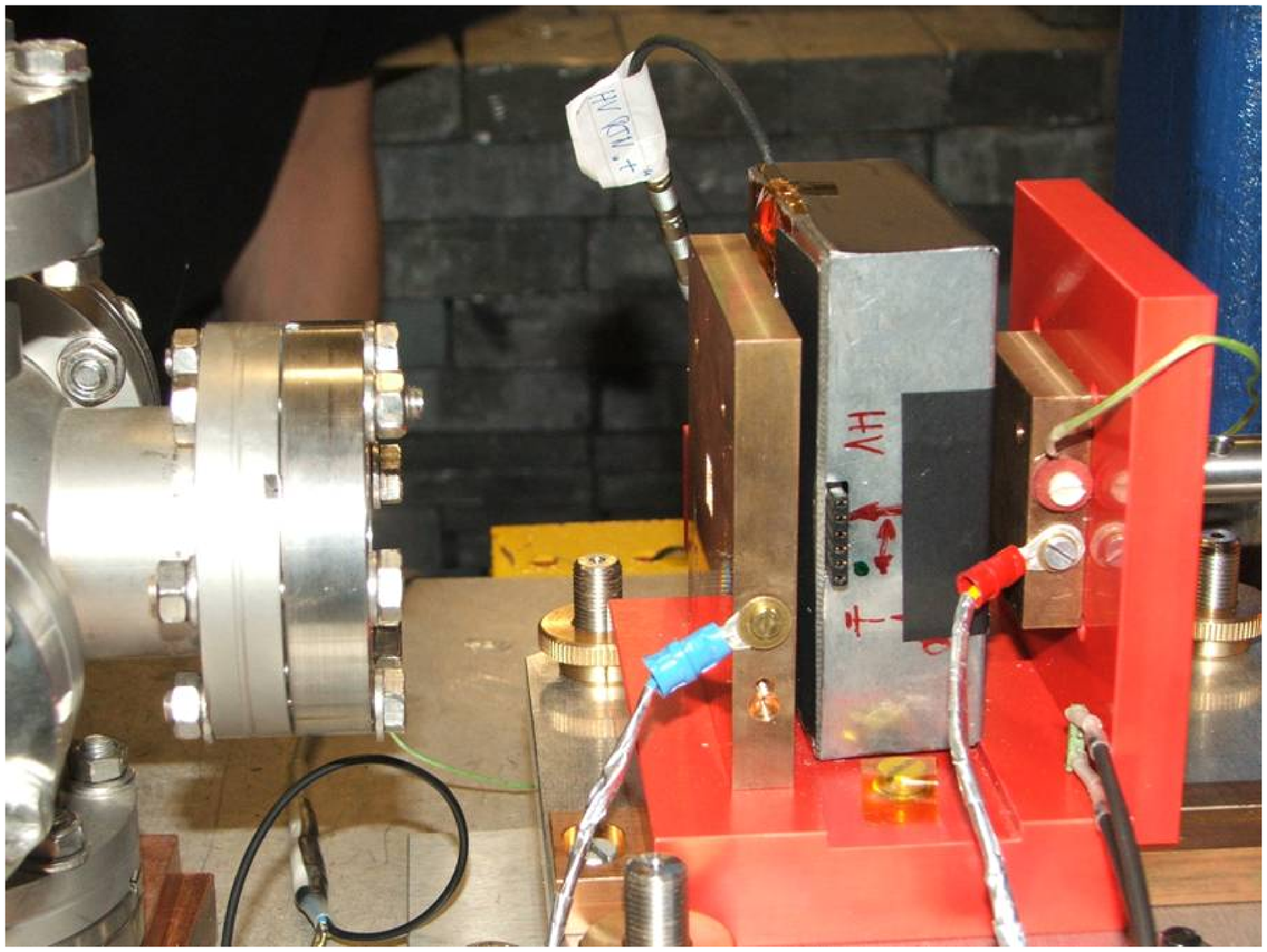}
\caption[Setup for sensor irradiation]{\label{fig:FCAL_tb} 
The setup for sensor irradiation. Left is the exit window of the 10 ~MeV electron beam.
The sensor is inside the grey PCB box. A brass collimator and  Faraday cap are used to measure the 
electron current crossing the sensor.
  }
\end{minipage}
\begin{minipage}{0.02\textwidth}
~~
\end{minipage}
\begin{minipage}{0.45\textwidth}
\includegraphics[width=\textwidth,height=0.7\textwidth]{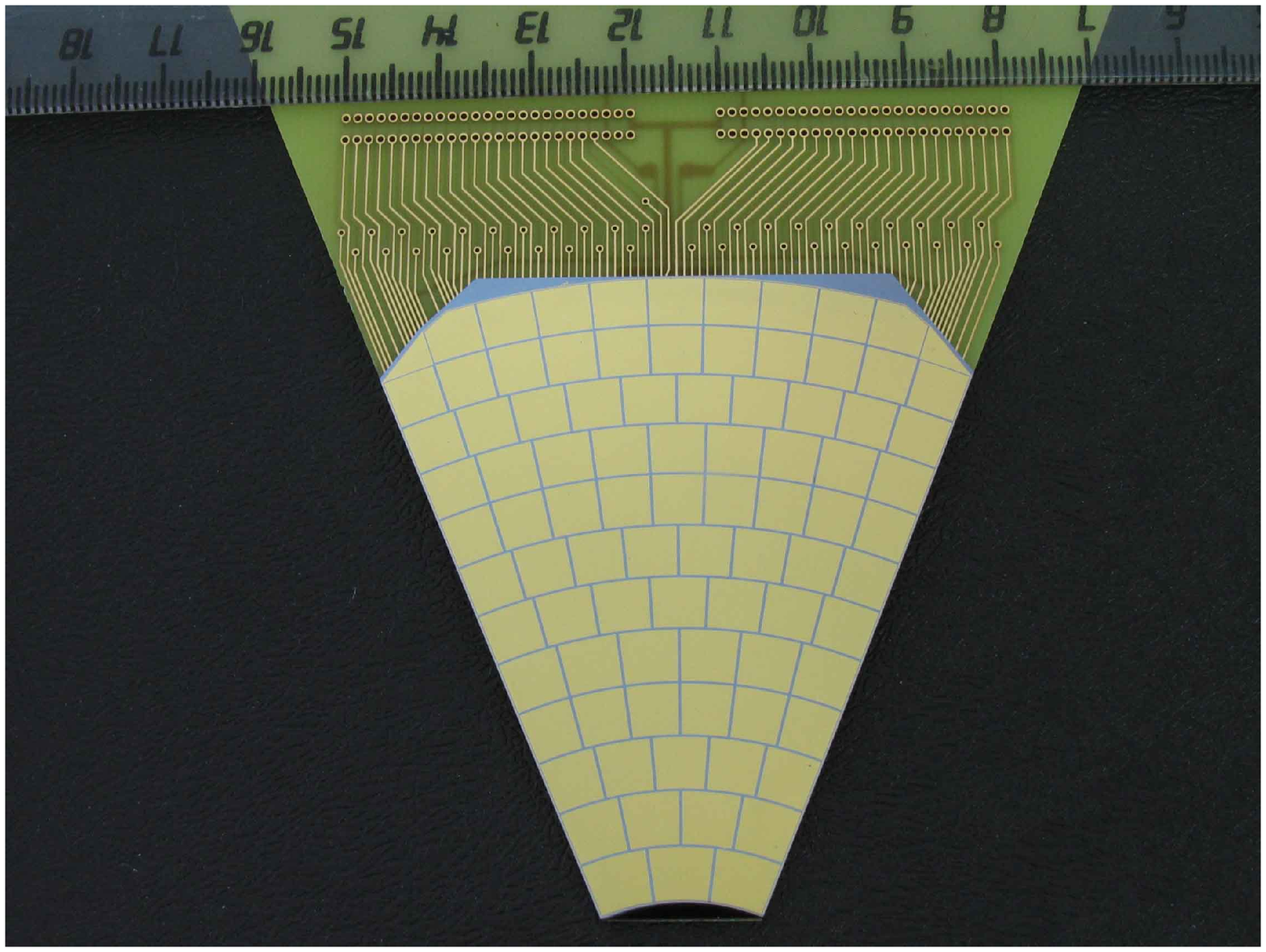}
\caption[GaAs sensor proposed for BeamCal]{\label{fig:GaAs} 
A Gallium-Arsenid (GaAs) sensor tile designed for BeamCal to be prepared for test measurements. 
  }
\end{minipage}
\end{figure} 
\begin{figure}[htb]
\begin{minipage}{0.45\textwidth}
\includegraphics[width=\textwidth,height=0.7\textwidth]{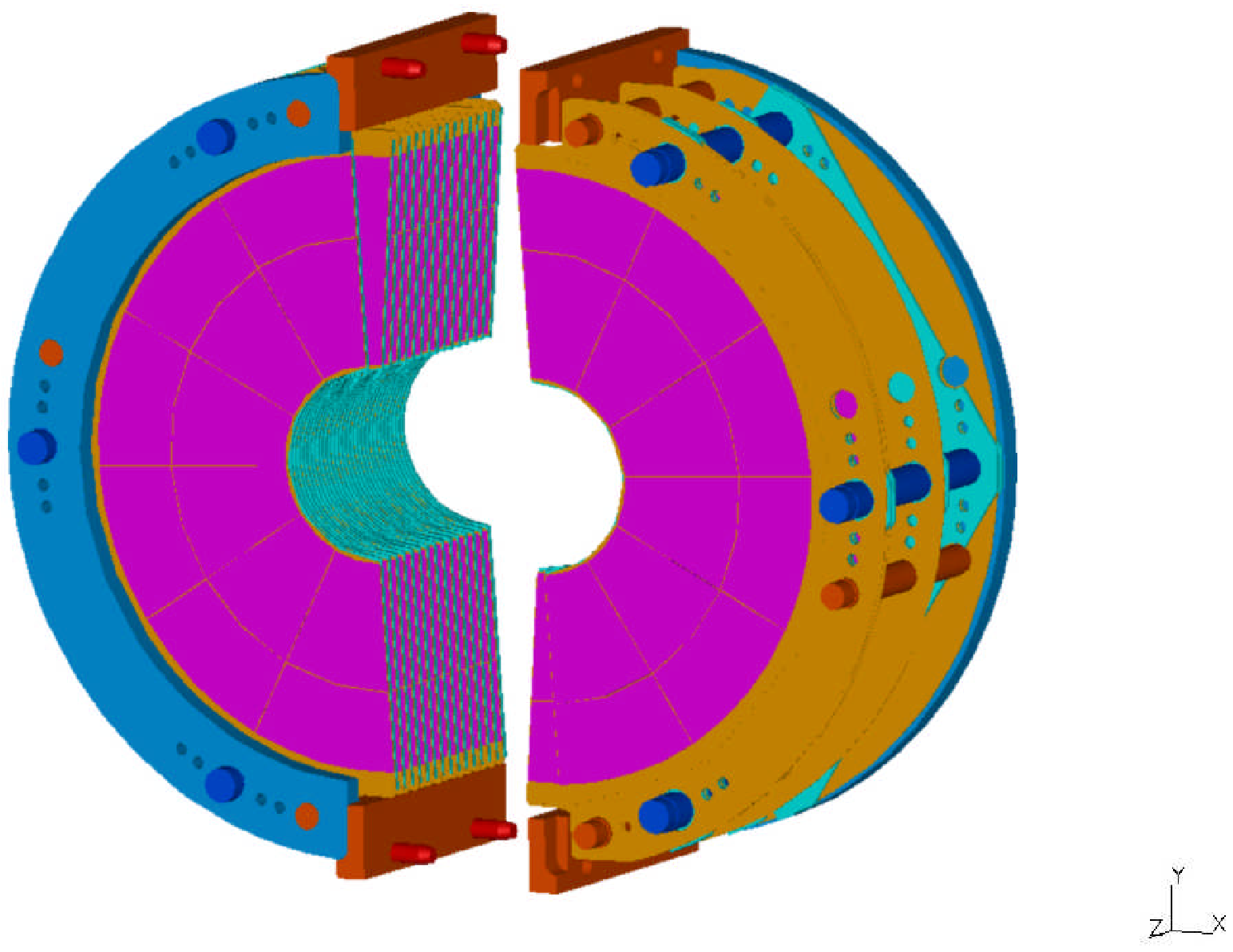}
\caption[Mechanical structure of the LumiCal]{\label{fig:lumi_mech} 
The mechanical structure of the LumiCal. Absorber disks are held by the blue bolts, the sensor
layers by the red bolts. The latter must ensure that sensor planes are positioned with 
$\mu$m accuracy.
  }
\end{minipage}
\begin{minipage}{0.02\textwidth}
~~
\end{minipage}
\begin{minipage}{0.45\textwidth}
\includegraphics[width=\textwidth,height=0.85\textwidth]{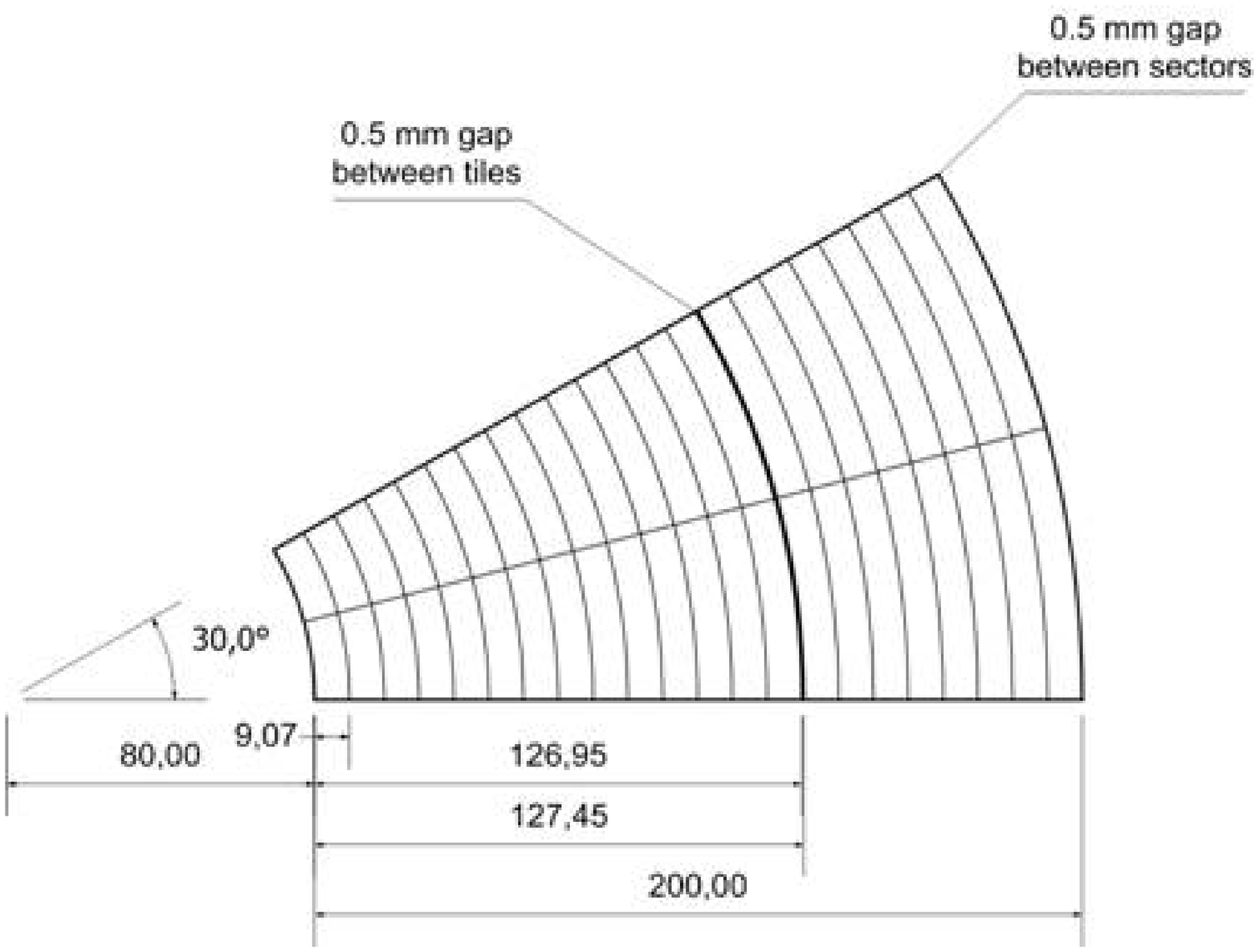}
\caption{\label{fig:lumi_sensor} 
The design of a pad structured silicon sensor for the LumiCal
  }
\end{minipage}
\end{figure} 

\subsubsection{Digital HCAL}

Digital HCAL designs use gaseous signal amplification in GEMs, Micromegas or
RPCs. Thin and large  
area chambers are interspersed between steel absorber plates. The chamber anode is 
segmented in small pads of about 1 cm$^2$ size, 
matching the granularity needed for the PFA application. 
Research work is done within the CALICE collaboration~\cite{digihcal}.

The structure of a digital HCAL with GEMs as sensors is shown in
Figure~\ref{fig:HCAL_gem_structure}. An electric field is created between
the cathode and the first kapton foil, forcing electrons created by ionization
to drift to the kapton foil.
The kapton foil is metalized on both sides
and perforated with holes of about $70 \mu$m diameter.
Applying high voltage to the metal layers on both sides of the foil results in a
strong electric field inside the holes and leads to gas amplification.
The use of several
layers  
ensures a sufficient signal size.
 The size of the
pads might be arbitrarily small.
 GEMs provide fast signals and recovery time. 
The pad signals are amplified and discriminated.
If the signal of the pad is above the threshold it is counted as a hit.

Test chambers of area 30 x 30 cm$^2$, as shown in Figure~\ref{fig:HCAL_gem_TC}, are 
operated with cosmics, a radioactive source and a high intensity 
electron beam.
A gas mixture of 80\% Ar/20\% CO$_2$ has been shown to work well with a gain of 10$^4$, 
an efficiency of about 95\% and a hit multiplicity
of 1.27. Chambers have been exposed to a high intensity electron beam. Even after
collecting $2 \times 10^{12}$ electrons/pad,  no decrease in gain was observed. 

A full-size test beam module, planned for completion in 2008, will be equipped 
with 100 x 30 cm$^2$ chambers. 
Major activities are ongoing for the mechanical aspects of large GEM-layer 
assembly and the fabrication of large area GEM foils in collaboration with an industrial company
in the US.

Micromegas function as illustrated in Figure~\ref{fig:micromega1}. Electrons liberated by a
 ionizing particle drift to a mesh,
are amplified when crossing the mesh 
and collected on the anode. The gap between the mesh and the anode can be made 
in the range of 100$\mu$m, leading to a small size avalanche and excellent spatial resolution
and potentially low pad multiplicity. Chambers based on Micromegas can be made very thin, about 4 mm,
allowing to built a very compact calorimeter. 
Test chambers as shown in  Figure~\ref{fig:micromega2} 
are of $50 \times 50$~cm$^2$ size. They are being 
prepared for test-beam measurements in 2007/ 2008. Provided the results
from these measurements are satisfactory, a 1~m$^2$ plane will be built for test-beam studies in 2008. 

The scheme of a glass RPC is shown in Figure~\ref{fig:scheme_RPC}. 
Thin glass plates enclose a volume filled with a suitable gas mixture.
The glass plates 
are covered outside with
a conductive layer. Applying high voltage a charged particle traversing the gas gap creates a 
local discharge.
Covering the chamber by an isolating foil 
with fine segmented pads, an image charge is induced on the pads around the
discharge position.    
 
The structure of the calorimeter will be the same as shown in Figure~\ref{fig:HCAL_gem_structure}
replacing the GEMs by RPCs.
A prototype of a RPCs of $100 \times 30$~cm$^2$ is shown in Figure~\ref{fig:HCAL_RPC}.

Tests have been carried out with cosmic rays, sources, and a particle beam. The gas mixture consists of Freon
(R134A), isobutane (5\%) and a small admixture of SF6. The efficiency, measured in a proton beam 
of 120 GeV at FNAL, is 
shown as a function of the high voltage in Figure~\ref{fig:HCAL_RPC_test}. It approaches nearly 100\% 
above 7~kV.
The pad multiplicity for a $1 \times 1$~cm$^2$ pads ranges between 1.1 and 1.6, depending
on the high voltage and on the particular design of the 
chamber. 
Further beam tests will start early in 2007 with eight fully equipped chambers 
interleaved with 20~mm steel absorber plates. If the results are satisfactory
the construction of a 1~m$^3$ prototype will be initiated.

Several ASIC chips are under development for the readout of digital calorimeters.
The HArDROC chip delivers a quasi-binary readout of 64 analog channels with two thresholds per channel.
A digital memory will save data during a bunch train and can be readout via one serial output.
The first version was submitted in September 2006.
The DCAL chip, also containing 64 analog channels,
is
under development in Argonne. A few prototypes have been tested
in the lab and fulfill the requirements on noise level, linearity, and time stamping.
A second version with decreased input sensitivity
was submitted  in summer 2006. 
Also a special version of the 
KPiX chip is under development. Both chips are to be beam tested in 2007.

\subsection{DREAM Calorimeter}

The DREAM collaboration~\cite{dream} follows a fundamentally different concept to 
improve the jet energy resolution.
Usual hadron sampling calorimeters are limited in the energy resolution due to
fluctuations induced by the different 
response from the electromagnetic and hadronic shower component.
DREAM uses a dual readout concept. The sensors inside the absorber are scintillation and clear fibers.
Scintillating fibers respond to all charged particles in a shower 
whereas clear fibers detect 
Cherenkov light induced mainly 
by electrons and positrons.
Because of the fine-grained spatial sampling also fluctuations in the density of local
energy  are accounted for.  
In addition, the detection of MeV neutrons in e.g. hydrogen enriched fibers, might further improve the
energy resolution.
A possible structure of a DREAM-type calorimeter is shown in   
 Figure~\ref{fig:DREAM}.

The DREAM Collaboration performs test-beam studies both with scintillating and clear 
fibers inside a copper absorber. Figure~\ref{fig:DREAM1} shows a module
operated in a CERN test-beam. 
The box at the end contains the photo-multipliers reading out the bundles
of clear and scintillating fibers.
Measurements are done with muons, electrons and hadrons. 
As an example, the measured resolution
for hadrons as a function of the hadron energy is shown 
in Figure~\ref{fig:DREAM2}. Using only the Cherenkov light output the resolution
follows a  86\%/$\sqrt E$ + 10\% dependence as a function of the energy.
Using the scintillator readout only the resolution can be parametrized as
49\%/$\sqrt E$ + 7\%. Combining both in a proper way using the measured ratio of
Cherenkov and  scintillator light the resolution can be improved 
to 41\%/$\sqrt E$ + 4.2\%. Further improvement of the latter result 
seems possible by reducing shower leakage, using e.g. a larger prototype, and to 
measure neutron depositions.

PbWO$_4$ crystals are under study, read out with several photosensors on their front and rear
sides. Filters enhance 
sensitivity to either scintillation light (in a relatively restricted 
range of wavelengths) and Cherenkov light (which covers the whole range,
albeit with $1/\lambda^2$ spectrum).
The different timing and the directionality of the Cherenkov light could also be exploited
to improve the shower energy measurement in crystals. 

DREAM promises an alternative to the particle flow concept. Studies are underway to characterize
the performance of a calorimeter based on the DREAM technology in the ILC environment,
e.g. in the reconstruction of multi-jet final states.

\subsection{Very Forward Calorimeters}
In the very forward region two calorimeters, BeamCal and LumiCal, 
are planned for a fast and precise measurement
of the luminosity and ensure detector hermeticity~\cite{FCAL}.
Recently a third calorimeter, GamCal, was proposed to support
the fast luminosity measurement and beam parameter optimisation. 
The first two calorimeters will be sampling calorimeters consisting of tungsten absorber disks
interspersed with pad-structured solid-state sensor planes. 
GamCal is still under design.

The BeamCal adjacent to the beam-pipe covers a polar angles down to about 5 ~mrad.
Electrons and positrons originating from beamstrahlung photon conversions deposit
several TeV per bunch-crossing in the BeamCal. The distribution of this energy will be measured
to assist in tuning the beams.
The expected dose collected is up to 10 ~MGy per year for nominal accelerator parameters
at 500 GeV center-of-mass energy.
Fine granularity and small Moliere radius is necessary to identify the
localized depositions from high-energy electrons on top 
of the broader spread of energy from beamstrahlung remnants. 
The requirements on the sensors are 
stable operation under high electromagnetic doses,
 very good linearity over a 
dynamic range of about 10$^4$, 
very good homogeneity, and fast response.
BeamCal has to be fully readout after each bunch-crossing requiring a specialized
fast FE electronics and data acquisition to be developed. 
Test-beam studies have are 
been done using samples of CVD diamond sensors of 1 cm$^2$ area and a few 100 $\mu$m thickness.
A reasonable linearity has been measured over a dynamic range of larger than 
10$^5$. 
The performance of several sensors as a function of the absorbed dose has been measured in a 10 ~MeV
electron beam, as shown in  Figure~\ref{fig:FCAL_tb} for doses up to 7~MGy. 
For the sensors 
produced so far we observe a drop of the signal to about 30\% and stable noise.
Studies in  close collaboration with the manufacturers are underway 
to improve performance. 
In addition, alternatives like
GaAs or special silicon are foreseen to be investigated.
An example of a GaAs sensor designed for BeamCal is shown in
Figure~\ref{fig:GaAs}. Several such sensors will be prepared for 
test-beam studies in 2007. 

The LumiCal is the luminometer 
of the detector and covers larger polar angle outside the reach
of beamstrahlung pairs. 
The goal is to measure the luminosity 
by counting Bhabha events with an accuracy better than
10$^{-3}$. A silicon tungsten calorimeter has been simulated
and e.g. requirements on the tolerances
of the mechanical frame, the sensor positioning and the position of the calorimeters 
relative to the beam have been estimated.
In particular the inner acceptance radius must be controlled at the $\mu$m level.

The mechanical design is shown in Figure~\ref{fig:lumi_mech}.
To avoid effects of gravitational sag 
the support for the absorber disks is decoupled from the one of the 
sensor planes.  
The sensor layers will consist of silicon sensors
made from 6-inch wavers and structured as 
sketched in Figure~\ref{fig:lumi_sensor}.
Prototypes of sensor planes will be available beginning of 2007
for test measurements.
The design of the FE electronics has just started.
  
\subsection{Conclusions}

The requirements on the performance of the calorimeters are physics driven. Potentially, all technologies 
pursued in the different collaborations and projects may match these requirements.
To rank the proposed technologies test-beam studies and full-system tests 
are necessary.

There is a large variety in the development level
of the projects. CALICE is taking data with a first prototype for an ECAL and an analog HCAL
in a test-beam and will be able to answer many questions using these data in the near term.
Other projects are going to built prototypes in the near future and test them in beams. Since the latter 
is a complex undertaking, sufficient infrastructure and person-power are needed to ensure success.
Smaller groups may find it advisable to combine their efforts or to join one of the 
larger collaborations.  
Tests of the DREAM concept continue. To demonstrate the feasibility
of this technology for the ILC, a design suitable for a collider detector should be worked out
which provides the performance demanded by ILC physics.

The special calorimeters to instrument the very forward region are still in a relatively 
early development phase. Ongoing sensor tests are necessary to make a suitable
choice. The FE electronics requirements are just worked out and the design started.
Construction of prototypes for beam tests will be the next important future step.   



\section{Superconducting Detector Magnets for ILC}
\label{sec-subs-magnet}

The use of magnetic fields is a fundamental method to analyze the momentum of charged particles.  In order to extend the energy 
range in particle physics, large-scale magnetic fields are inevitably required.  The basic relation between magnetic field strength, 
charged particle momentum, and bending radius is described with $p = \gamma mv = q \rho B$ where $p$ is the momentum, $m$ the mass, $q$ the charge, $\rho$ 
the bending radius, and $B$ the magnetic field.  The deflection (bending) angle, $\phi$, and the sagitta,$s$, of the trajectory 
are determined by;
\begin{equation}
   \phi	\approx L / \rho = q B L / p, \ \ \ {\rm and} \ \ \ 
      s \approx q B L^2 / 8 p   \label{eq-subs-magnet1}
\end{equation}
where $L$ is the path length in the magnetic field. For practical measurements in high energy colliding beam detectors, both field strength
and magnetic volume have been increased. A general-purpose detector consists of three major stages:  a central tracker 
close to the interaction point, a set of electromagnetic and hadronic calorimeter systems, and an external muon detector system.  
The momentum measurement is carried out in the region of the tracker, although it may be augmented for muons in the muon detector,
and requires a powerful magnetic field to achieve high resolution.

   The solenoid field has been widely used in many colliding experiments \cite{ref-mag1,ref-mag2,ref-mag3,ref-mag4,ref-mag5,ref-mag6}.  It features uniform field in the axial direction with self-supporting structure.  
The magnetic flux needs to return outside the solenoid coil, and in most cases, an iron yoke provides the flux return.  An external solenoid with the reversed field may also provide flux return, and it requires a much more sophisticated magnet system in terms of stored energy, quench protection and mechanical support. 

   The momentum analysis is usually performed by measurement of particle trajectories inside the solenoid, and the momentum resolution 
is expressed by
\begin{equation}
   \sigma_p/p \approx  p / (B • L^2) \label{eq-subs-mag2}
\end{equation}	
where $L$ corresponds to the solenoid coil radius, $R$.  Therefore, a larger coil radius may be an efficient approach to reach better 
momentum resolution, although overall detector cost considerations must also be taken into account. In this section progress in the 
design of solenoidal detector magnets 
is reviewed, and possible detector magnet design for 
the International Linear Collider (ILC) experiments are discussed.

\subsection{Progress in superconducting detector solenoid magnets} 
   Table~\ref{tab-mag1} lists progress of the superconducting solenoids in collider experiments~\cite{ref-mag1}. The CDF solenoid 
established a fundamental technology of the ``co-extruded aluminum stabilized superconductor"~\cite{ref-mag7} which has become a 
standard for the detector magnet based on the pioneer work for the colliding beam detector solenoid at ISR ~\cite{ref-mag8} and 
CELLO \cite{ref-mag9}.  The TOPAZ solenoid established the ``inner winding'' technique to eliminate the inner coil 
mandrel \cite{ref-mag10}.  The ALEPH solenoid demonstrated indirect and thermo-syphon cooling for stable cryogenics operation 
in large scale detector magnets~\cite{ref-mag11}. In the SDC prototype solenoid, the mechanical reinforcement of 
aluminum-stabilized superconductor was further developed \cite{ref-mag12}. In the LHC project at CERN, two large superconducting 
magnet systems have been developed for ATLAS and CMS with extensive efforts for the high-strength aluminum stabilized superconductor. 

   The magnetic field design of the ATLAS detector is composed of an axial field by using a solenoid coil in the central region and of an azimuthal field by using a set of toroidal coils~\cite{ref-mag13}.  Since the solenoid coil is placed in front of the liquid-argon calorimeter, it is required to be as thin and transparent as possible to achieve the best calorimeter performance. Therefore, the solenoid coil was designed with features (1) high-strength aluminum stabilized superconductor  uniformly reinforced, (2) pure-aluminum strip technique for uniform energy absorption and quench protection, and (3) a common cryostat with the LAr calorimeter to ultimately save magnet wall-material. Extensive efforts have been made to reinforce the aluminum stabilizer while keeping adequate low electrical resistivity as discussed below~\cite{ref-mag14}.  

   The CMS detector is designed as a single solenoidal magnet surrounded by the iron return yoke~\cite{ref-mag15, ref-mag16, ref-mag17}.  Special effort has been made to reinforce the aluminum stabilized superconductor in a hybrid configuration as discussed below~\cite{ref-mag18}.  

{\footnotesize{
\begin{table}[thb]
  \caption{\label{tab-mag1}Progress of detector solenoid magnets in high energy physics.}
\begin{footnotesize}
\begin{tabular}{ll|cccccc}
\hline
Name & Laboratory &B & R & L & E & X  & E/M  \\
           &     & [T]  & [m] & [m] &       [MJ]   & [$x_0$] & [kJ/kg] \\
\hline
CDF& Tsukuba/FNAL& 1.5&1.5&5.07&30&0.84&5.4\\
TOPAZ$^*$ & KEK & 1.2& 1.45& 5.4& 20& 0.70& 4.3\\
VENUS$^*$ & KEK& 0.75& 1.75& 5.64& 12& 0.52& 2.8\\
AMY$^*$ & KEK& 3& 1.29& 3& 40& $\#$ & \\ 
CLEO-II&Cornell&1.5&1.55&3.8&25&2.5&3.7\\
ALEPH$^*$&Saclay/CERN&1.5&2.75&7.0&130&2.0&5.5\\
DELPHI$^*$&RAL/CERN&1.2&2.8&7.4&109&1.7&4.2\\
ZEUS & INFN/DESY& 1.8& 1.5& 2.85&11&0.9&5.5\\
H1&RAL/DESY&1.2&2.8&5.75&120&1.8&4.8\\
BABAR& INFN/SLAC& 1.5& 1.5& 3.46& 27& $\#$ & 3.6\\
D0 & Fermi& 2.0& 0.6& 2.73& 5.6& 0.9& 3.7\\
BELLE&KEK&1.5&1.8&4&42&$\#$&5.3\\
BES-III$+$&IHEP&1.0&1.45&3.5&9.5&$\#$&2.6\\
ATLAS & & & & & & & \\
~~~Central&ATLAS/CERN&2.0&1.25&5.3&38&0.66&7.0\\
~~~Barrel &ATLAS/CERN&1&4.7-9.7&5&26&1080&  \\
~~~Endcap &ATLAS/CERN&1&0.825-5.35&5&2 $\times$ 250&-&\\
CMS$+$ &CMS/CERN&4&6&12.5&2600&$\#$&12\\
\hline
\multicolumn{8}{l}{$^*$ operation complete}\\
\multicolumn{8}{l}{$+$detector under construction}\\
\multicolumn{8}{l}{$\#$EM calorimeter inside solenoid, so small radiation length, X, not a goal}\\
\end{tabular}
\end{footnotesize}
\end{table}
}}

\subsection{Progress of Aluminum Stabilized Superconductor}
\begin{figure}
	\centering
		\includegraphics[width=12cm]{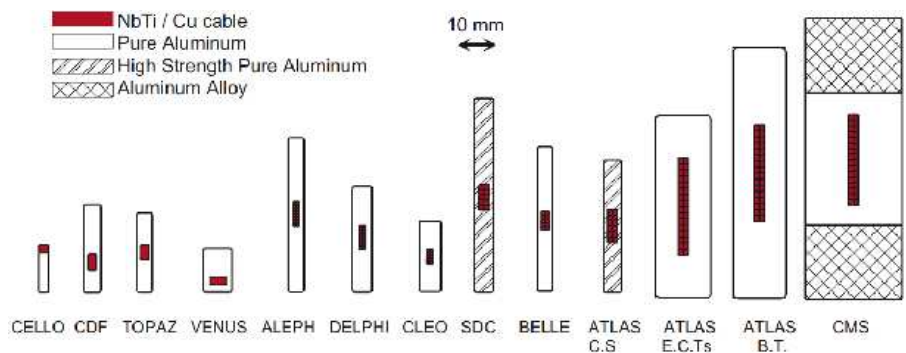}
	\caption[Development of aluminum stabilized superconductors in particle physics.]
{Cross sectional view of aluminum stabilized superconductors
which have been used for different collider detector magnets.
The conductors are arranged chronologically from left to right.
A scale is indicated by the double arrow in the center of the figure.}
	\label{fig-mag1}
\end{figure}

Aluminum stabilized superconductor represents a major technological advance in detector magnets. It has been developed to provide large-scale magnetic fields with minimum material~\cite{ref-mag5}.  In Figure~\ref{fig-mag1} cross-sectional vviews of aluminum stabilized superconductors are shown for various generations of detector magnets used high-energy physics experiments.  
Major progress on the mechanical properties of the conductor has been achieved by using a 
NbTi/Cu superconductor co-extruded with an aluminum stabilizer by diffusion bonding. One approach consists of a ''uniform reinforcement''~\cite{ref-mag5}, and the other consists of a ``hybrid configuration''~\cite{ref-mag18}. The uniform reinforcement was made, in the SDC and ATLAS solenoid, by ``micro-alloying'' followed by ``cold-work hardening''~\cite{ref-mag19}. By this the strength of the aluminum stabilizer could be much improved while the excellent electrical properties were retained without increasing the material.   Another technical approach for the reinforcement has been made, in the CMS solenoid, by using a hybrid (or block) configuration, which consists of pure-aluminum stabilized superconductor and high strength aluminum alloy (A6082) blocks at both ends fixed by using electron-beam welding.  The hybrid configuration is very useful and practical in large-scale conductors~\cite{ref-mag20, ref-mag21}.
          
\begin{figure}
	\centering
		\includegraphics[height=3cm]{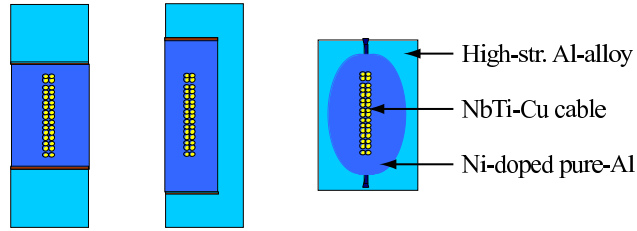}
		\caption[Possible hybrid magnet configurations]{Possible hybrid magnet conductor configurations based on technologies established for 
the LHC detector magnets.}
	\label{fig-mag2}
\end{figure}

   Based on these successful developments of both ``uniform reinforcement'' and ``hybrid configuration'' further improvement has been proposed, for future applications, in combining these efforts~\cite{ref-mag21,ref-mag22} as summarized in Table~\ref{tab-mag2}, and shown in Figure~\ref{fig-mag2}.  The central part of the CMS conductor partly composed of pure-aluminum stabilizer may be replaced by a Ni-doped high strength aluminum stabilizer developed for the ATLAS solenoid. This may result in a further reinforcement of the overall aluminum stabilized superconductor. It may be applicable in the ILC detector magnets especially for the high field magnet such as the SiD solenoid discussed later.  
{{   
\begin{table}[thb]
\caption[High-strength aluminum stabilized superconductors]{\label{tab-mag2}{Progress of high-strength aluminum stabilized superconductor and possible future upgrade.}}
\begin{center}
\begin{footnotesize}
\begin{tabular}{ll|lccc}
\hline
     & Reinforced  &Feature        &Aluminum        &Full cond       & Full cond \\ 
     &             &               &\multicolumn{2}{c}{yield strength} & RRR \\
\hline
\multicolumn{2}{l}{\underline{Progress at LHC}}&&&\\
ATLAS&  Uniform & Ni(0,5 \%)-Al    & 110      & 146      &        590 \\
CMS  & Hybrid   & Pure-Al\&A6082-T6& 26 / 428 & 258      & $\sim$ 1400 \\
\hline
\multicolumn{2}{l}{\underline{Improvements for ILC}}&&&\\
     & Hybrid   & Ni-Al\& A6082-T6 & 110 / 428 & $\sim$ 300 & $\sim$ 300\\
     & Hybrid   & Ni-Al\& A7020-T6 & 110 /677  & $\sim$ 400  & $\sim$ 300\\
\hline
\end{tabular}
\end{footnotesize}
\end{center}
\end{table}
}}

\subsection{$\bf E/M$ ratio as a Performance Measure}
The ratio of stored energy to cold mass ($E/M$) in a superconducting magnet is a useful performance measure. In an ideal solenoid with a perfect axial field, it is also expressed by a ratio of the (hoop) stress, $c_h$, to the average 
density, $d$~\cite{ref-mag1,ref-mag5}:
\begin{equation}
\label{eq-subs-mag3}
   E/M	\approx c_h / 2d.
\end{equation}

\begin{figure}
	\centering
		\includegraphics[height=8cm]{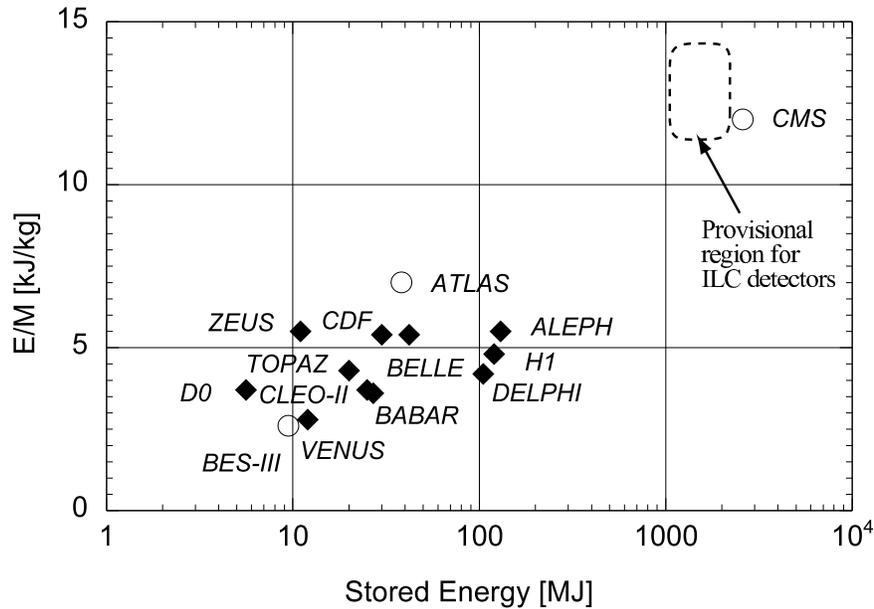}
	\label{fig-ratio}
	\caption{Ratio E/M of stored energy to cold mass for existing thin detector solenoids}
\end{figure}

Figure~\ref{fig-ratio} shows the E/M ratio in various detector magnets in high energy physics.  Assuming an approximate average density of $3\times10^3 $kg/m$^3$ for the conductor, and an E/M ratio of $10$~kJ/kg, the hoop stress level will be $\sim 60$~MPa. In the case of an iron free solenoid, the axial stress, $\sigma_z$, will be one half of the hoop stress, and the stress intensity $(c_h + c_z)$ then is around $90$~MPa. This has to be sufficiently lower than yield strength of the coil material, in a mechanically safe design.

	The E/M ratio in the coil is approximately equivalent to the enthalpy of the coil, $H$. It determines the average coil temperature rise after energy absorption in a quench:
\begin{equation}
\label{eq-subs-mag4}
   E/M = H (T_2) - H (T_1) \approx  H (T_2), 
\end{equation}
where $T_2$ is the average coil temperature after the full energy has been absorbed in a quench, and $T_1$ is the initial temperature. E/M ratios of 5, 10 and 20~kJ/kg correspond to $\sim 65$ , $\sim 80$, and $\sim 100$~K temperature rise, respectively. The E/M ratios in various detector magnets are shown in Figure~\ref{fig-ratio} as a function of the total stored energy. The CMS magnet has the 
currently largest E/M ratio at $12$~kJ/kg as well as the largest stored energy. In the case of a quench, 
the protection system at CMS has been designed to keep the temperature below $80$~K.

In developing a high field magnet, the limiting factor usually is the temeprature increase the 
coil can tolerate in case of a quench. This would favour high mass coils, contrary to 
the requirements on a light coil expressed by all experiments. 
The success of the CMS solenoid suggests that the E/M ratio of the detector magnet design in the ILC experiment as large as 12 kJ/kg or even slightly higher can be tolerated. 
It may help to design the higher field magnet with larger stored energy.  It should also be noted that a higher E/M ratio would be required to realize much higher field and/or much larger-scale superconducting magnets for the ILC detector magnets as discussed below.

\subsection{Detector Magnets at the ILC} 
According to the ILC detector outline documents~\cite{ref-GLD,ref-LDC,ref-SiD,ref-4th}, the detector magnet design requirements are 
listed in Table~\ref{tab-mag3}, and are compared with the LHC detector solenoids successfully commissioned.  

\begin{table}[tbh]
\caption[Detector Solenoids at LHC and ILC]{\label{tab-mag3}Superconducting detector solenoids for ILC compared with detector solenoids at LHC.}

\begin{footnotesize}
\begin{tabular}{ll|ll|lllll}
\hline
Parameters& unit & \multicolumn{2}{c}{LHC} & \multicolumn{5}{c}{ILC}\\
\hline
& & ATLAS& CMS& GLD& LDC& SiD& \multicolumn{2}{c}{4th}\\
& & \multicolumn{1}{c}{CS} & & & & & Inner & Outer \\ 
\hline
\multicolumn{2}{l|}{\bf Basic requirements}&&&&&&\\
\hline
Clear-bore radius      & m     &7    & 1.18  & 4.00  &3.00   & 2.5     & 3.0   &     \\
Central magnetic field & Tesla & 2   & 4     &3      & 4     & 5    &3.5   & 1.5 \\
\hline
\hline
\multicolumn{2}{l|}{\bf Design parameters}&&&&&&\\
\hline
Coil inner radius      & m     & 1.23 & 3.25  & (4.0) &	3.16  &2.65 & 3   & 5.4\\
Coil half length       & m     & 2.7  &6.25   & 4.43  & 3.3   & 2.5 & 4   & 5.5\\
Coil layers            &       & 1    &4      & 2     &4      & 6   &   6 & \\
Cold mass thickness    & m     & 0.04 & 0.3   & 0.4   &  0.3  & 0.4 & 0.3 & \\
Maximum field in coil  & Tesla & 2.6  & 4.6   & 3.5   &  4.6  & 5.8 & 5.8 & \\
Nominal current        & kA    & 7.73 & 20    &       &       & 1.8 & 2.0 & \\
Stored energy          & GJ    & 0.04 & 2.6   & 1.6   & 1.7   & 1.4 & 2.8 & \\
Cold mass weight       & ton   & 5.7  & 220   & 78    & 130   &     &     & \\
E/M                   & kJ/kg & 7    & 12.3  & 20    & 13    & 12   & 12.6& \\
\hline
Reference&  & \cite{ref-mag12}& \cite{ref-mag15} & \cite{ref-GLD} & \cite{ref-LDC} & \cite{ref-SiD,ref-mag27}  & \multicolumn{2}{c}{\cite{ref-4th}} \\
\hline
\end{tabular}
\end{footnotesize}
\end{table}

\begin{figure}
	\centering
		\includegraphics[height=8cm]{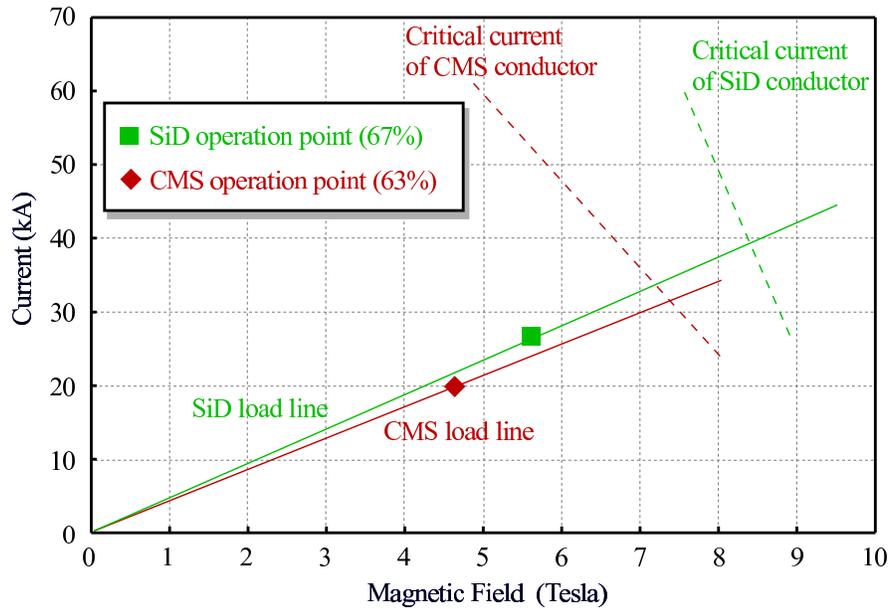}
	\caption[Load lines for different magnet designs.]{The load line (solid line) and the critical current as a function of
the magnetic field (dashed line) of the proposed SiD superconductor(green)
and of the CMS magnet(red).  The operation point
of the SiD magnet and the CMS magnets are shown by a green square symbol
and a red diamond symbol, respectively.  The numbers in \%
are loads relative to their (mechanically) critical loads.}
	\label{fig-mag4}
\end{figure}

The development of the ILC magnets will profit heavily from the LHC experiences. The CMS magnet 
concept is the basis for the proposed ILC magnets. The GLD detector solenoid will require a larger cold mass because of the larger coil 
radius resulting in more stored energy per length.  The LDC magnet design is most similar to the CMS solenoid design.  
The SiD detector solenoid will be most challenging with the high field of 5 T resulting in larger mechanical stress and increased stored 
energy.  The smaller coil radius may help to manage the mechanical stress~\cite{ref-mag27}. Figure~\ref{fig-mag4} shows the ratio between 
actual mechanical load and the critical load for the CMS and the SiD solenoids. For SiD, it is $67 \%$ of the critical load. 
Combining the advanced superconductor technology of ATLAS and CMS may help to make the GLD and SiD solenoids more stable and reliable. 
Advances in technology may also reduce the thickness 
of all proposed coils at the ILC.

Future advances in superconducting technology may make it possible to push the limits on E/M as 
high as $15 \sim 20$~hJ/kg, which could result in either more powerful or thinner 
magnets for the ILC. Further aggressive conductor development is crucial if 
these goals should be met.

\subsection{Summary and Outlook}
  	Based on the recent experience in the superconducting detector magnets at LHC, the ILC detector magnets will be designed as follows:

\begin{enumerate}
\item The magnet will use high-strength aluminum stabilized superconductor as used in the ATLAS and CMS solenoids. The magnet technology developed for the CMS solenoid can be applied and extended.   
\item The conductor mechanical strength may be further improved by combining both features of the high-strength aluminum stabilizer (Ni-doping and cold-work reinforcement) developed for the ATLAS solenoid and ``hybrid configuration'' reinforced by high-strength aluminum alloy placed at both ends of the coil developed for the CMS solenoid.   
\item High $E/M$ ratios in the cold mass has been reached by CMS and it does seem possible to achieve even higher E/M ratio up to 20~kJ/kg, depending on the required magnetic field and compactness of the magnet design. 
\item A highly redundant and reliable safety system needs to be an integral part of 
any magnet design, to protect the magnet in case of quench. Energy extraction should be the primary protection scheme and fast 
quench trigger, which initiates a heater induced quench, should be an important backup system.  
\end{enumerate}

	A special effort will be required to realize the 4th detector design. It will require major efforts in mechanical design as well
 as in the magnet safety design because of the extraordinary large electromagnetic force (and de-centering force), stored energy and 
the resulting mechanical complexity.  A sophisticated mechanical design and engineering work will be required. 	

	In the long range future, even higher field solenoids might be feasible by pushing $E/M > 20$~kJ/kg.  Such a design may be 
realized with further improvements in the conductor, with significantly larger yield strength. 
	
\section{Data Acquisition}

As outlined in all four detector concept studies \cite{ref-GLD,ref-LDC,ref-SiD,ref-4th} the data acquisition (DAQ) system of a 
detector at the ILC has to fulfill the needs of a high luminosity, high precision experiment without compromising on rare or yet 
unknown physics processes. Although the maximum expected physics rate, of the order of a few kHz, is small compared to that of 
recent hadron colliders, peak rates within a bunch train may reach several MHz due to the bunched operation.

In addition the ILC physics goals require higher precision in jet and momentum resolution and better impact parameter resolution 
than any other collider detector built so far. This improved accuracy can only be achieved by substantially increasing the number of 
readout channels.
Taking advantage of the bunched operations mode at the ILC, event building without a hardware trigger, followed by a software-based
event selection was proposed \cite{ref-TESLA:2001} and has been adopted by all detector concept studies. This will 
assure the needed flexibility and scalability and will be able to cope with the expected complexity of the physics 
and detector data without compromising efficiency.

The increasing numbers of readout channels for the ILC detectors will require signal processing and data compression already at the detector electronics level as well as high bandwidth for the event building network to cope with the data flow.
The currently built LHC experiments have up to 108 front-end readout channels and an event building rate of a few kHz, moving data with up to 500 Gbit/s \cite{ref-CMSDAQ}. The proposed DAQ system will be less demanding in terms of data throughput although the number of readout channels is likely to be a factor of 10 larger.

The rapid development of fast network infrastructures and high performance computing technologies, as well as the higher integration and lower power consumption of electronic components are essential ingredients for this data acquisition system.
Furthermore it turned out that for such large systems a restriction to standardized components is vital to achieve maintainability at an affordable effort, requiring commodity hardware and industry standards to be used wherever possible.

Details of the data acquisition system depend to a large extent on the final design of the different sub detector electronic components, most of which are not fully defined to date. 
Therefore the DAQ system presented here will be rather conceptual, highlighting some key points to be addressed in the coming years.

\subsection{Concept}
In contrast to currently operated or built colliders, such as HERA, Tevatron or LHC, which have a continuous rate of equidistant bunch crossings the ILC has a pulsed operation mode. For the nominal parameter set \cite{ref-raubenheimer} the ILC will have

\begin{itemize}
\item $\sim$ 3000 bunch crossings in about 1ms,
\item 300 ns between bunch crossings inside a bunch train and
\item $\sim$ 200 ms without collisions between bunch trains.
\end{itemize}

This operation mode results in a burst of collisions at a rate of $\sim$3MHz over 1~ms followed by 200~ms without any interaction. The integrated collision rate of 15 kHz is moderate compared to the LHC and corresponds to the expected event building rate for the LHC experiments.

The burst structure of the collisions at the ILC immediately leads to the suggested DAQ system:

\begin{itemize}
\item dead time free pipeline of 1 ms,
\item no hardware trigger,
\item front-end pipeline readout within 200 ms and
\item event selection by software.
\end{itemize}

The high granularity of the detector and the roughly 3000 collisions in 1 ms still require a substantial bandwidth to read the data in time before the next bunch train. To achieve this, the detector front end readout has to perform zero suppression and data condensation as much as possible. Due to the high granularity it is mandatory to have multiplexing of many channels into a few optic fibres to avoid a large number of readout cables, and hence reduce dead material and gaps in the detector as much as possible.

\begin{figure}
	\centering
		\includegraphics[height=8cm,bbllx=42mm,bblly=15mm,bburx=177mm,bbury=106mm]{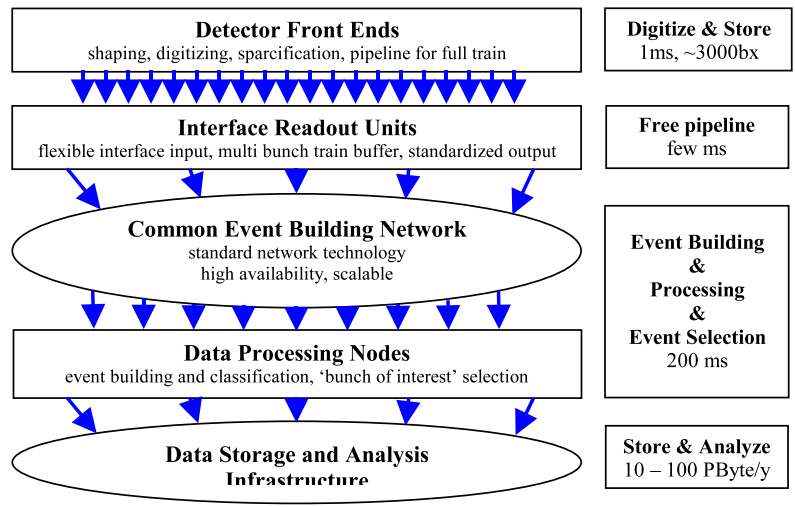}
	\caption{Conceptual diagram of the proposed data flow at the ILC}
	\label{fig-daq1}
\end{figure}

The data of the full detector will be read out via an event building network for all bunch crossings in one train. After the readout, the data of a complete train will be situated in a single processing node. The event selection will be performed on this node based on the full event information and bunches of interest will be defined. The data of these bunches of interest will then be stored for further physics analysis as well as for calibration, cross checks and detector monitoring. Figure~\ref{fig-daq1} shows a conceptual diagram of the proposed data flow.

The sub detector specific part is realized in the front-end readout units which receive the detector data via a fast serial link. The readout units will consist of three parts:

\begin{itemize}
\item a programmable interface to the front end readout,
\item the event data buffer which will allow storing data of several trains and
\item the standardized network interface to the central DAQ system.
\end{itemize}

The programmable interface should enable one common type of readout unit to adapt to the detector specific front end designs.
To allow for variations in the readout timing to more than 200 ms the readout units could be equipped with event data buffers with multiple train capacity.
The full event is built via the event building network into a single data processing node which will perform final data processing, 
extract and apply online calibration constants and will select the data for permanent storage.

In the data processing node the complete data of all bunch crossings within a train will be available for event processing. Distributing data of one train over several processing nodes should be avoided because sub detectors such as the vertex detector or the TPC will have overlapping signals from consecutive bunch crossings and unnecessary duplication of data would be needed.

Event selection is performed in these data processing nodes such that for each class of physics process a specific finder process will identify the bunch crossings which contain event candidates and mark them as ``bunches of interest''. All data for the `bunches of interest' will be fully processed and finally stored permanently for the physics analysis later on. By using software event selection with the full data available, a maximum event finding efficiency and the best possible flexibility in case of unforeseen conditions or physics processes is ensured. The best strategy for applying these finders and processing the data, depends on the topology of the physics processes to be selected and their background processes. This has to be further studied and optimized based on full Monte Carlo simulations.

Several trains will be built and processed in parallel in a farm of data processing nodes and buffering in the interface readout units will allow for fluctuations in the processing time.

Using commodity components like PCs and standardized network components allows for the scaling of the processing power or 
network bandwidth according to the demands. The use of off-the-shelf technology for the network and the computing units will 
ease maintainability and benefit from the rapid commercial development in this area. The DAQ system will also profit from the 
use of a common operating system, for example Linux, and high level programming languages already at the event building 
and event finding stage,  making the separation of on-line and off-line code obsolete and therefore avoid the need to rewrite, 
and debug, code for on-line or off-line purposes. This results in a more efficient use of the common resources.

\subsection{Detector front end electronics}
The detector front end readout is discussed in the specific chapters of the different detector components. A few common issues of 
particular relevance are summarized below.

The amount of data volume to be collected by the DAQ system is dominated by pair background from the machine. Simulations for the 
nominal ILC parameters \cite{ref-raubenheimer} at $E_{cm}$ = 500~GeV for the LDC \cite{ref-LDC} show in the vertex detector 455, 189 
and 99 hits per bunch crossing for layer 1, 2 and 3 respectively. In the TPC volume roughly 18000 hits are produced per bunch crossing. Similar studies for the other concepts confirm the high background near the beam pipe.

Except for the inner layers of the vertex detector the occupancy for a full train imposes no constraints onto the readout 
scheme. 
For the inner vertex detector layers the data has to be read out during the train to keep the hit density low enough so as 
not to compromise 
the tracking performance. In the SiD main tracker, associating hits with the bunch crossing which produced them, reduces
the background especially in the forward region.

For the SiW based ECAL systems, the high granularity requires large multiplexing on the front end detectors with an 
adequate multi-hit capability and efficient hit detection or zero suppression. Single chips with hit detection, 
charge and time digitization and multi-hit
 storage capacity for up to 2048 channels were proposed by several groups.

For the TPC novel readout technologies are developed with reduced ion feedback to allow for a gateless operation with 
sufficient gas amplification for a period of 1~ms.

The electronic noise of the front end systems or the detectors themselves is a third, possibly very dangerous, 
source of data volume in a triggerless system. It has to be sufficiently under control or it must be suppressed by the 
front end data processing. 

The high granularity of the detector systems and the increased integration of electronic at the detector front end, will result in large power dissipations. To avoid excessive cooling needs, all detector systems investigate the possibility of reducing the power at the front end electronics by switching power off between trains (power cycling). This has to be balanced against power up effects, the readout time needed between trains and the ability to collect data between trains for calibration purposes, e.g. cosmic muon tracks.

\subsection{Machine Interface}
The machine operation parameters and beam conditions are vital input for the high precision physics analysis and will therefore be needed alongside the detector data.
Since the amount of data and time structure of this data is similar, a common data acquisition system and data storage model should be used. Up to now very little has happened to integrate the DAQ for the beam delivery system into the physics data flow. It is mainly assumed that integration of parts or all of the machine parameters should be straight forward due to the programmable interface units and the network based structure of the DAQ system.

\subsection{Detector Control and Monitoring}
The data acquisition and its operation is closely coupled to the detector status and detector conditions, as well as the 
machine conditions. Hence it is proposed that the detector slow control and the detector monitoring are tightly linked 
to the DAQ system with an overall experiment control system.

For detector commissioning and calibration the DAQ system has to allow for partial detector readout as well as local DAQ runs for many sub components in parallel.
The DAQ system has to be designed such that parts of a detector component or complete detector components can be excluded from the readout or be operated in local or test modes without disturbing the physics data taking of the remaining parts.

The ILC as well as the detector will be operated by truly worldwide collaborations with participants around the world. 
The global accelerator network (GAN) and global detector network (GDN) have been proposed to operate both the machine 
and the detector remotely from the participating sites. This in turn requires that the data acquisition system, as well 
as the detector control, be designed with remote control and monitoring features built in from the start.

\subsection{Outlook and R\&D}
To benefit from the online software event selection an accurate online calibration is needed. Strategies for calibrating and monitoring the detector performance as well as efficient filter strategies have to be worked out. Simulation studies will be needed in the coming years to prepare this in more detail.

Although for the main DAQ system commodity components will be used, to be chosen at the time the DAQ has to be built, some R\&D is needed to prepare the decisions.
In addition, for the front end readout electronics and the interface to the DAQ system, decisions have to be made during the prototyping phase of the large detector components.

A DAQ pilot project is planned to serve as a frame for R\&D on the front end readout interface, the machine and detector DAQ interface, detector slow control issues, online calibration and event selection strategies.

Recent developments on technology (for example ATCA \cite{ref-ATCA}) should be followed and if possible explored to gain the necessary experience needed for the DAQ technology choice.

\section{Test Beams}
The intense detector R\&D program described earlier in this document will need support by significant test beam resources and test facilities. 
In this section a brief summary of the status and plans of the existing facilities at the time of writing of this document is presented.  

\subsection{Facilities}

Currently seven laboratories in the world are providing eight different beam test facilities: 
CERN PS, CERN SPS, DESY, Fermilab MTBF, Frascati, IHEP Protvino, LBNL and SLAC.  
In addition, three laboratories are planning to provide beam test facilities in the near future; 
IHEP Beijing starting 2008, J-PARC in 2009 and KEK-Fuji available in fall 2007.  Of these facilities, 
DESY, Frascati, IHEP Beijing, KEK-Fuji and LBNL facilities provide low energy electrons ($<$ 10 GeV). 
SLAC End Station-A facility provides a medium energy electron beam but the availability beyond 2008 is uncertain at this point. 
IHEP Protvino provides a variety of beam particles in 1- 45~GeV energy range, but the facility provides test beams only 
in two periods of one month each per year.  CERN PS and SPS facilities can provide a variety of beam particle species 
in energy ranges of $1 - 15$~GeV and $10 - 400$~GeV, respectively.  
Finally, the Fermilab Meson Test Beam Facility can provides a variety of particles in the energy range of $1 - 66$~GeV, thanks to a recent beam line upgrade, and protons up to 120~GeV.  This facility is available throughout the year for the foreseeable future.
Table~\ref{tab-testbeam} below summarizes the capabilities of these facilities and their currently known availabilities and plans.

\begin{table}[htbp]
	\caption{Summary of test beam facilities and their availabilities.}
	\centering
	{\small
		\begin{tabular}{lllll}
		\hline
		Facility & E (GeV) & Particle & N$_{\rm beams}$ & Availability and plans \\
		\hline
		CERN PS     & $1-15$   & e,h,$\mu$ & 4 & part of year availability\\
		CERN SPS    & $10-400$ & e,h,$\mu$ & 4 & part of year availability\\
		DESY        & $1-6.5$  & e         & 3 & $>$ 3 month/ year \\
		FNAL-MTBF   & $1-120$  & p,e,h,$\mu$ & 1 & continuous at 5\% duty factor \\
		Frascati    & $0.25-0.75$ & e      & 1 & 6 month/ year \\
		LBNL        & $1.5, < 0.06, < 0.03$ & e,p,n& 1 & continuous \\
		IHEP Protvino & $1-45$ & e, h, $\mu$ & 4 & 2 $\times $ 1 month/ year \\
		SLAC        & $28.5, 1-20$ & e, e, $\pi$, p& 1 & future after 2008 unclear\\
		\hline
		\multicolumn{4}{c}{{\bf {future facilities}}}& available\\
		IHEP Beijing & $1.1-1.5$ & e & 3 & March 2008 \\
					       & $0.4-1.2$ & e,$\pi$, p& 3 &  March 2008 \\
		J-PARC       & $< 3$ & e, h, $\mu$ & ? &  2009  \\
    KEK-Fuji     & $0.5-3.4$ & e & 1 & fall 2007, 8 month/ year\\
    \hline
		\end{tabular}}
	\label{tab-testbeam}
\end{table}

\subsubsection{Beam instrumentation/ machine-detector interface}
At the ILC beam instrumentation and the interface between the machine and the detector play a very important role. 
It is a very active field of R\&D, as described in the section on MDI in this document. 

The detectors require very specialized instrumentation in the very forward direction, to measure precisely luminosity and 
energy of the colliding beams. These devices need to be radiation hard under intense electromagnetic radiation, and 
at the same time precise and fast for the expected physics signals. Tests therefore are required of the 
radiation hardness, and of the actual instrumentation. The latter is done at some irradiation facilities, not 
listed in table~\ref{tab-testbeam}, while the former needs primarily high energy test beams as provided by 
CERN or FNAL to test the response of calorimeters to different types of beams. 

The machine requires very ambitious monitoring and control of the beams in the interaction region. A number of experiments are planned or under way to develop and test beam instrumentation for the ILC. These 
tests typically need high energy electron beams, to be able to test fast feedback systems, beam energy 
spectrometers, or high energy polarimeters. This is currently possible at SLAC and, for some applications, 
at the ATFII test facility at KEK. Beam size, bunch size and repetition rate of the beam are 
very important, and need to be matched to the actual ILC conditions as closely as possible. Thus, the needs for these activities can utilize accelerator test facilities which are independent of detector R\&D beam test facilities. The SLAC facility plays a 
central role in these tests, and its unclear future beyond 2008 present a major problem for the 
community.

\subsection{Tracking R\&D}
The R\&D plans of the different tracking groups have been summarized earlier in this document. The development 
work covers three different types of detectors: pixel Silicon detectors, Silicon-based strip detectors, and 
large volume time projection chambers. Of central concern for these groups is the availability of 
moderately high energy beams (to minimize the effects of multiple Coulomb scattering), to test and understand 
the response of the detectors, 
study the achievable resolutions, and develop algorithms for alignment and calibration. Test facilities 
like the one at DESY for low to medium energies make significant contributions. In particular for the gaseous 
detectors, tests with different particle species will eventually be needed, to understand the 
particle identification capabilities of the detector. The beams at CERN or at FNAL are well 
suited for these applications. 

A central problem for these tests is the availability of large bore high field magnets. Many tests 
can be performed however at lower fields. A 1 T magnet facility will become available within the EUDET program, 
initially at DESY from 2008 onwards, eventually at CERN or FNAL after 2009. Currently no facilities 
exist where high fields are available with a beam for larger detector volumes. A small scale 
high field test facility, without access to beams though, is available at DESY to the community. 

The studies would profit from a time structure in the beam similar to the one expected at the ILC. 
However most studies can be done also with different time structures in the beam. 

\subsection{Calorimeter R\&D}
The calorimeters are a central part of the different ILC detector concepts. They play a very important role in 
the concept of particle flow, as explained earlier in this document. The development work outlined in this 
document requires extensive tests under realistic beam conditions. 

The planned tests serve a dual purpose: Firstly the technologies proposed for the different calorimeters 
need to be tested and developed to a point where they can be proposed for an ILC detector. Secondly, 
in particular for the hadronic part of the shower development, little to no data exist currently 
for a detailed modeling of the shower. Therefore data taken with test calorimeters of sufficient size will 
be of great interest to the modeling and understanding of the hadronic shower. 

The calorimeter tests therefore require an extensive range of beam energies and particle types, 
from $<1$ GeV/particle to a few 100 GeV/particle. A
well understood beam is very important, with a good knowledge of the particle content and its energy, and with a flexible 
setup which allows the calorimeters to be scanned with beam under a wide range of conditions. For some studies 
it might be important to also model the time structure of the beam, though in most cases, collection of 
large data samples is probably more important than the study of detailed timing requirements. Given the broad 
range of proposed technologies, and the large step in performance needed compared to 
established technologies, significant beam time allocations will be needed. 

The facilities at CERN and FNAL are both well suited for these tests, if enough beam time 
can be allocated to the experiments. 

\subsection{Muon Detector R\&D}
Three detector concepts propose instrumented iron absorbers for muon detection. The muon system may also serve to catch
shower leakage from the main calorimeter. Tests of muon system components are therefore naturally
coupled with tests of the calorimeters. At the CALICE test beam in 2006 and 2007,
a significant tail catcher installation was installed behind the hadron calorimeter prototype module
and intensively used and tested with the calorimeters. 

The requirements for the beam are of course primarily muons at different energies, but also 
other hadron species to tests its capabilities as a tail catcher. Experiments at CERN and FNAL 
are well suited for this task. 

\subsection{Conclusion}
Test beams play a very central role in the ongoing detector R\&D for the ILC. The scarcity of beams around the world 
makes is imperative that these resources are efficiently used and optimally coordinated.


\section{Luminosity, Energy, and Polarization}

A crucial asset of an electron-positron collider is that the initial state is well known. 
However, the benefits of this advantage are not fully realized in a linear collider 
unless properties of the initial state -- luminosity, collision energy, and polarization 
(LEP) -- are measured. However, the unique collision dynamics at the ILC make 
these measurements particularly challenging, which requires some well-directed 
R\&D. In particular, beamstrahlung gives rise to a collision energy spectrum which 
depends strongly on the beam parameters, and hence will vary with time. 
Knowledge of the luminosity-weighted energy spectrum, or luminosity spectrum, is 
therefore a fundamental input for physics measurements at the ILC. It is well known 
that polarized beams provide a crucial ingredient for elucidating the fundamental 
electroweak structure of new physics processes. The control of the polarization 
state also provides an important experimental handle for separating competing 
processes from each other, or from backgrounds. The strategy for the polarization 
measurement will depend on the physics program and it will depend somewhat on whether 
only the electron beam is polarized, or if the positron beam is also polarized. 
In any case, it will be necessary to include the capability for
polarization measurements of unprecedented accuracy. 
Precision measurements will also require a state-of-the-art or better 
measurement of the integrated luminosity. Finally, the LEP 
instrumentation can potentially provide important feedback 
to the operating accelerator, in close to real time,
for optimization of the luminosity and reduction of backgrounds. 

Given the direct input of the LEP measurements to physics analyzes, it is
very likely that the development of the LEP instrumentation will eventually be
integrated closely with ILC detector collaborations. If there
are two interaction regions, each collaboration would presumably optimize the
LEP instrumentation to best fit their needs. In the case of a single interaction
region, the LEP development will need to ensure compatibility with both
detectors, as well as with the chosen beams crossing angle.

A critical input to the luminosity spectrum is the measurement of the beam 
energy, averaged over the beam populations, preferably both before and after the 
interaction point. An energy measurement of 200 ppm will suffice for most 
of the physics cases. 
However, a 100 ppm measurement would be required to ensure that this not limit a 
light Higgs mass measurement. If the program includes a very precise $W$ mass 
measurement or a Giga-Z program with positron polarization, then a 50 ppm 
measurement would perhaps be required. This is an accuracy which challenges 
conventional techniques. The leading technique is the magnetic spectrometer, 
either using the accelerator lattice itself (upstream of the interaction point) or an 
extraction line measurement. In the former case, the position measurement might 
be carried out using BPMs, while in the latter case other position-sensitive 
detectors can be used. In either case, R\&D is needed to ensure that viable 
solutions are available. One can hope to access the variable energy-loss spectrum 
by direct measurement of the beamstrahlung. At the SLC these measurements also 
provided important feedback on IP collision parameters. At the ILC, one might hope 
to avoid the high power in the forward hard photon beamstrahlung, opting to access 
the lower-energy parts of the spectrum. Other aspects of the luminosity spectrum 
determination will be carried out within the detectors themselves. These include the 
measurement of the a-collinearity distribution of Bhabha pairs, the measurement of 
radiative return events, the Bhabha scattering rate at large and small angles, and 
the direct measurement in very forward calorimeters (BeamCal) of low-energy pairs produced 
at the IP. The forward calorimeters which provide some of these measurements are 
included in the calorimetry section of this report.

For much of the physics of interest, a beam polarization measurement (of both beams, in
general) of about 0.5\% will be sufficient. 
For the the most demanding precision measurements, one would 
gain\cite{bib_LEP:ppol} by providing moderate positron polarization, along with 0.25\% polarization measurements. For the $A_{LR}$ measurement in
Giga-Z running, one could use the Blondel scheme with 0.25\% polarization measurements
as systematic consistency checks.
The use of Compton scattering of the beam electrons with a polarized 
laser beam was carried out successfully at the SLC and, with considerable effort,
provided a measurement of 0.5\% accuracy. 
However, the ILC presents greater challenges. Because there is significant 
depolarization at the IP, one hopes to make a Compton measurement both before 
and after the IP. R\&D is needed to ensure that 0.25\% to 0.5\% measurements 
(of both beams) can be carried out at the ILC. 

The machine-detector interface (MDI) is a catch-all term which includes not only 
the LEP measurements, but all aspects of interplay between the accelerator and 
the experiment, including the configuration of the beamline magnets and masking in 
the detector halls. An especially important issue is that of backgrounds -- their 
production mechanisms and transport to the detectors. Some of this work has been 
carried out as part of the accelerator design efforts. However, it is crucial that 
studies which simulate the appearance of backgrounds in the detectors be 
supported. As the detector concepts move closer to technical designs, the need for 
detailed background studies will increase. The coupling between accelerator and 
MDI also means that the requirements for MDI R\&D will evolve with the accelerator 
design, especially with respect to IP beam crossing angle configurations, beam 
parameters, or beam time structure.

\subsection{Current status and R\&D challenges}

\subsubsection{Luminosity, Energy, and Luminosity Spectrum}

The integrated luminosity can be determined by precision calorimeters (LumiCal)
placed at small scattering angle.
Following its successful application at LEP/SLC, many layers of silicon-tungsten
sandwich are being considered for these calorimeters, as described in the
calorimeter section of this report.

As discussed above, the average energy of each beam can be best measured with
magnetic spectrometers, upstream or downstream of the IP, or both. LEP/SLC
spectrometers provided resolutions of $\approx 200$ ppm. Not only does this
miss the requirement for ILC by about a factor of two, the conditions at ILC
are more challenging. Hence, R\&D is required to demonstrate the required performance.

The upstream spectrometer is based on beam-position monitors (BPMs) for the
position measurements to deduce the bend angle in the spectrometer, as used at LEP. 
A collaboration
of Notre Dame, UC Berkeley, Royal Holloway, Cambridge, DESY, Dubna, SLAC, and UC London
are developing this technique. Prototypes have been successfully tested in beam lines
at SLAC (ESA) and KEK (ATF). The next major step in the R\&D 
involves installation of an interferometer-based
metrology grid on the BPM structure, followed by additional beam tests. This is crucial,
as individual elements are performing at the level of the 100 ppm requirement, but the
full system is not. Challenges include BPM electronics stability, mechanical stability,
magnetic field tolerance, and insensitivity to beam parameters. The last point is critical,
since the installed system should measure energy independent of the luminosity. Future
progress will depend on the availability of appropriate test beams, such as ESA.

The downstream spectrometer has, in principle, more possible implementation options.
A collaboration of Oregon and SLAC is basing its design on the measurement of the
distance between two synchrotron stripes, one produced before the spectrometer bend, and one after. This technique was used at SLC\cite{bib_LEP:espec}. 
The R\&D is focusing on the development of a
viable detector of the synchrotron stripes. A preliminary test of a quartz fiber
detector read out by multi-anode PMTs was performed in the SLAC ESA beam. While the
downstream spectrometer is more susceptible to backgrounds, it has the advantage that
the dispersion of the stripe separation is sensitive to the IP luminosity spectrum.

The combination of integrated luminosity and beam energy measurements do not provide what is 
directly related to the physics -- the distribution of $e^+e^-$ collision energies at the IP, 
typically known as the luminosity spectrum. The final states of these collisions can, 
of course, offer observables which are directly related to the luminosity spectrum.
As mentioned above, these include Bhabha a-collinearity, radiative returns (to the $Z$), or
even the reconstruction of the dimuon invariant mass. The consideration of these
processes, as well as their interplay with the beam measurements, are important topics for
near term simulation R\&D. An additional simulation topic for the beam energy
measurements is the systematic difference between the average beam energy and the
average energy from the luminosity spectrum.

\subsubsection{Polarimetry}

While it is possible to extract beam polarization information from the 
physics final states, we assume that these will not serve as the primary
polarization measurement, but rather as consistency checks. Polarimetry for the
individual beams can be carried out either before or after the interaction point,
or, if possible, both. In either case, a Compton scattering IP is
provided by directing a laser with known polarization across the charged beamline
some tens (or more) of meters from the $e^+e^-$ IP. Either the scattered
electrons (positrons) or photons can be analyzed, since various Compton
observables are strongly polarization dependent, allowing the charged beam polarization
to be extracted.

A collaboration of SLAC, DESY, Orsay, Tufts, and
Oregon has been developing detailed designs for polarimeter measurements both 
upstream and downstream of the interaction point using Compton scattering. This 
R\&D effort has so far focused on the design of the Compton interaction region and 
the measurement chicanes\cite{bib_LEP:pol1}. It has been assumed so far that the detection 
of the Compton-scattered electrons will be functionally very similar to the systems
used at the SLC\cite{bib_LEP:pol2}.

Beam chromaticity can lead to a beam polarization which varies across the
spatial profile of the beam. Therefore, since the electron (positron) beam will in general
sample the positron (electron) beam at the IP
differently than does the laser at the Compton IP, a systematic shift will be present
between the measured polarization and that which applies to the $e^+e^-$ collisions.
One of the main challenges for
precision polarimetry is to minimize and quantify these differences. Another systematic shift
results from depolarization at the IP as one beam passes through the field of the other.
Again, such shifts need to be measured, for example by comparing measurements before and
after the IP or by periodically taking beams out of collision. Finally, it is important
to develop the instruments for the Compton measurements. 

\subsubsection{Other MDI Instrumentation}

There are several types of beamstrahlung monitor which are under consideration.
The FCal collaboration\cite{bib_LEP:fcal}
is designing a beamstrahlung monitor, called
GamCal, for the extraction lines. One option under consideration for detecting the
high-power flux is to convert a fraction of it using a gas jet target. A group
at Wayne State is investigating the detection of the visible part of the 
beamstrahlung spectrum, which is emitted at larger angles. The FCal
collaboration is developing the technology for the BeamCal instrument, which
would surround the beamlines in the far forward region of the detectors. The front section
of the BeamCal would measure the pair production resulting from beamstrahlung.
The instruments which measure the beamstrahlung directly (e.g. GamCal and visible)
are designed primarily to provide fast feedback to the accelerator
controls for luminosity optimization. 
For this reason, even though these devices
share close physical proximity to some LEP instrumentation, 
they more logically should be folded into accelerator R\&D, as is the case, for example,
for the FONT (fast beam feedback) collaboration. The BeamCal, on the other hand,
also provides important direct information to the physics analyzes (electron veto). 
And it is physically connected to the detector proper. 
Hence, we have included it in the calorimeter section of this report.

\subsection{Milestones}

A critical near term goal, common to all LEP R\&D, is to understand the implications
for LEP of all of the configurations under consideration for the interaction region. 
At a minimum
this requires that simulation software be run on each configuration. For the next
one to two years, the focus should be on the specific technologies and methods
for a given IR configuration. Each technique has particular requirements for
space, background tolerance, and beam parameters. And likewise, each LEP instrument
has interactions with the final focus, the other LEP
instrumentation, or the detectors, which must be understood. Therefore, it is not
possible, in most cases, to develop any element of LEP R\&D in isolation. These issues
must be largely settled before a reasonably definite technical design can be produced.





\cleardoublepage

\chapter{Sub Detector Performance}
\label{subdetector_performance}
\section{Introduction}

The performance goals for ILC detectors require advancing detector  
designs and technologies
beyond the current state of the art. The detector subsystems have been 
designed accordingly, and promise the high performance required for ILC 
physics.  Evaluating and characterizing subsystem 
performance accurately and believably has required
going beyond simple estimations, to careful studies with 
full simulation and reconstruction programs.
In the full simulation programs, effects such as 
particle interactions with detector materials and 
shower development in calorimetric detectors have been 
taken into account, because they may well have a non-negligible impact on detector performance.
In the reconstruction programs, codes have been developed
to perform pattern recognition and track fitting on
simulated data in the trackers, and ``particle flow'' algorithms have been 
developed to assess the performance of the particle flow concept
and its impact on detector design. These are labor intensive approaches,
and uncommon at this stage in the development of detector concepts, but 
they are seen as necessary to establish the credibility of the new detector designs. 
  
For these studies, the concept teams (SiD, LDC, 
GLD, and 4th) have developed GEANT4-based simulation packages, respectively SLIC~\cite{ref-SLIC}, Mokka~\cite{ref-MOKKA}, 
Jupiter~\cite{ref-JUPITER}, and ILCRoot~\cite{ref-ILCRoot}.  At this stage of development, 
simplified sub-detector geometries and averaged densities 
for detector materials are typically used in the 
detector descriptions, but some attempt is made to represent
the dead material associated with support structures, readout
electronics, and other services. 
In the reconstruction programs, simplifying 
assumptions may be used for the less central aspects of the simulation,
e.g. the tracking reconstruction is assumed to be perfect when
evaluating the calorimeter response for Particle Flow Algorithms.  
It is recognized that full reconstructions are needed, and they are close 
to being realized. 
Of course, full Monte Carlo studies
are only as good as the  models of particle 
interactions implemented in the simulation programs.  
Future test beam experiments must confirm that present hadronic shower codes 
adequately describe calorimeter simulations, or new codes must supplant them,
before the detector design are finalized.  

The results presented below
should be considered as a snap shot of the current understanding 
of sub-detector performance. This understanding is evolving rapidly. 
The aim is to illustrate what performance can be achieved for the various subsystems
by drawing on examples from all the concepts, and to demonstrate that present
designs largely meet the ILC performance goals.

\section{Material in the Tracking Volume}

In designing the ILC detectors, particular attention has been given 
to minimizing the material budget in the vertex detector and tracking volumes.  
This is crucial to achieve 
good momentum resolution even for low momentum tracks, to preserve
excellent electron ID, to guarantee efficient tracking in the forward region, 
to improve track and calorimeter cluster matching for particle flow, and 
to minimize the impact of conversions and interactions on the calorimetry.

\begin{figure}[htb]
\begin{center}
\includegraphics[height=7cm]{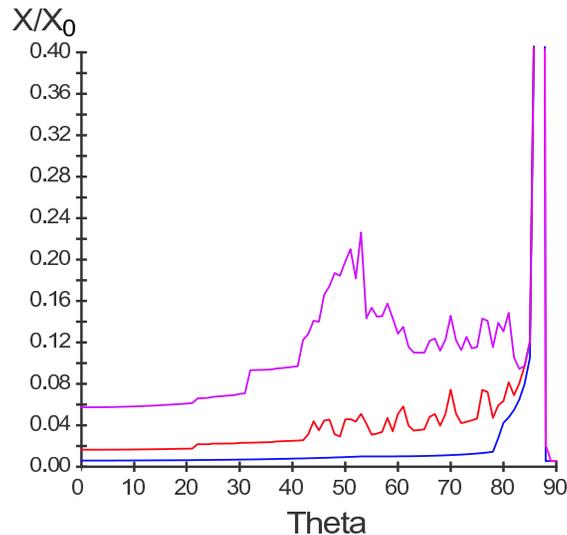}
\end{center}
\caption[Material budget in the tracker] {The material budget of the tracker(purple), vertex detector(red) and 
beam pipe(blue)
in radiation lengths, as a function of the polar angle, 
as modeled in the SiD Monte Carlo}
\label{fig-sec6-MaterialBudget-SID}
\end{figure}

The material budget as modeled in the SiD Monte Carlo is shown in 
Figure \ref{fig-sec6-MaterialBudget-SID}, as an example.
In this figure, the lowest curve shows the contribution from the beam pipe
and the readout for the vertex detector.  The material corresponding 
to the various readout elements has conservatively been assumed to be uniformly 
distributed in the tracker volume. The following two curves indicate 
the additional material due to the active vertex detector 
elements and the supports, respectively.  The outer curve
gives the amount of material of the tracker as a whole,
that is, the sum of the vertex detector and the outer tracker 
including the anticipated dead material in the tracking volume.

The total material in front of  the calorimeters for the other detector concepts 
is comparable in the central region, and ranges up to 20\% $X_0$ in the forward region.  
The GLD, with four layers of 
silicon tracker in addition to the vertex detector, has slightly more material
than the LDC case.  The material in the endcap region of the TPCs is thought to be
dominated by the readout system and 10\% $X_0$ was assumed in the performance study of GLD.

\section{Vertexing performance}

\subsection{Impact Parameter Resolution}
The impact parameter is the distance between the interaction point 
and the trajectory of the charged particle.  A non-zero value of the 
impact parameter indicates that the particle is a decay product of 
a parent particle, which has traveled some distance away from the IP before decaying.   
Measuring impact parameters with high resolution is the key to identifying heavy 
particle decays, and thus heavy flavor, in $e^+e^-$ jets.

\begin{figure}[htb]
\begin{center}
\begin{tabular}{c c}
\includegraphics[width=7.5cm]{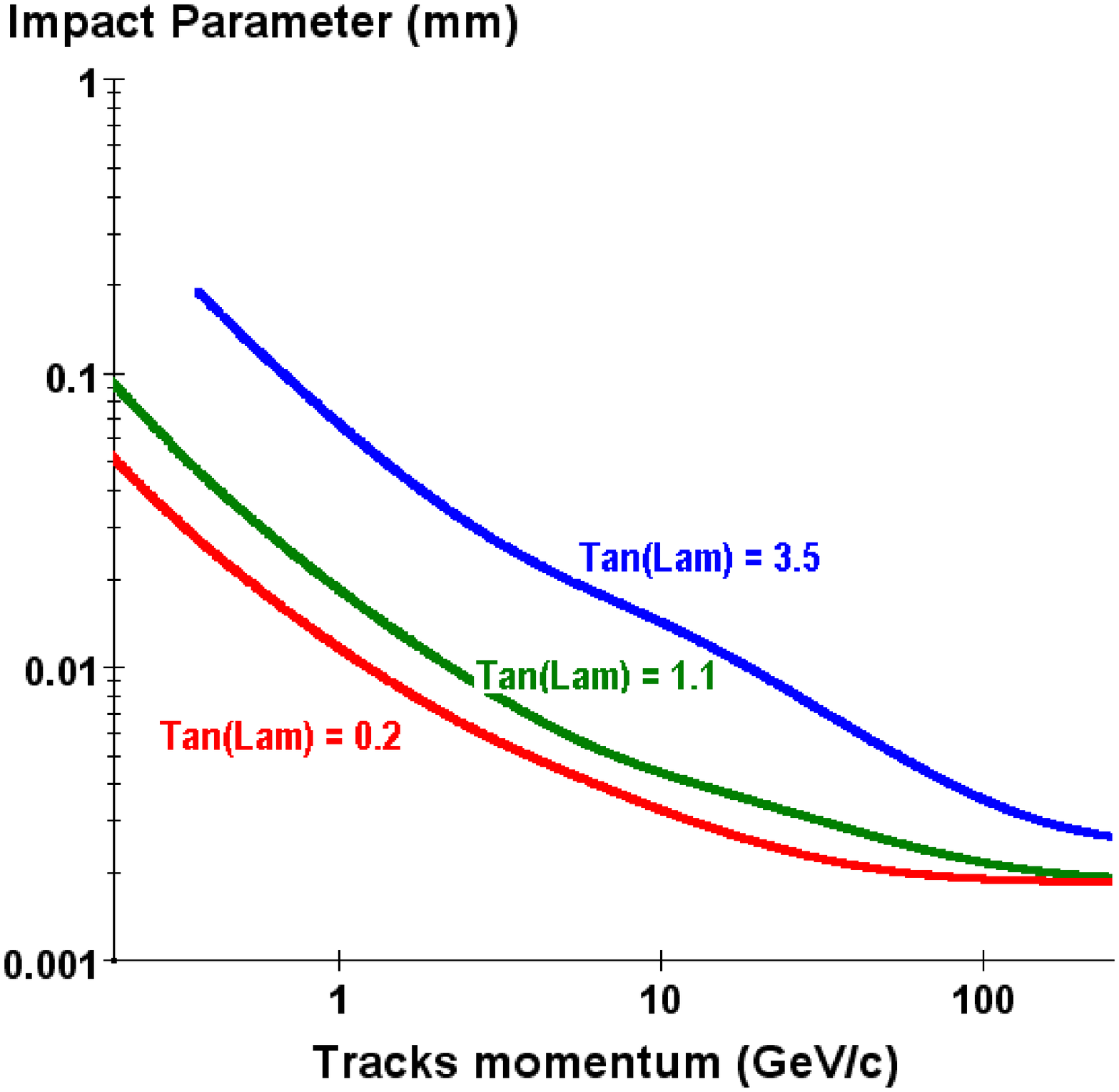}
&
\includegraphics[width=7.5cm]{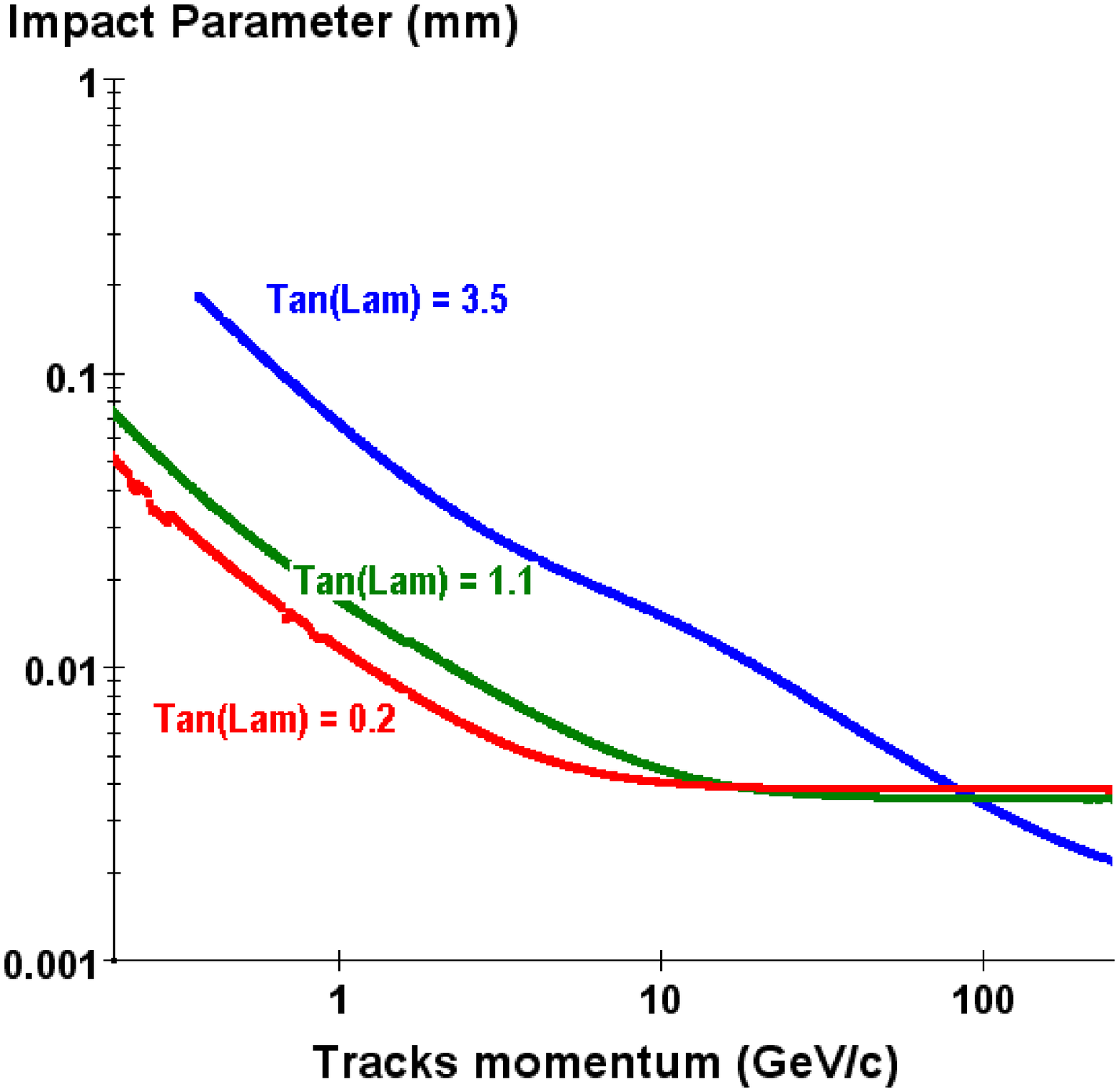}\\
\end{tabular}
\end{center}
\caption[Impact parameter resolution]{Track $r-\phi$ (left) and $r-z$ (right) impact parameter 
resolution as a function of track total momentum, 
for track dip angle $\tan\lambda$ values of  0.2 (red),  
1.1 (green) and 3.5 (blue) respectively. }
\label{fig-sec6-vtxip-vs-momentum}
\end{figure}

\begin{figure}[htb]
\begin{center}
\begin{tabular}{c c}
\includegraphics[width=7.5cm]{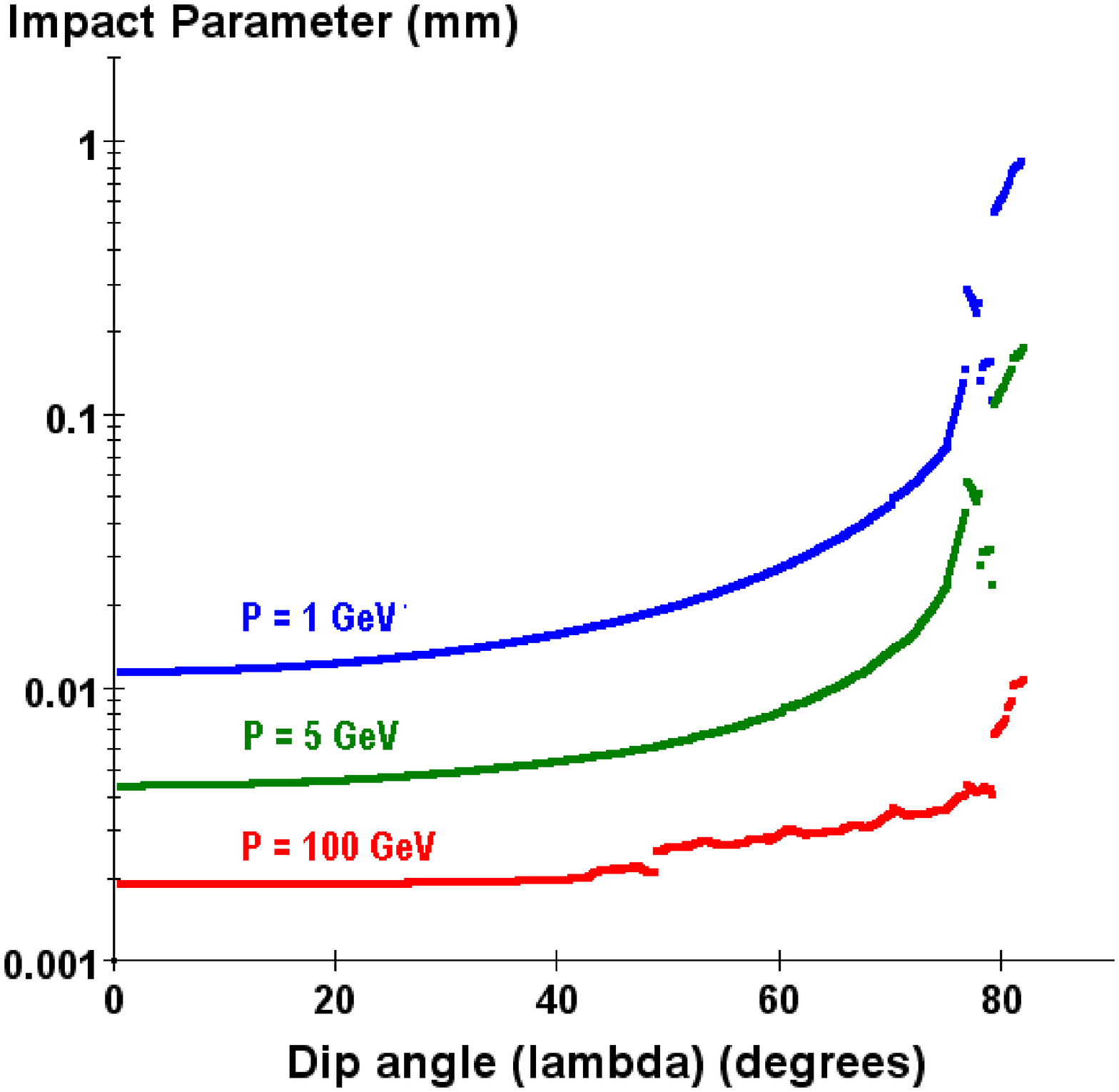}
&
\includegraphics[width=7.5cm]{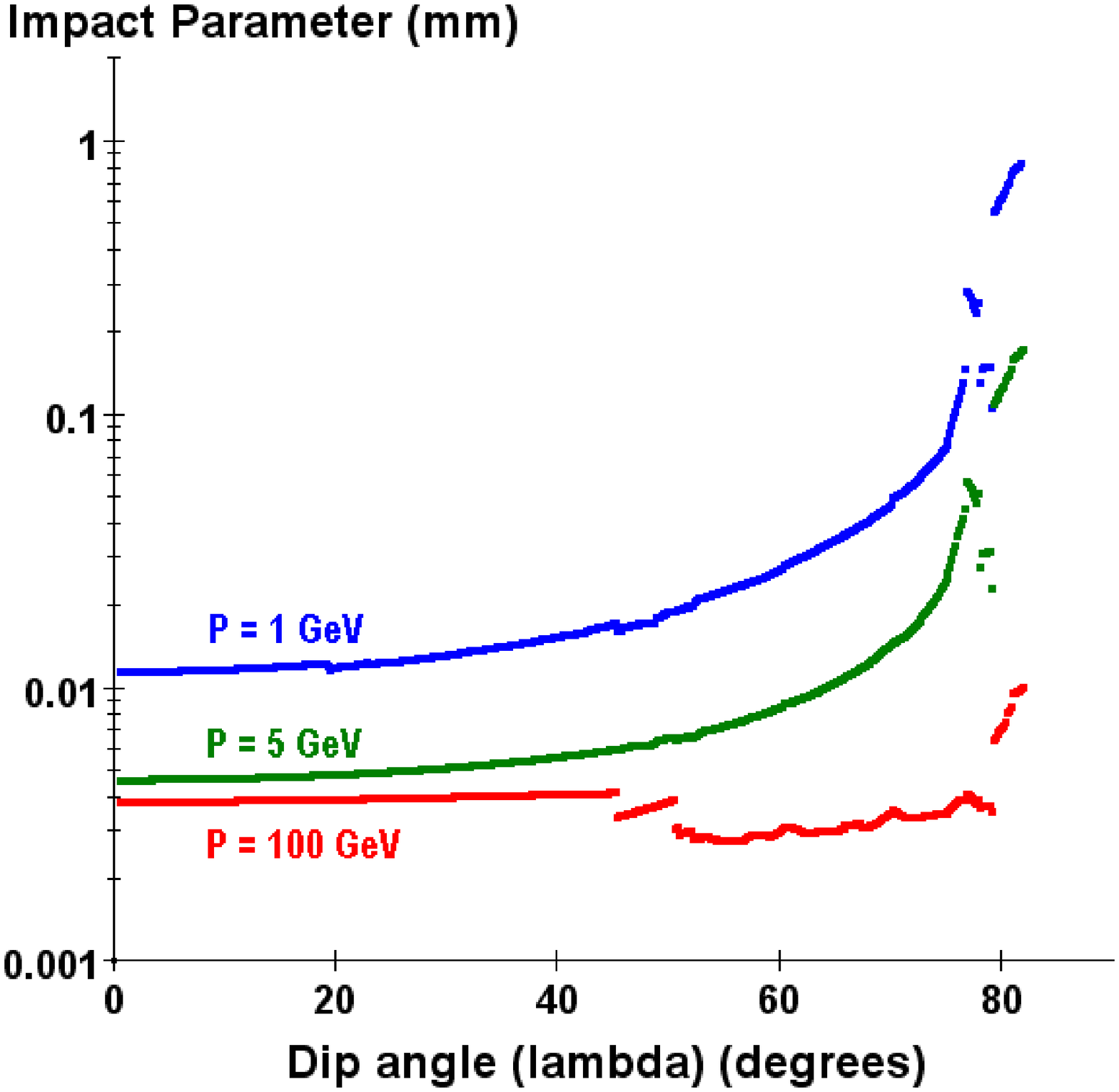}\\
\end{tabular}
\end{center}
\caption[Impact parameter resolution]{Track $r-\phi$ (left) and $r-z$ (right) impact parameter 
resolution as a function of 
track dip angle $\lambda$, for track momentum values of  100 GeV/c (red),  
5 GeV/c (green) and 1 GeV/c (blue) respectively.
}
\label{fig-sec6-vtxip-vs-angle}
\end{figure}

Typical $r-\phi$ and $r-z$ impact parameter resolutions 
as a function of the track momentum for a few characteristic 
polar angles are shown Figure~\ref{fig-sec6-vtxip-vs-momentum},
and those as a function of the track polar angle for a few different 
momentum values are shown in  Figure~\ref{fig-sec6-vtxip-vs-angle},
taking SiD as an example. 
In this study the impact parameter resolution was analyzed 
from the track-parameter error matrix 
taking into account both spatial resolution and the detailed GEANT material 
description. The spatial resolution per hit was assumed to be 
3.5 $\mu$m. In addition, the degradation of the spatial resolution in $Z$ due 
to signal broadening
is represented by a 4 $\mu$m error. An excellent impact parameter resolution of 
$< 10 \mu$m is generally achieved for 
the whole barrel region down to track momentum of about 1 GeV/c, while the asymptotic 
resolution for very high momentum tracks is expected to be about 2 to 3 $\mu$m. 
The other concepts have similar performance.
The resolution in the endcap region degrades somewhat due to the effect 
of extra material for support and detector services. However, for the high 
momentum tracks, good impact parameter resolution is maintained all the way to 
about 80$^\circ$ ( $\cos\theta \sim 0.988$ ).

\subsection{$\bf{b/c}$ Quark Tagging}
As already pointed out in the description of the vertex detector, it is important to
be able to tag decays with bottom and charm quarks in the final state with excellent
efficiency and purity. Most tracks have relatively low momentum, so good impact parameter resolution
down to small ($\sim$ 1 GeV/c) momenta are important. In addition, due to the large
average boost of heavy flavor hadrons in the case of energetic jets, 
decay vertices can be as far as a few cm away from
the primary vertex, and therefore can be outside the innermost vertex detector layer.
Therefore the overall detector must be flexible enough to cope with these
high boost events as well. The topological vertexing as pioneered by SLD has the
potential to allow efficient reconstruction of secondary and tertiary vertices for a very
large range of situations, and can help in tagging quark charge as well as quark flavor.

\begin{figure}[htb]
\begin{center}
\includegraphics[width=12cm]{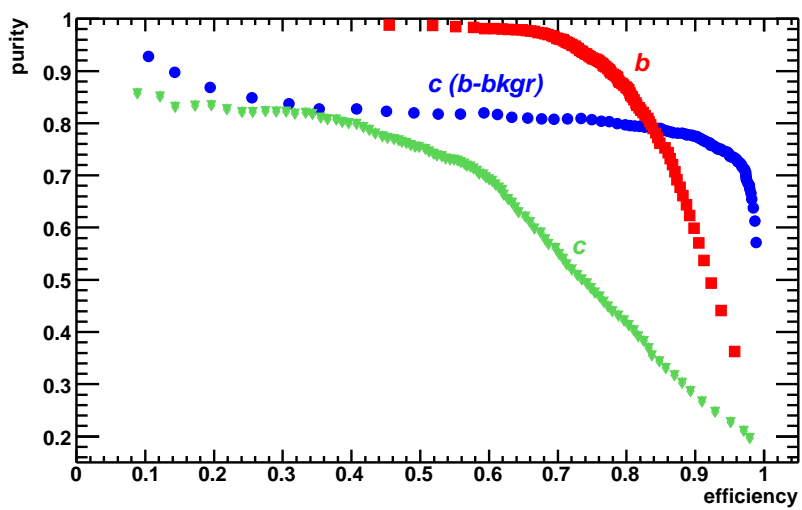}
\end{center}
\caption[Efficiency - purity for bottom and charm tagging]{Efficiency and purity for 
tagging a $b$-quark(red square) and
$c$-quark(green triangle) jets in Z decays, using a full simulation.
The blue-circle points indicate the further improvement in 
performance of the charm tagging in events with only bottom 
background is relevant for example to the measurement of 
Higgs branching ratios. }
\label{fig-sec6-vtx-tageff}
\end{figure}

A topological vertexing program for ILC detector has been developed.
Studies of its performance using a full detector simulator 
started recently using $Z \rightarrow q\bar{q}$ process
at $Z$ pole energy as a bench mark of $b/c$ quark tagging.
A typical initial result for the LDC detector is shown 
in the Figure~\ref{fig-sec6-vtx-tageff}\cite{ref-sonia-LCFIVertex}.
The obtained purity and efficiency using a realistic detector resolution is promising. 
These studies are currently being extended to higher energies, where tagging 
performance is influenced by tracks from hadronic interactions with detector 
materials. Refinements of the algorithms to account for this effect are underway.


\section{Tracker performance}
The tracking devices are designed to provide excellent momentum resolution 
and efficient reconstruction over a large range in polar angle, 
$\theta$.  To achieve this end, LDC, GLD, and 4th 
concepts use a Time Projection Chamber inside a solenoidal 
magnet with a magnetic field of 3 to 4 Tesla as a central tracking device, possibly augmented with 
intermediate and forward trackers. A different approach, using all silicon tracking
and a somewhat higher field, is adopted by SiD.

\begin{figure}[h]
\begin{center}
\includegraphics[width=10cm]{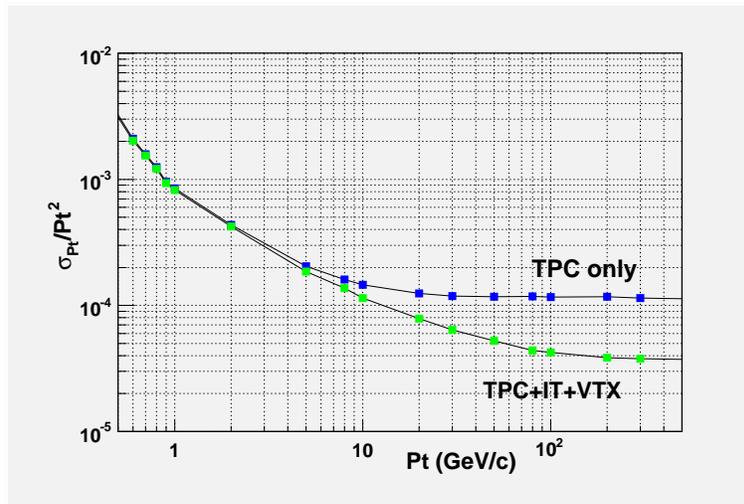}
\end{center}
\caption[Momentum Resolution for the GLD concept]{A typical momentum resolution of tracking device.
Shown is the case for the GLD tracking system.}
\label{fig-sec6-TPCResol}
\end{figure}

A typical momentum resolution in the case of GLD is shown in 
Figure \ref{fig-sec6-TPCResol}.  For this study, 
a muon particle was generated at a polar angle of 90$^\circ$.  The azimuthal spatial 
resolution was taken to be  150 $\mu$m, independent of 
the drift length, and simulated signals were fitted with a Kalman fitter
program.
The momentum resolution of the TPC in conjunction with the intermediate tracker 
and vertex detector, is better than $5 \times 10^{-5} p_t$ (GeV/c) at high momentum, thus
meeting the ILC momentum resolution goal.

\begin{figure}[h]
\begin{center}
\includegraphics[width=9cm]{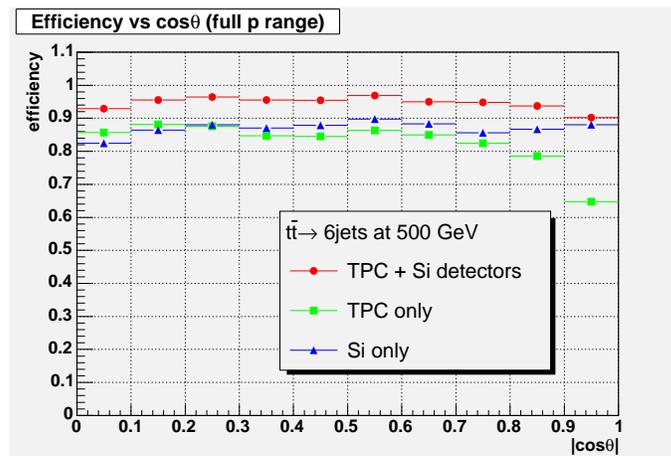}
\end{center}
\caption[Track reconstruction efficiency for LDC]{A typical TPC track reconstruction efficiency for $Z \rightarrow t\bar{t}$ events
taken for LDC.}
\label{fig-sec6-LDC-fig22}
\end{figure}

Pattern recognition and track reconstruction in a TPC is relatively 
straightforward, even in an  environment 
with a large number of background hits, thanks to the dense, three dimensional nature 
of the information recorded by the chamber. 
The efficiency to reconstruct tracks in the LDC TPC, is shown in Figure~\ref{fig-sec6-LDC-fig22}. 
In the central region, which is covered by the TPC, 
the track reconstruction efficiency is better than 99\%. 
The reconstruction efficiency in the forward 
region needs further study, and will be improved by including an algorithm which utilizes
the forward intermediate tracker hits.

In the ILC environment, several effects may influence the quality 
of space point measurements in a TPC. For example, the use of
a Dipole in Detector (DID) corrector magnet
will degrade the magnetic field uniformity and complicate reconstruction.
Positive ions, created in the amplification
process at the endcaps, will distort 
the electric field uniformity as they drift back through the TPC to the cathode.
The presence of through-going muons, generated upstream in the beam collimation section,
spiraling Compton electrons, produced when MeV photons scatter in the gas, and low
energy neutron interactions in the gas, will add to chamber backgrounds, but are not expected
to pose problems for the pattern recognition.  
The other effects are expected to be correctable, but 
studies of reconstruction efficiencies taking them and the additional backgrounds into account 
are yet to be done.

\begin{figure}[ht]
\begin{center}
\includegraphics[width=15cm]{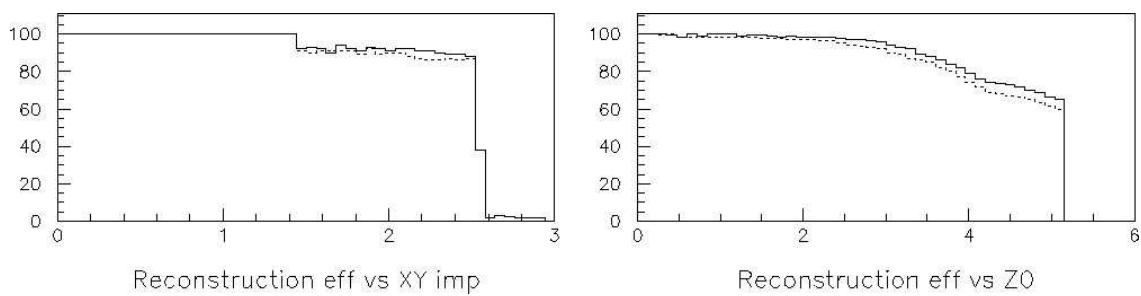}
\end{center}
\caption[Track finding efficiency in SiD]{Reconstruction efficiency of the vertex detector seeded 
track finding in SiD as a function of track impact parameters. Reconstruction cuts 
are set at 3.0~cm for the XY impact parameters and 5.0~cm for Z.  Solid 
lines correspond to high Pt($>1$ GeV), dashed to low Pt ($<0.5$ GeV) tracks. }
\label{fig-sec6-sideff}
\end{figure}

In the SiD detector, the central tracker consists 
of five layers of silicon microstrip detectors, but the vertex detector
and electromagnetic calorimeter play important roles in tracking as well. Making an efficient 
use of three dimensional information from the pixel vertex detector, 
the standard track finding algorithm for the SiD
detector is an ``inside-out" track finding algorithm. That is, pattern recognition 
begins in the vertex detector, and progresses by extrapolating tracks into the main tracker. 
Studies of the track finding efficiency have used a full Monte Carlo simulation of the vertex detector
raw data, and realistic cluster finding and coordinate determination codes. Tracker hit positions have
been smeared with the expected tracker resolution.   
The pattern recognition in the vertex detector begins by selecting hits in three of the five different
layers. A reconstructed track is required to have 
5 associated hits at least, including those in the central tracker. To reduce combinatorics
and reconstruction time, 
tracks are required to originate close to the interaction point 
and have transverse momentum exceeding 200 MeV/c.
The reconstruction efficiency for this algorithm for single tracks is shown in 
Figure~\ref{fig-sec6-sideff}. The present algorithm is fully efficient for 
tracks with small impact parameters. The tracking efficiency in the core of a jet has been studied 
using $q\bar{q}$ Monte Carlo events at $\sqrt{s}=500$ GeV, and is found
to be above 95\%.  In order to focus on 
the reconstruction efficiency in the fiducial volume of the central tracking system,
events were required to have $|\cos\theta_{thrust}|\leq 0.5$ with a thrust magnitude of 0.94 
or greater.  For these events, tracks with $|\cos\theta|\leq 0.5$ were found to be 
reconstructed with 94.3\% efficiency. Nearly all the inefficiency is due to tracks 
that originated outside the vertex detector, and consequently couldn't
be found with the vertex-seed algorithm. If tracks are required to originate within 
1 cm of the origin, the track finding  efficiency was approximately 99\% for $q\bar{q}$ events.

In order to reconstruct tracks originating outside the vertex detector, SiD uses 
Silicon Tracker Standalone Tracking and Calorimeter-Assisted Tracking.
For the Silicon Tracker Standalone Tracking, a simple pattern recognition 
algorithm that uses circle fits to all valid three-hit combination has been 
studied in the barrel tracker.  For single high-$p_t$ muons, 
the tracking efficiency of 99\% was achieved.  For 
$t\bar{t}$ events, a track finding efficiency of 94\% is achieved 
so far.  Further study and refinement of this technique is expected to 
yield improved efficiencies for tracks that originate beyond 
the vertex detector.

Calorimeter Assisted Tracking relies on the very fine segmentation of the 
EM calorimeter; the passage of minimum ionizing particle (MIP) through the EM 
calorimeter look track-like, thanks to the high granularity of the calorimeter.
The MIP stub found in the EM calorimeter is extrapolated back 
to the main tracker to find the associated hit, and identified 
as a track when certain criteria are satisfied.  In a proof of principle demonstration using
the simulated Z pole events, this algorithm 
reconstructed 61\% of all charge pions with $p_t > 1 $ GeV/c, 
produced by $K_S^0$ decays. Significant improvements are expected
with further refinements of the code.

\section{Calorimeter performance}
The performance of the electro-magnetic 
and hadron calorimeters have been studied with GEANT4-based simulations.

SiD, LDC and GLD all utilize sampling calorimeters,
whose energy resolution is essentially determined by 
the sampling fraction, and all aim at achieving a  jet energy resolution of $30\%/\sqrt{E(\mathrm{GeV})}$. 
The expected energy resolutions
of the electromagnetic calorimeters are similar concept to concept, as
are the energy resolutions of the proposed hadron calorimeters.
Other details differ, however, including the proposed transverse
segmentation, hadronic calorimeter depth, absorber materials, and choice of sensors. 
Since these three concepts adopt the particle flow
approach to calorimetry, single particle energy resolution is hardly the whole story; 
the ability to discriminate the energy deposited in the calorimeter 
by charged tracks from that deposited by photons or primary
neutral hadrons, becomes at least equally important. Jet energy resolution, or
even di-jet mass resolution, become the relevant figures of merit.

\begin{figure}[htb]
\begin{tabular}{c c}
\includegraphics[width=7.5cm]{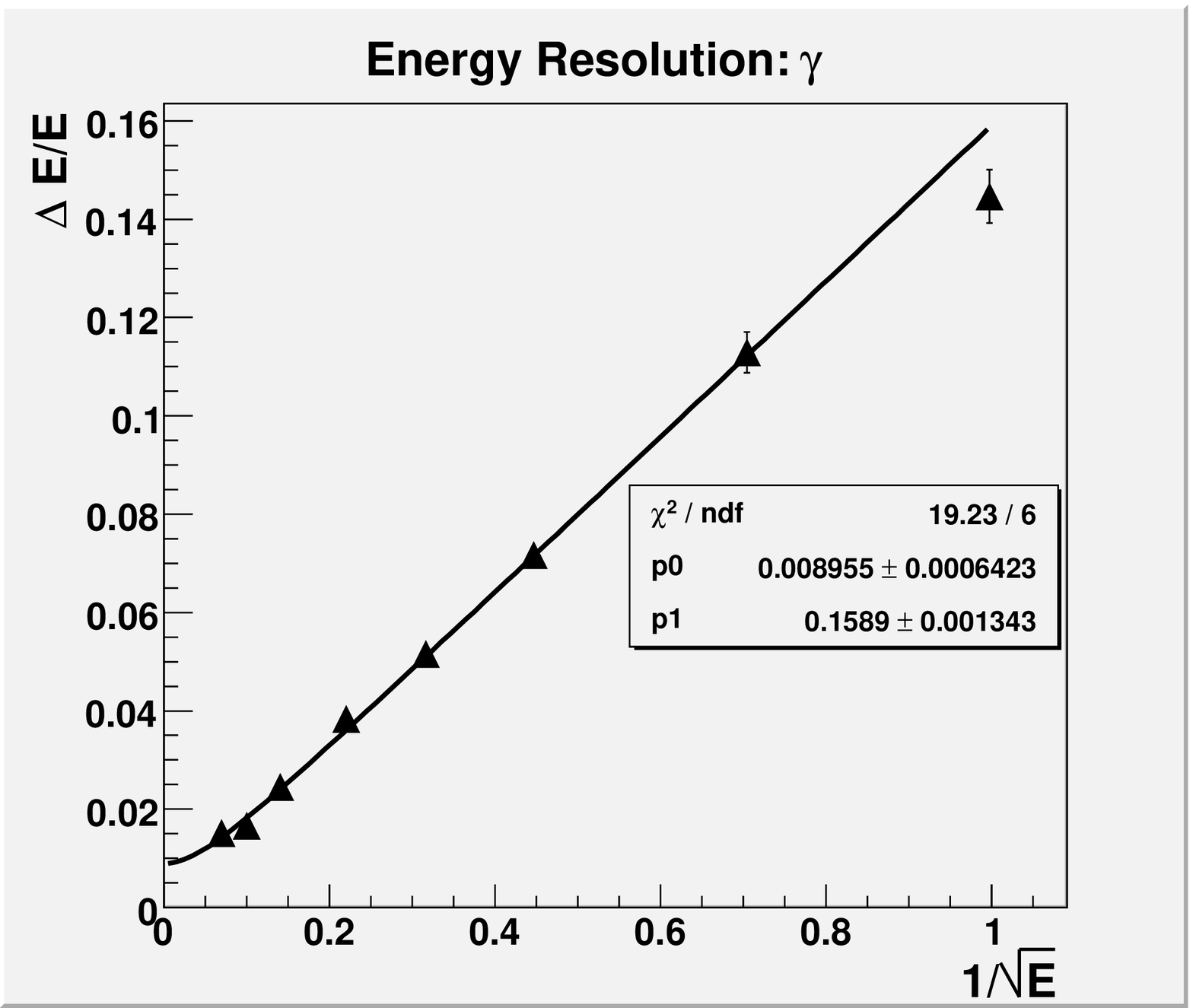}
&
\includegraphics[width=7.5cm]{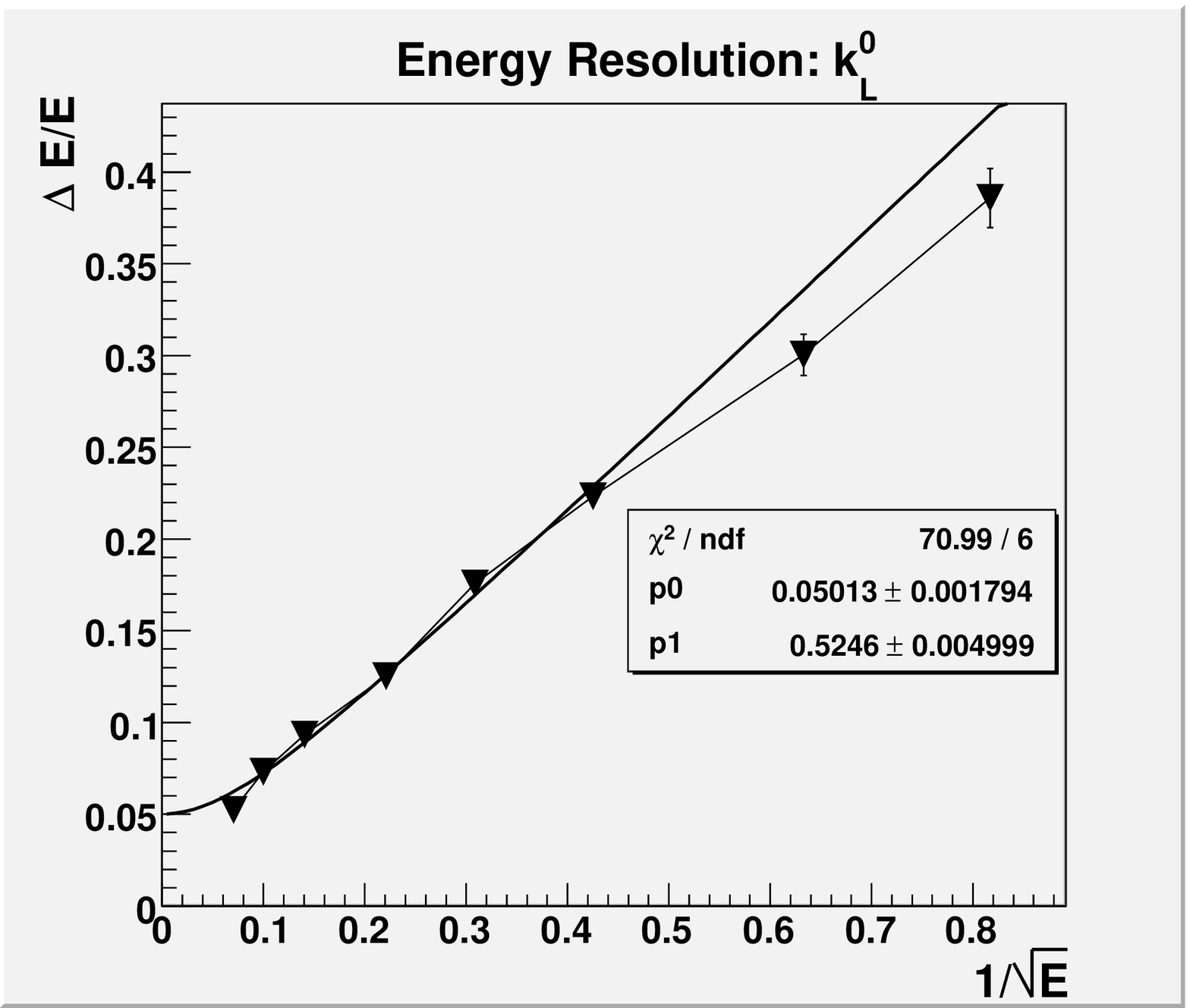}
\end{tabular}
\caption[Photon energy resolution in GLD]{The left figure is the energy resolution of photons in the angular
region of $|\cos\theta| < 0.8$ in the GLD as a function of the energy.
The resolution was derived 
from Gaussian fits to the peak of the response
distribution.  The right figure is the resolution of the hadron 
calorimeter}
\label{fig-sec6-CAL_Resol}
\end{figure}

The energy resolution of the electromagnetic calorimeters proposed for the various concepts
ranges from 14 to 17$\%/\sqrt{E}$ for the stochastic term 
and is about 1\% for the constant term. A typical 
energy resolution as a function of the photon energy 
is shown in Figure~\ref{fig-sec6-CAL_Resol} in the case of GLD.
The energy resolution of the hadron calorimeter of SiD,
LDC, and GLD is in the range of 
50 to $60\%/\sqrt{E}$ for the stochastic term 
and between 3 to 10\% for the constant term, depending 
on the absorber, readout detector, and particle type.
It should be noted that these resolutions have been estimated 
solely with the GEANT4 simulation, since they characterize new
designs and untested detectors. Clearly these results need
confirmation in test beam experiments in the coming years.

The fine segmentation of the electromagnetic calorimeters 
makes it possible to separate electromagnetic energy deposited by photons from
the energy deposited by incident tracks. The high granularity
also allows an accurate determination of the direction of photons.  
The measurement of the direction of photons is important, for example, 
in GMSB SUSY scenarios involving long-lived decays of heavy particles, where 
photons from the decay can point to a decay far from the IP.
In the case of LDC with $5 \times 5$ mm$^2$ readout 
cells, the angular resolution of the ECAL is estimated to be 
55 mrad$/\sqrt{E(\mathrm{GeV})}$. The position resolution of 
the EM cluster is estimated to be 0.9 mm$/\sqrt{E(\mathrm{GeV})}$. 
These features will also make it possible to fix the relative alignment of the tracker and the ECAL 
with high energy electrons.

Distinct from the other concepts, 4th uses a dual-readout, compensating calorimeter system. 
It reads out quartz and scintillating fibers, which are embedded in an absorber, with photon detectors.
The quartz fibers
are sensitive to \v{C}erenkov light coming primarily from electromagnetic energy deposits, and the 
scintillating fibers respond to the total ionization energy. Measuring the electromagnetic and 
ionization energy deposits separately allows software compensation, and delivers high resolution. 
The fibers are interleaved in an absorber made of Copper, in a fully projective 
geometry consisting of towers with cross-sectional area $2 \times 2$ cm$^2$.

At present, the 4th concept has implemented the Hadron Calorimeter, without a special
electromagnetic section, in their simulation program. The conversion of the energy into the number 
of Scintillation and \v{C}erenkov photons is handled by specific routines 
taking into account factors such as angles between the particle and 
the fiber as well as a Poisson statistics of produced photons\cite{fourth_jet_algorithm}.
Effects such as the response function of electronics, non-constant quantum efficiency, etc., 
have not yet been implemented. 

\begin{figure}[htb]
\begin{center}
\includegraphics[width=7.5cm]{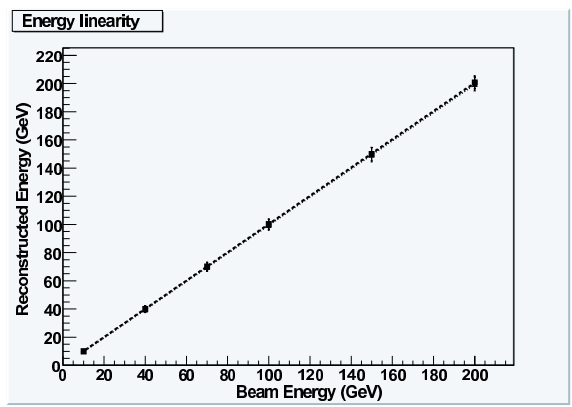}
\end{center}
\caption[Reconstructed vs beam energy in the 4th concept calorimeter.]{Reconstructed vs beam energy in the Hadronic Calorimeter 
for single pions for the 4th concept.}
\label{fig-sec6-fourth-hcal-linearity}
\end{figure}

To measure the energy of an incident particle in the calorimeter, the strengths of the signals 
from the the \v{C}erenkov fibers and Scintillation fibers \cite{fourth-etac-etas} are 
appropriately weighted.
The weighting factors, $\eta_C$ and $\eta_S$, are known to be 
independent of the incident particle energies and are obtained by 
simulating the response to 40 GeV electrons. The linearity of the calorimeter
response to pions is shown in Figure~\ref{fig-sec6-fourth-hcal-linearity}, and
indicates that compensation occurs at all energies with a unique set of calibration constants.
The energy resolution for hadronic showers ($\sigma_E / E$) obtained was 
$36 \sim 38\%/\sqrt{E}$, depending of the pattern recognition of calorimeter. 

\section{Jet Energy Resolution}
\label{sec-subdetector-PFA}

The majority of the interesting physics processes at the ILC involve multi-jet final
states. The reconstruction
of the invariant mass of two or more jets will provide a powerful tool both for event
reconstruction and identification. 
As described in Chapter~\ref{Challenges}, one of the goals of the ILC detector performance is to 
be able to separate $W$ and $Z$ in their hadronic decay modes.
In order to achieve this goal, the jet energy resolution of detectors ($\sigma_E/E$) 
is required to be as good as 
$30\%/\sqrt{E(\mathrm{GeV})}$ for a lower energy jet or less than 3\% for a higher energy jet.
This is a factor two better than the best jet energy resolution achieved at LEP.
To this end, GLD, LDC and SiD are equipped with a finely segmented 
calorimeter optimized for particle flow analysis (PFA).
The 4th concept is equipped with a high resolution dual-readout calorimeter 
and measures the jet energy precisely without PFA.

\subsection{Particle Flow Based Jet Energy Measurement}
A promising strategy for achieving the ILC goal of the jet energy resolution is the particle flow
concept which, in contrast to a purely calorimetric measurement, requires the
reconstruction of the four-vectors of all visible particles in an event. 
Present particle flow algorithms work best when the energies of the individual particles 
in a jet are below about 100 GeV. In this regime, the momentum of the charged particles is reconstructed in
the tracking system with an accuracy which exceeds the energy and angle
measurements in the calorimeters. Hence, in order to attain the best
reconstruction of events, the charged particle measurement must be solely based on
the tracking information, while the reconstruction of photons and neutral hadrons is
performed with calorimeter system. The crucial step of the particle flow algorithm is
the correct assignment of calorimeter hits to the charged particles and the efficient
discrimination  of close-by showers produced by charged and neutral particles.

\subsubsection{Algorithm}
The development of particle flow algorithms for the ILC
detector concepts is still at a relatively early stage. 
However, given that three of the concepts are designed for particle flow
calorimetry this is an active area of research. It should not be forgotten that the jet energy
resolution obtained is a combination of detector and reconstruction software.
The output of any particle flow algorithm is a list of reconstructed particles, 
termed particle flow objects (PFO). Ideally  these would correspond to the particles produced in the interaction.
Several programs have been developed, as described in the Detector Outline Documents
\cite{ref-GLD,ref-LDC,ref-SiD}. While the algorithms are distinct there are 
a number of features which are common. Only the general features of these algorithms are described here.
First, charged particle tracks are reconstructed in the 
tracking detectors.  Identification of neutral vertices, such as 
$K_s \rightarrow \pi^+\pi^-$ decays, and kinks from electron bremsstrahlung in the tracker 
material improves the performance slightly, by replacing a calorimetric measurement with
information from the tracker.

The next step is pattern recognition in the electromagnetic and hadron
calorimeters. The goal of the calorimeter clustering is to identify every cluster
resulting from single particles and to separate nearby showers.
Calorimeter reconstruction may be performed independently of the track 
reconstruction, or tracks may be used to guide the calorimeter clustering. 
The algorithms differ significantly in details of how calorimeter clusters are formed but all utilize the high granularity and tracking ability of the proposed calorimeters.  
Charged particle PFOs are formed from the tracks and those clusters associated with them.
The four-momenta of charged PFOs are determined solely with the
reconstructed track parameters and the results of any particle identification procedure.
Calorimeter clusters which are not  associated with tracks are considered as neutral PFOs and
may be identified as either photons or neutral hadrons. 
The reconstruction of the  four-momenta  of neutral objects is based 
on calorimetric energy and position measurements and particle identification
from the shower profiles.

\subsubsection{PFA Performance}
The results presented here
represent the current status of  the particle flow
algorithms. As the algorithms are further developed significant improvements are anticipated. For these initial studies the performance has been evaluated 
by summing the entire energy for hadronic events at the Z pole. These simulated events provide a clean environment for evaluating PFA performance since uncertainties associated with jet finding and the association of particles with the decaying bosons are avoided. PFA performance can be straightforwardly quantified
in terms of the resolution of the total reconstructed energy and visible mass. 
Studies in a multi-jet environment are at a relatively early stage.
\begin{figure}[h]
\begin{center}
\includegraphics[width=13cm]{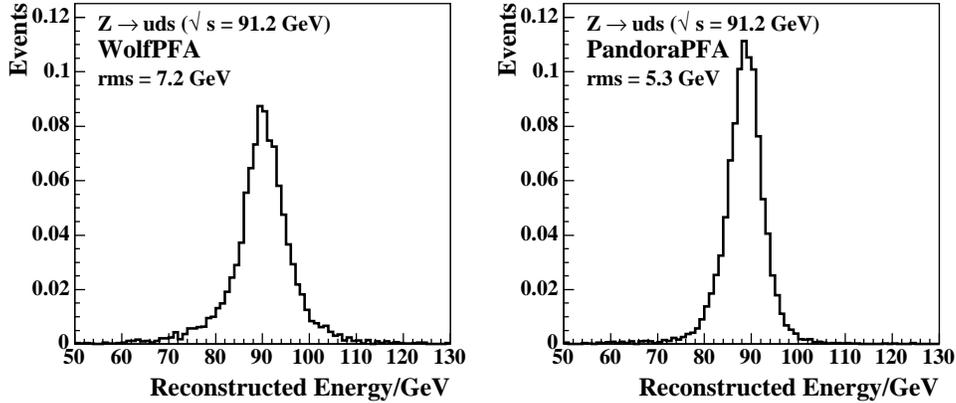}
\end{center}
\caption[Reconstructed energy in light flavour events]{Distributions of reconstructed energy for $Z \rightarrow q\bar{q}$ (uds only) events 
at $\sqrt{s} = 91.2$ GeV
obtained using WolfPFA and PandoraPFA for a GEANT4 simulation of the LDC detector.}
\label{fig-sec6-PFA_Resol}
\end{figure}

Figure~\ref{fig-sec6-PFA_Resol} shows a typical reconstructed energy distribution of $Z$ decays to $u$, $d$ and $s$ jets
(avoiding the need to account for unobserved neutrinos) which were generated without initial state radiation. 
These results come from the LDC,
using two different algorithms, WolfPFA~\cite{ref-WolfPFA} and PandoraPFA~\cite{ref-PandoraPFA}.
The distribution of measured energy is characterized by a narrow core
and a wider tail, which results from the failure to detect some low momentum particles
and those forward particles which miss the detector fiducial volume, and the imperfect 
subtraction of charged track energy from the calorimeter signal.
In order to quote a figure of merit for particle flow performance, $\sigma_{90}$is defined to be the root 
mean square of that part of the distribution  
that contains 90\% of the jets, because the usual rms is highly sensitive to tails of the distribution.  
Using $\sigma_{90}$ has the advantage that the effects of tails are suppressed and the quoted resolution reflects that for the majority of the events.
The significance of 10\% of tail events will depend on the
signal-to-noise ratio of the process if interest, and it should be further studied
using physics processes.
\begin{figure}[h]
\begin{center}
\includegraphics[width=7cm]{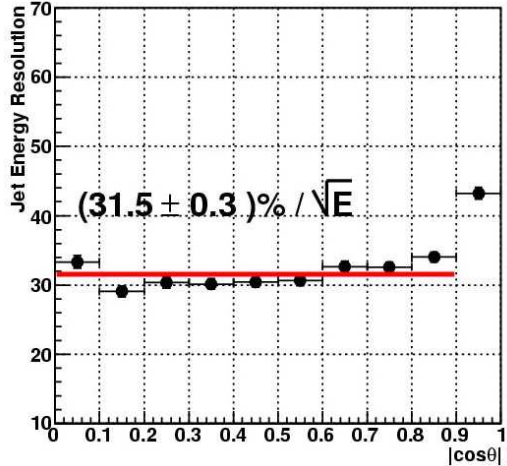}
\end{center}
\caption[The jet energy resolution, $\alpha$, as a function of the $|\cos\theta_q|$]{The jet energy resolution, $\alpha$,  as a function of the $|\cos\theta_q|$ in the case of 
$e^+e^- \rightarrow q\bar{q}$ (light quarks only)
events at $\sqrt{s} = 91.2$ GeV of the GLD detector.}
\label{fig-sec6-PFA_Resol_costheta}
\end{figure}                     

Figure~\ref{fig-sec6-PFA_Resol_costheta} shows the jet energy resolution, $\alpha\equiv \sigma_{90}/\sqrt{E}$,  as a function of 
the production angle of the jet ($|\cos\theta_q|$) for the GLD concept. 
In the barrel region of the detector (i.e. $|\cos\theta|<0.9$), the averaged
jet energy resolution is  $31.5\%/\sqrt{E}$. LDC and SiD 
obtained similar values in the range between about $(30 - 35\%)/\sqrt(E)$.

\begin{figure}[h]
\begin{center}
\includegraphics[width=10cm]{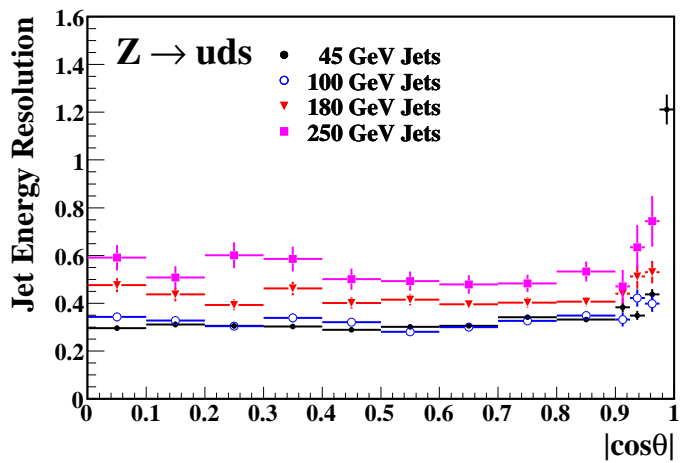}
\end{center}
\caption[$\cos\theta_q$ dependance of the jet energy resolution at different energies]
{The jet energy resolution, $\alpha$, as a function of the $|\cos\theta_q|$ 
for jets of energies from 45 GeV to 250 GeV.}
\label{fig-sec6-DEjet-HE-Costh}
\end{figure}

\begin{figure}[h]
\begin{center}
\includegraphics[width=6cm]{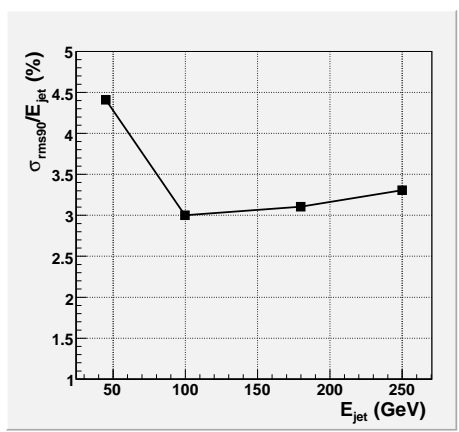}
\end{center}
\caption[The energy dependance of the relative jet energy resolution of PandoraPFA.]
{The relative jet energy resolution, $\sigma_{90}/E_{jet}$,
of PandoraPFA averaged in the region, $|\cos\theta_{jet}|<0.7$, as a function of the jet energy.}
\label{fig-sec6-DEjet-over-E}
\end{figure}

For higher energy jets the opening angles between particles decreases due to the larger Lorentz boost. 
This makes the separation of clusters in the calorimeter more challenging.
Recently, PandoraPFA has introduced an iterative re-clustering method to improve
cluster separations and cluster-track association\cite{ref-thomson-LCWS2007}. Accordingly, 
the jet energy resolution for higher energy jets improves significantly as seen in 
Figures~~\ref{fig-sec6-DEjet-HE-Costh} and \ref{fig-sec6-DEjet-over-E}.
In this study, 
$e^+e^- \rightarrow q\bar{q}$ (light quarks only)
events were generated to study jet energy resolution using 
the Tesla detector configuration\cite{ref-TESLA:2001} which is similar to LDC.

As seen in these figures, for jets of energy up to 100~GeV, 
PandoraPFA has achieved the required ILC jet energy 
resolution of $30\%/\sqrt{E}$. Further improvement of performance is anticipated.
Studies using perfect PFA, which uses Monte Carlo truth information for clustering indicate that
improvements in resolution of up to 30\,\% may be achievable.

\begin{figure}
\begin{center}
\begin{tabular}{c}
\includegraphics[width=10cm]{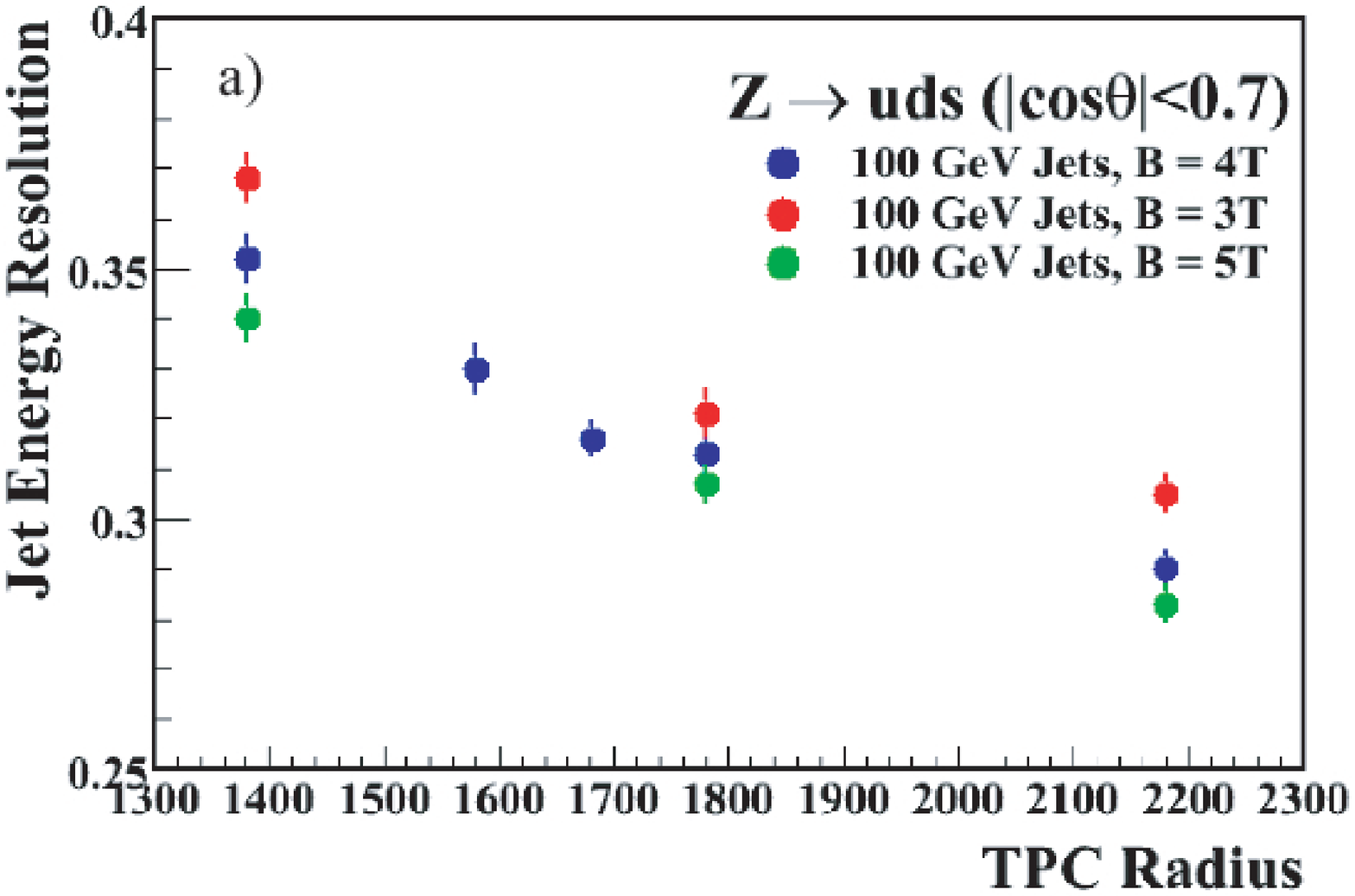} \\
\includegraphics[width=10cm]{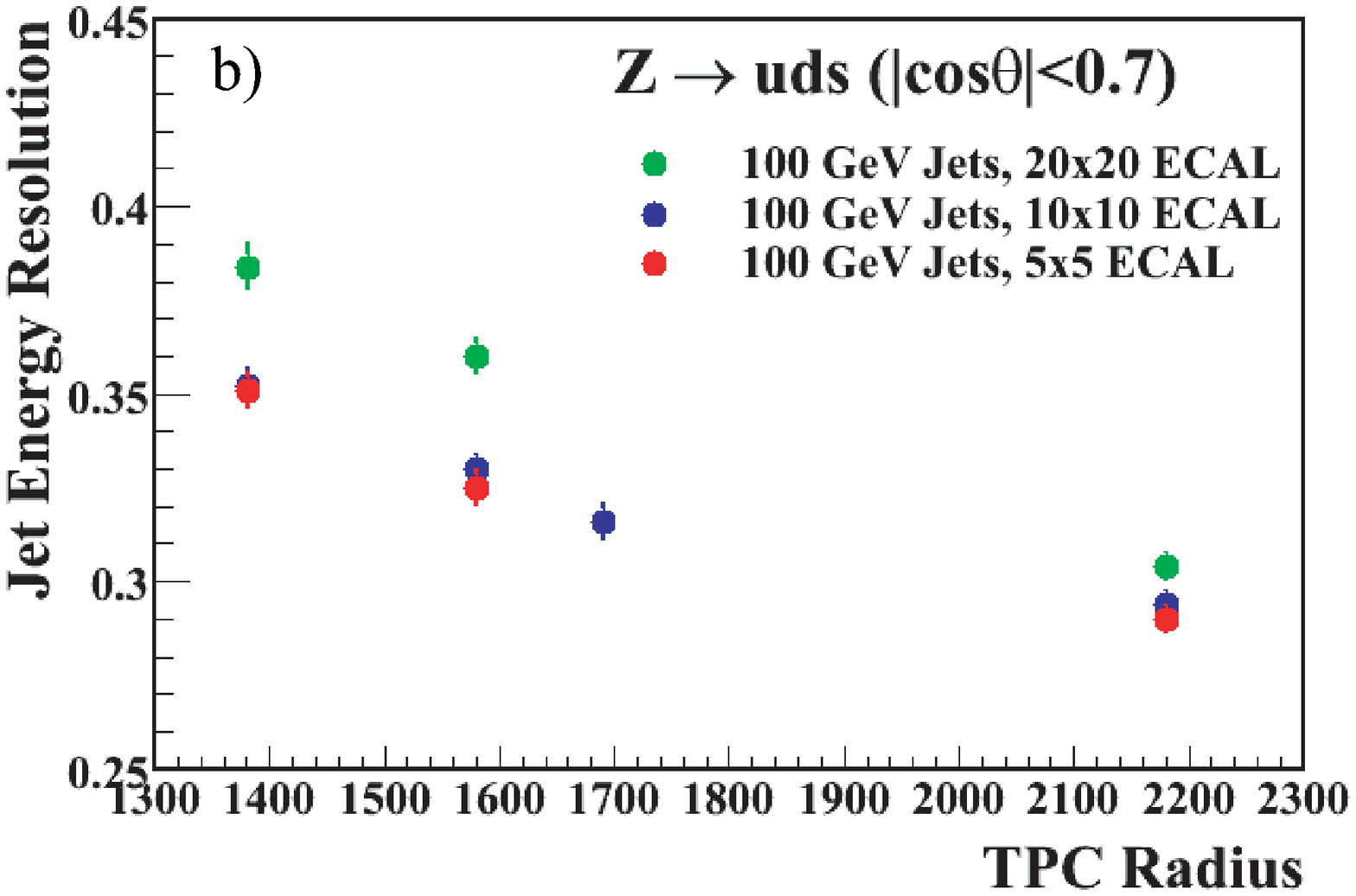}
\end{tabular}
\end{center}
\caption[PFA performance for several detector configurations]{{a) Jet energy resolution obtained with PandoraPFA 
and the Tesla TDR detector model plotted as a function of TPC outer radius 
(which is almost the same as the ECAL inner radius) and magnetic field.
 b) The jet energy resolution obtained with PandoraPFA and the 
Tesla TDR detector model plotted as a function of TPC outer radius and ECAL 
transverse segmentation (mm$^2$) for a magnetic field of 4T. 
For both plots jet energy resolution is defined as the $\alpha$ 
assuming the expression $\sigma_E/E=\alpha/\sqrt{E(\mathrm{GeV})}$.} 
\label{fig-sec6-figure3}}
\end{figure}
A number of detector optimization studies have recently been performed
with the PandoraPFA particle flow algorithm\cite{ref-thomson-LCWS2007}. 
For example, Figure~\ref{fig-sec6-figure3}a shows how the jet energy
resolution depends on the TPC radius and magnetic field. As expected, the resolution
improves with increasing radius and increasing magnetic field (both of which
increase the mean transverse separation of particles at the front face of the
ECAL). Larger calorimeter radii and 
stronger magnetic fields result in increased separation between the 
particles in a jet, thus they are preferred for better PFA performance.
In order to achieve the PFA performance goal with a reasonable 
detector cost, SiD adopts the highest magnetic field and 
smallest radius (5~T and 1.3~m), while GLD has 
the weakest field and largest radius (3~T and 2.0~m). The LDC
lies in between these extremes (4~T and 1.5~m).
The performance difference among three parameter choice 
is small with the current version of the PandoraPFA algorithm,
but the results suggest that the larger radius is more important than the stronger B-field.
Figure~\ref{fig-sec6-figure3}b shows how the jet energy
resolution depends on the transverse segmentation of the electro-magnetic
calorimeter (ECAL) for a number of different TPC outer radii. Again this study is based on the simulation of the Tesla TDR detector. As expected, higher granularity gives better resolution and it is 
apparent that a transverse segmentation of $20 \times 20$\,mm$^2$ is insufficient
in the case of smaller TPC radii.
The improvement in going from $10 \times 10 $~mm$^2$ segmentation to $5 \times 5$~mm$^2$ is not particularly large because for 100~GeV jets the confusion of clusters in the ECAL does not contribute significantly to the overall jet energy resolution in
either case.   

\subsubsection{Particle ID in a Jet Environment}
Track and cluster association done in particle flow analysis 
naturally identifies the charge of calorimeter clusters, and can provide particle ID even within jets.
Clusters in the electromagnetic calorimeter which are unassociated with tracks can be associated
with photons, or occasionally, neutral hadrons.
EM clusters whose position coincides with a charged track, and whose energy matches the track's momentum, 
are identified as electrons. A track-like cluster of small energy depositions, consistent with those 
expected from a minimum ionizing particle, 
is a muon candidate.  Thanks to the high granularity 
of the ILC calorimeter, charged particles leave identifiable tracks in the calorimeter.  

According to a study by the GLD group, 
about 94\% of the photon energy in the jet of $Z^0$ to the light quark pair decay is successfully 
identified as neutral electromagnetic energy. 87 \% of the identified photons are genuine. 
Photon identification proceeds by selecting energy clusters which are unassociated with 
tracks, matching the expected longitudinal shower shape, accounting for the energy 
deposited per calorimeter cell, and taking into account other variables. 

\subsection{Jet Energy Reconstruction in Non-PFA Calorimeters}

The calorimeter of the 4th concept aims to achieve good jet energy resolution via
compensating calorimetry, without the particle flow ansatz. It uses two jet finder 
algorithms, the UA1 cone type algorithm\cite{UA1Cone} 
and a modified Durham jet finder algorithm.
First, the tracks and V0's with $p_t > 10 $ GeV are input into a jet cone finding 
algorithm to find the number of jets and their angles.  Calorimeter clusters are then added to the identified jets until 
no further clusters are found or the maximum aperture of the cone reaches 
$60^\circ$.  An additional algorithm attaches  isolated clusters, low $p_t$ tracks, and muons 
to the jets.  For details, see ref.\cite{fourth_jet_algorithm}.

\begin{figure}[htb]
\begin{center}
\includegraphics[width=10cm]{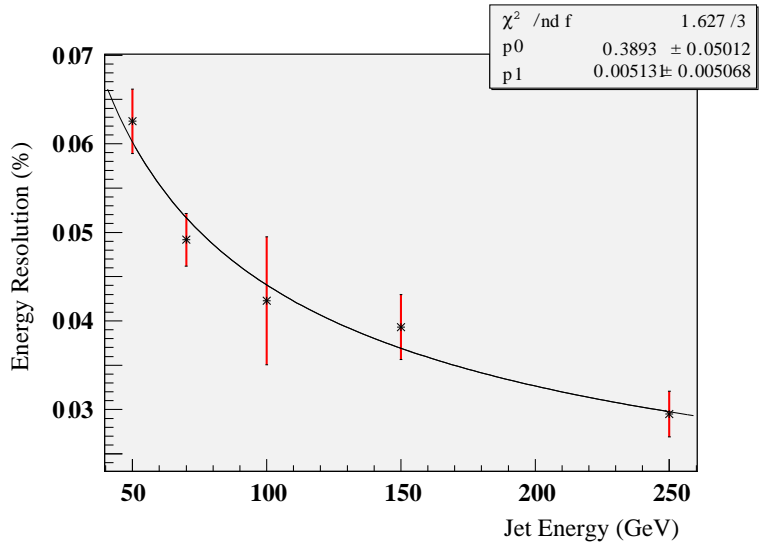}
\end{center}
\caption[Jet energy resolution for the 4th concept]{A preliminary performance of single jet energy resolution in 
$e^+e^- \rightarrow q\bar{q}$ for the 4th concept.}
\label{fig-sec6-fourth-ejet-vs-e}
\end{figure}

The performance of the jet energy reconstruction was studied 
using light quark pair production events by  $e^+e^-$ 
annihilation.  The energy resolution ($\sigma_E/E$) of about 3\%
is achieved for a jet of 250 GeV energy.  It is shown as a function of 
the jet energy in Figure~\ref{fig-sec6-fourth-ejet-vs-e}.

\section{Muon ID performance}
SiD, LDC and GLD have thick iron return yokes.
Tracking devices interleaved in the iron return yoke
serve to identify muons, augmenting muon ID
in the finely segmented hadron calorimeter.
The ILC detectors typically have strong solenoidal fields of 3 to 5 Tesla 
and appreciable material in the calorimeters (4-6 nuclear 
interaction lengths), so only energetic
muons even reach the barrel muon detector.

\begin{figure}[h]
\begin{center}
\includegraphics[width=9cm]{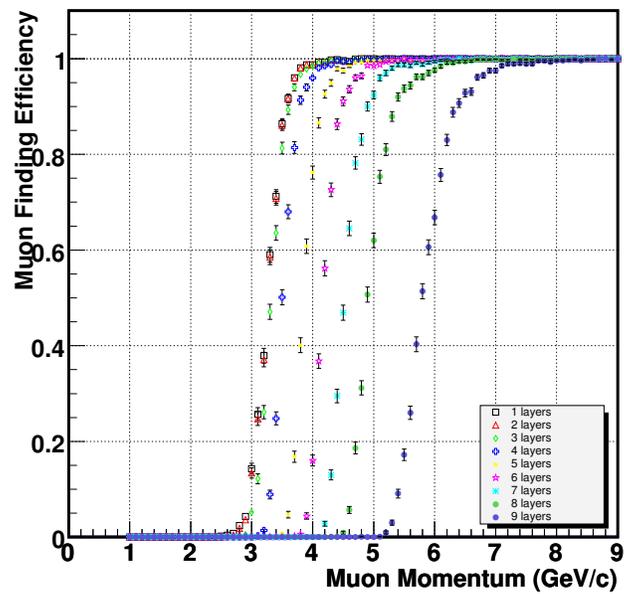}
\end{center}
\caption[Muon detection efficiency]{Muon detection efficiencies as a function of the muon energy in the baseline GLD
configuration. Muons were generated in 90$^\circ$ from the origin. The efficiency
threshold is found to be 3.5 GeV requiring a hit in the first layer, or 6 GeV requiring hits 
in all layers.}
\label{fig-sec6-muid}
\end{figure}

The GLD group studied the momentum acceptance of its muon detector in the baseline 
GLD configuration, using GEANT4-based full simulation.  
The muon was generated at 90$^\circ$.
As seen in Figure~\ref{fig-sec6-muid}, the
muon momentum has to exceed 3.5 GeV/c to reach 
the first layer of the muon detector, and 6 GeV/c 
to pass through the outer most muon detector.
The muon misidentification probability was estimated 
for the LDC design to be below 1\%\cite{ref-LDC}.

The momentum of the muon is measured well by 
the main tracker. Matching tracks found in the main tracker 
with those in the muon detector and the intervening
calorimeters is yet to be studied.

The $p_t$ resolution of isolated muons reconstructed in the 
muon spectrometer was also studied by the 4th concept. 
The muon spectrometer of the 4th concept 
utilizes proportional aluminum tubes of diameter 4.6~cm in the region between the solenoids.
The barrel part consist of 3 staves, each containing 20 layers of plane tubes of 4 meter long
and placed between the outer and the inner solenoid. The point resolution of 
$\sigma_{r\phi} = 200 \mu$m and $\sigma_z = 3$~mm was assumed in the analysis.
The tracks which had
been reconstructed by the combination of the TPC 
and the Vertex Detector were projected to the inner layer of the muon spectrometer, 
at which point the track parameters were estimated.   
Tracks which have originated in the Hadron Calorimeter 
and those which have released  an appreciable
amount of energy after exiting the TPC, are expected to fail
a track matching criterion, but this has not yet been implemented. Note that the muon spectrometer 
itself has a momentum resolution of  $\sigma(1/p_t) = 1.6\times 10^{-3}$ 
at high momentum, while for lower
momentum tracks it is dominated by the multiple scattering
in the aluminum tubes. Track matching thus involves comparing the TPC tracks with those reconstructed 
in the muon spectrometer, in position, direction, and momentum. The reconstruction efficiency is 
$94\%$ for muons with momentum above 7 GeV and not entering the cracks of the detector.
%
%



\clearpage

\cleardoublepage

\chapter{Integrated Physics Performance}
\label{physics_performance}
In this section the performance of the detector in a few selected
physics reactions is summarised. The purpose of this section is to
illustrate the level of maturity of both the understanding of the
detectors and of the reconstruction and analysis algorithms. 
The scope of these analyses is rather limited, and does not cover the 
full physics potential of the ILC. In particular analyses looking for 
physics beyond the Standard Model have not yet been studied in enough 
detail with realistic simulations to be included in this section. 
For
this reason, only channels where a complete simulation has been
done, based on detailed Monte Carlo, and analysed with realistic
algorithms, are shown. It should be pointed out that it is not the
intention of this chapter to illustrate the full physics program at
the ILC - for this the reader is referred to the volume describing
the physics program.

\section{Tools used in the Analyses}
Over the last years significant progress has been made in the development of complete simulation
and reconstruction software system for the ILC. A number of different approaches have been proposed,
and are available through a number of software repositories~\cite{ref-MARLIN, ref-JAS, ref-JUPITER, ref-ILCRoot}.

The detectors propose a tracking system composed of a number of
different sub-systems. Algorithms have been developed which do high
efficiency tracking in the individual sub-systems, and combine then
the results from all tracking detectors. Using realistic algorithms,
and including a simulation of the expected background rates, track
reconstruction efficiencies close to $99\%$ have been demonstrated,
with momentum resolutions around $\sigma (p_t) / {{p_t}^2} < 1
\times 10^{-4}{\rm GeV} ^{-1}$.

At least for energies below 1~TeV the best event reconstruction
resolution is believed to result from a particle flow algorithm, as
has been discussed in~\ref{sec-subdetector-PFA}. A number of
software packages are available which implement this approach, and
reach jet-energy resolutions which at least at moderate jet energies
up to around 100~GeV are close to the goal of $30\%/\sqrt{E}$
\cite{ref-PandoraPFA, ref-WolfPFA}.

While the tracking reconstruction codes have reached a fair level of maturity, 
development of the particle flow algorithms is still advancing rapidly. 
Therefore results presented in the following should be
interpreted as a snapshot of an ongoing development, where significant further improvements
can be expected over the next few years.

\section{Higgs Analyses}
The study of the properties of the Higgs boson - if it exists - will be a major undertaking at the ILC.
It also provides an excellent demonstration to illustrate the interplay between the detectors proposed for the
ILC and the physics to be done at the accelerator.

\subsection{Higgs Recoil Analyses}

One of the most challenging reactions for the tracking system of the
detector is the measurement of the Higgs mass using the technique of
recoil mass. The recoil mass technique allows a precise measurement of the 
Higgs boson mass and an essentially model-independent determination of 
the $ZHH$ coupling. 

In this method, the Higgs is analysed through the
reconstruction of a Z-boson produced in the decay of a virtual Z
into a ZH. Assuming that the center of mass energy of the collider
is known with sufficient accuracy the mass of the Higgs can then be
deduced from the measurement of the Z decay: $m_H^2 = s + m_Z^2 - 2
E_Z \sqrt {s}$, where $s$ is the center of mass energy, $m_Z$ the
mass of the $Z^0$ and $E_Z$ the reconstructed energy of the $Z^0$. 
Only the leptonic decay modes of the $Z$ are used.

\begin{figure}
    \centering
\includegraphics[bbllx=8mm,bblly=18mm,bburx=204mm,bbury=154mm,height=8cm]{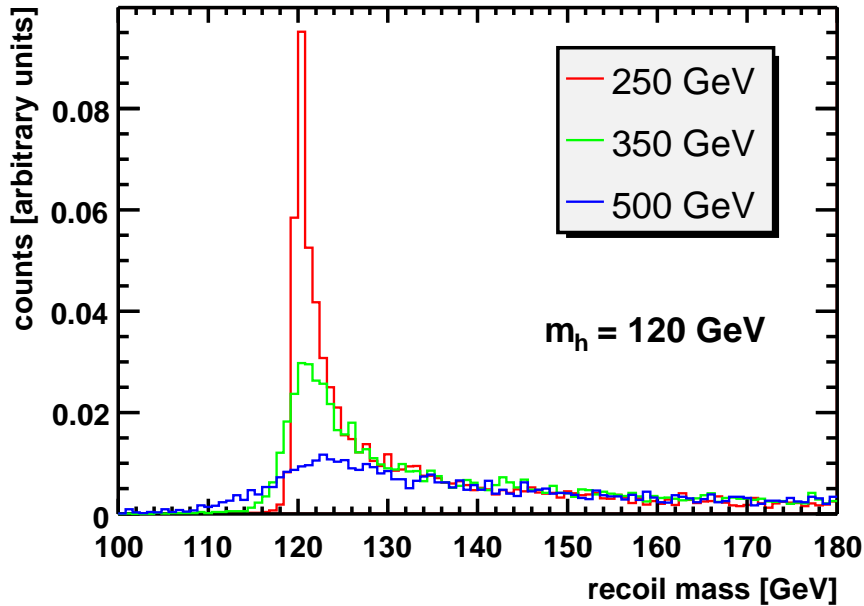}
    \caption[Recoil mass spectrum at 250, 350 and 500~GeV]{Recoil mass spectrum for a 120~GeV Higgs at 250, 350 and 500~GeV, without backgrounds, for $H$ decays into electrons and muons.}
    \label{fig:Higgs_recoil_ECMS}
\end{figure}

For a given mass of the Higgs boson the reconstruction of the invariant Higgs mass through the recoil
technique depends heavily on the center of mass energy at which the
experiment is performed. In figure~\ref{fig:Higgs_recoil_ECMS} the
recoil mass spectrum (with no background) is shown for running the
accelerator at 250~GeV, 350~GeV and at 500~GeV, for a Higgs of mass 120~GeV.
The improvement in the width of the signal is obvious.

As a test case the reconstruction of a hypothetical Higgs boson of mass $120$~GeV is studied at a 
center-of-mass energy of $250$~GeV. 
A central part of this analysis is the identification of the lepton
pair, into which the $Z$ decays. The analysis presented in
\cite{ref:ZH_lohmann} and done in the context of the LDC detector is based on a 
full GEANT simulation of the detector, and a complete track and shower reconstruction program. 

A likelihood method is used to separate
electrons, muons and pions from their signals left in the
calorimeter. The purity and contamination after the ID procedure is
shown in table~\ref{tab:PID}.

\begin{table}[htb]
\centering
\begin{tabular}{cccc}
\hline
 & electron & muon & pion \\
\hline
electron & 99.5\% & 0.0\% & 0.5\%\\
muon     & 0.3\% & 93.6\% & 6.1\%\\
\hline
\end{tabular}
\caption[Purity and contamination of particle ID]{\label{tab:PID}Table of purity and contamination of an electron and a muon sample after
running the particle identification likelihood method described in the text.}
\end{table}

The most important backgrounds to this analysis are Standard Model processes. The following
reactions have been studied and simulated: $e^+e^- \to ZZ \to llX$, $e^+e^- \to \mu^+\mu^-$, $e^+e^- \to W^+W^-$, $e^+e^- \to e^+e^-(\gamma)$. Not included yet is the background $e^+e^- \to \tau^+\tau^-$.
Events of these processes are generated with the MC generators Sherpa, BHWIDE and Pythia, 
and processed through the simulation and reconstruction step as the signal samples. 

\begin{figure}
    \centering
        \includegraphics[bbllx=14mm,bblly=12mm,bburx=187mm,bbury=139mm,height=8cm]{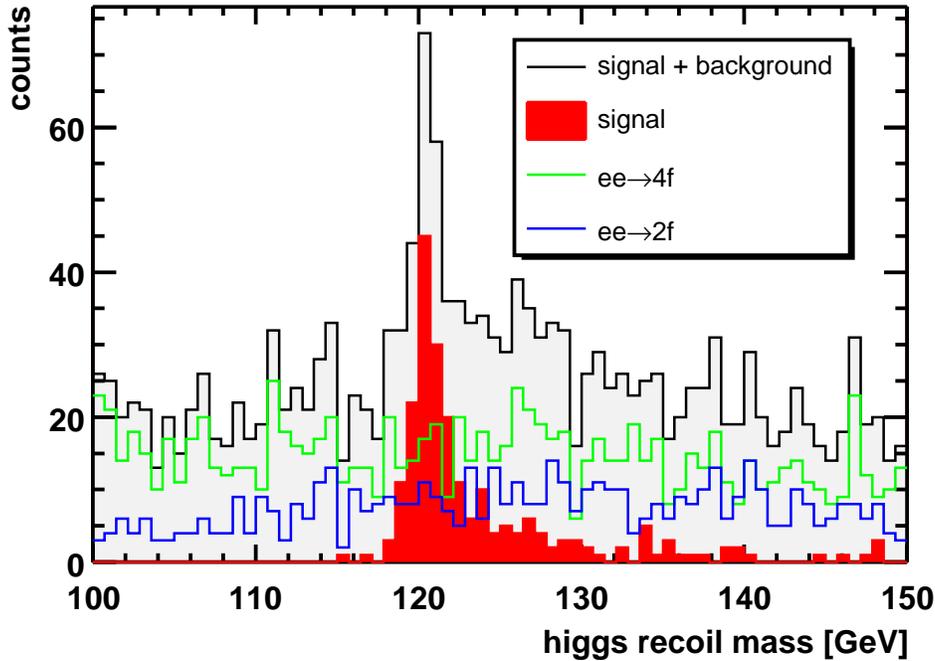}
    \caption[Recoil mass spectrum for a 120~GeV Higgs]{Recoil mass spectrum reconstructed for a 120~GeV Higgs, with full background simulation, at a centre-of-mass energy of 250~GeV. $Z$ decays into electroncs and muons are considered. The background 
    from four-fermion final state contains the pair production of heavy gauge bosons.}
    \label{fig:Higgs_recoil}
\end{figure}

Backgrounds are reduced by applying the particle ID code, and by simple cuts on the mass of
the invariant lepton system, and the angle relative to the beam line. After cuts, 
around $50\%$ of the $H \mu \mu$ final state, $40\%$ of the $H ee$ final state, are 
reconstructed. 
Based on a data sample equivalent to a luminosity of $50$~fb$^{-1}$, a clear signal from the
Higgs could be reconstructed, over a small background, as shown in figure~\ref{fig:Higgs_recoil}.
From this analysis the mass of the Higgs has been reconstructed 
using a simple fit to the mass distribution 
with an error of $\approx 70$~MeV, 
and the cross section with a relative error of $8\%$. Further improvements of this analysis 
are expected by applying a more sophisitcated likelihood method for the determination of the 
mass of the Higgs.

A similar analysis has been performed in the context of the SiD detector concept, at a center of mass energy
of the collider at 350~GeV. This analysis is based on a cut based event selection and background rejection. The
general flow of the analysis is very similar to the one described above. Only the dominant background source
from $e^+e^-\to ZZ$ decays has been simulated so far.

While in the previously mentioned LDC analysis the machine
backgrounds have been taken into account through a parametrised
approach, in the SiD analysis fully simulated machine background
events have been included. One event from each of the machine
backgrounds (GuineaPig pairs,$\gamma\gamma \rightarrow$~hadrons, and
$\gamma\gamma \rightarrow \mu^+\mu^-$) has been added to each of the
physics events. The events have been combined at the Monte Carlo hit
level, prior to digitization. The readout technologies envisioned
for the silicon tracking detectors are expected to provide
single-bunch timing capabilities. Extensions to this study will
investigate the impact of integrating over larger numbers of beam
crossings.

\begin{figure}[htb]
    \centering
    \includegraphics[height=8cm]{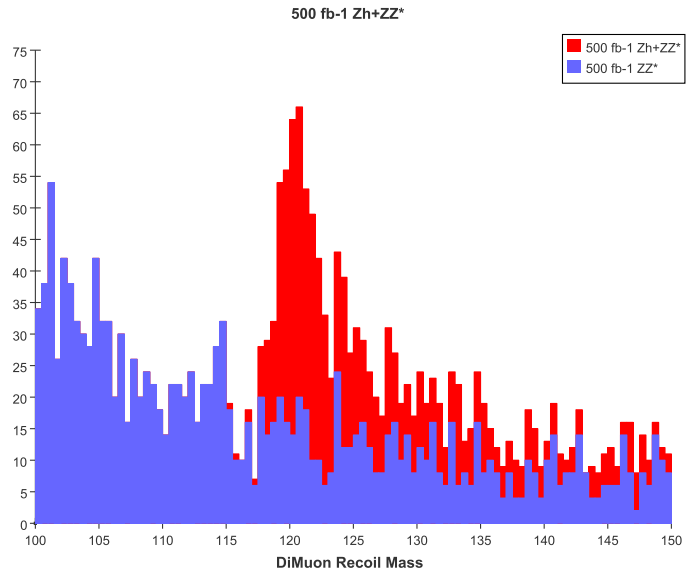}
    \caption[Dimuon recoil mass in ZH events.]{Dimuon recoil mass for $ZZ^*$ background (blue) and $ZH$ signal
    plus background (red) for centrally produced muons. The event sample
    corresponds to an  integrated luminosity of $500fb^{-1}$ at 350~GeV cms.}
    \label{fig:dimuonRecoilMass}
\end{figure}

The SiD analysis proceeds by looping over all the reconstructed
particles (charged and neutral) in the event and requiring two muons
with momentum greater than 20~GeV. Having found two high-momentum
muons, the invariant mass of the system is calculated and required
to be consistent with that of the $Z$ boson.
Figure~\ref{fig:dimuonRecoilMass} shows the recoil mass distribution
for the $ZZ^*$ background in blue and $ZH$ signal plus background in
red.

The precision of the Higgs mass from this measurement, based on a comparison between
the mass distribution reconstructed and template Monte Carlo distributions, is
estimated to be about 135~MeV. Taking into account the larger center-of-mass energy of 350~GeV, this is compatible with the results from the previous analysis. 

%
\subsection{The process, $\bf{e^+e^- \rightarrow \nu {\bar \nu} b {\bar b}}$}
The Higgs decay into bottom quarks is of particular interest since
it is the dominant decay mode of the Higgs boson if its mass is less than about 140~GeV.
A study has been performed using the Higgs-strahlung process, where $Z$ decays invisibly and
the Higgs decays hadronically.
The measured rate of the process provides information on the Yukawa coupling to the bottom quark.
The invariant mass of the measured particles is
the mass of the Higgs, since all visible particles stem from the Higgs decay, and
there is no ambiguity of the mass measurement due to an exchange of colored
particles in the final state as is the case in the four-jet mode of the Higgs-strahlung process.
Thus this process is considered as a benchmark for the capability of the detector
and the reconstruction performance. An excellent vertex detector is also a key element
for an efficient separation of the bottom quark jets from backgrounds.

In the GLD analysis presented here~\cite{ref-yoshioka-lcws2007} the
events were generated with Pythia 6.3. In the event generation,
beamstrahlung effect was taken into account together with
bremsstrahlung.  The nominal ILC parameter set, but at a beam
energy of 175~GeV, was used for the generation of the beamstrahlung
spectrum. The events were passed through a full simulation program,
Jupiter, using the GLD detector model, and reconstruted with the GLD
version of the particle flow algorithm. The study was performed for
a Higgs mass of 120~GeV. This study is based on a Monte Carlo
event sample of 200~fb$^{-1}$ .  The $e^{+}e^{-} \rightarrow ZZ$
process is the dominant source of physics background and was
included in the study.

The interesting events are characterized by missing energy and missing $p_t$ due to neutrino productions.
Bottom quark jets tagged through their secondary vertex are another signature.
In order to select these events, the following selection cuts were applied; the total visible energy was between 90~GeV and 200~GeV;
the total missing $p_t$ was greater than 20~GeV; the cosine of the jet axis was between -0.8 and 0.8;
and the event contained more than 4 tracks whose closest distance to the interaction point (IP)
was more than three sigma away from the interaction point.  In addition, the missing mass of the event,
calculated assuming the initial center of mass energy being equal to twice of nominal beam energy, was
required to be within 60~GeV of the $Z$ mass.

\begin{figure}[tbh]
    \centering
\includegraphics[height=8cm]{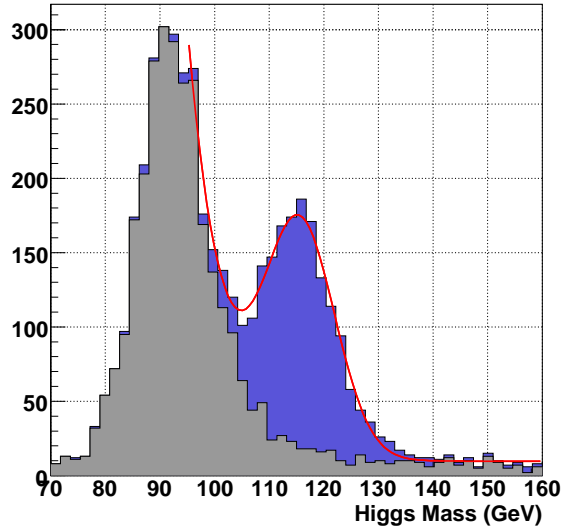}
    \caption[Reconstructed mass spectrum for Higgs candidates (120~GeV)]{Reconstructed mass spectrum for Higgs candidates (120~GeV) in the $ZH\rightarrow \nu \bar \nu b \bar b$ decay.}
    \label{fig:physics_ZHnnbb}
\end{figure}

The resulting mass spectrum is shown in
Figure~\ref{fig:physics_ZHnnbb}. 
In this study, the mass scale was calibrated using the position of the $Z^0$ resonance.

The reconstructed mass of the Higgs is lower than the input value. This is believed to be
due to the energy loss by neutrinos in $b$ decays and/or incomplete
correction for energy not properly identified in the current version of the 
particle flow algorithm. Further studies
with an improved PFA algorithm is needed.
The statistical error of the event rate is about $3.2\%$, which is consistent
with the previous analysis using a fast simulation~\cite{ref-GLD}.

\subsection{$\bf e^+e^- \rightarrow ZHH \rightarrow$ 6~jets}

\begin{figure}[tbh]
    \centering
        \includegraphics[height=6cm,bbllx=0,bblly=0,bburx=182mm,bbury=80mm]{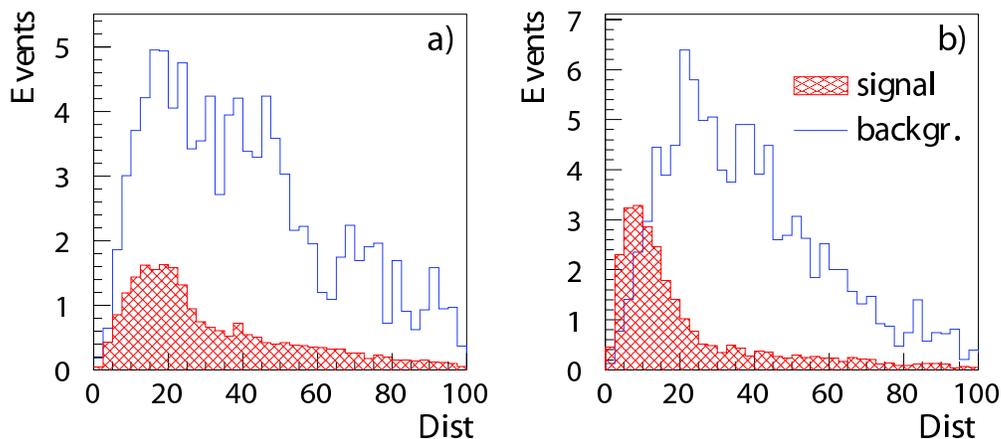}
    \caption[Distance variable for ZHH topology]{Distance variable for signal and background assuming
    a) $\Delta E/E = 60\%(1+|\cos{\theta_{jet}}|/\sqrt{E}$, or b) $30\%/\sqrt{E}$.}
    \label{fig:physics_zhh}
\end{figure}

Superior dijet mass resolution is necessary to identify intermediate
resonances, such as in the process $e^+e^- \rightarrow ZHH$, which is sensitive to the trilinear Higgs coupling. The cross section for this process is
at the sub-femtobarn level making identification above background
difficult.  A study~\cite{Castanier:2001sf} of ZHH decay into 6 jets
at $\sqrt{s} = 500$~GeV for $m_{H} = 120$~GeV finds that
conventional jet energy resolution (i.e., LEP experiments) is not
sufficient to identify a signal above background.  In this analysis,
a distance variable $Dist =
\sqrt{(m_{12}-m_{H})^{2}+(m_{34}-m_{H})^{2}+(m_{56}-m_{H})^{2}}$ is
used to characterize signal and background, as shown in
figure~\ref{fig:physics_zhh}.

\section{Top Analyses}
The measurement of properties of the top quark are an important part of the measurement program at the ILC.
The precise knowledge of its properties, its mass and width and its couplings to other particles, are
sensitive inputs to the overall constraint on the Standard Model.

The reconstruction of the top at the ILC can profit from the clean and well known environment at this collider. About
44\% of the top decays are expected to go into fully hadronic final states, which are reconstructed in the
detector as six jets. The fully hadronic top decay therefore is an excellent laboratory to
investigate and test the event reconstruction and algorithms. The drawback of the fully hadronic mode is
that there are a number of effects known which affect the final state: final state interactions, color rearrangements, Bose Einstein correlations, etc..
The extraction of the top mass from this channel has many theoretical difficulties, though in recent years
significant progress has been made in understanding them and showing solutions to some of them.

In this analysis~\cite{Chekanov:2003cp} $e^+e^- \rightarrow t \bar
t$ are studied in its fully hadronic decay mode into six jets.
Events at 500~GeV are generated using the Pythia event generator.
Six jets are reconstructed with the $k_\perp$ ~\cite{Catani:1991hj}
alogorithm. Full tracking and particle flow reconstruction are then
applied based on the BRAHMS software system with the SNARK particle
flow implementation~\cite{Chekanov:2003cp}.

Hadronic events are selected based on the total visible energy in the event, which should be close to the
event energy. The momentum imbalance along
the beam and perpendicular to the beam direction should both be small. Only events which have six well separated
jets are accepted, to clean up the sample. Events which have a well identified lepton in it are removed from the sample.

\begin{figure}
    \centering
        \includegraphics[height=8cm]{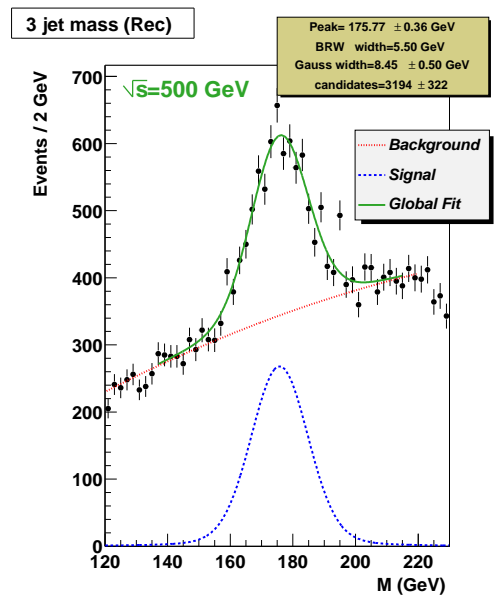}
    \caption[Invariant mass distribution of three-jet groups]{Invariant mass distribution of three-jet groups, after all cuts applied. The dashed line indiciates the background from other Standard Model processes.}
    \label{fig:physics_top1}
\end{figure}

In a next step the six jets are grouped into two groups of three
jets each. The total four-momenta of the three jet groups are
calculated. The best grouping of jets into three jet groups is then
selected with the constraint that the invariant masses of the two
groups should be similar, and by imposing total energy and momentum
conservation. The two groups should be produced approximately back
to back. Additionally the sample can be further cleaned up by
imposing a positive bottom tag on some of the jets, and by testing
whether two out of three jets in each group are consistent with
coming from the W decay. The invariant mass of the three jet groups
is shown in Figure~\ref{fig:physics_top1}.

The analysis includes physics backgrounds, but no beam-beam related
backgrounds. For technical reasons not all physics background
channels have been fully simulated. For the most part the events
were generated using the same tools as the signals, but were not
processed through the full simulation chain. Instead they were
passed through a fast smearing level Monte Carlo, before however
being fully reconstructed by the same program as the signal sample.
From this study a statistical uncertainty of the top mass
determination of $100 MeV$ has been found, for an integrated
luminosity of $300 fb^{-1}$. The mass resolution found is $5.5$~GeV
at $500$~GeV, which is approximately compatible with the goal of
$30\%/\sqrt E$.

\cleardoublepage

\chapter{The case for two Detectors}
\label{detector_2detectors}


The ILC's scientific productivity will be optimized with two
complementary detectors operated by independent international
collaborations, time-sharing the luminosity.  This will ensure the
greatest yield of science, guarantee that discoveries can be confirmed
and precision results can be cross-checked, provide the efficiency of
operations, reliability, and insurance against mishap demanded for a
project of this magnitude, and enable the broadest support and
participation in the ILC's scientific program.

\section{Complementary and Contrasting Detectors}
 
The two detectors will be designed to measure the physics events with
different approaches.  Ideally, given the unknowns of the experimental
environment at future colliders, the program must be prepared with two
detector philosophies in order to provide complementary sensitivity to
physics, backgrounds, and fake effects.  There is no unique, optimal
design for an ILC detector, because it is not known what will be
discovered, what physics will prove to be the most important, or what
the most significant backgrounds will be. Having two experiments
allows some level of aggressiveness in experimental design.  For
similar reasons, the LEP/SLC detectors were designed with different
strengths and weaknesses, arising from different assumptions on
physics and technical advantages; their complementarity broadened the
coverage.  At the Tevatron, the top quark discovery benefited from the
different detector approaches of CDF and D0.  ATLAS and CMS at the LHC
will provide this complementarity.  It is important for the ILC
detectors to provide similar breadth in detector response.

Experience with operating experiments at a linear collider is limited
to Mark II and SLD at SLC.  This experience raised unexpected issues
with beam halos, 'fliers', beam-related EMI, and other effects.
It is prudent to anticipate additional surprises related to operating
at the much higher currents and energy of the ILC.  The design of the
ILC will, of course, profit from the SLC experience, and be able to
avoid many of these problems.  But for a new machine, one must expect
new effects; having two complementary detectors will add flexibility
in dealing with such technical uncertainties.

\section{Broad Participation and Scientific Opportunity}

Having two complementary detectors will encourage the broadest
possible participation of the world HEP community in ILC physics.  A
worldwide financial and technical effort must be mounted to realize
the ILC.  The scale of this effort is unprecedented in the history of
particle physics, even exceeding that mounted for the Large Hadron
Collider.  In fact, no international scientific project of this
magnitude has yet been completed by any collaboration.  The number of
physicists in the world who will be interested, and must be enlisted
in order to justify the size of the enterprise, is very large.  One
detector effort will not satisfy the interest, or the need.

The level of financing for the project, and specifically that for the
detector efforts, will be determined by the size of the interested
community.  Having two detectors will generate significantly greater
scientific interest in the project throughout the world. This fact
must be considered when the potential cost saving of reducing to one
detector is evaluated.

The ILC will be a research facility for decades of exploration, and it
must provide the opportunities for more than a generation of particle
physicists.  Two detectors mounted by two collaborations double the
possibilities for meaningful contributions to the experimental
program, and accommodate the research interests of twice as many
physicists.  With two detectors employing complementary technical
solutions, the development and training opportunities, especially
those for young scientists and engineers, will be greatly enhanced.

\section{Efficiency, Reliability, Insurance}

Having two independent detector collaborations will yield highly
efficient, reliable data taking, with the insurance to deal with
unexpected problems.  The efficiency of operation will benefit from
time-sharing the luminosity, since the maintenance of one detector can
be carried out while the other is accumulating data.  Furthermore,
unexpected problems or the failure of one detector will not stop the
operations of the collider. There are risks associated with operating
large and complex detector systems. A major failure could disable the
program for a long time if a second detector were not available.

The competition between two detectors collaborations will drive the
scientific productivity of both experiments, as has been demonstrated
frequently in the past.  This important force in the scientific
enterprise results in a more effective utilization of the program's
resources, and more rapid progress.

\section{Confirmation, Cross-checks and Scientific Redundancy}

Only by having two detectors can there be genuine scientific
confirmation of new discoveries, or critical cross-checks of precision
measurements.  Indeed, the ILC is expected to make major discoveries
about the nature of the universe.  Such discoveries will be accepted
and integrated into the scientific paradigm only with sound
confirmation.  Two complementary experiments, with differing detector
approaches, will provide the required cross-checks on discoveries.
While discoveries require confirmation; precision measurements require
redundancy.  Two collaborations will develop independent analyses
which will be characterized by separate data sets and different
systematic errors. Having two detectors will ensure the most accurate
assessment of new physics found by discoveries or by precision
measurements. Furthermore, the fact that one collaboration's results
are subject to confirmation or refutation by the other is an important
protection against false conclusions.

For important results, we can expect each detector collaboration to
develop two or more competing analyses.  However, having two analysis
chains within the same detector collaboration does not create the
level of competition, redundancy, and independence needed for optimal
scientific outcomes.  There are many examples in particle physics of
important scientific results not being properly resolved by parallel
analyses within a single experimental collaboration.  The degree of
autonomy enjoyed by each such analysis effort within a collaboration
is fundamentally limited by the collaboration's goal of finding a
common answer. On the other hand, having two experimental
collaborations naturally results in truly independent analyses, which
may reach alternative conclusions, preventing confirmation of
incorrect results.

Confirmation and redundancy have been necessary for progress in
high-energy physics in the past, as demonstrated by many fixed-target
and collider experiments \cite{Con_SM01:2001}. For decades the ILC
will be at the cutting edge of the unknown, where cross-checks are
imperative for a rapid and thorough understanding of the data and the
physics.  In fact, confirmation and redundancy are an indispensable
part of science, a principle understood broadly.




\clearpage

\cleardoublepage

\chapter{Costs}
\label{detector_cost}

Three detector concepts, GLD, LDC, and SiD, have estimated the costs of their respective detectors. 
Although the methodologies employed differed somewhat, all three used a complete work breakdown 
structure, and attempted to identify all the significant costs associated with their various subsystems, 
as well as costs associated with assembly and installation. These cost estimates have been made in 
light of the GDE costing rules, but have included contingency at a level of $\approx$ 35\%. 
Costs below are quoted in year 2007 
dollars (\$) without escalation. To include inflation effects, a rate of 3\%/year can be applied. For example, the cost 
evaluated in 2014 dollars, a year that could see the middle of construction, would be higher by 23\%.  
To get a common basis for the costs in different regions, the following assumptions about conversions 
between dollars and euros and yen were employed: 1 Yen = 0.00854 \$ and 1 Euro = 1.20 \$. Clearly, 
some uncertainty arises because of the inconstancy of these conversion factors. 

Costs have been divided into those for materials and supplies (M\&S), and those for in-house manpower, 
which is given in man-years and then converted to dollars depending on local labor rates. Because of  
regional accounting differences, rather different amounts are assigned to these two categories by the 
different concepts, but the sum of the two is relatively constant region to region.

The cost drivers for the M\&S budgets are the calorimeters and the solenoidal magnet and flux return iron. 
Costs for common materials, like silicon detectors, or tungsten, or steel, are estimated differently by 
the different concepts, occasionally leading to rather large differences on individual detector parts. 
These differences are assumed to average out over the entire detector.  Costs for such materials are 
estimated with various methods, sometimes from one or preferably more industrial quotations, sometimes 
from the actual expenses borne in building previous detectors.  When comparing the estimates concept to 
concept, most items which appeared in one accounting, but not in another, were accounted for, and added 
in where absent. Integration, transportation, 
and computing have been included. Indirect costs associated with both M\&S and labor have also been included.

\begin{figure}
\begin{center}
\includegraphics[height=8cm]{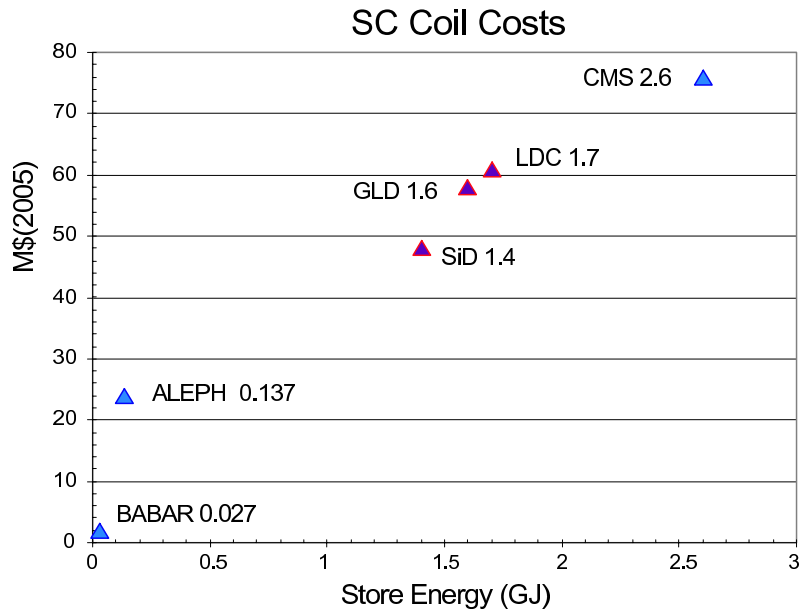}
\caption{Estimated cost of superconducting coils as a function of the stored energy}
\label{fig-DCR-coil_costs}
\end{center}
\end{figure}

Overall, there is reasonable agreement among the three concept estimates.  Explicit comparison of some 
of the major items, like the magnet coil, return yoke, and calorimeters, have been considered in some 
detail, and discrepancies understood.
Coil costs present an interesting example. Figure~\ref{fig-DCR-coil_costs} shows the estimated costs of 
the coils for each of the concepts as a function of the stored energy. Costs for the BaBar, 
Aleph, and CMS coils are included for reference. Costs include the manpower for design and  fabrication. 
The dependence of the individual estimates on stored energy looks reasonable, and the present estimates 
look in line with the reference points. Costs associated with the Detector Integrated Dipole are at the 
level of a few percent of the main coil cost.

\begin{figure}
\begin{center}
\includegraphics*[height=9cm]{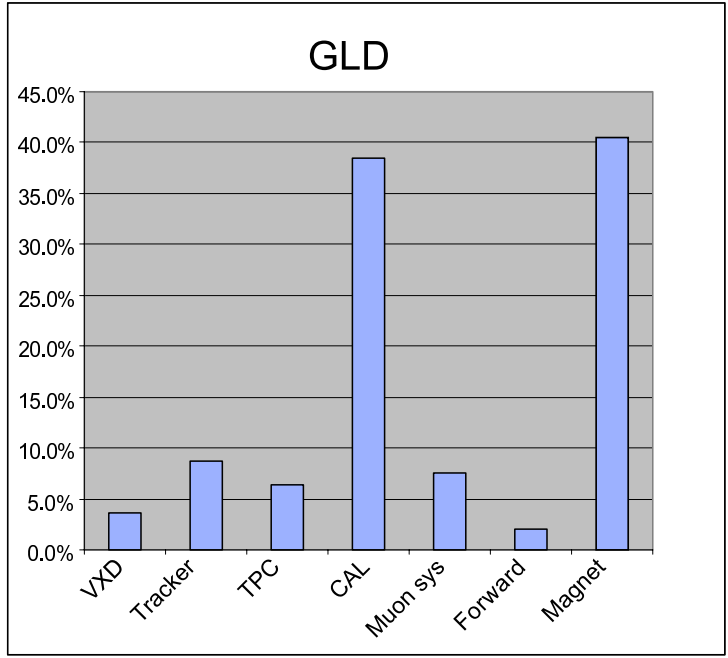}
\caption{Relative subsystem costs for GLD}
\label{fig-GLDrelsystemcosts}
\end{center}
\end{figure}

\begin{figure}
\begin{center}
\includegraphics*[height=9cm]{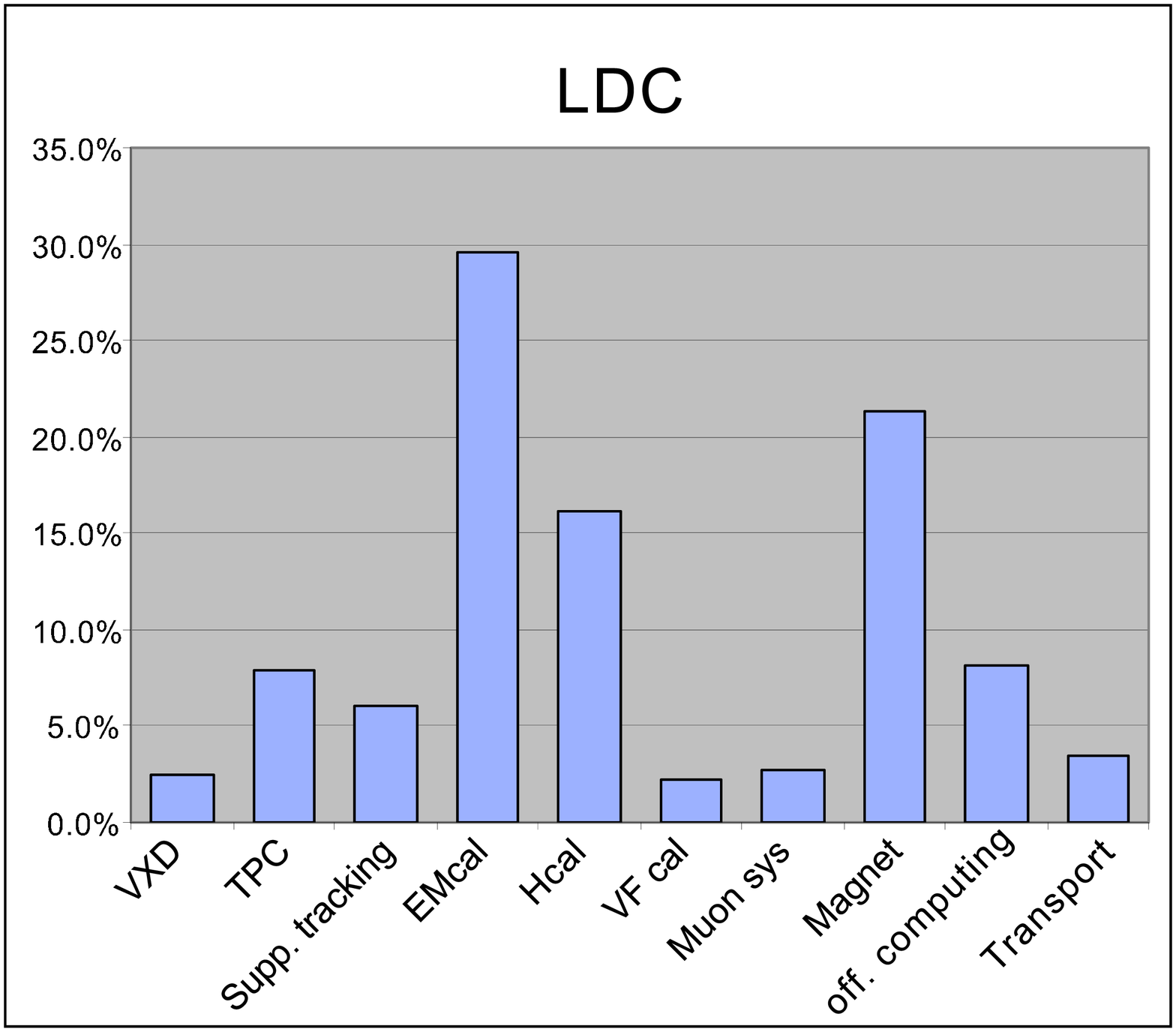}
\caption{Relative subsystem costs for LDC}
\label{fig-LDCrelsystemcosts}
\end{center}
\end{figure}

\begin{figure}
\begin{center}
\includegraphics*[height=9cm]{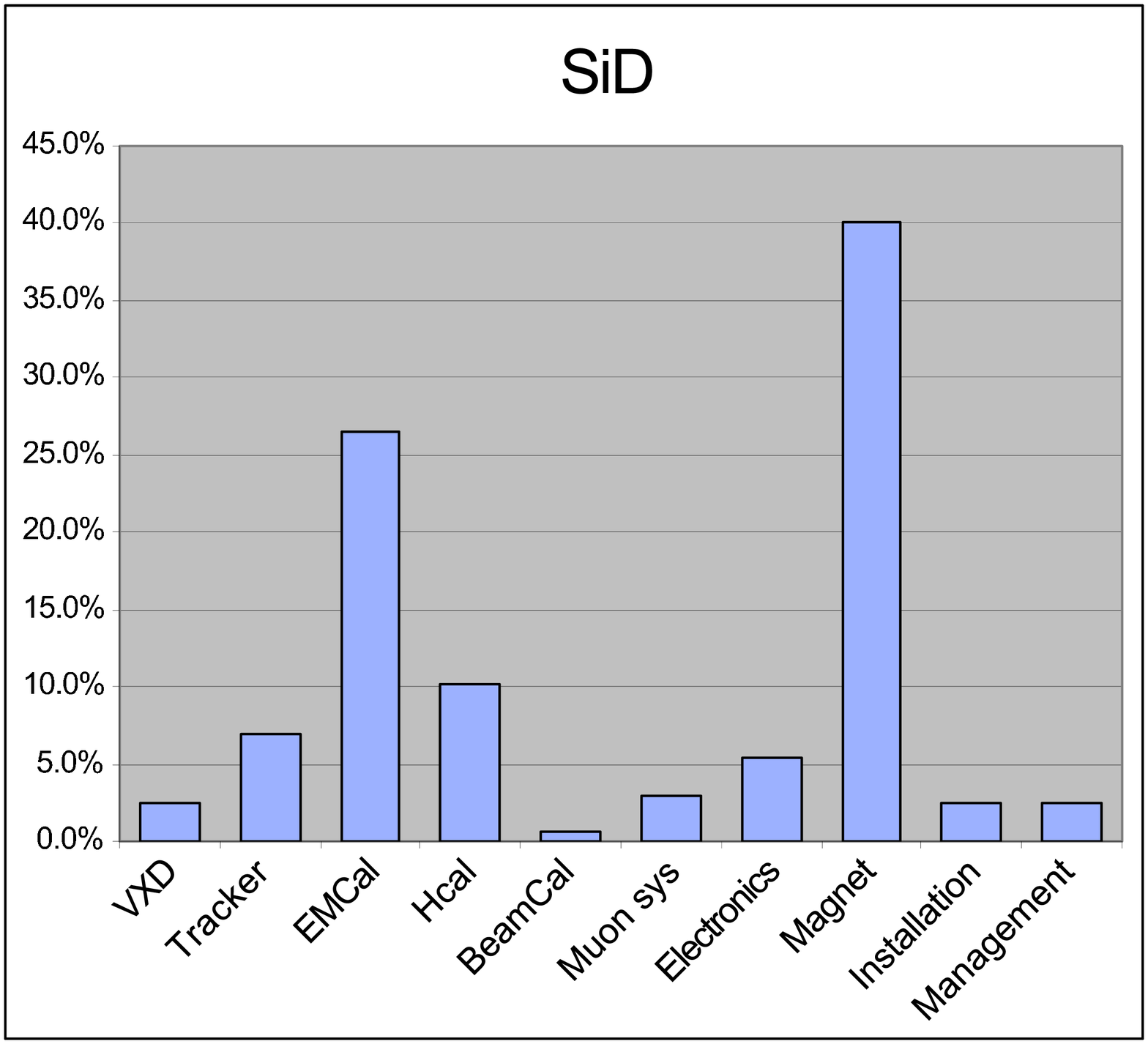}
\caption{Relative subsystem costs for SiD}
\label{fig-SiDrelsystemcosts}
\end{center}
\end{figure}

Figures~\ref{fig-GLDrelsystemcosts}, \ref{fig-LDCrelsystemcosts} and \ref{fig-SiDrelsystemcosts} show the cost breakdowns across detector subsystems for each concept. The figures make clear that the detailed categories for costing differ concept to concept.  For example, SiD has costed electronics, installation, and management as separate items whereas LDC and GLD have embedded these prices in the subdetector prices. In another example, GLD chooses to cost both hadron and electromagnetic calorimeters as a single item, since the detectors used are similar. LDC and SiD have separated these expenses, because the detection techniques are quite different. The prominence of costs associated with the magnet, which is here taken to be the sum of coil and flux return, and the calorimeter is obvious from the figures. Inevitably, some costs have not been treated equally in the different concepts at this stage in the cost estimation process. For example, LDC has costed the transportation independently and provided an estimate for off-line computing. GLD and SiD have not provided these costs.

The total detector cost lies in the range of 460-560 M\$ for any of the detector concepts, including contingency.
For SiD and LDC, M\&S costs lie in the range 360-420 M\$ and manpower is estimated to be 1250-1550 person-years, again
with contingency included. The GLD estimate includes most of the manpower with the M\&S, but as mentioned before, its total cost is 
comparable to that of SiD or LDC.

\chapter{Options}
\label{detector_options}
The baseline for experimentation at the ILC is a 500 GeV collider for electrons and positrons. A number of 
options exist to expand the scope of the collider by colliding different particles, or 
by slightly modifying the layout. These options in general are connected with additional costs, 
which are not estimated in this document. Nevertheless since they represent a significant 
extension of the physics capabilities of the facility, their physics motivations and 
impact on detector design are discussed in this section. 

The simplest option, which does not require any significant detector upgrades, is the operation of the 
collider as an $e^-e^-$ collider. A significant body of literature exists for this option, both 
describing its physics program, and its possible realization within a linear collider like the ILC. 

The GigaZ program requires running the collider at an energy corresponding to the Z pole. The ILC 
could reach very high luminosities at the Z, and thus become a very powerful laboratory for advancing the tests
of the Standard Model performed at LEP/SLC to a new level of accuracy. The physics program of this option is 
summarized in some detail in section~\ref{sec-options-gigaz}. 

The largest modifications to both the accelerator and the detector are required by the photon collider option,
described in section~\ref{sec-option-gg}. Here a discussion of both the highlights of the anticipated 
physics program and the technological challenge for the experiment are described. 
\section{GigaZ}
\label{sec-options-gigaz}
\newcommand {\stl}  {\sin^2 \theta_{\rm eff}^l}
\newcommand{\ALR}    {A_{\mathrm{LR}}}

The name ``GigaZ'' denotes the possibility to run the ILC back on the
Z resonance and, if needed, at the W-pair threshold to measure the W
mass. If all other parameters of the accelerator are kept unchanged the luminosity is 
${\cal L} = 4 \cdot 10^{33} {\rm cm}^{-2} {\rm s}^{-1}$ at the Z peak and
${\cal L} = 8 \cdot 10^{33}{\rm cm}^{-2} {\rm s}^{-1}$ at the W-pair
threshold. This corresponds to $10^9$ hadronic Z decays in less than a
year of running, and $10^6$ W-pairs close to threshold in the same time.

\subsection{Physics motivation}

The main objective of Z-pole physics is to measure the axial-vector ($g_{\rm A,f}$)
and vector coupling ($g_{\rm V,f}$) of the Z to fermions, where the best
precision, theoretically and experimentally can be obtained for
leptons. The ratio of the two is sensitive to the weak mixing angle
$\sin^2 \theta$ (where $g_{\rm V,f}/g_{\rm A,f} = 1 - 4 q_f \sin^2 \theta$). 
If there is no new physics in fermion production on the
Z-pole at the Born level all deviations of $g_{\rm A,f},\ g_{\rm V,f}$ from their Born
level Standard Model predictions can be absorbed in two effective
parameters, $\Delta \rho$ and $\sin^2 \theta_{\rm eff}$ (where $g_{\rm A,f} = \sqrt{\Delta \rho} a_{f,{\rm Born}}$ and  
$g_{\rm V,f}/g_{\rm A,f} = 1 - 4 q_f \sin^2 \theta_{\rm eff})$). In principle these
parameters still depend on the fermion flavor, however for $f\ne b$
the difference between the flavors does not contain additional
information, so that usually the values for leptons are given. Only
the b-quark is interesting on its own since it is the isospin partner
of the heavy top quark and in some models, like the little Higgs
models, the $(b,t)$ doublet is different from the other isospin
doublets. In case new physics enters directly via the exchange of a
new vector boson, $Z'$, the $Z$ observables are sensitive to the mixing of
the Standard Model $Z$ with the $Z'$.

$\stl$ can be measured with extremely good precision
at GigaZ. It depends only on a ratio of couplings which can be
obtained from asymmetry measurements. For this reason many systematic
errors like efficiency and luminosity drop out in the calculation so
that the full statistics at GigaZ can be used. The most precise
determination of $\stl$ can be obtained from the
left-right asymmetry with a polarized electron beam
\[
   \ALR =  \frac{1}{\cal P}\frac{\sigma_L - \sigma_R} 
                                   {\sigma_L + \sigma_R} 
        =  \frac{2 g_{\rm V,l} g_{\rm A,l}}{g_{\rm V,l}^2+g_{\rm A,l}^2}
\]
where $\sigma_L$ ($\sigma_R$) denotes the cross section with left-
(right-) handed beam polarization and ${\cal P}$ the polarization
vector.  If both beams can be polarized, the polarization can be
unfolded internally and a precision of $\Delta \ALR = 10^{-4}$ is
possible corresponding to $\Delta \stl = 0.000013$ \cite{GigaZ}. This
corresponds to an improvement of a factor of ten compared to the LEP/SLD combined
value of $\stl$.

To measure $\Delta \rho$ absolute cross section measurements as well
as the total width of the Z, which has to be measured from a scan, are
needed. In both quantities several systematic uncertainties enter so
that here an improvement is much more difficult. Under optimistic
assumptions $\Delta \rho = 5 \cdot 10^{-4}$ can be achieved which
corresponds to a factor two improvement with respect to LEP.

The W-mass can be measured with a precision of $\Delta m_W = 6~{\rm MeV}$
from a scan of the W threshold corresponding to an
improvement of a factor six to the present value and a factor three to
the projected LHC precision.

All models of new physics, once they are calculable, have to
predict the size of the loop corrections or of new Born level
processes for electroweak processes at or below the Z. In this sense
the GigaZ option is interesting in all possible cases. However the
number of new particles in the ILC energy range and corresponding
thresholds or peaks to scan varies largely between the models. It thus
has to be decided at a later stage if there is time available for Z
and W-threshold running.

\begin{figure}[htbp]
  \centering
  \includegraphics[width=0.6\linewidth]{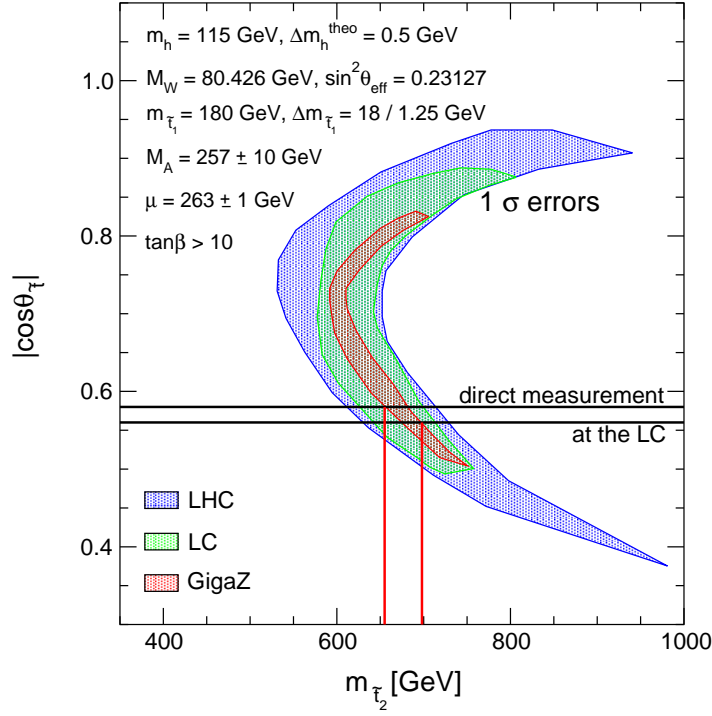}
  \caption[Constraints in the $\tilde{t}_2-\cos \theta_{\tilde{t}}$ now and with GigaZ]{Constraints in the $\tilde{t}_2-\cos \theta_{\tilde{t}}$
    plane from the electroweak precision data now and with GigaZ.}
  \label{fig:gigazsusy}
\end{figure}

As an example for the use of GigaZ in supersymmetry, 
Figure~\ref{fig:gigazsusy} shows the indirect constraint in the
$\tilde{t}_2-\cos \theta_{\tilde{t}}$ plane from the electroweak precision
data now and with GigaZ when the other relevant SUSY parameters are known.

Experimentally the situation is more challenging if nature has chosen
a scenario in which the (t,b) isospin doublet plays a special role. In
this case GigaZ can also provide fundamental measurements on the
b-sector like the normalised b-cross section on the peak, $R_b$, or the
forward backward asymmetry for b-quarks with polarized beam
$A_{LR,FB}$ which measures the couplings of the Z to b-quarks. These
measurements require a pure b-tagging with very good knowledge of the
background and, in the case of the asymmetries, in addition efficient
b-charge tagging.

\begin{figure}[htbp]
  \centering
  \includegraphics[width=0.6\linewidth]{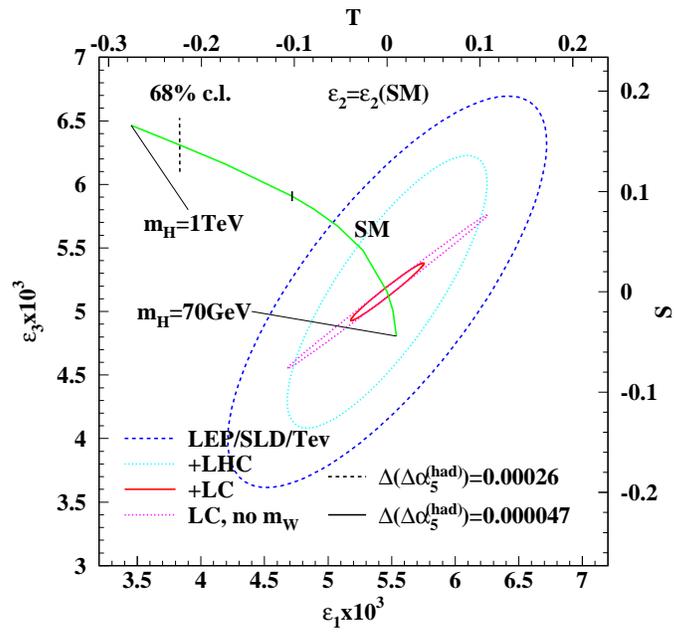}
  \caption{S-T ($\varepsilon_1-\varepsilon_3$) parameters now, after
    LHC and after GigaZ}
  \label{fig:gigazst}
\end{figure}

If a Higgs is found with a mass incompatible with the current precision data
or no Higgs is found, GigaZ is needed to confirm the old data with
higher precision and to determine where the discrepancy comes from.
Figure~\ref{fig:gigazst} shows the present and possible future precision
data in the STU and $\varepsilon_{1,2,3}$ representations. 
In many models it is easy to modify T ($\varepsilon_1$) which depends
on the mass splitting in the isospin-doublets.
Due to the correlation between the two parameters a change in the Higgs
mass can be compensated by a change in T. To separate the two effects
the precise measurement of the W-mass is thus extremely important in
this case.

\begin{figure}[htbp]
  \centering
  \includegraphics[width=0.6\linewidth]{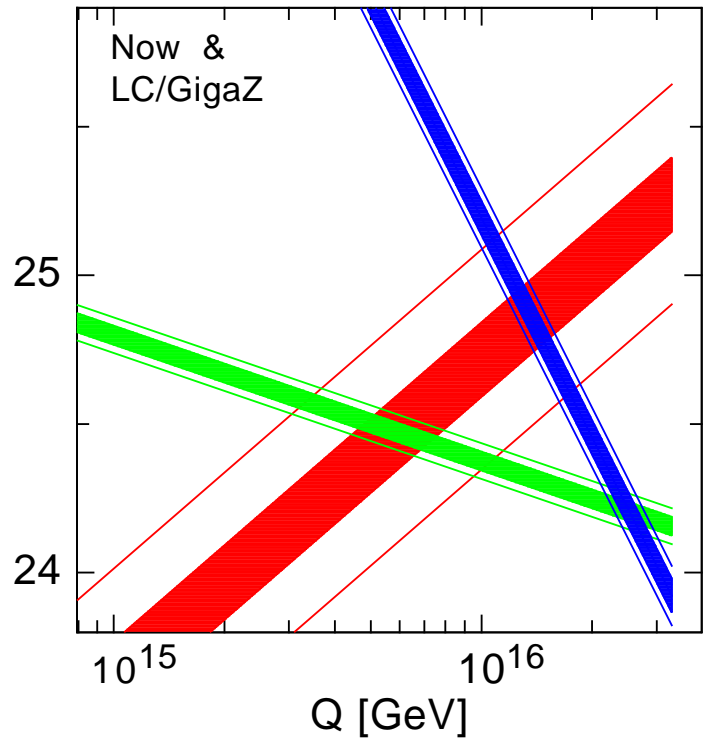}
  \caption{Unification of couplings now and after GigaZ}
  \label{fig:gigazas}
\end{figure}

Another task at GigaZ is the measurement of the strong coupling constant
$\alpha_s$ which can be obtained from the ratio of hadronic to leptonic Z
decays to a precision of 0.0005 - 0.0007. Tests of grand unification are
limited by the knowledge of the strong coupling constant (see
Figure~\ref{fig:gigazas}). Since some models, e.g.~within string theory, predict
small deviations from unification, this measurement turns out to be very
important.

\subsection{Experimental challenges}

For the detector GigaZ seems not very problematic. The event rate is high,
about 30 events per bunch train. However this is compensated by the much
smaller rate of two-photon events and the about one order of magnitude smaller
background from beamstrahlung compared to 500\,GeV. About 1\% of the Z-events
contain a second and $10^{-4}$ a third Z-event in the same bunch crossing. For
Z-counting, which is needed in the $\ALR$ measurement this should be no
problem. A slight challenge might be flavor tagging in this case, however one
can exclude these events from the analysis and correct for the very small bias
this introduces.

If only electron polarization is available, it has to be measured to 
$\Delta {\cal P}/{\cal P} = 7 \cdot 10^{-4}$ which seems hopeless. However
if polarized positrons are also available and the sign of the polarization can be
changed rapidly, no absolute polarimetry is needed. Only relative
measurements are needed to track time dependencies and differences
between the two helicity states.

The real challenge of GigaZ is the beam energy measurement. $\ALR$
depends strongly on the center of mass energy due to $\gamma$-Z interference. A
beam energy measurement of $\Delta E_b /E_b < 3 \cdot 10^{-5}$
relative to the Z mass is needed so as not to limit the precision on the weak mixing
angle. To improve knowledge of the Z-width,
the beam energy must be very well known, 
$\Delta E_b /E_b < 10^{-5}$. In this case the beamstrahlung and
the beam energy spread also have to be measured to a few percent. These requirements 
are significantly more aggressive than those for the $500$~GeV ILC. 

For a scan of the W threshold the detector requirements are more
relaxed because the event rate is much lower. However this measurement also
requires the beam energy
to be known to $\Delta E_b /E_b < 3 \cdot 10^{-5}$
relative to the Z mass.

\newlength{\figwidth}
\newlength{\figheight}
\newlength{\twofigheight}
\newlength{\twofigwidth}
\setlength{\figheight}{0.5\textwidth}
\setlength{\figwidth}{0.7\textwidth}
\setlength{\twofigheight}{0.35\textwidth}
\setlength{\twofigwidth}{0.45\textwidth}
\newcommand{\ttbs}{\char'134}
\newcommand{\AmS}{{\protect\the\textfont2  A\kern-.1667em\lower.5ex\hbox{M}\kern-.125emS}}
\newcommand{\epem}{\ensuremath{e^+e^-}}
\newcommand{\gamgam}{\ensuremath{\gamma\gamma}}
\newcommand{\emem}{\ensuremath{e^-e^-}}
\newcommand{\cms}{\ensuremath{cm^{-1}s^{-1}}}

\hyphenation{author another created financial paper re-commend-ed Post-Script}


\section{Photon Collider}
\label{sec-option-gg}

\begin{figure}[t]
  \begin{center}
  \includegraphics*[width=\figwidth]{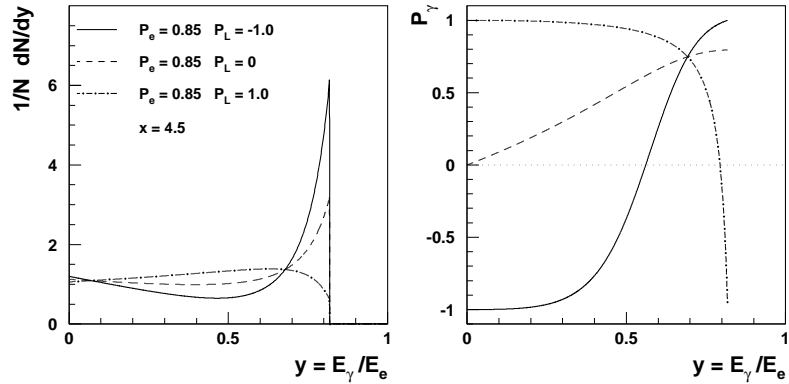}
  \end{center}
  \vskip -0.5cm
  \caption[Energy and polarization of photons from the Compton back-scattering]{Energy distribution (left plot) and polarization (right plot)
           for photons from the Compton back-scattering, for fixed
           electron beam polarization  $P_e$=85\% and different laser
           polarizations: $P_L=1,0,-1$.
           Parameter $x=4.5$ corresponds to laser wave length
           of 1.06 $\mu m$ and primary electron beam energy of 250 GeV.
           }
  \label{fig:compton}
\end{figure}

The elegant idea~\cite{ginzburg-telnov} to convert an \epem\ collider 
to a \gamgam\ collider can expand the physics reach of the ILC.  
%
The Photon Linear Collider (PLC) denotes both the $\gamma \gamma $ and
$e \gamma$ options of the ILC. In order to produce high energy photon
beams the electron beams of the ILC, running in the $e^-e^-$, mode
are used. Just a few millimeters before reaching the interaction point(IP),
the focused electron bunches collide with a very intense laser
beam. 
In the process of Compton backscattering, most of the electron
energy can be transferred to the final photon, moving in the
direction of the initial electron.
With a proper choice of electron beam and laser polarization, one can produce
a peak of high energy photons with a high degree of polarization, 
as shown in Figure~\ref{fig:compton}.
By converting both electron beams, a study of $\gamma \gamma$
interactions is possible in the energy range up to $\sqrt{s_{\gamma\gamma}}\sim
0.8 \cdot \sqrt{s_{ee}}$, whereas by converting one beam the
$e\gamma$ processes up to $\sqrt{s_{e\gamma}}\sim 0.9 \cdot
\sqrt{s_{ee}}$ can be studied.

In the $\gamma \gamma $ or $e \gamma$ modes it is possible to reduce the
emittance of the electron beams and apply stronger beam focusing in
the horizontal plane.
The luminosity is not limited by beamstrahlung and beam-beam
interactions, therefore  for nominal electron beam energy of 250~GeV
the geometric luminosity $L_{geom}=12\cdot 10^{34} cm^{-2}s^{-1}$, 
about four times larger than the expected $e^+ e^-$ luminosity.
However, due to the high intensity of the electron and laser beams, higher
order processes  become important and the beams will be dominated by
low energy photons.
Even so, the luminosity in the high energy $\gamma \gamma$  peak (see
Figure~\ref{fig:spectra}) corresponds to about $1/3$ of the
nominal  $e^+ e^-$ luminosity. For a nominal electron beam energy of
250~GeV it is expected that $L_{\gamma \gamma}(\sqrt{s_{\gamma\gamma}}> 0.65
\cdot \sqrt{s_{ee}})$
of about 100~fb$^{-1}$ per year (400~fb$^{-1}$ for a whole energy
range).
In first approximation, the luminosity of the photon collider is
proportional to the electron beam energy.


\subsection{Physics Reach}
 The PLC is an ideal observatory for the scalar sector of the Standard Model
and beyond, leading to important tests of the EW symmetry breaking mechanism
which are in many cases complementary to the $e^+e^-$ ILC case.
In addition the PLC is also a natural place to study in detail
hadronic interaction of photons
\cite{Brodsky:2004wx,Badelek:2001xb,Boos:2000ki,Ginzburg:2000cq,Hagiwara:2000bt,Telnov:1999vz}.
The most important aspects of physics of the PLC, illustrated by
some examples, are listed below.
\begin{itemize}
\item At a $\gamma \gamma$ collider the neutral C=+ parity resonances can be produced,
in contrast to C=$-$ resonances in the $e^+e^-$ collision. The
lowest spin of a resonance allowed  is zero, as for a Higgs boson,
while spin 1, dominating at the $e^+e^-$, is forbidden.

\begin{figure}[ht]
  \begin{center}
  \includegraphics*[width=\figwidth]{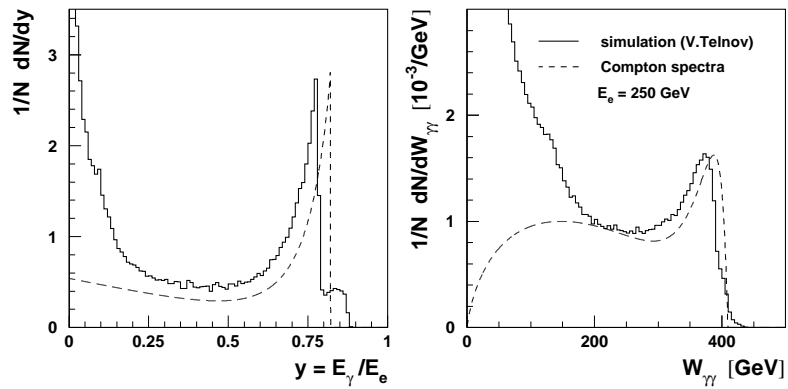}
  \end{center}
  \caption[Distributions of the photon energy and the $\gamma\gamma$ center-of-mass energy]{Energy distribution for photons (left plot)
           and the $\gamma \gamma$ center-of-mass energy
           distribution (right plot) from  simulation of
           the PLC luminosity spectra by V.Telnov,
           compared to the ideal (i.e. the lowest order QED) Compton spectra.
           }
  \label{fig:spectra}
\end{figure}

\item The  s-channel resonance production of  C=+ particles permits
precise measurements of their properties. For example, the precision of the
cross section measurement for the SM Higgs decaying into the $b\bar b$
final state is between 2 to 3 \% for Higgs masses between 120 and 155 GeV (Figure~\ref{fig:phys_h1});
for Higgs masses between 200 and 350 Gev, and decays into the WW final state, the accuracy is
between 3 and 8 \%.  Using both linearly
and circularly polarized colliding photons one can select CP-even
and CP-odd states. Study of the CP nature of the Higgs bosons (both
for the case of CP conservation, and of CP violation in the Higgs
sector), is feasible even by using only the initial polarization
asymmetries \cite{Grzadkowski:1992sa}. Additional information can
come by from measurements of the final state.

\begin{figure}[htb]
  \begin{center}
  \includegraphics*[width=\twofigwidth,height=\twofigheight]{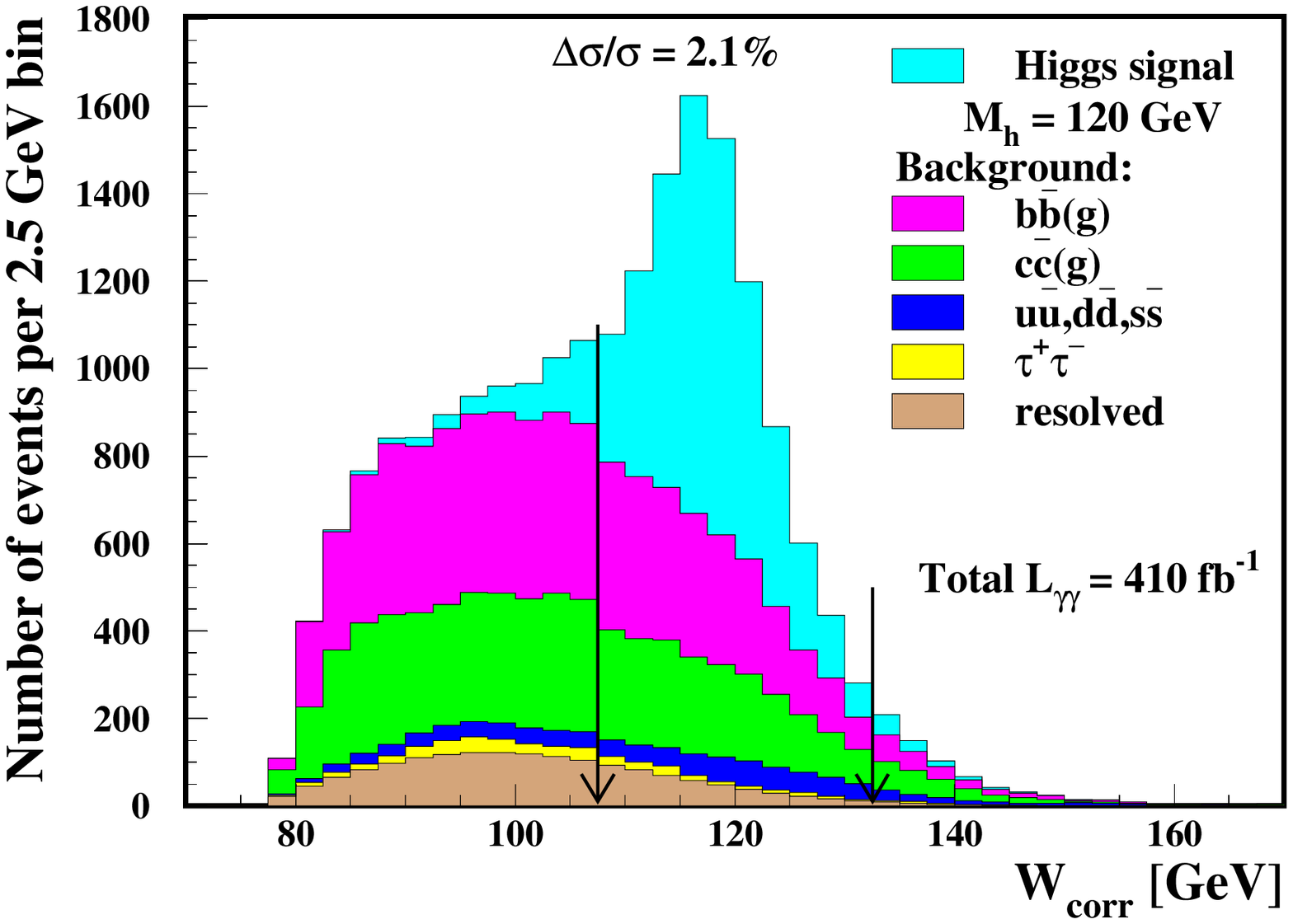}
     \includegraphics*[width=\twofigwidth,height=\twofigheight]{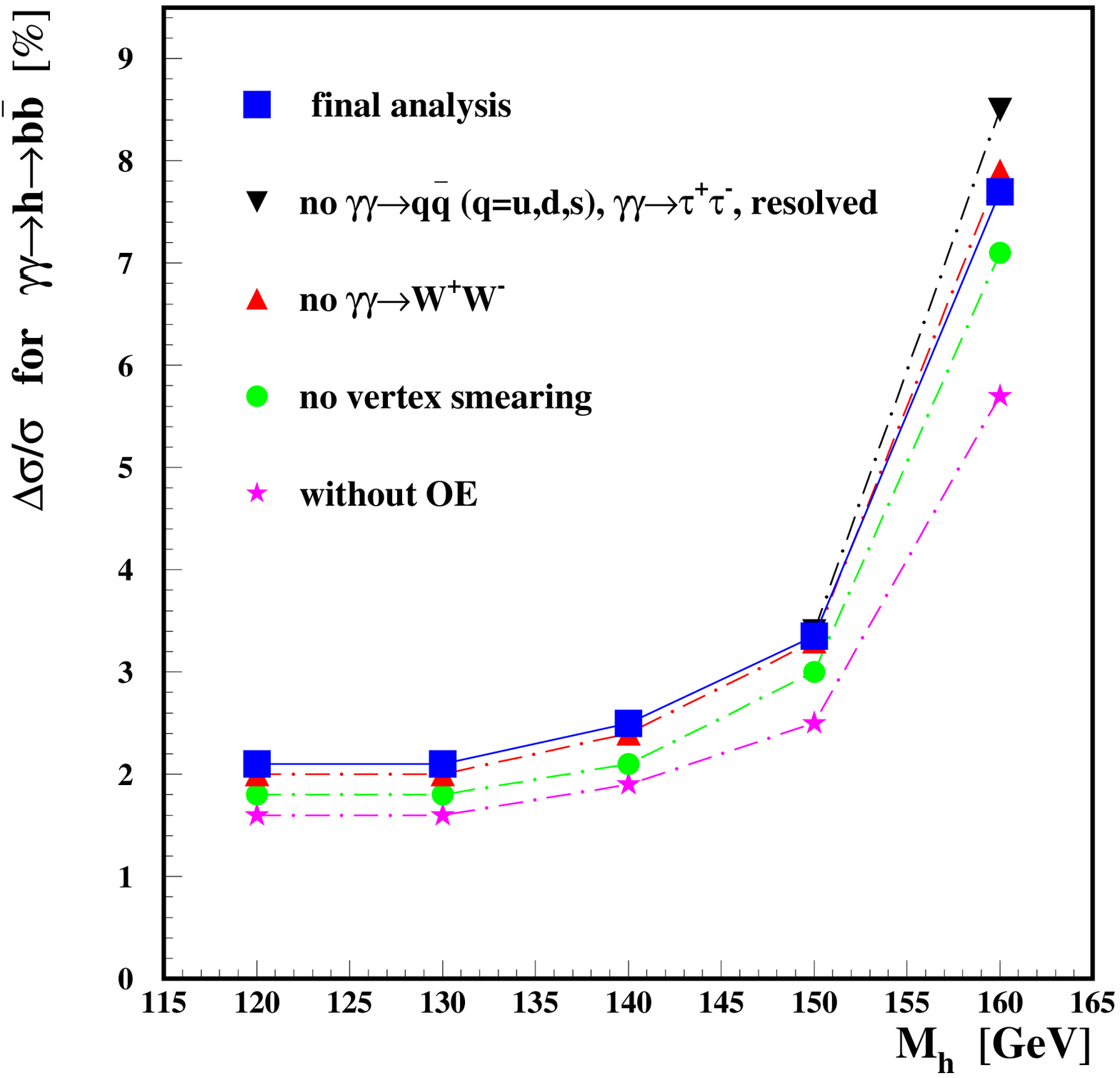}
  \end{center}
 \caption[Distribution of invariant mass and achievable precision for SM Higgs]{
Left: distributions of the corrected invariant mass, $W_{corr}$, for
selected $b \bar{b}$ events; contributions of the signal, for $M_h =
$ 120~GeV, and of the  background processes, are shown separately.
Right: statistical precision of $\Gamma ( h \rightarrow \gamma
\gamma) BR(h \rightarrow b \bar{b})$ measurement for the SM Higgs
boson with mass 120 $\sim$ 160~GeV for various stages  of analysis
\cite{Niezurawski:2005cp}.
 }
 \label{fig:phys_h1}
 \end{figure}

\item Neutral resonances couple to photons via loops involving
charged particles. The Higgs $\gamma \gamma$ coupling is dominated by
loops involving those heavy charged particles which couple strongly to the Higgs.
Therefore the $\gamma \gamma$ partial width  is
sensitive to the contributions of particles with masses 
beyond the energy of the $\gamma \gamma$ collision. By combining the
production rate for $\gamma \gamma \rightarrow Higgs\rightarrow b
\bar b$ with the measurement of the $Br(h\rightarrow bb)$ at
$e^+e^-$ ILC, with accuracy ~1~\%, the width $\Gamma(h\rightarrow
\gamma \gamma)$ can be determined with high accuracy 2 \%, for a Higgs mass
of 120 GeV. This allows discriminating between various models for the Higgs.
For example, in the 2 Higgs Doublet Model, which has all couplings of  neutral Higgs bosons 
identical to those in the SM, the
contribution of the $H^+$ with mass 800 GeV, leads to 10 \%
suppression in the $h$ decay width, for $M_h$ around 120 GeV
\cite{Ginzburg:2001ph}. Also the effect of new heavy
particle with mass around 1~TeV, as suggested in the Littlest Higgs
model, should be seen at PLC. In some cases it is
possible to measure not only the absolute value of the $h \gamma \gamma$
amplitude but also its phase, due to the interference with
non-resonant background. By combining $WW$ and $ZZ$ channels for the SM
Higgs boson, accuracy of the phase measurements is between 30 and
100~mrad \cite{Niezurawski:2002jx}. A similar conclusion was obtained
for the $t \bar t$ channel \cite{Asakawa:2003dh}.

\item Since particles can be produced singly at a $\gamma \gamma$ collider,
it is possible to produce high mass neutral Higgs bosons which would be
inaccessible at the parent $e^+e^-$ ILC,
where they are typically produced in pairs or associatively with other heavy
particles. PLC can play an important role in covering the so-called LHC
wedge, which appears in the MSSM for the intermediate $\tan
\beta$. In the wedge region,  LHC and ILC may not be able to
discover other Higgs particles beside the lightest SM-like Higgs
boson $h$. But at the PLC observation of heavy (degenerate) A and H bosons,
with masses above 200 GeV, would be possible (Figure~\ref{fig:susyhiggswedge})
\cite{Asner:2001ia,Niezurawski:2005cr,Spira:2006aa}.
%

%

\item
In $\gamma \gamma$ collisions, any kind of charged particles
(scalars, fermions and vectors) with  masses  below the kinematic
limits, can be directly produced in pairs, through lowest
order QED. Moreover their cross sections are typically larger than
the corresponding cross sections in $e^+e^-$. Especially important
for a $\gamma \gamma$ collider is the production of pairs of charged
Higgs bosons and charged sfermions and charginos. The $e \gamma$
option allows study of the associated production of heavy sfermions and
light charginos/neutralinos, when the $e^+e^-$ energy will be not
high enough for the sfermion pair production \cite{Datta:2002mz}.

 \begin{figure}[ht]
\begin{minipage}{.4\textwidth}
\includegraphics[width=\textwidth]{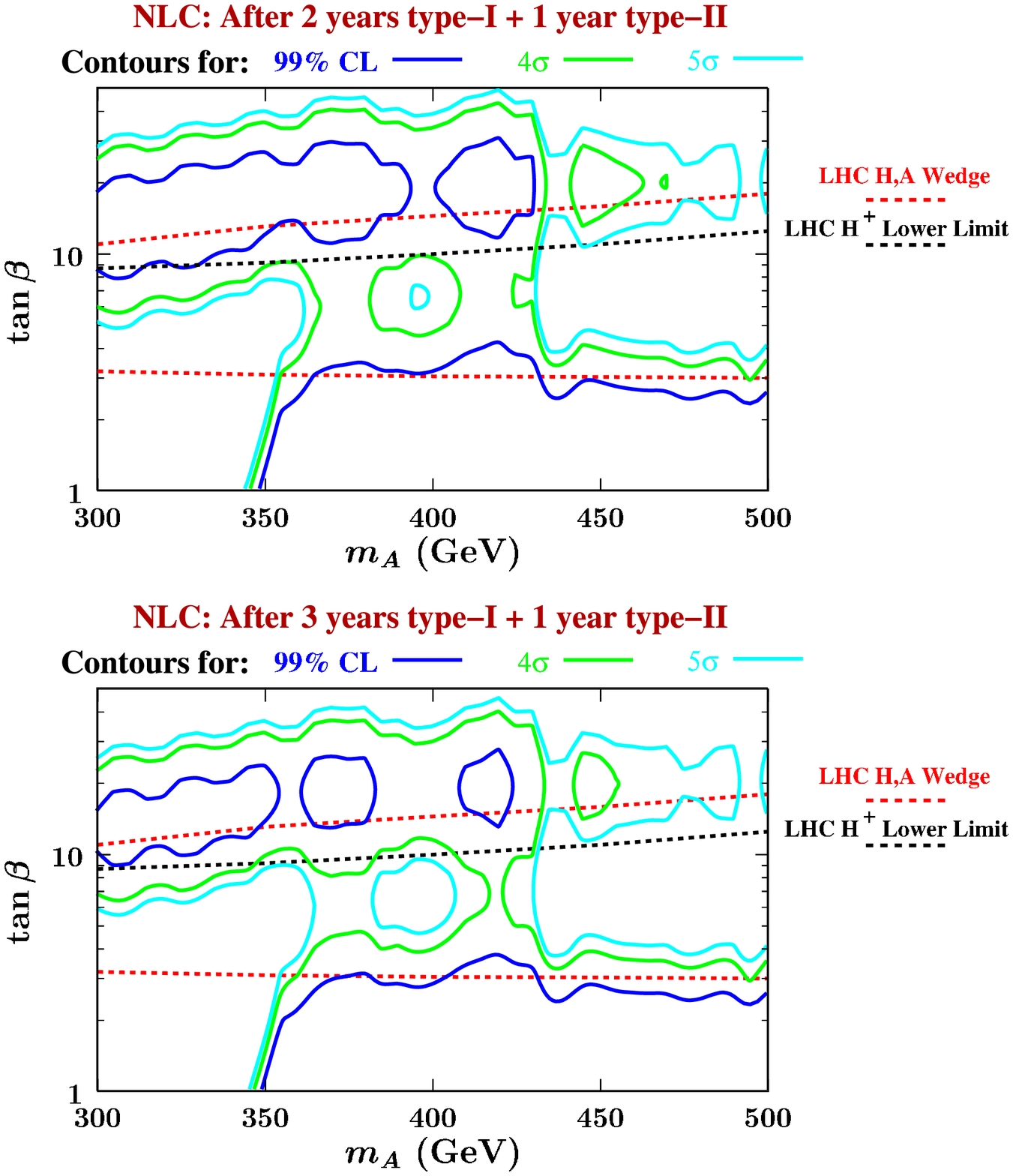}
\end{minipage}
\begin{minipage}{.59\textwidth}
\includegraphics*[width=\textwidth]{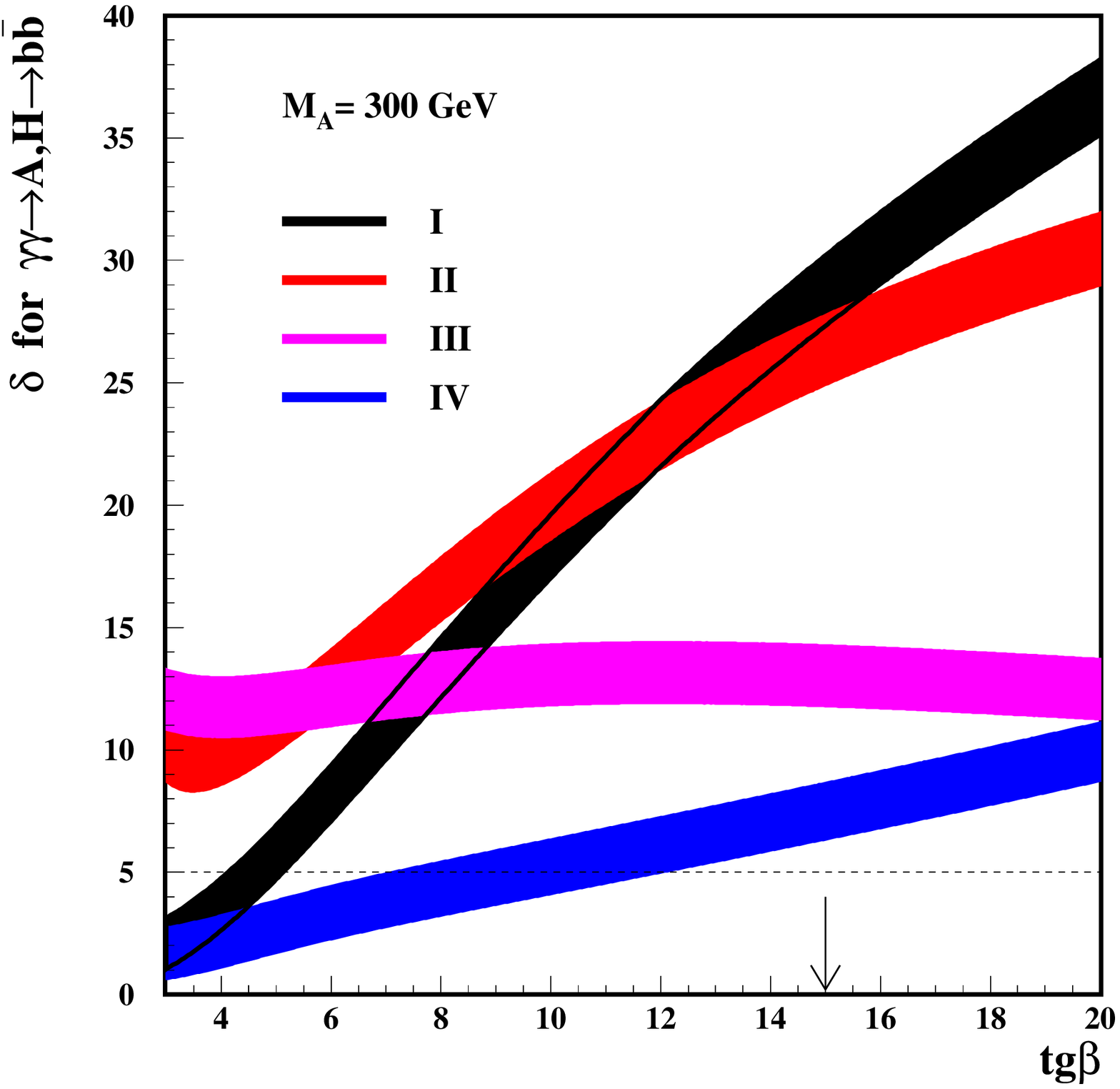}
\end{minipage}
\caption[Production of A and H at the $\gamma \gamma$ collider]{Production of  A and H, with  parameters  corresponding to
 the LHC wedge, at the $\gamma \gamma$ collider.
 Left: Exclusion and discovery limits obtained at the linear collider for
$\sqrt {ee} =$  630 GeV, after 3 or 4 years of operation (using the
broad and peaked energy spectra) \cite{Asner:2001ia}; Right:
statistical significance of the Higgs-boson production measurement
as a function of $\tan\beta$, for $M_A = $ 300~GeV
\cite{Niezurawski:2005cr}.} \label{fig:susyhiggswedge}
\end{figure}

\item
The huge cross sections for the $\gamma \gamma \rightarrow W^+W^-$
and $e^- \gamma \rightarrow \nu W^-$ processes permit study of the
anomalous $WW\gamma$ coupling, with an accuracy similar to that of the
$e^+e^-$ collider. (See Volume II, Section 3.2.) Because of the very
clean production mechanism and cross section, the PLC can provide
precision top quark measurements, and searches for anomalous top
couplings. Here the sensitivity is large, because the $t\bar t$ production rate
depends on the 4th power of the $ht\bar t$ coupling.
Note, that at PLC the $\gamma t \bar t$ and $ Zt \bar t$ couplings
are separated. Single top production at an $e\gamma$ collider is
the best option for measuring the $Wtb$ coupling.

\item
 Detailed studies of neutral gauge boson scattering processes,
$\gamma \gamma \rightarrow \gamma \gamma / \gamma Z / ZZ$, which
appear only atn one-loop level in the Standard Model,  constrain
new physics contributions, which could affect these
channels either at the tree-level or through additional loop
contributions \cite{Gounaris:2005pq}.

\item The production of pairs of neutral Higgs bosons at the $\gamma \gamma$ collider
proceeds, in contrast to pairs of charged Higgs bosons,
via  box and triangle loop diagrams
\cite{Belusevic:2004pz}. It is sensitive to the trilinear Higgs
self-coupling, which must be measured in order to
reconstruct the Higgs potential.

\item At a PLC two photons can form a $J_z=0$ state with either even
or odd CP parity.  Testing the CP nature of the Higgs bosons can
be performed by using the polarization asymmetries and/or the
observation of correlations among the decay products. For the ZZ and WW decay channels,
the angular distribution of the secondary WW and ZZ decay products
can be used \cite{Niezurawski:2004ga}. In $\gamma \gamma
\rightarrow Higgs \rightarrow \tau\bar \tau /t \bar t$, one can
perform a model independent study of CP-violation, exploiting
fermion polarization (Figure~\ref{fig:phys_h2})
\cite{Asakawa:2000jy,Asakawa:2003dh,Godbole:2006eb}. 
In addition $\gamma \gamma \rightarrow Higgs \rightarrow \tau\bar \tau
$ can be used  \cite{Godbole:2006eb} to look for a light CP-violating Higgs, 
which may escape discovery at the  LHC.
\item
The cross sections for Higgs boson and SUSY particle production at the
PLC depend on different combinations of couplings than the corresponding
processes at other machines. Therefore combination of precision
measurements at $pp$, $e^+e^-$ and $\gamma\gamma$ collisions can
give us useful additional information, and can be used to
differentiate between models both with and without CP violation.

\begin{figure}[htb]
  \begin{center}
 \includegraphics*[width=\twofigwidth]{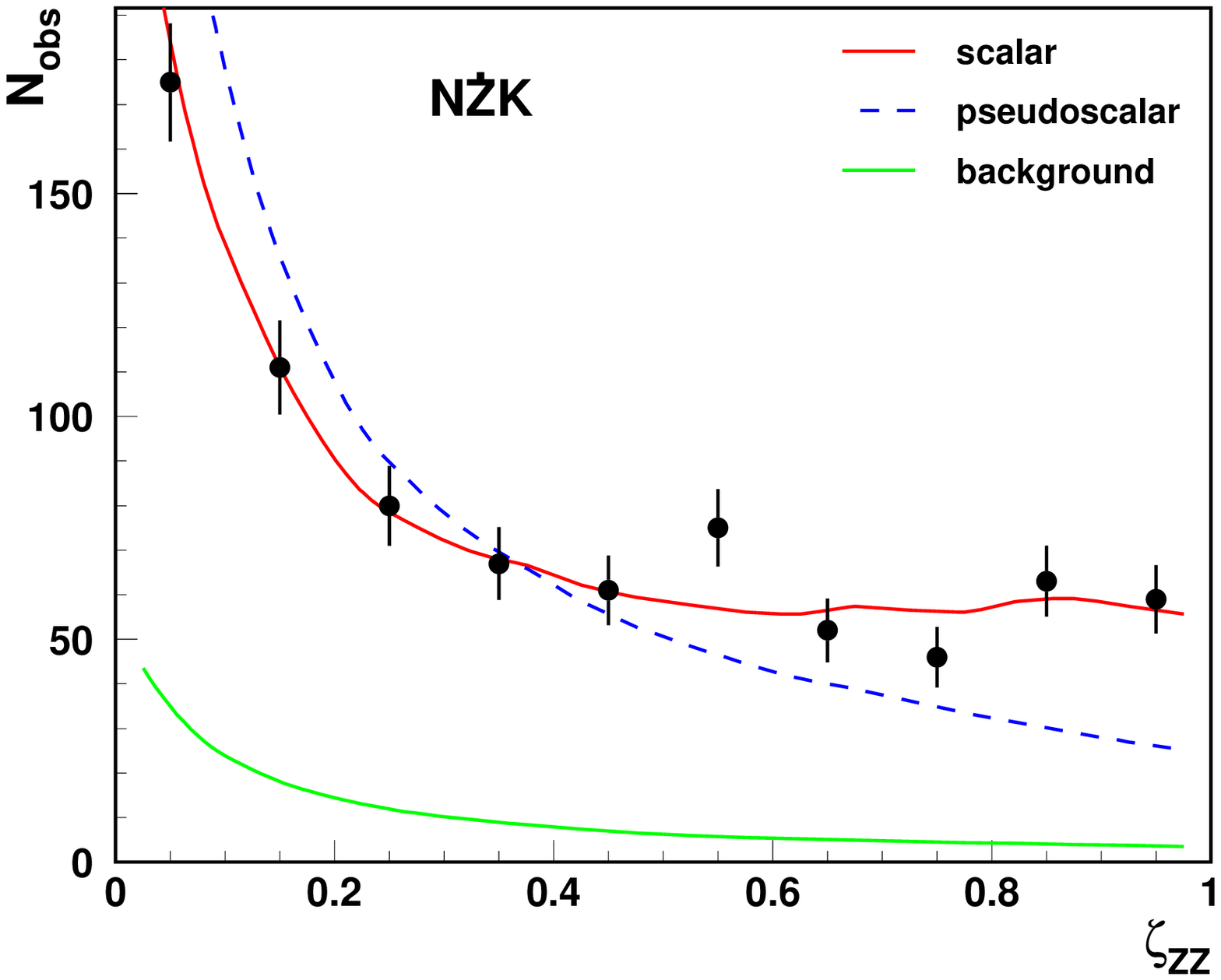}
  \includegraphics*[width=\twofigwidth,height=\twofigheight]{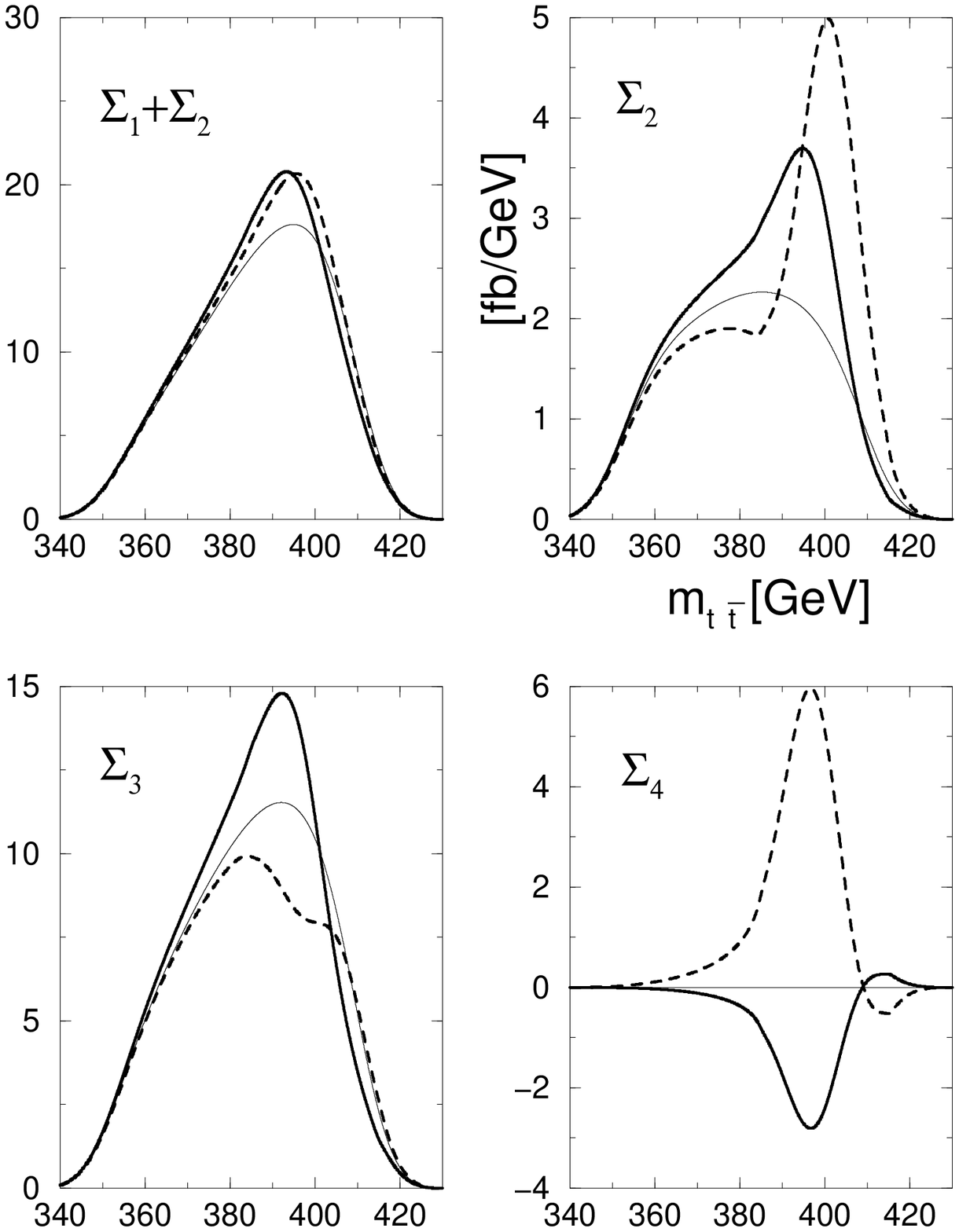}
  \end{center}
 \caption[Higgs analysis in $\gamma \gamma$ collisions]{
Left:   Measurement
        of the variable $\zeta_{ZZ}$ calculated
        from the polar angles of the $Z\rightarrow l^+ l^-$
        and  $Z\rightarrow j j $  decays
        for  $Z Z \rightarrow l^+ l^- j j $ events.
         Signal and background calculations are performed for
         primary electron-beam energy of 152.5~GeV and
         the Higgs-boson mass of 200~GeV \cite{Niezurawski:2004ga}.
Right: cross section contribution $\Sigma_2$ extracted from the
measurement of $\gamma \gamma \rightarrow t\bar{t}$ events, as a
function of the reconstructed invariant mass for the scalar (dashed)
and pseudo scalar (thick solid) Higgs-boson with mass of 400 GeV
\cite{Asakawa:2003dh}.
 }
 \label{fig:phys_h2}
 \end{figure}
\end{itemize}

\subsection{Detector and Beam Line Modifications}

No modifications to the main accelerator are required for \gamgam\ 
running as long as the accelerator can support \emem\ running.

\subsubsection{Crossing Angle}

The outgoing electron beam has a large energy and angle spread 
after the Compton backscattering.  An exit aperture of 
$\pm 10$ mrad must be provided so that the disrupted beam 
can avoid hitting the detector.  The exit aperture must also be 
shielded from the magnetic field of the final focusing quad.
Concepts for a final focus quad have been developed~\cite{BNL-FF}
which require a minimum crossing angle of $25$ mrad.  
This requires the Beam Delivery tunnel layout to support the 
$25$ mrad crossing angle.  Either one interaction point must
be designed for initial operation at $25$ mrad or 
additional conventional infrastructure
to support a conversion to $25$ mrad must be provided.



\subsubsection{Extraction Line}

The energy spread of the outgoing beam makes any attempt at 
steering likely to lose excessive amounts of beam.  The preferred 
design for the \gamgam\ extraction line is a field-free
vacuum tube following the $\pm 10$ mrad stay clear of the beam.

\begin{figure*}[tb]
\centering
\includegraphics*[width=\textwidth]{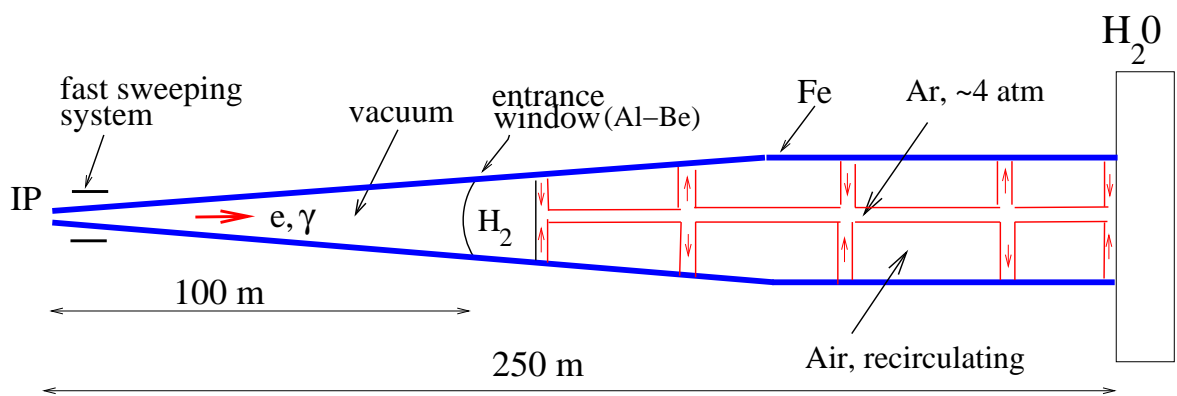}
\caption[A conceptual layout of the beam dump]{A conceptual layout of a beam dump~\cite{Telnov-dump} 
with a gas filled region to disperse the energy of the photon beam.}
\label{fig:dump}
\end{figure*}

The \gamgam\ beam dump will have to be designed to handle the 50\% 
of the beam power which is in the photon beam.  For the standard 
design this could lead to boiling of the water since the photon 
beam cannot be steered or smeared out.  A gas based beam dump has been 
proposed~\cite{Telnov-dump} to deal with this.  A conceptual layout is
shown in Figure~\ref{fig:dump},  
it would require a longer tunnel for the extraction line.

\subsubsection{Final Focus}

The \epem\ beam is designed to be flat in order to minimize disruption.  
This is not required for \gamgam\ operations and changes to the 
final focus magnet strengths can reduce the spot size in $x$, 
increasing the luminosity.  This would have no impact on the 
\epem\ operation.

\subsubsection{Detector Modification}

\begin{figure*}[tb]
\centering
\includegraphics*[width=.75\textwidth]{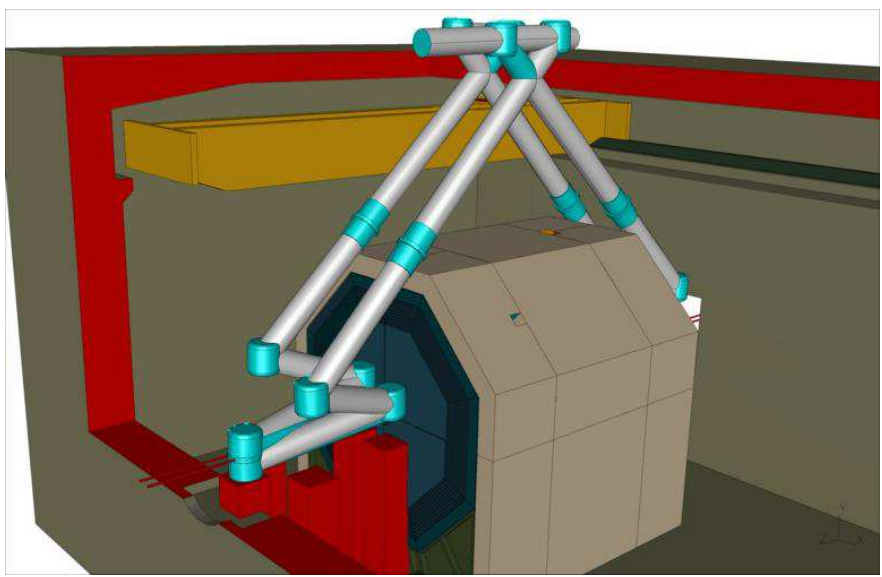}
\caption[Schematics of the laser cavity system]{The laser cavity~\cite{Klemz} has a path length equal to the bunch
spacing of the accelerator.  This makes it a natural fit to circulate
the laser light around the outside of the detector.  Two cavities are
required, one for each accelerator arm.}
\label{fig:layout}
\end{figure*}

\begin{figure*}[tb]
\centering
\begin{minipage}{.69\textwidth}
\includegraphics*[width=\textwidth]{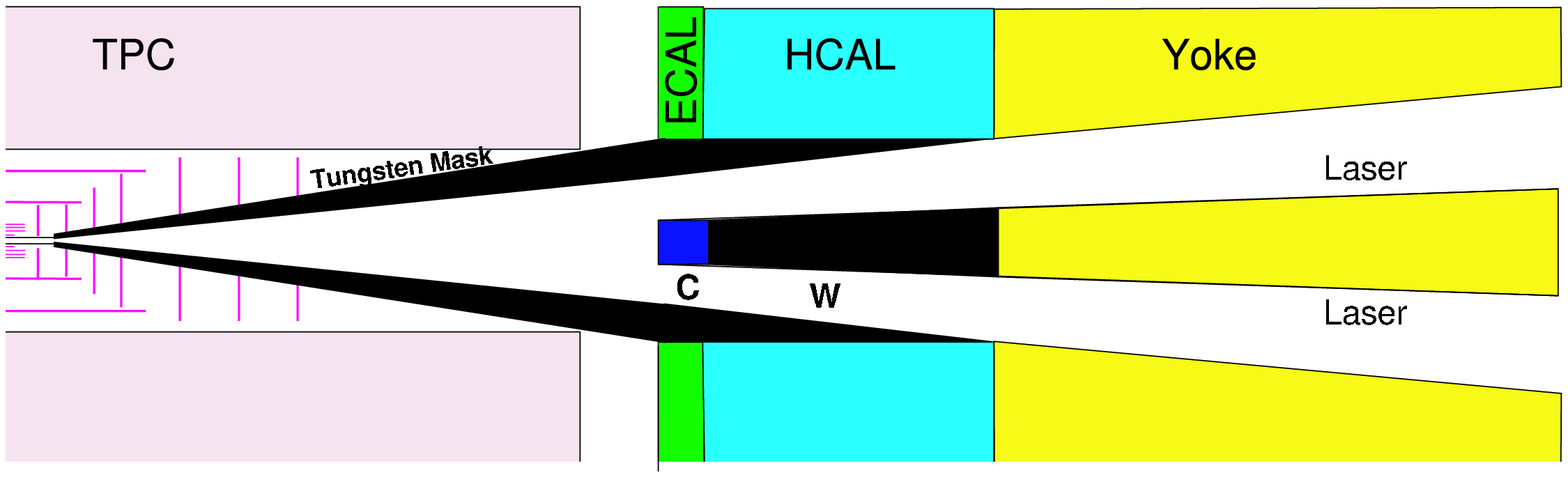}
\end{minipage}
\begin{minipage}{.3\textwidth}
\includegraphics*[width=\textwidth]{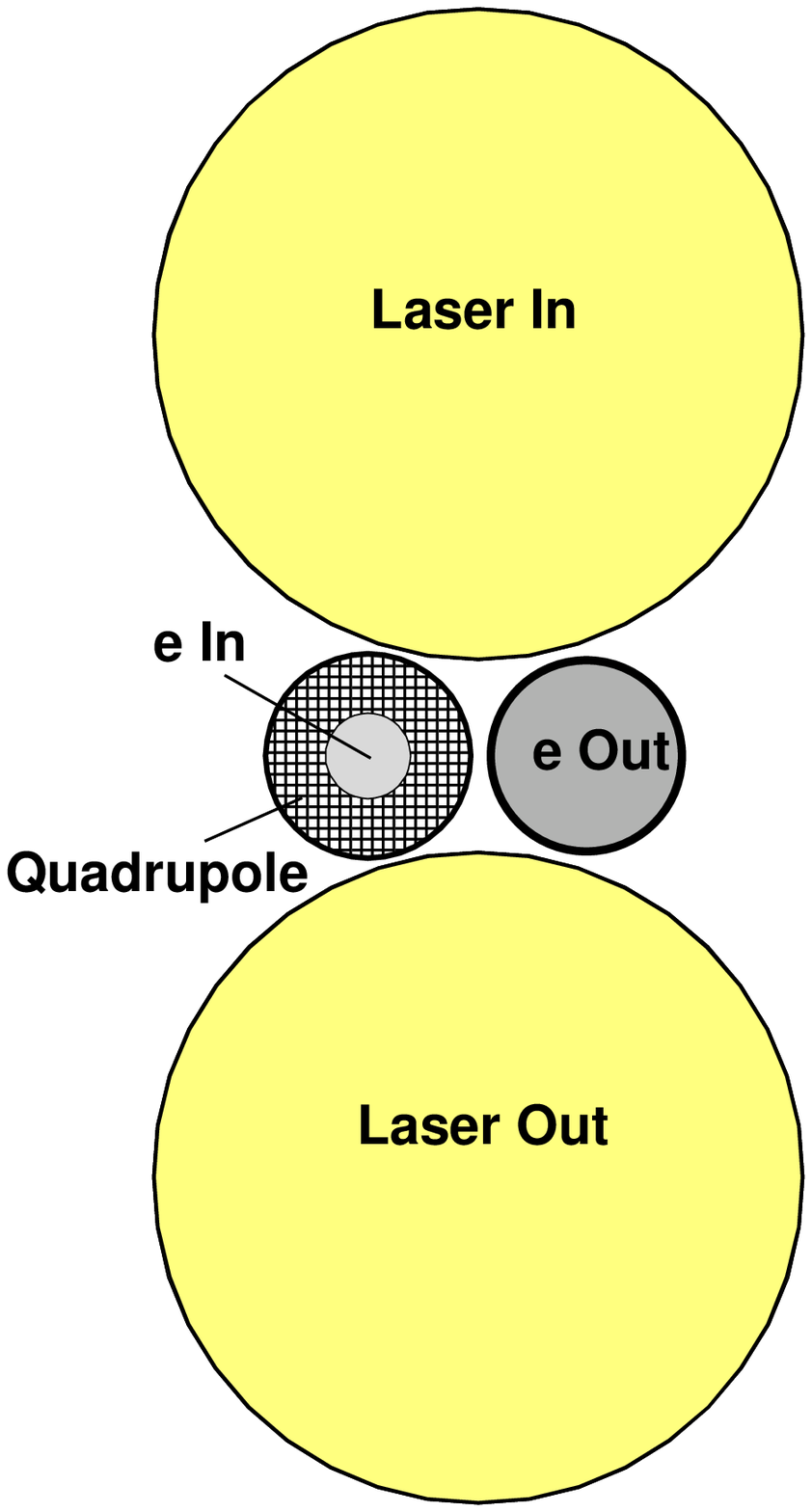}
\end{minipage}
\caption[Light path for the laser in the IP]{Focusing mirrors direct the light pulses into the
detector to collide with the electron beam.  An unobstructed
path from the mirror to the IP must be provided.  The left figure
is a concept for the modifications to the endcap and beam pipe
region needed to accommodate this.  The right figure is an end view
looking down the beam pipe from the IP~\cite{Moenig:det}.}
\label{fig:cross}
\end{figure*}

The detector modifications required are mainly restricted to the 
area around the beam-pipe and the beam input and extraction lines.  
A collaboration from MBI and DESY~\cite{Will:ha,Klemz} has developed
a conceptual design for a recirculating cavity that would greatly reduce
the average laser power required for a photon collider.
Space must be provided in the detector hall, as shown in 
Figure~\ref{fig:layout} to support the optical cavity and the 
source laser for each arm.  As shown in Figure~\ref{fig:cross}, 
a line-of-sight from the IP to the outside of the end cap must be 
provided for the laser light to traverse the cavity and collide with the
electron beam.  This will require modification to the endcap 
calorimeter and possibly any forward tracking that exists in that area.  
No optical hardware is located within the detector.

The increased aperture in the extraction line will increase the 
radiation load seen by the vertex detector from the beam dump.  
Initial estimates are that the fluence from the beam dump is 
$10^{11}$ neutrons/$cm^2$/year.  This is well within the capabilities 
of existing technologies but is a tighter requirement than for 
standard \epem\ running.

\subsubsection{Change-over}

It is expected that operation of the laser cavities will have been 
demonstrated off-site before change-over to \gamgam\ running is 
contemplated.  A shutdown will be required to install the laser hardware 
and configure the IP for $25$ mrad crossing angle.  During the 
shutdown one would:
\begin{itemize}
\item{Remove the detector components around the beam pipe and replace 
them with one configured for $25$ mrad crossing angle.}
\item{Install the laser and optics hardware.}
\item{Either, move the detector to the $25$ mrad IP};
\item{or, if already at the $25$ mrad IP replace the 
\epem\ extraction line with the \gamgam\ extraction line and beam dump.}
\end{itemize}

\subsection{Conclusion}

The \gamgam\ option adds significantly to the physics reach 
of the ILC.  In order to maintain this option the ILC design should 
include a capability to run the detector with a $25$ mrad 
crossing angle.  The detector should also be designed so that
the area around the beam pipe can be easily replaced with one 
configured for $25$ mrad running. Space in the detector 
hall should be reserved for the laser and optics installations.


\chapter{Conclusions}
\label{detector_conclusions}

Experiments at the ILC can profoundly advance particle physics. Ensuring that advance requires the design and development of a new 
generation of particle physics detectors. The machine environment imposes constraints on the design and  boundary conditions on the 
viable detector technologies. ILC physics requires detector performance well beyond the present state of the art.  Satisfying these 
constraints, achieving the needed detector performance, and integrating subdetector systems in a way which maximizes overall physics 
performance have stimulated a world wide effort to design, research, and develop ILC detectors.

This Report has summarized how far these designs and technologies have been developed over the past years. 
It has summarized the 
challenges posed by the ILC machine environment, and by the physics itself, and it has described integrated detector designs 
that can do this physics, and are within reach technologically. The physics performance goals for these detectors are ambitious. 
Assessing whether the proposed detector concepts work has required a high level of detail in the simulation codes which model 
their performance. Full Monte Carlo analyses have been used to characterize subsystem performance and are beginning to be used 
to benchmark integrated detector physics performance as well. 

Significant progress on subsystem design and technological development is reported here.
Two of the major technical challenges, developing fast readout schemes for highly pixellated vertex detectors and developing 
the calorimeters and reconstruction codes capable of greatly improved jet energy resolution, have engaged world wide R\&D. 
Both efforts have reported important progress. Work on the charged particle trackers, which have unparalleled momentum 
resolution, and the far-forward calorimeters, that must survive the intense radiation generated in the collision process, 
shows comparable progress. Technological proofs of principle are not yet completed, but the outstanding technical questions 
are under intense study, and answers should be available within the next few years. Designs for the detectors themselves, 
summarized in Detector Outline Documents from the four existing detector concept studies, and progress in proto-engineering 
for the machine-detector interface, the experimental halls, surface assembly, and possible push-pull operations, record 
progress toward realistic and realizable designs for the ILC experiments.

The claim that detectors can be built which do justice to ILC physics rests on more than technical arguments. It has a 
financial component as well. The DCR has presented a first comprehensive look at the costs of ILC detectors, based on the 
separate evaluations for three of the detector concepts. The total cost for the two detectors called out in the ILC 
baseline will be approximately $10\%$ of the cost of the machine. This is an appropriate level of investment for delivering 
ILC physics. Having two detectors will allow new results and new discoveries to be confirmed (or refuted) independently. 
It will guarantee productive data taking even if there is mishap with one of the detectors. Two complementary designs 
will better adapt to the full range of ILC background and physics unknowns. Two collaborations will double the world's 
involvement in this physics, double the base to support it, and double the opportunities for young physicists to contribute. 
Competition between these two will deliver the best science for the best value at the earliest time. Two is the right number.

What's next? Detector development is as crucial to the sucess of the ILC program as the 
accelerator development.
The GDE plans to have an Engineering Design Report for the accelerator 
completed by 2010. The detector R\&D and integrated detector design 
efforts must keep pace with 
progress on the ILC. The detector R\&D program, which has already developed over many years, includes efforts in all regions, 
with inter-regional collaboration in some cases, and inter-regional coordination in all cases. The R\&D is reviewed within 
the global context by the World Wide Study. This R\&D is critical to the success of the ILC experimental program.

To focus integrated detector design efforts over the next few years, the current studies for four distinct concepts are expected
to be 
concentrated into two engineering design efforts, in time for the submission of
detector EDRs on the same time scale as the ILC machine EDR. 
The next steps are still being developed by the ILCSC, but will include appointing 
a central coordinator,
who will be responsible for coordinating the ILC experimental program, 
together with appropriate international review and control mechanisms.
The resulting detector designs are expected to have complementary and contrasting 
strengths, as well as broad international participation, and can serve
as the basis for the ILC experimental program once the project has been approved.
 


\section*{Acknowledgment}
The editors wish to thank their many colleagues whose work has been reported in this Volume 4, Detectors, of the ILC Reference Design Report. In particular they would like to acknowledge 
the help of the following persons who contributed to the text of this volume: T.~Barklow, J.~Brau, P.~Burrows, K.~B\"usser, G.~Eckerlin, R.~Frey, J.~Hauptmann, D.~Karlen, W.~Lohrmann, T.~Markiewicz, R.~Partridge, A.~Savoy-Navarro, M.~Thomson, H.~Videau, A.~Yamamoto and J.~Yu.

\newpage
\addcontentsline{toc}{chapter}{Bibliography}




\backmatter

\clearpage
\listoffigures %

\clearpage
\listoftables

\typeout{Lastpage = \arabic{page}}
\end{document}